\documentclass[11pt, pdftex]{cmu_ece_dissertation_hyos}
\pdfoutput=1

\usepackage{amsmath}
\usepackage{multirow}
\usepackage{multicol}
\usepackage{ragged2e}
\usepackage{color}
\usepackage{url}
\usepackage{bigfoot}
\usepackage{setspace}
\usepackage[subrefformat=parens,labelformat=parens]{subfig}
\usepackage{verbatim}

\usepackage{amsthm}
\usepackage{amssymb}
\usepackage[noend]{algorithmic}
\usepackage{algorithm}
\usepackage{array}
\usepackage{balance}
\usepackage{booktabs}
\usepackage{paralist}
\usepackage{tabularx}
\usepackage{dblfloatfix}
\usepackage{fixltx2e}
\usepackage{textcomp}
\usepackage{csquotes}

\ifx\pdfoutput\undefined
  \usepackage{graphicx}
\else
  
  \usepackage[pdftex]{graphicx}
  \usepackage{epstopdf}
  \usepackage[protrusion=true,expansion=true]{microtype}
\fi

\usepackage[numbers, square, comma, sort]{natbib}

\usepackage[T1]{fontenc}
\usepackage{textcomp}

\usepackage[letterpaper,twoside,inner=1.3in,outer=1.3in,bottom=1.3in,top=1.3in]{geometry}

\usepackage[%
  hypcap=true,
  font=small,
  labelfont=bf
]{caption}
\usepackage[%
  pdftex,             
  pageanchor=true,
  plainpages=false, 
  pdfpagelabels, 
  bookmarks,
  bookmarksnumbered,
  linkcolor=black,
  citecolor=black,
  pdftitle={},
  pdfauthor={},
  pdfsubject={PhD thesis},
  pdfkeywords={},
  pdfproducer={pdfLaTeX},
]{hyperref}

\newcommand{\tableref}[1]{Table~\ref{tab:#1}}
\newcommand{\figref}[1]{Figure~\ref{fig:#1}}
\newcommand{\figsubref}[1]{Figure~\subref*{fig:#1}}

\newcommand\T{\rule{0pt}{2.3ex}}
\newcommand\B{\rule[-1.1ex]{0pt}{0pt}}
\renewcommand\footnoterule{\kern-3pt \hrule width 2in \kern 2.6pt}

\DeclareMathOperator*{\argmin}{argmin}

\newcommand{\VS}[1]{}
\newcolumntype{C}[1]{>{\centering\arraybackslash}p{#1}}
\newtheorem{definition}{Definition}

\begin{document} 

\frontmatter

\pagestyle{empty}

\title{\bf Towards Predictable Real-Time Performance on Multi-Core Platforms}
\author{Hyoseung Kim}
\date{June}
\Year{2016}
\trnumber{}

\degrees{
B.S., Computer Science, Yonsei University, Seoul, Korea\\
M.S., Computer Science, Yonsei University, Seoul, Korea\\
}

\support{}
\disclaimer{}


\maketitle


\begin{keywordspage}
\end{keywordspage}
\pagestyle{plain} 

\clubpenalty=9999
\widowpenalty=9999

\doublespacing

\begin{abstract}

Cyber-physical systems (CPS) integrate sensing, computing, communication and actuation capabilities to monitor and control operations in the physical environment. A key requirement of such systems is the need to provide {\em predictable real-time performance}: the timing correctness of the system should be analyzable at design time with a quantitative metric and guaranteed at runtime with high assurance. This requirement of predictability is particularly important for safety-critical domains such as automobiles, aerospace, defense, manufacturing and medical devices. 

The work in this dissertation focuses on the challenges arising from the use of modern multi-core platforms in CPS. Even as of today, multi-core platforms are rarely used in safety-critical applications primarily due to the temporal interference caused by contention on various resources shared among processor cores, such as caches, memory buses, and I/O devices. Such interference is hard to predict and can significantly increase task execution time, e.g., up to 12$\times$ on commodity quad-core platforms. To address the problem of ensuring timing predictability on multi-core platforms, we develop novel analytical and systems techniques in this dissertation. Our proposed techniques theoretically bound temporal interference that tasks may suffer from when accessing shared resources. Our techniques also involve software primitives and algorithms for real-time operating systems and hypervisors, which significantly reduce the degree of the temporal interference. Specifically, we tackle the issues of cache and memory contention, locking and synchronization, interrupt handling, and access control for computational accelerators such as general-purpose graphics processing units (GPGPUs), all of which are crucial to achieving predictable real-time performance on a modern multi-core platform. Our solutions are readily applicable to commodity multi-core platforms, and can be used not only for developing new systems but also migrating existing applications from single-core to multi-core platforms.

\end{abstract}

\begin{acknowledgments}

This dissertation would have been impossible without the help and support of many people. First and foremost, I would like to thank my advisor, Prof. Raj Rajkumar. I was lucky to work with Raj. His guidance and expertise have made me a better thinker, writer, and researcher. Raj gave me opportunities to participate in exciting projects, demonstrate my research results, and mentor other students, all of which led me to become an independent researcher and to pursue an academic career.

I am grateful to the members of my thesis committee, Prof. Onur Mutlu, Prof. Anthony Rowe, and Dr. Shige Wang for their time, effort and inputs in completing this dissertation. Thanks to Onur for his insights on various aspects of my work. I learned a lot from Onur on computer architectures, which was a great asset for my research. Thanks to Anthony for his feedback and advice, even since my very early days at CMU. I enjoyed lively conversations with Anthony and liked to hear his view on cyber-physical systems. Thanks to Shige for his giving me many inputs and motiving me with various practical examples. Working with Shige was a great pleasure to me.

I would like to thank my research colleagues at the Software Engineering Institute (SEI): Dio de Niz, Bj{\"o}rn Andersson, and Mark Klein. Our weekly meeting was an excellent opportunity to share lots of interesting discussions and do some good collaborative work. I also would like to thank Prof. John Lehoczky for his keen insight and wisdom during our meetings at SEI.

I wish to thank the members of the CMU's autonomous driving team: Prof. John Dolan, Jongho Lee, Tianyu Gu, Chiyu Dong, Adam Werries, Zhiding Yu, and all other former members. Their passion and efforts made me proud of being part of the team and contributing to our autonomous car.

A special thanks to General Motors (GM), National Science Foundation (NSF), and the Fulbright association for funding my research.

Most of my time during my doctoral studies was spent at the Real-Time and Multimedia systems Lab (RTML). Thanks to all the members of RTML who shared their time with me: Gaurav Bhatia, Karthik Lakshmanan, Arvind Kandhalu, Junsung Kim, Reza Azimi, Alexei Colin, Young-Woo Seo, Anand Bhat, Sandeep D'souza, and Shunsuke Aoki. Also, I would like to thank Toni M. Fox for her kind support on administrative work. 

Besides the RTML members, I am grateful to my friends at CIC: Max Buevich, Niranjini Rajagopal, Oliver Shih, Adwait Dongare, Donghyuk Lee, Sang Kil Cha, Gihyuk Ko, and Soo-Jin Moon. I am grateful to my Korean friends who I met in Pittsburgh: Sungwon Yang, Jaesok Yu, Yongjune Kim, Min Suk Kang, Minhee Jun, and Kiryong Ha. Without these people, I could not have fully enjoyed my time at CMU. I would like to thank my old buddies who are currently geographically far from me but always on my side: JongMan Koo, Jaehun Ha, Kwangkyu Park, Hwan Lee, Jungho Kim, San Yoon, Junoh Jeon, Jungmyung Kim, and Woongjung Do. I am also very grateful to Wonwoo Jung, Shinyoung Yi, Jongho Rim, and Youngbin You, for their being always supportive of me. 

My family has given me their endless love and support. Thanks to my parents for being my parents. My immeasurable gratitude is due to them. Thanks to my parents-in-law for their understanding me during the long years of my studies. Thanks to my brother-in-law, Taegon Lee, for his encouragement. Lastly, my thanks go to my wife, Whayoung Lee. She has been the greatest source of warmth, love and support since I met her. I would never have completed this dissertation without her.

\end{acknowledgments}

\tableofcontents
\listoffigures
\listoftables
\listofalgorithms

\mainmatter


%
%
%
%
%

\chapter{Introduction}

Cyber-physical systems (CPS) are increasingly used in safety-critical application domains including automotive, aerospace, defense, manufacturing and medical devices. Since many CPS directly impact human safety and the environment, they must sense, process and react to external events with stringent timing requirements. Any transient violation of the timing requirements may lead to system failures, resulting in catastrophic consequences. Hence, a cyber-physical system for safety-critical applications should provide {\em predictable real-time performance}. The timing correctness of the system should be analyzable at design time with a quantitative metric and be guaranteed at runtime with high assurance.

The conventional approach to developing a system for safety-critical applications is to use single-core processor platforms. This is because, although ensuring real-time predictability is a challenging issue, existing theoretical foundations and real-time operating systems (OSs) make it achievable on only single-core platforms. Unfortunately, these approaches have limitations in meeting the ever-increasing computational demands for additional functionalities in safety-critical applications. For example, in the automotive domain, some recent cars such as Lexus LS430 already have more than a hundred of processors each~\cite{Navet_RTS10}, and adding advanced automotive technologies like adaptive cruise control, pedestrian detection and collision avoidance is becoming harder due to space and cost requirements. 

Modern multi-core processors are therefore receiving much attention as promising candidates for the development of next-generation CPS for  safety-critical applications. 
The use of multi-core platforms gives an opportunity to consolidate multiple applications onto a single hardware platform. Such consolidation leads to a significant reduction in space requirements while also reducing installation, management and production costs by reducing the number of processor chips and wiring harnesses among them. However, providing real-time predictability on a multi-core platform is substantially different from doing so on a single-core platform. Tasks executing in parallel on different cores may contend with each other to access shared resources, e.g., a last-level cache, main memory, and I/O devices. 
This contention causes temporal interference among tasks and may result in significant delay, e.g., up to $12\times$ increase in task execution time~\cite{Kim_RTAS14}, which can easily jeopardize the timing predictability of the entire system. For this reason, government regulations and certification standards for safety-critical systems, e.g., DO-178C by the U.S. Federal Aviation Administration (FAA), still do not advise the use of multi-core platforms.

In this dissertation, we focus on the challenges arising from the use of modern multi-core platforms in CPS. Specifically, we develop novel analytical and systems techniques that address the predictability issues associated with shared resources in multi-core platforms. 
Our techniques theoretically bound temporal interference among tasks in the presence of contention on the shared resources. Also, our techniques reduce the interference by complementary software techniques and algorithms for real-time OSs and virtualization. With these techniques, we provide predictable real-time performance on accessing each type of shared resource, e.g., caches, main memory, sensor, I/O devices and GPUs, and ensures the predictability of the entire system in an efficient way.
The main thesis supported by this dissertation is as follows:

\begin{displayquote}
	\textbf{Thesis Statement:} Novel primitives in systems software combined with analytical techniques yield timing predictability on a multi-core platform by bounding and significantly reducing temporal interference from shared platform resources.
\end{displayquote}

The remainder of this chapter provides context for this dissertation. First, we describe the scope of this work. Secondly, we discuss the challenges associated with each type of shared resource. Thirdly, we present the contributions of this dissertation, along with forward references to later chapters. Lastly, we describe the organization of this dissertation.

\section{Scope of This Work}
\label{scope_of_this_work}

We give a brief description of shared resources and task execution environments considered in this dissertation. The detailed system model used in this work will be described in Chapter~\ref{system_model}.

\subsection{Multi-Core Platform and Shared Resources}

The work in this dissertation considers a computing platform equipped with a single-chip, homogeneous multi-core processor. 
All the cores of the processor are identical to each other, in terms of clock speed, instruction execution performance, and access time to platform resources.
This corresponds to many of today's multi-core processors, such as Intel Core i7, AMD FX, ARM Cortex-A15, and NXP QorIQ processors. Although there also exist other types of multi-core processors, such as IBM Cell and ARM big.Little architectures, we focus on the former type of processors in this work.

The computing platform has various resources shared among all processor cores, as illustrated in \figref{platform_resources}. We categorize those shared resources into the following three types:

\begin{figure}[t]
	\centering
	\subfloat{
		\includegraphics[width=0.8\columnwidth]{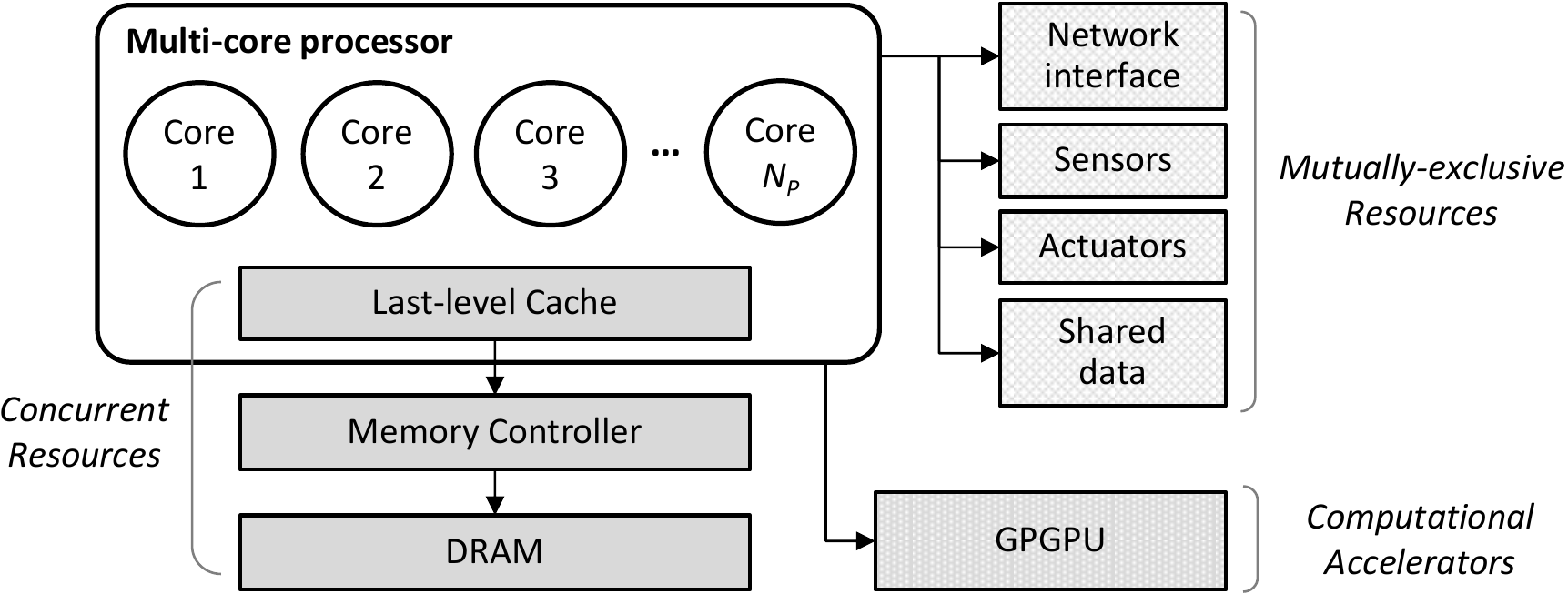}
	}
	\caption{Multi-core platform and shared resources}
	\label{fig:platform_resources}
\end{figure}

\begin{itemize}
	\item \textbf{Concurrent Resources:} Shared resources that allow concurrent access from multiple tasks are referred to as concurrent resources. The resources in the multi-core memory hierarchy, such as a last-level cache, a memory controller, and DRAM, belong to this type. 
	
	\item \textbf{Mutually-Exclusive Resources}: 
	Shared resources that require no more than one task to access them at a time are referred to as mutually-exclusive resources. 
	Any access to mutually-exclusive resources should obey the requirement of mutual exclusion to prevent data corruption and/or unexpected behavior. I/O devices, such as sensors, actuators and network interfaces, typically belong to this category. Also, shared data regions are considered as mutually-exclusive resources. 	
	
	\item \textbf{Computational Accelerators}: Shared resources that supplement the computational capacity are referred to as computational accelerators. GPGPUs (general-purpose graphics processing units), DSPs (digital signal processors), and FPGAs (field-programmable gate arrays) fall into this category, but in this work, we specifically focus on GPGPUs. In the rest of this dissertation, we will use the terms ``GPGPU'' and ``GPU'' interchangeably. 
\end{itemize}

\subsection{Tasks and Task Execution Environments}

CPS applications are typically composed of a set of recurrent tasks with timing constraints. Hence, we consider the {\em sporadic task} model~\cite{SporadicTask} to represent CPS applications in an analyzable way. Under the sporadic task model, each task repeatedly releases a workload, called a {\em job}, with a minimum time interval. The {\em response time} of a task is the time duration from the release of a job of the task to the completion of the job execution. Each task has a timing constraint, called a {\em relative deadline}, and the timing constraint of a task is deemed to be satisfied if the worst-case response time of the task is smaller than or equal to its relative deadline. A task is called {\em schedulable} if it satisfies its timing constraint. A set of tasks (taskset) is schedulable if all tasks in the set satisfy their timing constraints. 

\begin{figure}[t]
	\centering
	\subfloat[Non-hierarchical scheduling in a native environment]{
		\label{fig:non_hierarchial_scheduling}
		\includegraphics[width=0.3\columnwidth]{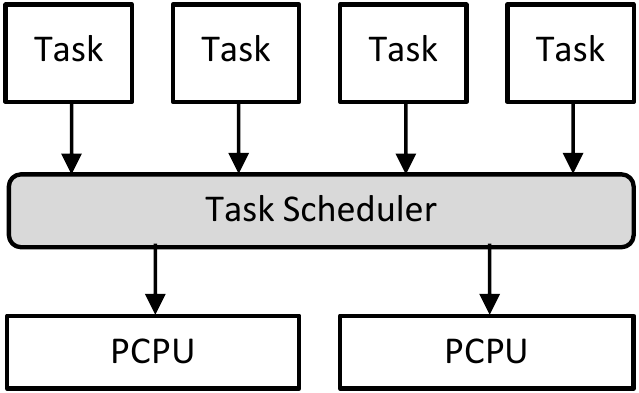}
	}
	\hspace{15pt}
	\subfloat[Hierarchical scheduling in a virtualized environment]{
		\label{fig:hierarchical_scheduling}
		\includegraphics[width=0.6\columnwidth]{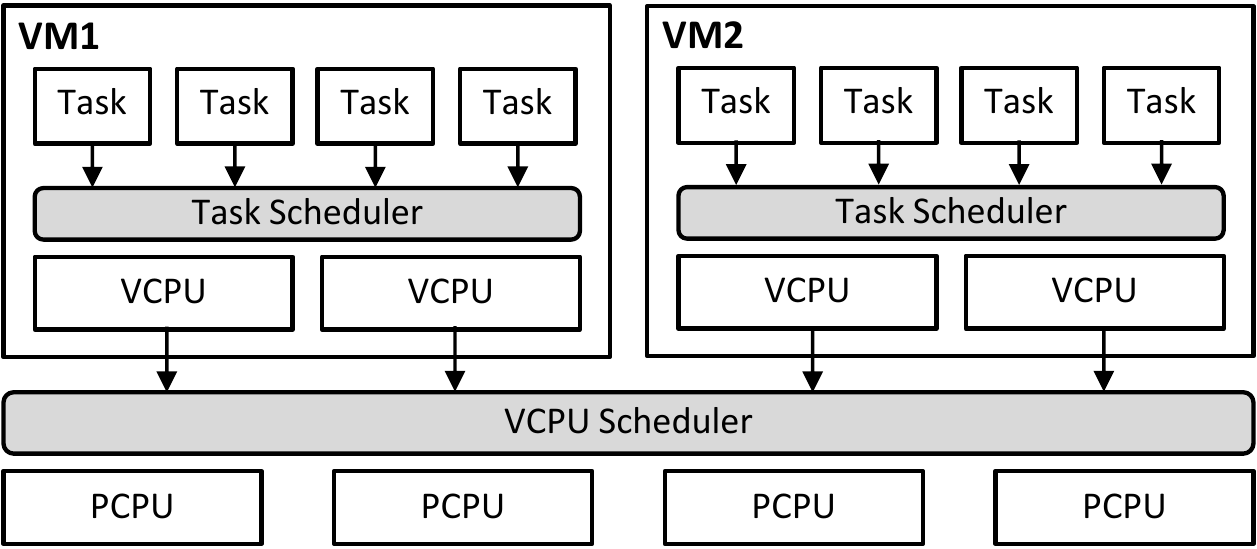}
	}
	\caption{Task execution environments and scheduling structures}
	\label{fig:scheduling_structures}
\end{figure}

Task execution environments considered in this work fall into two categories: {\em native} and {\em virtualized} environments. In a native environment, the system runs an OS that schedules tasks on physical CPU cores (PCPUs)\footnote{We will use the terms ``cores'' and ``PCPUs'' interchangeably in the rest of this dissertation.}, as shown in \figsubref{non_hierarchial_scheduling}. This is referred to as a non-hierarchical scheduling structure. In a virtualized environment, the system runs a hypervisor providing a two-level hierarchical scheduling structure as shown in \figsubref{hierarchical_scheduling}. The hypervisor hosts multiple guest virtual machines (VMs), each of which has one or more virtual CPUs (VCPUs). The tasks of a VM are scheduled on the VCPUs of that VM by the guest OS. Each VCPU is a scheduling entity to the hypervisor, meaning that the hypervisor schedules VCPUs on PCPUs. Each VCPU has an execution budget and a budget replenishment period, and the tasks of a VCPU can only execute when the VCPU has spare budget. The VCPU budget replenishment policy used in the system, such as a periodic server~\cite{periodic_server}, a deferrable server~\cite{deferrable_server} and a sporadic server~\cite{sporadic_server}, determines when and how to refill the budget of VCPUs. The characteristics of these policies will be described in Section~\ref{system_model}.

There are two approaches to schedule tasks on multiple processing cores: {\em partitioned} and {\em global}. Partitioned scheduling statically assigns each task to a core and always executes the task on that core. Under partitioned scheduling, finding an optimal task allocation can be modeled as a bin-packing problem. Global scheduling, on the other hand, allows tasks to migrate from one core to another at runtime. In this dissertation, we focus on {\em partitioned fixed-priority preemptive scheduling} for both OSs and hypervisors due to the following reasons:
\begin{enumerate}[(i)]
	\item It is widely supported in many commercial real-time embedded OSs and hypervisors such as OKL4~\cite{OKL4} and PikeOS~\cite{PikeOS}.
	\item It does not introduce task migration costs.
	\item It can benefit from the well-established uniprocessor theoretical framework.
\end{enumerate}

\section{Challenges with Shared Resources}
Shared resources on a multi-core platform cause different challenges depending on their types.
We briefly discuss the specific challenges associated with each type of shared resource. 

\subsection{Concurrent Resources}

\noindent\textbf{Cache Interference:} Many of today's multi-core processors incorporate a large last-level shared cache to improve the performance and efficiency of the system. The shared cache can efficiently bridge the performance gap between memory access latency and processor clock speed by backing up small private caches. Each of the cores can access the entire shared cache so that a better cache hit ratio can be statistically achieved. However, the uncontrolled use of the shared cache introduces significant worst-case timing penalties in task execution, due to {\em cache interference} among tasks.
It has been shown in~\cite{Kim_TR2013} that cache interference on a quad-core processor increases task response time by up to 40\%, compared to when the task runs alone in the system with no cache interference from other tasks. As the number of cores increases, the negative impact of cache interference becomes more significant. 
In this dissertation, we develop OS-level and hypervisor-level techniques to provide predictable cache performance to tasks executing in native and virtualized environments, respectively.

\smallskip
\noindent\textbf{Memory Interference:} Main memory is another major shared resource among processor cores. A task running on one core can be delayed by other tasks running simultaneously on different cores due to interference in the shared main memory system, which is referred to as {\em memory interference}. Memory interference delay can be large and highly variable, thereby posing a significant challenge for the design of predictable systems. Specifically, in modern systems, commercial-off-the-shelf (COTS) DRAM systems have been widely used as main memory to cope with high performance and capacity demands. The DRAM system contains multiple hardware components such as a memory controller, DRAM banks, and buses. Each of these components has different timing characteristics, making it difficult to analyze memory interference. 
In addition, as memory-intensive applications are becoming more prevalent in CPS, the reduction of this interference is critical to making effective use of multi-core platforms. The work in this dissertation presents techniques to bound and reduce memory interference.

\subsection{Mutually-Exclusive Resources}
\label{challenges_mutually_exclusive_resources}

\noindent\textbf{Synchronization:} Consolidating multiple tasks onto a single hardware platform inevitably introduces sharing of mutually-exclusive resources, e.g. shared data regions for inter-task communication, network interfaces, and I/O devices. Those resources are typically protected by mutually-exclusive locks to avoid race conditions. When a task requests access to such a resource, the resource can be granted to the task only if it is not held by another task. Otherwise, the task is blocked until the requested resource is released. Hence, for the timing predictability of tasks, we need a synchronization mechanism that provides bounded blocking times. 
The sharing of mutually-exclusive resources and task synchronization issues have been intensively studied in the context of native environments.  However, prior approaches can lead to unbounded blocking times in a virtualized environment due to the hierarchical scheduling structure. In this dissertation, we present a novel synchronization scheme to address such timing penalties in a virtualized environment.

\smallskip
\noindent\textbf{Interrupt Handling:} I/O devices like sensors and actuators use interrupts to notify events in the physical environment to the computing system. Hence, in addition to synchronization, interrupt handling and resulting execution flows should be carefully designed for predictable access to I/O devices. 
We have identified two strong requirements for the interrupt handling scheme of CPS: (i) providing responsive and bounded interrupt handling time while ensuring the schedulability of tasks, and (ii) enforcing interrupts to protect task execution from interrupt storms\footnote{An interrupt storm is a condition where a system receives interrupts at an unexpectedly high rate and the processing of those interrupts takes the majority of the CPU time. It is also known as the receive livelock problem~\cite{Mogul_ACM97}.}. The issues of responsive and enforced interrupt handling have been mainly studied for a native execution environment. However, these requirements are not satisfied by prior work in a virtualized environment. In this dissertation, we develop an analyzable interrupt handling scheme to address the aforementioned requirements in a virtualized environment.

\subsection{Computational Accelerators}

\noindent\textbf{GPGPU Management:} The high computational demands of complex algorithmic tasks used in recent CPS pose substantial challenges in guaranteeing their timeliness. 
For example, the CMU's autonomous vehicle~\cite{Wei_IV13} executes perception and motion planning algorithms along with running tasks for data fusion from tens of sensors equipped within the vehicle. Since each of these tasks is computation intensive, it becomes harder to satisfy their timing requirements when they execute on the same hardware platform. 
Fortunately, many of today's embedded multi-core processors, such as NXP i.MX6 and NVIDIA Tegra K1, have an on-chip GPGPU, the use of which can greatly help in addressing the timing challenges of computation-intensive tasks by accelerating their execution. However, today's COTS GPU hardware and device drivers are not designed with predictability as a primary concern. First of all, execution on a GPU is non-preemptive. While a lower-priority task is using a GPU, GPU execution requests from higher-priority tasks are delayed until the current GPU execution finishes. In addition, GPU device drivers do not consider the scheduling policy used in the system. Hence, GPU requests from lower-priority tasks may be handled earlier than those from higher-priority tasks, which negatively impact the schedulability of tasks. In this dissertation, we present techniques to control GPU access in a timely and efficient manner.

\section{Contributions}

The overarching contribution of this work is the development of novel analytical and systems techniques to yield timing predictability on a multi-core platform. Our techniques address cache and memory interference, synchronization, interrupt handling, and GPGPU management issues, all of which are crucial to achieving predictable real-time performance on modern multi-core platforms.

\subsection{Analytical and Systems Support for Concurrent Resources}

The following is a brief description of our contributions to concurrent resources. 
Details on these contributions are described in Chapters~\ref{chapter_coordinated_cache_management}, \ref{chapter_bounding_and_reducing_memory_interference}, and \ref{chapter_cache_management_for_virtualization}.

\begin{itemize}
\item \textbf{Coordinated Approach for Predictable Cache Management:} We develop a coordinated OS-level cache management scheme to address cache interference. Our scheme provides predictable cache performance through tight coordination of cache reservation, reserved cache sharing, and cache-aware task allocation. This approach also mitigates the two major problems of the conventional software cache partitioning technique: (i) the memory co-partitioning problem, which results in page swapping or waste of memory, and (ii) the availability of a limited number of cache partitions, which causes degraded performance.\footnote{For additional details, please see Section~\ref{literature_review_cache_interference}.} We have implemented and evaluated our scheme in Linux/RK running on a quad-core platform. 
Experimental results indicate that, compared to the traditional approaches, our scheme is up to 39\% more memory space efficient and consumes up to 25\% fewer cache partitions while preserving timing predictability. Our scheme also yields a significant utilization benefit that increases with the number of tasks. 

\item \textbf{Bounding and Reducing Memory Interference:} 
We present techniques to reduce memory interference and find an upper bound on the worst-case memory interference on a multi-core platform with DRAM-based main memory. We explicitly model major resources in the DRAM system, including banks, buses, and the memory controller. By considering their timing characteristics, we analyze the worst-case memory interference delay imposed on a task by other tasks running in parallel. Experimental results show that our approach provides an upper bound very close to our measured worst-case interference. From our analysis, we find that memory interference can be significantly reduced by (i) partitioning DRAM banks, and (ii) co-locating memory-access-intensive tasks on the same processing core. Based on these observations, we develop a memory interference-aware task allocation algorithm for reducing memory interference. Our memory interference-aware task allocation algorithm provides a significant improvement in task schedulability over previous work, with as much as 96\% more tasksets being schedulable.

\item \textbf{Predictable Cache Management for Virtualization:} 
In addition to OS-level techniques, we develop a predictable cache management framework for a virtualized environment. Our framework introduces two hypervisor-level techniques, vLLC and vColoring, that enable the allocation of cache partitions to individual tasks running in a virtual machine (VM), which is not achievable by prior work. Our framework also provides a cache management scheme that determines cache allocation to tasks, designs VMs in a cache-aware manner, and minimizes the aggregated utilization of VMs to be consolidated. As a proof of concept, we implemented vLLC and vColoring in the KVM hypervisor running on x86 and ARM multi-core platforms. Experimental results with three different guest OSs, namely Linux/RK, vanilla Linux and MS Windows Embedded, show that our techniques can effectively control the allocation of cache partitions to tasks in VMs. Experimental results also show that our cache management scheme yields a significant utilization benefit compared to other approaches.

\end{itemize}

\subsection{Analytical and Systems Support for Mutually-Exclusive Resources}

The following is a brief description of our contributions to mutually-exclusive resources. 
Details on these contributions are described in Chapters~\ref{chapter_synchronization} and \ref{chapter_interrupt_handling}.

\begin{itemize}

\item \textbf{Synchronization for Multi-Core Virtual Machines:} 
We develop vMPCP, a synchronization framework for tasks executing in a virtualized environment. vMPCP exposes the execution of critical sections of tasks in a guest virtual machine to the hypervisor. Using this approach, vMPCP reduces and bounds blocking time on accessing mutually-exclusive resources shared within and across virtual CPUs (VCPUs) assigned to different physical CPU cores.
vMPCP supports various VCPU budget replenishment policies, with an optional budget overrun to reduce blocking times. 
We provide the VCPU and task schedulability analyses under vMPCP, with different VCPU budget replenishment policies, with and without budget overrun. 
The case study using our hypervisor implementation shows that vMPCP yields significant benefits compared to a virtualization-unaware multi-core synchronization protocol, with 29\% shorter response time on average.

\item \textbf{Responsive and Enforced Interrupt Handling:} 
We develop a novel interrupt handling scheme for a multi-core virtualization environment, called vINT. vINT provides a pseudo-VCPU abstraction dedicated for interrupt handling, which overcomes the limits imposed by the timing parameters of virtual CPUs in an analyzable way. vINT also accounts for and enforces interrupt handling and resulting execution flows within a guest virtual machine. 
We analyze interrupt handling time as well as VCPU and task schedulability, with and without vINT. 
Our experimental results indicate that vINT achieves timely interrupt handling while providing as good task schedulability as when it is not used. Our case study based on a prototype implementation on the KVM hypervisor shows that vINT yields significant benefits in reducing interrupt handling time and in protecting tasks against interrupt storms permeating into the virtual machine.

\end{itemize}

\subsection{Analytical and Systems Support for Computational Accelerators}

The following is a brief description of our contributions to computational accelerators, specifically general-purpose GPUs.
Details on these contributions are described in Chapter~\ref{chapter_predictable_gpgpu_management}.

\begin{itemize}
\item \textbf{Predictable GPGPU Access Control:} We develop a server-based GPU access control approach to manage a GPU in a predictable manner. Our proposed approach introduces a dedicated server task that handles GPU requests from other tasks with respect to their priority order. Although we focus on a GPU in this work, our approach can be used for other types of computational accelerators, such as DSPs.
Our server-based approach also addresses the main limitations of an existing real-time synchronization-based GPU access control approach, which will be discussed in Section~\ref{GPU_synchronization_approach}. 
Experimental results indicate that our server-based approach yields significant improvements in task schedulability over the synchronization-based approach. For example, a quad-core system with our approach schedules 66\% more randomly-generated tasksets than the same quad-core system that uses the synchronization-based approach with the multiprocessor priority ceiling protocol~\cite{MPCP,MPCP2}. 
	
\end{itemize}

\section{Organization}

The rest of this dissertation is organized as follows. Chapter~\ref{chapter_background} reviews the background and related work, and Chapter~\ref{system_model} describes the system model used in this work. Chapters~\ref{chapter_coordinated_cache_management}, \ref{chapter_bounding_and_reducing_memory_interference}, and \ref{chapter_cache_management_for_virtualization} present our work for concurrent resources. Chapters~\ref{chapter_synchronization} and \ref{chapter_interrupt_handling} present our work for mutually-exclusive resources. Chapter~\ref{chapter_predictable_gpgpu_management} presents our work for computational accelerators. Chapter~\ref{chapter_guidelines_for_future_computer_architectures} discusses guidelines for future computer architecture designs.  Chapter~\ref{conclusions} concludes this dissertation.

\chapter{Background and Related Work}
\label{chapter_background}

This chapter presents the background and related work on the following five issues: cache interference, memory interference, synchronization, interrupt handling, and GPGPU management. Each section of this chapter reviews relevant systems software techniques and/or hardware components, and discuss related prior work.

\section{Cache Interference}
\label{literature_review_cache_interference}

Many researchers have recognized and studied the problem of cache interference in order to use a shared cache in a predictable manner. Among a variety of approaches, software cache partitioning, called {\it page coloring}, has been considered as an appealing approach to address this issue. Page coloring prevents cache disruptions from other tasks by assigning exclusive cache partitions to each task. It does not require any hardware support beyond what is available on most of today's multi-core processors. In this section, we describe the page coloring technique and discuss its problems. We then review related work on cache interference.

\begin{figure}[t]
	\centering
	\VS{-8pt}
	\subfloat{
		\includegraphics[width=0.7\textwidth]{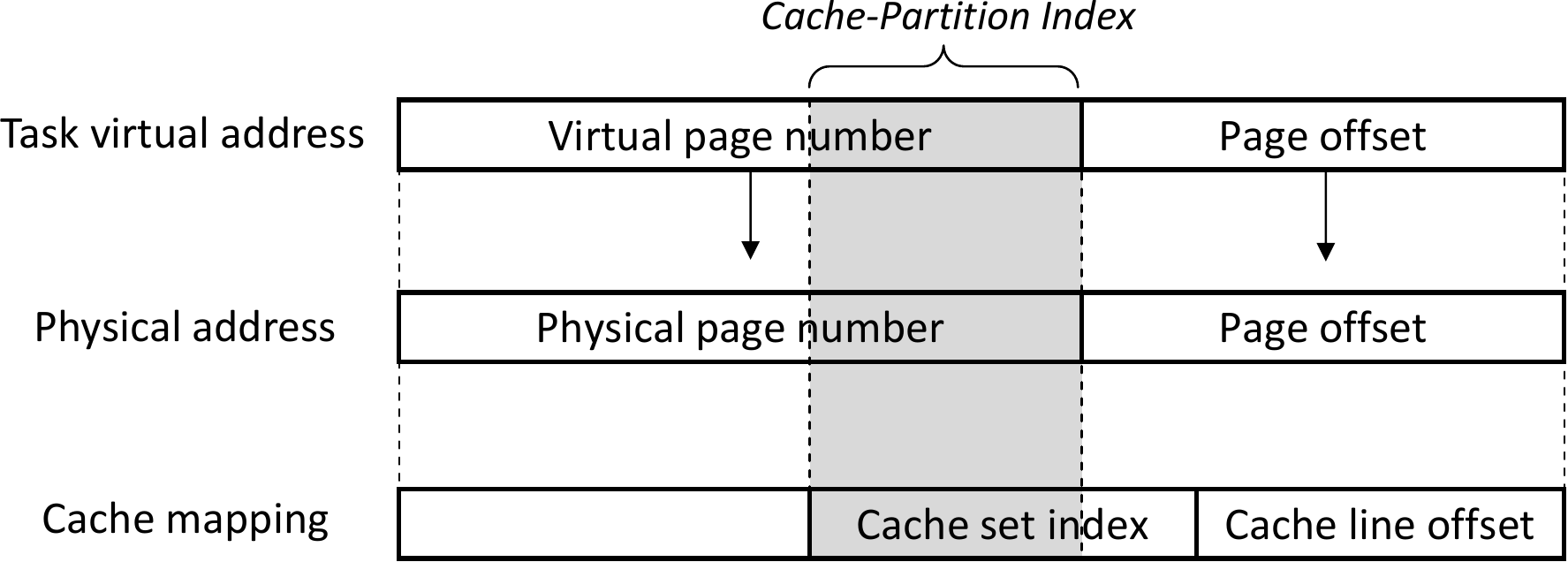}
	}
	\VS{-7pt}
	\caption{Memory address to cache mapping and page coloring}
	\label{fig:memory_address_and_page_coloring}	
\end{figure}

\subsection{Page Coloring}
\label{page_coloring}

Page coloring is a software technique used to control a physically-indexed set-associative cache, which is the case for most shared caches on modern processors. On a physically-indexed cache, page coloring uses the mapping between physical addresses and cache set indices. 
As shown in \figref{memory_address_and_page_coloring}, there are overlapping intersection bits between the physical page number and the cache set index. Page coloring uses those intersection bits as a {\em cache-partition index} which partitions the cache into $n$ cache partitions. Simultaneously, the cache-partition index co-partitions the entire physical memory into $n$ memory partitions. In other words, physical memory pages with the same cache-partition index are grouped into a memory partition, and each physical memory partition corresponds to a cache partition with the same cache-partition index. Since the OS has direct control over the mapping between physical pages and the virtual pages of an application task, it can allocate specific cache partitions to a task by providing the task with physical pages in the corresponding memory partitions. 

The number of cache partitions available in the system is calculated as follows: $n=S/(W \times P)$, where $n$ is the number of cache partitions, $S$ is the cache size, $W$ is the number of ways of the cache, and $P$ is the size of a page frame and is typically 4KB. Hence, if $S=\text{256KB}$, $W=\text{16}$ and $P=\text{4KB}$, the number of cache partitions $n$ is 4. One implicit assumption in page coloring is that the number of cache sets is a power of two. In some architectures like Intel Sandy Bridge and Haswell, the last-level cache consists of cache slices, the number of which is equal to that of physical cores~\cite{Intel_HC25,Intel_HC23}. As shown in \cite{Kim_ECRTS13,Ye_PACT14}, although the mapping between physical addresses and cache slices is not publicly known, page coloring on such architectures can be implemented on a per cache-slice basis. This results in the number of cache partitions equal to $n=S/(W\times P\times N_P)$, where $N_P$ is the number of physical cores.

\subsection{Problems with Page Coloring}
\label{problems_with_page_coloring}

There are several challenging problems to be solved before page coloring can be used widely in multi-core systems. The first problem is the memory co-partitioning problem~\cite{Liedtke_RTAS97,Lin_HPCA08}. Page coloring simultaneously partitions the entire physical memory into the number of cache partitions. If a certain number of cache partitions is assigned to a task, the same number of memory partitions is also assigned to that task. However, a task's memory usage is not necessarily related to its cache usage. If a task requires more number of memory partitions than that of cache partitions, the required memory partitions should be assigned to the task despite its small cache usage. Otherwise, the task would suffer from page swapping. If a task requires more number of cache partitions than that of memory partitions, some of the assigned memory would be wasted. 

The second problem is the availability of a limited number of cache partitions. As the number of tasks increases, the amount of cache that can be used for an individual task becomes smaller and smaller, resulting in degraded performance. Moreover, the number of cache partitions may not be enough for each task to have its own cache partition. This second problem also unfortunately applies to hardware-based cache partitioning schemes.

Page coloring was originally developed for a native environment. In a virtualized environment, there is one more problem: page coloring implemented in a guest OS running in a VM can no longer map a task's virtual page to a specific cache partition. This is because there is an additional address translation layer at the hypervisor, which is to spatially isolate VMs from each other. One simple approach to consider is to implement page coloring in the hypervisor and assign cache partitions to VMs, as proposed in \cite{Li_VEE14, Ma_JSA13}. However, this approach cannot allocate cache partitions to individual tasks running in a VM. In other words, all tasks within the same VM share the cache partitions assigned to that VM and will suffer from cache interference. In this dissertation, we address these problems.

\subsection{Related Work}

With cache partitioning, the system performance is largely dependent on how cache partitions are allocated to tasks. Yoon et al.~\cite{Yoon_RTSS11} formulated cache allocation as an MILP problem to minimize the total CPU utilization of Paolieri's new multi-core architecture~\cite{Paolieri_ISCA09}. Fu et al.~\cite{Fu_ECRTS11} proposed a sophisticated low-power scheme that uses both cache partitioning and DVFS. 
Paolieri et al.~\cite{Paolieri_RTAS11} proposed a task and cache allocation algorithm for a system using non-preemptive partitioned scheduling. 
These approaches, however, assume hardware cache partitioning support, which is not yet widely available in current commodity processors~\cite{AMDDevDoc, IntelDevDoc, POWER4_Arch}. 

Software cache partitioning or page coloring is an alternative to hardware cache partitioning support. Wolfe \cite{Wolfe_JCSE94} and Liedtke et al. \cite{Liedtke_RTAS97} used page coloring to prevent cache interference in a single-core system. Bui et al. \cite{Bui_RTCSA08} focused on improving the schedulability of a single-core system with page coloring. Page coloring also has been studied for multi-core systems in \cite{Cho_MICRO06,Tam_WIOSCA07,Zhang_EuroSys09}. Guan et al. \cite{Guan_EMSOFT09} proposed a non-preemptive scheduling algorithm for a multi-core real-time system using page coloring. 
Lin et al.~\cite{Lin_HPCA08} conducted a comparative study on various multi-core cache partitioning schemes by implementing them with page coloring. 
Mancuso et al.~\cite{Mancuso_RTAS13} proposed the Colored Lockdown technique that combines page coloring and cache lockdown to better keep the frequently accessed pages of tasks in a cache. 
Ye et al.~\cite{Ye_PACT14}~developed COLORIS that supports both static and dynamic cache partitioning based on page coloring. 
Ward et al.~\cite{Ward_ECRTS13} focused on cache management issues in multi-core mixed-criticality systems and proposed cache locking and scheduling techniques that use page coloring. 
Zhang et al. \cite{Zhang_EuroSys09} proposed a {\em hot-page} coloring approach that assigns cache partitions only to a small set of frequently accessed pages. However, since they use on-line page access monitoring and page migration, it may not be suitable for time-critical systems. 

Cache interference also happens in single-core systems due to task preemption, which causes the eviction of the cache contents of a preempted task. 
Such cache interference penalties are bounded by accounting them as cache-related preemption delays while performing schedulability analysis. Altmeyer et al.~\cite{Altmeyer_RTSS11}, Lunniss et al.~\cite{Lunniss_RTAS13} and Lee et al.~\cite{CGLee_IEEE01} focused on reducing the cache penalties by using static cache analyses. However, they do not consider cache partitioning that can prevent the cache penalties by assigning exclusive cache partitions to tasks. 
Busquets-Mataix et al. \cite{Busquets_ECRTS97} proposed a hybrid technique of cache partitioning and schedulability analysis for a single core system, but it cannot be directly applied to a shared cache of a multi-core processor.
Xu et al.~\cite{Xu_RTSS13} extended multi-core compositional analysis to incorporate cache interference delay caused by private caches, assuming that there is no shared cache. Lunniss et al.~\cite{Lunniss_RTNS14} extended CRPD analysis to a single-core hierarchical scheduling environment. However, none of these approaches focuses on a shared cache in a multi-core platform.

There also exist some research efforts to address cache interference in a virtualized environment. Previous work on software-based cache management in a virtualization environment~\cite{Li_VEE14, Ma_JSA13} proposed to implement page coloring in the hypervisor and to allocate cache partitions to virtual machines (VMs). This approach, however, cannot be used to address cache interference among tasks running within a VM due to an additional address translation layer at the hypervisor. Kim et al.~\cite{Kim_IEEE14} proposed a hardware-based solution to enable page coloring implemented in a guest OS to work. However, hardware modification required by this approach does not allow the use of commodity multi-core processors. In addition, if a guest OS does not have page coloring support, tasks running on that guest OS cannot get any benefit. 

\section{Memory Interference}

\label{MEM_background}
Memory interference in a DRAM system is largely affected by two major components: (i) the DRAM chips where the actual data are stored, and (ii) the memory controller that schedules memory read/write requests to the DRAM chips. In this section, we provide a brief description of these two components. Our description is based on DDR3 SDRAM systems, but it generally applies to other types of COTS DRAM systems. For more information, interested readers may refer to \cite{Rixner_ISCA00, Nesbit_MICRO06, Moscibroda_2007, Mutlu_MICRO07, Lee_HPCA13}.

\begin{figure}[t]
	\centering
	\subfloat{
		\includegraphics[width=0.7\textwidth]{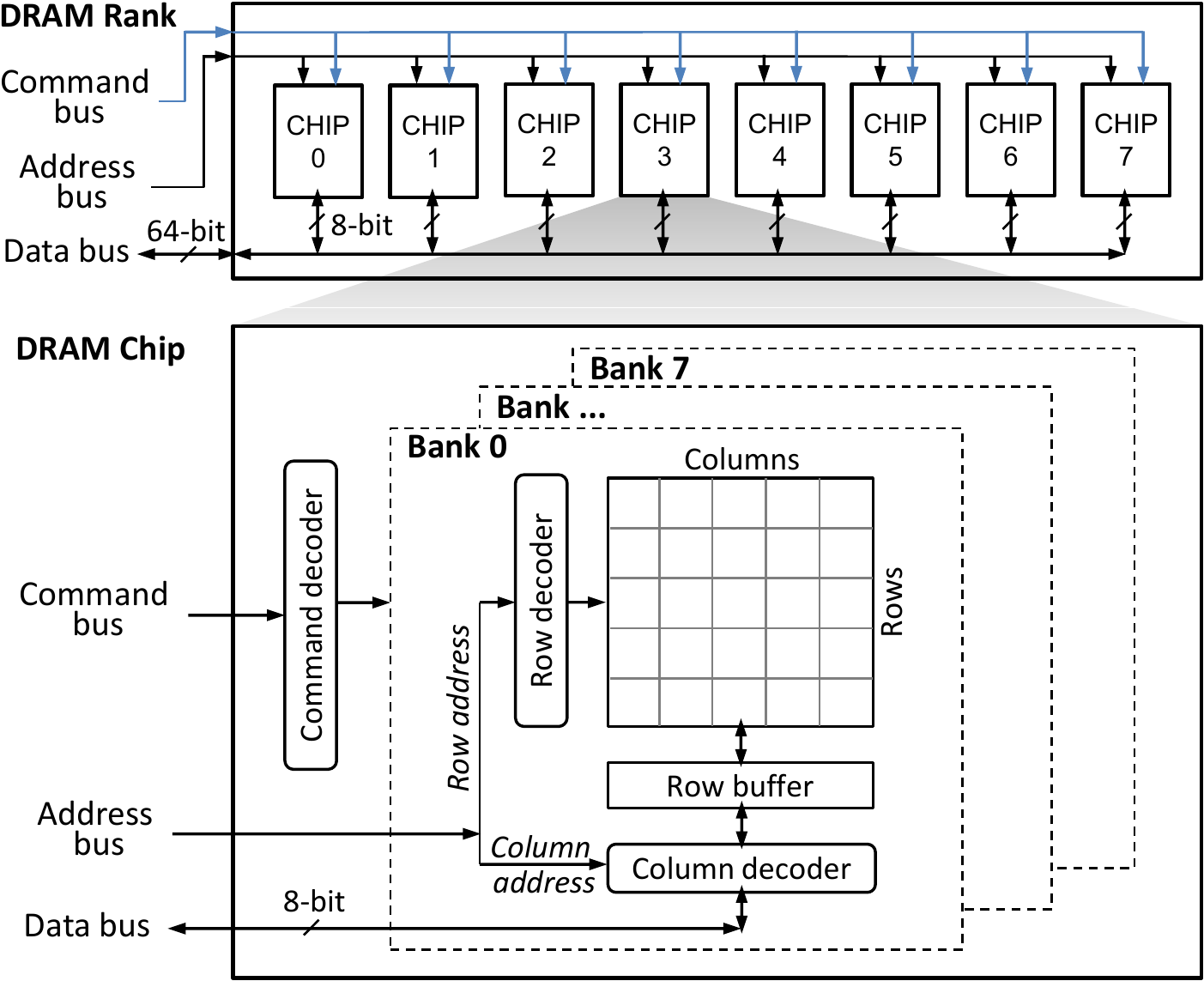}
	}
	\caption{DRAM device organization}
	\label{fig:MEM_dram-organization}	
\end{figure}

\subsection{DRAM Organization}
A DRAM system as shown in \figref{MEM_dram-organization} is organized as a set of {\em ranks}, each of which consists of multiple DRAM chips. Each DRAM chip has a narrow data interface (e.g. 8 bits), so the DRAM chips in the same rank are combined to widen the width of the data interface (e.g. 8 bits/chip $\times$ 8 chips = 64 bits data bus). A DRAM chip consists of multiple DRAM {\em banks} and memory requests to different banks can be serviced in parallel. Each DRAM bank has a two-dimensional array of rows and columns of memory locations. To access a column in the array, the entire row containing the column first needs to be transferred to a {\em row-buffer}. This action is known as {\em opening} a row. Each bank has one row-buffer that contains at most one row at a time. The size of the row-buffer is therefore equal to the size of one row, which is 1024 or 2048 columns in a DDR3 SDRAM chip \cite{JEDEC_DDR3}.

The DRAM access latency varies depending on which row is currently stored in the row-buffer of a requested bank. If a memory request accesses a row already in the row-buffer, the request is directly serviced from the row-buffer, resulting in a short latency. This case is called a {\em row hit}. If the request is to a row that is different from the one in the row-buffer, the currently open row should be closed by a {\em precharge} command and the requested row should be delivered to the row-buffer by an {\em activate} command. Then the request can be serviced from the row-buffer. This case is called a {\em row conflict} and results in a much longer latency. In both cases, transferring data through the data bus incurs additional latency. The data is transferred in a burst mode and a {\em burst length (BL)} determines the number of columns transferred per read/write access.

\begin{figure}[t]
	\centering
	\subfloat{
		\includegraphics[width=0.7\textwidth]{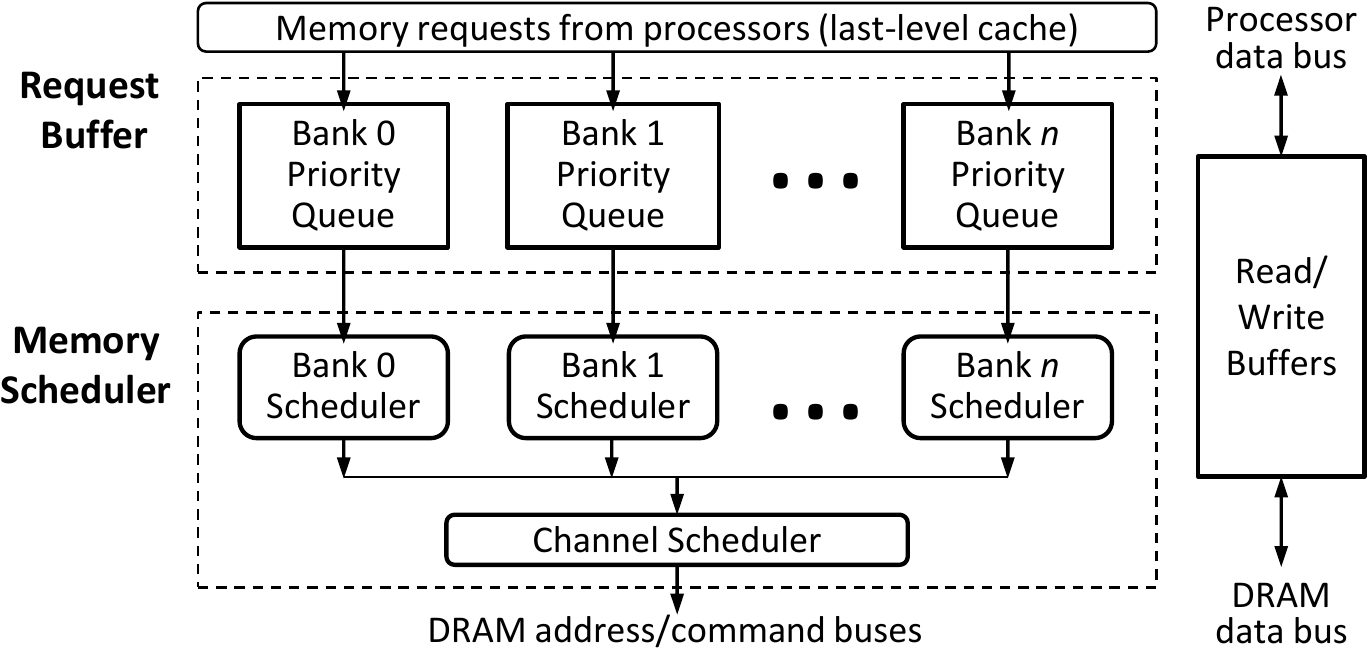}
	}
	\caption{Logical structure of a DRAM controller}
	\label{fig:MEM_dram-controller}	
\end{figure}

\subsection{Memory Controller}

\figref{MEM_dram-controller} shows the structure of a memory controller in a modern DRAM system. The memory controller is a mediator between the last-level cache of a processor and the DRAM chips. It translates read/write memory requests into corresponding DRAM commands and schedules the commands while satisfying the timing constraints of DRAM banks and buses. To do so, a memory controller consists of a request buffer, read/write buffers, and a memory scheduler. The request buffer holds the state information of each memory request, such as an address, a read/write type, a timestamp and its readiness status. The read/write buffers hold the data read from or to be written to the DRAM. The memory scheduler determines the service order of the pending memory requests. 

The memory scheduler typically has a two-level hierarchical structure.\footnote{The physical structure of priority queues, bank schedulers, and the channel scheduler depends on the implementation. They can be implemented as a single hardware structure \cite{Nesbit_MICRO06} or as multiple decoupled structures~\cite{Mutlu_MICRO07,Mutlu_ISCA08,Ausavarungnirun_ISCA12}.} As shown in~\figref{MEM_dram-controller}, the first level consists of per-bank {\em priority queues} and {\em bank schedulers}. When a memory request is generated, the request is enqueued into the priority queue that corresponds to the request's bank index. The bank scheduler determines priorities of pending requests and generates a sequence of DRAM commands to service each request. 
The bank scheduler also tracks the state of the bank. If the highest-priority command does not violate any timing constraints of the bank, the command is said to be {\em ready} for the bank and is sent to the next level.
The second level consists of a {\em channel scheduler}. It keeps track of DRAM commands from all bank schedulers, and monitors the timing constraints of ranks and address/command/data buses. Among the commands that are {\em ready} with respect to such channel timing constraints, the channel scheduler issues the highest-priority command. Once the command is issued, the channel scheduler signals ACK to the corresponding bank scheduler, and then the bank scheduler selects the next command to be sent.

\smallskip
\noindent\textbf{Memory Scheduling Policy:} 
Scheduling algorithms for COTS memory controllers have been developed to maximize the data throughput and minimize the average-case latency of DRAM systems. Specifically, modern memory controllers employ First-Ready First-Come First-Serve (FR-FCFS) \cite{Rixner_ISCA00, Nesbit_MICRO06} as their base scheduling policy. FR-FCFS first prioritizes ready DRAM commands over others, just as the two-level scheduling structure does. At the bank scheduler level, FR-FCFS re-orders memory requests as follows: 
\begin{enumerate}
	\item Row-hit memory requests have higher priorities than row-conflict requests.
	\item In case of a tie, older requests have higher priorities. 
\end{enumerate}
Note that, in order to prevent starvation, many DRAM controllers impose a limit on the number of consecutive row-hit requests that can be serviced before a row-conflict request~\cite{Mutlu_MICRO07,Subramanian_ICCD14}. We will discuss such a limit in Section~\ref{MEM_analysis}.
At the channel scheduler level, FR-FCFS issues DRAM commands in the order of their arrival time. Therefore, under FR-FCFS, the oldest row-hit request has the highest priority and the newest row-miss request has the lowest priority.

\begin{figure}[t]
	\centering
	\subfloat{
		\includegraphics[width=0.8\textwidth]{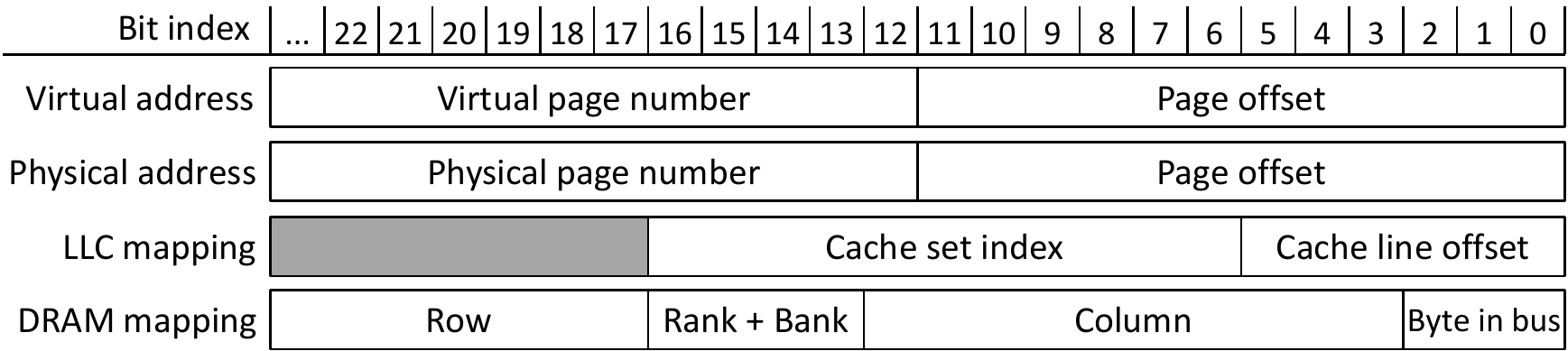}
	}
	\caption[Memory address to cache and DRAM mapping]{Task address to cache and DRAM mapping (Intel i7-2600)}
	\label{fig:MEM_address-mapping}	
\end{figure}

\subsection{Bank Address Mapping and Bank Partitioning}
\label{MEM_background_bank_partitioning}
In modern DRAM systems, physical addresses are interleaved among multiple banks (and ranks) to exploit bank-level parallelism for average-case performance improvement. The granularity of address interleaving is typically equal to the size of one row, because mapping adjacent addresses to the same row may provide better row-buffer locality. This strategy is called a {\em row-interleaved} address mapping policy and it is widely used in many COTS systems. As an example, \figref{MEM_address-mapping} shows the address mapping of the system equipped with the Intel i7-2600 processor which follows the row-interleaved policy.\footnote{The DRAM mapping of \figref{MEM_address-mapping} is for the single-channel configuration in this system.} 
In this system, bits 13 to 16 of the physical address are used for the rank and bank indices.

The row-interleaved policy, however, can significantly increase the memory access latency in a multi-core system~\cite{Muralidhara_MICRO11, Liu_PACT12, Jeong_HPCA12}. For instance, multiple tasks running simultaneously on different cores may be mapped to the same DRAM banks. This mapping can unexpectedly decrease the row-buffer hit ratio of each task and introduce re-ordering of the memory requests, causing significant delays in memory access. 

Software bank partitioning~\cite{Liu_PACT12, Suzuki_ICESS13, Yun_RTAS14, Kim_RTAS14, Kim_RTS16, Xie_HPCA14} is a technique used to avoid the delays due to shared banks. By dedicating a set of specific DRAM banks to each core, bank partitioning prevents both (i)~the unexpected close of the currently-open row and (ii)~the negative effect of request re-ordering. Therefore, with bank partitioning, bank-level interference among tasks simultaneously executing on different cores can be effectively eliminated. Similar to software cache partitioning discussed in Section~\ref{page_coloring}, bank partitioning can be implemented by exploiting the mapping between physical addresses and rank-bank indices. If a task is assigned only physical pages with a specific rank-bank index~$b$, all the memory accesses of that task are performed on the rank-bank~$b$. By controlling the physical page allocation in the OS, the physical memory space can be divided into {\em bank partitions} and a specific bank partition can be assigned to a core by allocating corresponding physical pages to the tasks of the core. 
Each bank partition may comprise one or more DRAM banks. If the memory requirement of the tasks of a core is larger than the size of one DRAM bank, each bank partition can be configured to have multiple DRAM banks to sufficiently satisfy the memory requirement with a single bank partition. However, due to the resulting smaller number of bank partitions, 
it may not be feasible to assign a dedicated bank partition to each core. In our work, we therefore consider not only dedicated DRAM banks to reduce memory interference delay but also shared banks to cope with the limited number of banks.

\subsection{Related Work}
\label{MEM_related_work}

Several prior studies have developed special non-COTS memory components to achieve predictable memory access time. The Predator memory controller~\cite{Akesson_predator_CODES07} uses credit-based arbitration 
and closes an open row after each access. The AMC memory controller \cite{Paolieri_ESL10} spreads the data of a single cache block across all DRAM banks so as to reduce the impact of interference by serializing all memory requests. The PRET DRAM controller~\cite{Reineke_CODES11} hardware partitions banks among cores for predictability. 
A memory controller that allows different burst sizes for different memory requests has been proposed \cite{Li_ECRTS14}. A memory controller that partitions the set of banks so that a single memory access can fetch data from multiple banks (bank interleaving) within a partition of banks has been proposed and it uses the open-row policy \cite{Krishnapillai_ECRTS14}.
Researchers have also proposed techniques that modify a program and carefully set up time-triggered schedules
so that there is no instant where two processor cores have outstanding memory operations
\cite{rosen07}.

We have heard, however, a strong interest from practitioners in techniques
that can use COTS-based multi-core platforms and existing applications without requiring modifications. Therefore, this has been the focus of our work.
In this context, 
some previous work considers the entire memory system
as a single resource such that a processor core holds the memory system exclusively
until the requested data are delivered to the core~\cite{Pellizzoni_DATE10,Andersson_SIGBED10,Dasari_11,Schliecker_DATE10,lv_RTSS10}.
They commonly assumed that each memory request takes a constant service time and memory requests from multiple cores are serviced in the order of their arrival time. However, these assumptions may lead to overly pessimistic or optimistic estimates in modern COTS DRAM systems, where the service time of each memory request varies and the memory controller re-orders the memory requests~\cite{Moscibroda_2007}.

Instead of considering the memory system as a single resource, recent work \cite{Wu_RTSS13} makes a more realistic assumption about the memory system, where the memory controller has one request queue per DRAM bank and one system-wide queue connected to the per-bank queues.
That analysis, however, only considers the case where each processor core is assigned a private DRAM bank. Unfortunately, the number of DRAM banks is growing more slowly than the number of cores, and the memory space requirement of a workload in a core may exceed the size of a single bank. 
Due to this limited availability of DRAM banks, it is necessary to consider sharing of DRAM banks among multiple cores. With bank sharing, memory requests can be re-ordered in the per-bank queues, thereby increasing memory request service times. The work in \cite{Wu_RTSS13} unfortunately does not model this request re-ordering effect. In addition, the work assumes a special non-COTS memory controller. In this dissertation, we address these limitations. 

In the field of task allocation, the problem of finding an optimal allocation of tasks to cores is known to be NP-complete~\cite{Johnson_74}. Hence, many near-optimal algorithms based on the bin-packing heuristics have been proposed as practical solutions to the task allocation problem~\cite{deNiz_06,Lakshmanan_ICDCS10,Lakshmanan_ECRTS09}. The IA$^3$ algorithm~\cite{Paolieri_RTAS11} is the first approach to take memory interference into account when allocating tasks. IA$^3$ pessimistically assumes that the amount of memory interference for a task is only affected by the number of cores used, and does not consider the actual number of interfering memory requests generated by other tasks that run in parallel. Motivated by this, we develop a memory interference-aware allocation algorithm that reduces memory interference by considering the memory access intensity of each task.  Chapter~\ref{chapter_bounding_and_reducing_memory_interference} will present details on our algorithm and its experimental results.

Finally, there has been recent work in the computer architecture community on the design of memory controllers and memory systems that can dynamically estimate application slowdowns~\cite{Subramanian_HPCA13,Moscibroda_2007,Ebrahimi_ASPLOS10,Subramanian_MICRO15}. These designs, however, do not aim to provide worst-case bounds and may under-estimate memory interference. There also exist research efforts on designing memory controllers for heterogeneous systems (e.g.,~\cite{Usui_TACO16,Kayiran_MICRO14}) and nonvolatile memory (e.g.,~\cite{Zhao_MICRO14}).
Future memory controllers might incorporate ideas like batching and thread prioritization (e.g.,~\cite{Mutlu_ISCA08, Kim_MICRO10, Kim_HPCA10, Subramanian_ICCD14, Subramanian_IEEE16}), which would raise interesting research questions regarding predictability.

\section{Synchronization}
\label{background_synchronization}

From a scheduling perspective, shared mutually-exclusive resources, such as I/O devices and shared data regions, are categorized into two types: {\em global} and {\em local} resources. Global resources are the resources shared among tasks executing on different physical CPU cores (PCPUs) in a native environment, or on different virtual CPUs (VCPUs) in a virtualized environment. The critical sections corresponding to the global resources are referred to as global critical sections. Conversely, local resources are shared among tasks executing on the same PCPU or VCPU. The corresponding critical sections are local critical sections. 

In this section, we characterize timing penalties that arise from the two types of shared mutually-exclusive resources in native and virtualized environments. Then, we review related prior work.

\subsection{Timing Penalties from Mutually-Exclusive Resources}
\label{SYNC_shared_resource_penalties}

Timing penalties caused by accessing mutually-exclusive resources in a multi-core platform can be categorized into {\em local blocking} and {\em remote blocking}. Local blocking time is the duration for which a task needs to wait for the execution of lower-priority tasks assigned on the same core. Uniprocessor real-time synchronization protocols like PCP~\cite{PCP} can bound the local blocking time to at most the duration of one local critical section. Remote blocking time is the duration that a task has to wait for the executions of tasks of any priorities assigned on different cores. If a task tries to access a resource held by another task on a different core, task $\tau_i$ suspends by itself until the resource-holding task finishes its corresponding critical section. Multiprocessor real-time synchronization protocols such as MPCP~\cite{MPCP} are proposed to bound and minimize the duration of remote blocking.

Unlike local blocking, remote blocking causes additional timing penalties even though a multiprocessor synchronization protocol like MPCP is used~\cite{Lakshmanan_RTSS09}: 
\begin{itemize}
	\item {\bf Back-to-back execution}: If a task suspends by itself due to remote blocking, its self-suspending behavior can cause a back-to-back execution phenomenon~\cite{Rajkumar_91}, resulting in additional interference to lower-priority tasks.
	\item {\bf Multiple priority inversions}: Whenever a medium-priority task suspends due to remote blocking, lower-priority tasks get a chance to execute and issue requests for local or global resources. In case of local resources under PCP, every normal execution segment of a medium-priority task can be blocked at most once by one of the lower-priority tasks executing their local critical sections with inherited higher priorities. In case of global resources under MPCP, every normal execution segment of a task can be preempted at most once by each of the lower-priority tasks executing global critical sections. Consequently, multiple priority inversions caused by remote blocking increase the local blocking time.
\end{itemize}

In a virtualized environment, the length of remote blocking time may become even significantly longer due to:
\begin{itemize}
	\item {\bf Preemptions by higher-priority VCPUs}: Consider a task $\tau_i$ in a VCPU $v_j$ waiting on a global resource held by another task in a VCPU $v_k$ assigned on a different physical core. If the VCPU $v_k$ is preempted by higher-priority VCPUs on its core, the remote blocking time for the task $\tau_i$ is increased by the execution times of those higher-priority VCPUs. 
	\item {\bf VCPU budget depletion}: Tasks in a VCPU are scheduled by using their VCPU's budget. When the VCPU budget of a resource-holding task is depleted, a task waiting remotely on that resource needs to wait at least until the start of the next replenishment period of the resource-holding task's VCPU.
\end{itemize}

\subsection{Related Work}
\label{SYNC_related_work}

Synchronization issues in multi-core and multiprocessor systems have been intensively studied in the non-hierarchical scheduling context. MPCP (Multiprocessor Priority Ceiling Protocol)~\cite{MPCP2,MPCP} provides bounded remote blocking time on accessing global shared resources under partitioned fixed-priority scheduling. MPCP uses the uniprocessor PCP~\cite{PCP} for accessing local resources. Recently, a new schedulability analysis for MPCP is proposed in \cite{Lakshmanan_RTSS09}. MSRP (Multiprocessor Stack-based Resource Policy)~\cite{Gai_RTAS03} is an extension of the uniprocessor SRP~\cite{SRP} for resource sharing under partitioned EDF scheduling. A comparison of MPCP and MSRP is also provided in \cite{Gai_RTAS03}. FMLP (Flexible Multiprocessor Locking Protocol)~\cite{Block_RTCSA07} is the first protocol that supports both partitioned and global EDF scheduling. MSOS (Multiprocessors Synchronization for real-time Open Systems)~\cite{Nemati_ECRTS11} is designed for resource sharing among independently-developed systems where each processor uses different scheduling algorithms. All these protocols, however, are designed for non-hierarchical scheduling, so they may cause indefinite remote blocking time under the hierarchical scheduling of virtualization environments.

In the hierarchical scheduling context, much research has been conducted on the schedulability analysis of independent tasks on uniprocessors~\cite{Davis_RTSS05, Saewong_ECRTS02, Shin_RTSS03, Shin_ACM08} and multiprocessors~\cite{Leontyev_RTS09,Shin_ECRTS08}. For tasks with shared resources, HSRP (Hierarchical Stack Resource Policy)~\cite{Davis_RTSS06} is the first synchronization protocol proposed in the context of uniprocessor hierarchical scheduling. HSRP uses budget overrun and payback mechanisms to limit priority inversion. SIRAP (Subsystem Integration and Resource Allocation Policy)~\cite{Behnam_EMSOFT07} uses the idea of self-blocking to bound delays on accessing shared resources without knowing the timing parameters of other subsystems. RRP (Rollback Resource Policy)~\cite{Asberg_RTAS13} uses a rollback mechanism to avoid a lock-holding task to be blocked while holding a lock. However, none of these protocols has been extended to the multi-core hierarchical scheduling context. 

In~\cite{Nemati_09}, the authors propose to group tasks sharing a resource into a single component and to use the hierarchical scheduling model to schedule the tasks and the component. The purpose of this approach is to avoid global resource sharing in a multi-core system, but it limits the sum of the utilization of tasks sharing a resource to be less than one. 

The virtualization of real-time and cyber-physical systems have recently received much attention. RT-Xen~\cite{Lee_RTAS12,Xi_EMSOFT11} is the first hierarchical real-time scheduling framework for the Xen hypervisor. RT-Xen implements a suite of fixed-priority servers for the VCPU budget replenishment policy. The work in \cite{Bruns_ECRTS10} investigates the real-time performance of the L4/Fiasco microkernel-based hypervisor~\cite{L4/Fiasco}. However, these approaches have not considered the synchronization issues. 

In this dissertation, our goal on mutually-exclusive resources is to minimize the remote blocking in a multi-core virtualized environment. Another goal is to bound the remote blocking time of a task as a function of the duration of global critical sections of other tasks (and the parameters of VCPUs having those tasks when overrun is not used), and {\em not} as a function of the duration of normal execution segments or local critical sections.

\section{Interrupt Handling}
\label{background_interrupts}

Interrupt handling and resulting execution flows are indispensable for many systems that interact with the physical environment in a lower latency compared to polling. As discussed in Section~\ref{challenges_mutually_exclusive_resources}, techniques for predictable interrupt handling have been intensively studied in a native environment, but not in a virtualized environment. Therefore, we focus on predictable interrupt handling in a virtualized environment.

In a virtualized environment, a {\em physical} interrupt generated by a sensor or network interface is first handled by the interrupt service routine (ISR) of the hypervisor, and then delivered to the corresponding VCPU in the form of a {\em virtual} interrupt. Once that VCPU is scheduled, the virtual interrupt is handled by the ISR of the guest OS while consuming the VCPU's budget. Finally, the interrupt triggers the execution of any task responsible for reacting to that interrupt. 
In this section, we describe problems with interrupt handling in virtualization, and then review related prior work.

\subsection{Problems with Virtual Interrupts} 
\label{INTR_interrupt_handling_problems}
The main difference between interrupt handling in virtualized and non-virtualized environments is the presence of virtual interrupts. Here, we detail two major problems associated with virtual interrupts. 

\begin{itemize}
\item {\bf Timing penalties to virtual interrupt handling:} Once a virtual interrupt is injected into a VCPU, it is handled by using the priority and budget of the VCPU. Virtual interrupt handling time is thus affected by the following two factors. First, when a virtual interrupt is delivered to a VCPU, the budget of the VCPU $v_i$ might have been completely consumed by other tasks within $v_i$. Hence, the handling of the virtual interrupt may be delayed until the start of the next replenishment period of the VCPU $v_i$. Second, although a VCPU $v_i$ has an enough budget to handle a virtual interrupt, the handling of that virtual interrupt may be delayed by the execution of any task on higher-priority VCPUs that can preempt the VCPU $v_i$. 
\item {\bf Virtual interrupt storms:} Previous work proposed to address interrupt storms in a native environment~\cite{Danish_RTAS11, Lewandowski_RTAS07, Parmer_RTSS08} uses a dedicated aperiodic server, e.g., a deferrable server or a sporadic server, for interrupt handling. When the server budget is depleted, the associated interrupt is not handled until the start of the next replenish period of the server. By doing so, the impact of an interrupt storm on CPU time is limited to the amount of the budget assigned to the associated server. 

While previous work can be applied to the hypervisor to address physical interrupt storms, it may not be used for virtual interrupt storms in a full-virtualization scenario, where an unmodified guest OS is used and it is unaware of being virtualized. In general, OSs measure the passage of time by reading and comparing two clock values, e.g., $t_1 - t_0 =$ elapsed time from $t_0$ to $t_1$. Under full virtualization, an unmodified guest OS can check the passage of {\em physical time} in this manner. However, the guest OS cannot use the same manner to check the passage of {\em virtual time}, which is the actual CPU time used by the guest VCPU. This is because the guest OS is unaware of when and how much VCPU-level preemptions are caused. In other words, when previous work is used for virtual interrupts under full virtualization, it may result in significant errors in the accounting of virtual interrupt handling. 
\end{itemize}

In this dissertation, our goals on interrupt handling are twofold: (i) minimize and bound interrupt handling time in a virtualized environment, and (ii) account for virtual interrupt handling and protect tasks from virtual interrupt storms without any modifications to the guest OS.

\subsection{Related Work}
\label{INTR_related_work}

Previous work on interrupt handling in a native environment commonly uses a {\em split interrupt handling} model to execute deferrable work within a task context~\cite{Lewandowski_RTAS07, Manica_10, Steinberg_ECRTS05}. Specifically, Zhang and West~\cite{Zhang_RTSS06} proposed the Process-Aware Interrupt (PAI) mechanism that schedules and accounts Linux bottom halves with the highest priority of the tasks waiting on the corresponding interrupt. Palmer and West~\cite{Parmer_RTSS08} proposed to use deferrable servers to handle interrupts in order to minimize the receive livelock problem~\cite{Mogul_ACM97}. Danish et al.~\cite{Danish_RTAS11} proposed a Priority Inheritance Bandwidth-Preserving (PIBP) policy to handle interrupts and I/O requests with the budget and priority of the associated task. All these schemes, however, are designed for a native system and cannot address the problems of virtual interrupt handling in a virtualized system, discussed in the previous subsection.

There also exist many research efforts attempting to address other aspects of interrupts in a native environment. 
Leyva-del-Foyo et al.~\cite{Leyva_ACM12} proposed an integrated task and interrupt management model. By using a very short ISR that only activates a task corresponding to the interrupt, the proposed model could reduce the interference from interrupts associated with lower-priority tasks. Elliott and Anderson~\cite{Elliott_ECRTS12} focused on the priority inversion problem caused by the interrupts of GPU asynchronous I/O in a multi-core system using global scheduling.
Brandenburg et al.~\cite{Brandenburg_JSA11} investigated various interrupt accounting mechanisms for multi-core systems using global EDF scheduling.

To overcome the limitations of hierarchical scheduling in virtualization, approaches based on {\em paravirtualized scheduling}~\cite{Kiszka_RTLWS09, Lackorzynski_EMSOFT12, Ma_JISE13} have been studied.
All of these approaches require modifications to the scheduler of a guest OS to let the hypervisor know the currently-executing task within the VM. Using this information, the hypervisor increases the priority of the corresponding VCPU so that the VCPU is not preempted by other VMs executing lower-priority tasks. 
However, none of these approaches bounds the worst-case interrupt handling time. They also do not enforce virtual interrupt handling. Specifically, the work in \cite{Lackorzynski_EMSOFT12} proposes to assign a separate budget and priority to a subset of tasks and interrupts of a VCPU, but does not consider virtual interrupt storms and does not show how the separate budget and priority values can be determined. 

Beckert et al.~\cite{Beckert_DAC14} proposed an interrupt handling scheme for virtualization. However, their approach has several limitations: (i) the hypervisor is assumed to use TDMA to schedule VCPUs, which does not conform to the latest research efforts on real-time system virtualization, (ii) virtual interrupts may be handled while consuming the budgets of unrelated other VCPUs, meaning that each VCPU is not guaranteed to use its assigned budget for its own purpose, and (iii) task schedulability in the presence of virtual interrupts is not considered. In this dissertation, we address these limitations.

\section{GPGPU Management}
\label{backgroun_gpu_management}

A GPGPU or simply GPU is a computational accelerator. Tasks running on CPU cores can offload some of their workloads to the GPU to reduce their response times and to save CPU utilization. The use of a GPU, of course, causes a different execution pattern compared to when it is not used. The characteristics of the GPU hardware and device driver also affects the timing behavior of the task execution. In this section, we first describe the execution pattern of a task using a GPU, and then review prior work on predictable GPU management.

\begin{figure}[h]
	\centering
	\subfloat{
		\includegraphics[width=0.7\textwidth]{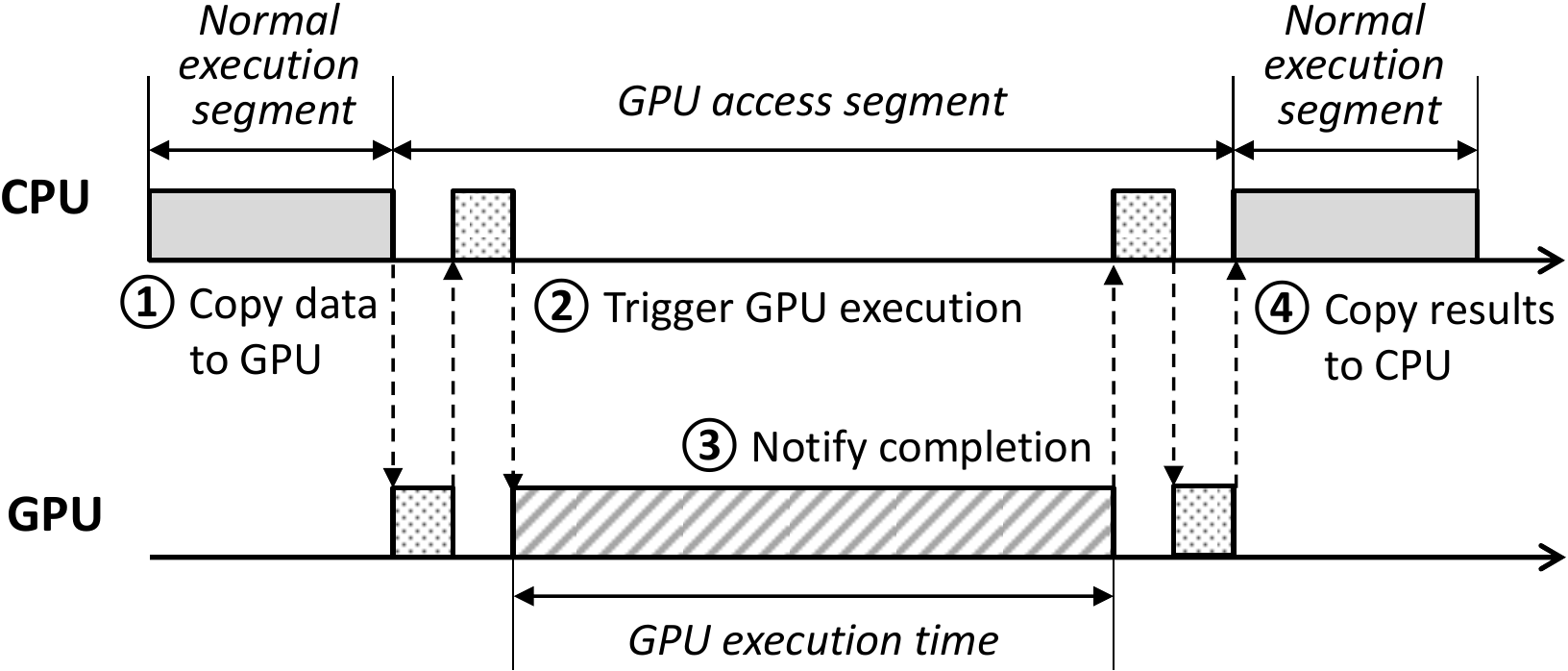}
	}
	\caption{Execution pattern of a task accessing a GPU}
	\label{fig:GPU_execution_pattern}	
\end{figure}

\subsection{GPU Execution Pattern}
\label{backgroun_gpu_execution_pattern}

The execution time of a task using a GPU can be decomposed into normal execution segments and GPU access segments. Normal execution segments run entirely on CPU cores and GPU access segments involve GPU executions. \figref{GPU_execution_pattern} depicts an example of a task having one GPU access segment. In the GPU access segment, the task first copies data needed for GPU execution, from CPU memory to GPU memory. Then, the task triggers the GPU execution and waits until the GPU execution finishes. During this time, the task may suspend or busy-wait, depending on the implementation of the GPU device driver and the configuration used. The task is notified when the GPU execution finishes, and it copies the results back from the GPU to the CPU. Finally, the task continues its normal execution segment. 

There are several issues we need to consider for the use of a GPU in a predictable manner. First, today's COTS GPUs do not support a preemption mechanism, and GPU execution requests from multiple tasks are handled in a sequential, non-preemptive manner. This is primarily due to the high overhead expected on GPU context switching~\cite{Tanasic_ISCA14}. Second, COTS GPU device drivers do not respect task priorities and the scheduling policy used in the system. Hence, in the worst case, the GPU access request of the highest-priority task may be delayed by the requests of all lower-priority tasks in the system, which causes possibly unbounded priority inversion. These issues have motivated the development of a predictable GPU management scheme to ensure task timing constraints while achieving performance improvement.

\subsection{Related Work}
\label{GPU_related_work}

Many software techniques have been developed to utilize a GPU as a predictable, shared computing resource. 
TimeGraph~\cite{Kato_ATC11} is a real-time GPU scheduler that schedules GPU access requests from tasks with respect to task priorities. This is done by modifying an open-source GPU device driver and monitoring GPU commands issued by tasks at the driver level. TimeGraph also provides a resource reservation mechanism that accounts for and enforces the GPU usage of each task, with posterior and apriori enforcement techniques. RGEM~\cite{Kato_RTSS11} is another real-time GPU scheduler implemented as a user-level library. Hence, RGEM can be used with proprietary, closed-source GPU device drivers. RGEM provides similar features to TimeGraph, such as scheduling of GPU requests in task priority order. In addition, RGEM allows splitting a long data-copy operation into smaller chunks, reducing blocking time on data-copy operations. Gdev~\cite{Kato_ATC12} is similar to TimeGraph and RGEM in the GPU scheduling perspective, but provides common APIs to both user-level tasks and the OS kernel to use a GPU as a computing resource. GPES~\cite{Zhou_RTAS15} is a software technique to break a long GPU execution segment into smaller sub-segments, allowing preemptions at the boundaries of sub-segments. 
While all these techniques can mitigate the limitations of today's GPU hardware and device drivers, they have not considered the schedulability of tasks. In other words, GPU requests from tasks are handled in a predictable manner under those techniques, but the timing behavior of tasks on the CPU side, especially on a multi-core CPU, has not been studied as a primary concern. 

Kim et al.~\cite{Kim_RTSS13} focused on the schedulability analysis of tasks using hardware accelerators like GPUs. They found that conventional real-time scheduling analysis requires tasks not to suspend while accessing GPUs, which may significantly waste CPU utilization. As a solution to this problem, they proposed a new scheduling policy, called {\em segment-fixed priority scheduling}, which assigns different priorities and phase offsets to each segment of tasks. Since the determination of the optimal priorities and offsets for individual segments is NP-hard in the strong sense, they developed several heuristics for priority and offset assignment. However, their approach is limited to single-core systems.

Elliott et al.~\cite{Elliott_RTS12, Elliott_RTSS13} modeled GPUs as mutually-exclusive resources and developed GPUSync, a software framework based on real-time synchronization protocols to access GPUs. This synchronization-based approach has many benefits. First, it can schedule GPU requests from tasks in a predictable manner, without making changes to GPU device drivers. Second, it allows the task schedulability analysis originally developed for multi-core synchronization protocols to be directly used for analyzing the tasks accessing GPUs in a multi-core environment. However, this approach requires the GPU access segments of tasks to be treated as {\em critical sections}, meaning that tasks cannot suspend during GPU executions. Also, the use of real-time synchronization protocols for GPUs may unnecessarily delay the executions of high-priority tasks due to the priority-boosting mechanism employed in such protocols. In this dissertation, we develop a new approach to address the limitations of the synchronization-based approach for GPU management. Section~\ref{chapter_predictable_gpgpu_management} will discuss more details on the use of the synchronization-based approach under partitioned fixed-priority scheduling, present our proposed approach, and compare the performance characteristics of these two approaches.

\chapter{System Model}
\label{system_model}

\begin{figure}[h]
	\centering
	\subfloat{
		\includegraphics[width=0.9\columnwidth]{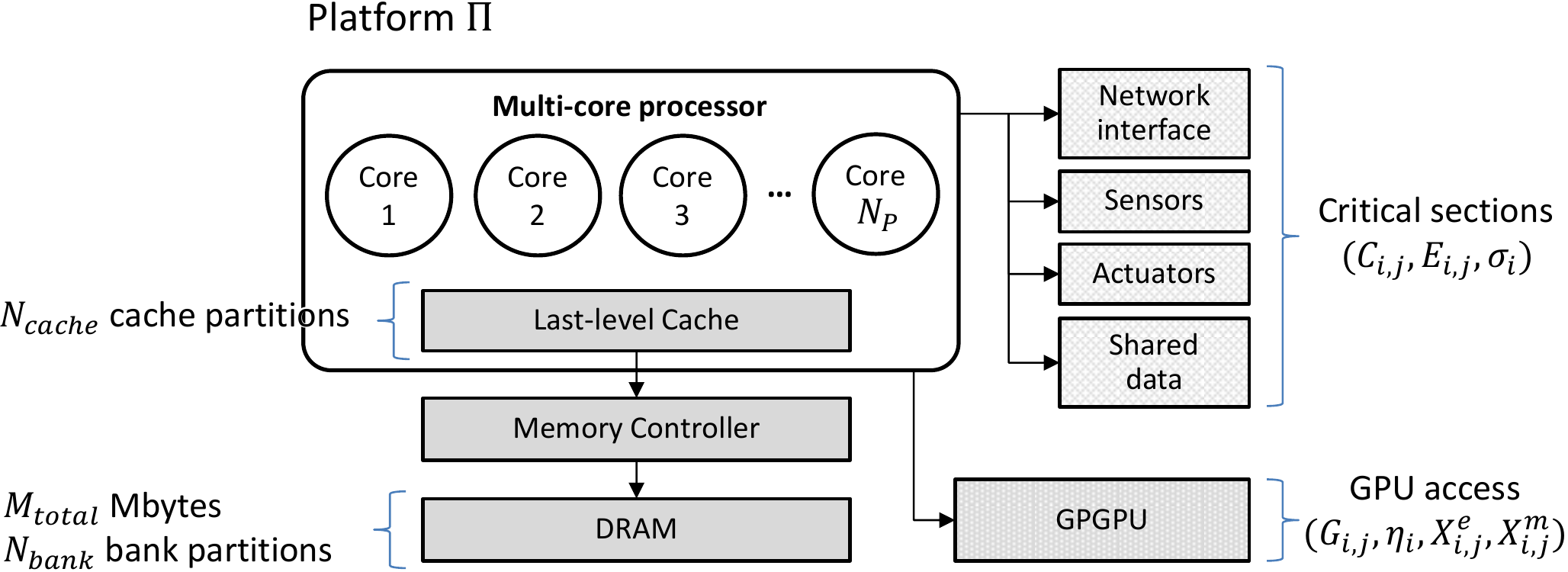}
	}
	\caption{Multi-core platform considered in this work}
	\label{fig:platform_details}
\end{figure}

This chapter describes the system model used throughout this dissertation. As briefly described in Section~\ref{scope_of_this_work}, we consider a multi-core platform equipped with three types of shared resources: concurrent resources, mutually-exclusive resources, and computational accelerators. \figref{platform_details} illustrates the computing platform, denoted as $\Pi$, considered in this work and the parameters associated with the shared platform resources. Tasks are executed with access to the shared platform resources of $\Pi$ in either native or virtualized environments. The entire system parameters we use, including the platform, tasks, and virtual machines, are summarized in \tableref{system_model_params}. Detailed explanations are given in the following sections.

\begin{table}[t]
\centering
{
	\footnotesize
	\caption{Summary of system model parameters}\label{tab:system_model_params}
	\begin{tabular}{c|c|l}
	\hline
	Type 		& Params 	& Descriptions\\
	\hline
	Platform 	& $N_P$			& Number of physical CPU cores\\
	($\Pi$)		& $M_{total}$ 	& Total memory size in Mbytes\\
				& $N_{cache}$ 	& Number of cache partitions\\
				& $N_{bank}$	& Number of bank partitions\\
	\hline
	Task		& $C_i(k)$		& WCET of task $\tau_i$, when $k$ cache partitions are assigned to it\\
	($\tau_i$)	& $C_i$ 		& Simplified form of $C_i(k)$\\	
				& $T_i$			& Minimum inter-arrival time of each job of $\tau_i$\\
				& $D_i$			& Relative deadline of $\tau_i$\\
				& $M_i$			& Required physical memory size in Mbytes\\
				& $H_i(k)$		& Max. DRAM requests of $\tau_i$, when $k$ cache partitions are assigned to it\\
				& $H_i$ 		& Simplified form of $H_i(k)$\\				
				& $G_i$ 		& Maximum accumulated GPU access time of $\tau_i$\\						
	\hline
	Critical	& $C_{i,j}$		& WCET of $j$-th normal execution segment of $\tau_i$\\
	section		& $E_{i,j}$		& WCET of $j$-th critical section segment of $\tau_i$\\
				& $\sigma_i$			& Number of critical section segments of $\tau_i$\\
	\hline
	GPU			& $G_{i,j}$		& Maximum length of $j$-th GPU access segment of task $\tau_i$\\
	access		& $\eta_{i}$	& Number of GPU access segments of $\tau_i$\\
				& $X^e_{i,j}$		& Worst-case GPU execution time in $j$-th GPU access segment of $\tau_i$\\
				& $X^m_{i,j}$		& WCET of miscellaneous operations in $j$-th GPU access segment of $\tau_i$\\				
	\hline
	Virtual		& $N_{vcpu}$	& Number of VCPUs in a VM \\
	machine		& $C_i^v(k)$	& Execution budget of a VCPU $v_i$, when $k$ cache partitions are assigned to it\\
	(VM)& $C_i^v$		& Simplified form of $C_i^v(k)$\\
				& $T_i^v$		& Budget replenishment period of a VCPU $v_i$\\
	\hline
	\end{tabular}
	}
\end{table}

\section{Platform Model}
\label{platform_model}

We consider a computing platform $\Pi$ equipped with a single-chip multi-core processor and $M_{total}$ Mbytes of DRAM as main memory. The processor has $N_P$ identical cores running at a fixed clock speed. 
In this work, we assume that each core has a fully timing-compositional architecture as described in~\cite{Wilhelm_IEEE09}. This means that each core is in-order with one outstanding memory access request and any delay from shared resources are additive to task response time. 

\smallskip
\noindent\textbf{Shared Cache:}
The multi-core processor has a unified last-level cache (LLC) shared among all cores. We use page coloring to manage the shared cache in software. Page coloring is implemented in the OS in a native environment, and in the hypervisor in a virtualized environment. With page coloring, the LLC is divided into $N_{cache}$ cache partitions. Each cache partition is represented as a unique integer in the range from 1 to $N_{cache}$. 

\smallskip
\noindent\textbf{DRAM System:}
We assume the DDR SDRAM system described in Section~\ref{MEM_background}. The memory controller uses the FR-FCFS policy, and the arrival times of memory requests are assumed to be recorded when they arrive at the memory controller. 
DRAM consists of one or more ranks.
The memory controller uses an {\em open-row} policy which keeps the row-buffer open. We assume that the DRAM is not put into a low-power state at any time.

The LLC and the DRAM system are connected by a single memory channel.
We assume that all data fetched from the DRAM system are stored in the LLC. A single memory request can fetch one entire cache line from the DRAM because of the burst-mode data transfer. The addresses of memory requests are aligned to the size of {\it BL} (burst length). We limit our focus on memory requests from CPU cores and leave DMA (Direct Memory Access) as our future work. 

Bank partitioning is considered to divide DRAM banks into $N_{bank}$ partitions. Each bank partition comprises one or more DRAM banks that are not shared with other bank partitions, and is represented as a unique integer in the range from 1 to $N_{bank}$. It is assumed that the number of DRAM banks in each bank partition and the number of bank partitions assigned to a task do not affect the task's worst-case execution time. 

\smallskip
\noindent\textbf{GPGPU:} We assume that the platform $\Pi$ is equipped with a single, general-purpose GPU device. 
The GPU has the characteristics described in Section~\ref{backgroun_gpu_management}. Hence, although the GPU can be shared among multiple tasks, GPU requests from tasks are handled in a sequential, non-preemptive manner. The GPU has its own memory space, which is assumed to be sufficiently enough for the GPU memory usage of tasks. We also assume that the data copy request of a GPU-using task from the main memory to the GPU memory, and vice versa, is handled by the memory controller, just like normal memory requests. Analyzing the effects of using DMA for GPU data copy remains as our future work.

\section{Task Model}
\label{task_model}

We consider sporadic tasks with constrained deadlines. Tasks are scheduled by {\em partitioned fixed-priority preemptive scheduling}. Thus, each task is statically assigned to a single physical core in a native environment, and to a single virtual CPU (VCPU) in a virtualized environment. Any fixed-priority assignment can be used for tasks, such as Rate-Monotonic~\cite{Liu_Layland}. Task $\tau_i$ is represented with the following parameters:
$${\tau}_i:=(C_i(k), T_i, D_i, M_i, H_i(k), G_i)$$
\begin{itemize}
	\item $C_i(k)$: the worst-case execution time (WCET) of task ${\tau}_i$, when it runs alone in a system with $k$ cache partitions assigned to it
	\item $T_i$: the minimum inter-arrival time of each job of ${\tau_i}$
	\item $D_i$: the relative deadline of each job of $\tau_i$ ($D_i \le T_i$)
	\item $M_i$: the size of required physical memory in Mbytes, which should be assigned to $\tau_i$ to prevent swapping
	\item $H_i(k)$: an upper bound on the number of DRAM requests generated by any job of $\tau_i$, when $k$ cache partitions are assigned to it
	\item $G_i$: the maximum accumulated GPU access time of $\tau_i$
\end{itemize}
We assume that $C_i(k)$ is monotonically decreasing with $k$. This is a common assumption in the literature: the actual WCET function may not be monotonic, but this assumption can be easily satisfied by monotonic over-approximations of WCETs with insignificant pessimism~\cite{Altmeyer_ECRTS14}.
Each task $\tau_i$ has a unique priority $\pi_i$. An arbitrary tie-breaking rule can be used to achieve this under fixed-priority scheduling. 
Note that no assumptions are made on the memory access pattern of a task (e.g., access rate). 

Parameters $C_i(k)$ and $H_i(k)$ can be obtained by either measurement-based or static-analysis tools. When a measurement-based approach is used, $C_i(k)$ and $H_i(k)$ need to be conservatively estimated. Especially in a system with a write-back cache, $H_i(k)$ should take into account dirty lines remaining in the cache. We assume that $C_i(k)$ and $H_i(k)$ parameters remain the same in both native and virtualized environments.\footnote{Capturing the overhead of virtualization in those parameters is beyond the scope of our work. However, we believe this does not limit the practicality of our work because it is relatively small (e.g., more than 99\% of native performance can be achieved in full-virtualization mode with recent hardware virtualization support~\cite{Steinberg_EuroSys10}).} 

In the rest of this dissertation, $C_i$ and $H_i$ may be used instead of $C_i(k)$ and $H_i(k')$, respectively, when each task is assumed to have been already assigned its cache partitions. 

\smallskip
\noindent\textbf{Tasks with Critical Sections:} Mutually-exclusive resources considered in this work are protected by suspension-based mutex locks. Tasks access shared resources in a non-nested manner, meaning that each task can hold only one resource at a time. 
If a task $\tau_i$ accesses such resources, the WCET $C_i$ can be decomposed into an alternating sequence of normal execution segments and critical section segments as follows:
$$C_i:=(C_{i,1}, E_{i,1}, C_{i,2}, E_{i,2}, ..., E_{i,\sigma_i}, C_{i,\sigma_{i}+1})$$
\begin{itemize}
  \item $C_{i,j}$: the WCET of the $j$-th normal execution segment of task $\tau_i$
  \item $E_{i,j}$: the WCET of the $j$-th critical section segment of $\tau_i$  
  \item $\sigma_i$: the number of critical section segments of $\tau_i$
\end{itemize}
We use $E_i$ to denote the sum of the WCETs of the critical section segments of $\tau_i$. Hence,
$$E_i=\sum_{j=1}^{\sigma_i}E_{i,j}\text{, and  } C_i=\sum_{j=1}^{\sigma_i+1}C_{i,j}+\sum_{j=1}^{\sigma_i}E_{i,j}$$

\smallskip
\noindent\textbf{Tasks with GPGPU Accesses:} 
As presented in Section~\ref{backgroun_gpu_execution_pattern}, a task using a GPU has one or more GPU access segments. We use $\eta_i$ to denote the number of GPU access segments of task $\tau_i$, and $G_{i,j}$ to denote the maximum length of the $j$-th GPU access segment of $\tau_i$. Hence, 
$$G_i=\sum_{j=1}^{\eta_i}G_{i,j}$$
The $j$-th GPU access segment of $\tau_i$ can be decomposed as follows:
$$G_{i,j}:=(X^e_{i,j}, X^m_{i,j})$$
\begin{itemize}
	\item $X^e_{i,j}$: the worst-case GPU execution time in the $j$-th GPU access segment of $\tau_i$
	\item $X^m_{i,j}$: the WCET of miscellaneous operations, including data copies and notifications, in the $j$-th GPU access segment of $\tau_i$
\end{itemize}

\section{Virtual Machine Model}

When virtualization is used, the system runs a hypervisor hosting multiple guest virtual machines (VMs). To distinguish it from VMs, the system is also referred to as a {\em host machine} in a virtualized environment. Each VM has one or more VCPUs. The VCPUs are scheduled on the physical CPU cores (PCPUs) of the host machine by {\em partitioned fixed-priority preemptive scheduling}. Hence, each VCPU is statically assigned to a single PCPU, and any fixed-priority assignment can be used for VCPUs. Each VM  is represented as follows:
$$\text{VM}:=(v_1, v_2, ..., v_{N_{vcpu}})$$
where $v_i$ is a VCPU and $N_{vcpu}$ is the number of VCPUs in the VM. We represent a VCPU $v_i$ as follows: 
$$v_i:=(C_i^v(k), T_i^v)$$
\begin{itemize}
	\item $C_i^v(k)$: the execution budget of a VCPU $v_i$, represented as a function of the total number of cache partitions ($k$) assigned to $v_i$\VS{-5pt}
	\item $T_i^v$: the budget replenishment period of a VCPU $v_i$\VS{-3pt}
\end{itemize}
Since task execution time is affected by the number of assigned cache partitions, it is obvious that the required budget of a VCPU is also affected by the number of cache partitions to be used by its tasks. With this model, the computational demand of each VM can be presented to the hypervisor and other VMs, without revealing its task attributes. We will show in Section~\ref{VCACHE_virt_cache_algo} how to find the budget of each VCPU with respect to the number of cache partitions. For brevity, $C_i^v$ may be used instead of $C_i^v(k)$, when each VCPU is assumed to have been assigned its cache partitions.

For the VCPU budget supply and replenishment policies, we consider {\em periodic server}~\cite{periodic_server}, {\em sporadic server}~\cite{sporadic_server}, and {\em deferrable server}~\cite{deferrable_server} variants, because they have been widely used in real-time virtualization~\cite{Xi_EMSOFT11, Li_VEE14,Kim_RTSS14, Kim_RTCSA15}. Under the periodic server policy, each VCPU becomes active periodically and executes its tasks that are ready to be executed until the VCPU's budget is exhausted. When a VCPU has no task ready to execute, the VCPU cannot preserve its budget; the budget is idled away. Under the deferrable server policy, a VCPU can preserve its budget until the end of its current period. Hence, the tasks of the VCPU can execute any time while the VCPU's budget remains. The budget-preserving feature of the deferrable server policy causes a jitter equal to $T^v-C^v$~\cite{Bernat_RTSS99}. Under the sporadic server policy, a VCPU can preserve its budget, but only the amount of budget used is replenished $T_i^v$ units after the start of the use of that amount, yielding a zero release jitter. 

\section{Other Assumptions}

We further make the following assumptions in this dissertation:
\begin{itemize}
	\item We assume that the multi-core processor considered in this work uses neither simultaneous multithreading (SMT)~\cite{Tullsen_ISCA95} nor dynamic voltage and frequency scaling (DVFS)~\cite{Burd_IEEE00}. This assumption is made to minimize timing uncertainties possibly caused by such techniques. Although there has been work on utilizing SMT (e.g.,~\cite{Cazorla_IEEE06}) and DVFS (e.g.,~\cite{Colin_RTCSA14,Saewong_RTAS03}) in predictable systems, we leave them as our future work.
	\item We assume that tasks do not use dynamic memory allocation, since it is typically prohibitively expensive when real-time predictability is important~\cite{Mancuso_RTAS13}. Also, we assume that tasks do not experience page swapping and they have been allocated their required physical memory size ($M_i$).	
	\item We assume that each VM has been allocated a sufficient number of host physical pages and that page swapping does not happen at run-time. This is a reasonable assumption in CPS virtualization scenarios because, unlike in server virtualization, memory underprovisioning is considered to be harmful to timing predictability~\cite{Kiszka_RTLWS09}. Also, this assumption can be easily achieved by VM admission control at the hypervisor.
\end{itemize}

\chapter{Coordinated Approach for Predictable Cache Management}
\label{chapter_coordinated_cache_management}
Cache interference in multi-core systems can be categorized into two types: {\em inter-core} and {\em intra-core}. Inter-core cache interference happens when tasks running on different cores access the last-level shared cache (LLC) simultaneously. Since the execution of a task may be potentially affected by memory accesses of {\em all} tasks running on other cores, the accurate analysis of inter-core cache interference is extremely difficult~\cite{Guan_EMSOFT09}. Intra-core cache interference, in contrast, occurs within a core. When a task preempts another task, the preempting task may evict the cache contents of the preempted task. Moreover, while a task is inactive, other tasks can corrupt its cache. 

In this chapter, we introduce a novel OS-level cache management scheme to address inter-core and intra-core cache interference in a multi-core platform. 
Our scheme provides predictable cache performance and addresses the problems of page coloring discussed in Section~\ref{problems_with_page_coloring}, through tight coordination of {\em cache reservation}, {\em cache sharing}, and {\em cache-aware task allocation}.
Cache reservation ensures the exclusive use of a certain amount of cache for individual cores to prevent inter-core cache interference. Within each core, cache sharing allows tasks to share the reserved cache, while providing a safe upper bound on intra-core cache interference. Cache sharing also significantly mitigates the memory co-partitioning problem and the limitations on the number of cache partitions. By using cache reservation and cache sharing, cache-aware task allocation determines efficient task and cache allocation to schedule a given taskset. 
Our scheme does {\em not} require special hardware cache partitioning support or modifications to application software. Hence, it is readily applicable to commodity processors such as the Intel Core i7. Our scheme can be used not only for developing a new system but also for migrating existing applications from single-core to multi-core platforms.

The detailed contributions of our scheme are as follows. First, we introduce the concept of sharing cache partitions under page coloring to counter the memory co-partitioning problem and the limited number of cache partitions. We show how pages are allocated when cache partitions are shared, and provide a condition that checks the feasibility of sharing while guaranteeing the allocation of the required memory to tasks. Second, we provide a response time test for checking task schedulability when cache partitions are shared among tasks. Our approach is independent of the specific cache analysis used and allows estimating the worst-case execution time (WCET) of a task in isolation from other tasks. Third, our cache-aware task allocation algorithm reduces the number of cache partitions required to schedule a given taskset, while meeting both the task memory requirements and the task timing constraints. We also show that the remaining cache partitions after the allocation can be used to save the total CPU utilization. Forth, we have implemented and evaluated our scheme by extending the Linux/RK platform~\cite{LinuxRK, ResourceKernel} running on the Intel Core i7 quad-core processor. The experimental results on a real machine demonstrate the effectiveness of our scheme. 

For simplicity, our analysis provided in this chapter considers cache interference only. Interference delays from other shared resources will be analyzed in later chapters. However, it is worth noting that our analysis provided in this chapter can be easily combined with those in other chapters, since any delay from shared resources is additive to task response time in a fully timing-compositional architecture~\cite{Wilhelm_IEEE09} we assume. An example of combining cache and memory interference analyses will be provided in Section~\ref{combining_with_cache_interference_analysis}.

The background and related prior work on cache interference have been discussed in Section~\ref{literature_review_cache_interference}. The system model including assumptions and notation for a shared cache and tasks can be found in Chapter~\ref{system_model}. 

The rest of this chapter is organized as follows. Section~\ref{CACHE_coordinated_cache_mgmt} presents our coordinated cache management scheme. A detailed evaluation of our scheme is provided in Section~\ref{CACHE_evaluation}. Section~\ref{CACHE_conclusions} summarizes this chapter.

\section{Coordinated Cache Management}
  \label{CACHE_coordinated_cache_mgmt}
\begin{figure}[t]
\centering
\vspace{-10pt}
  \includegraphics[width=0.7\textwidth]{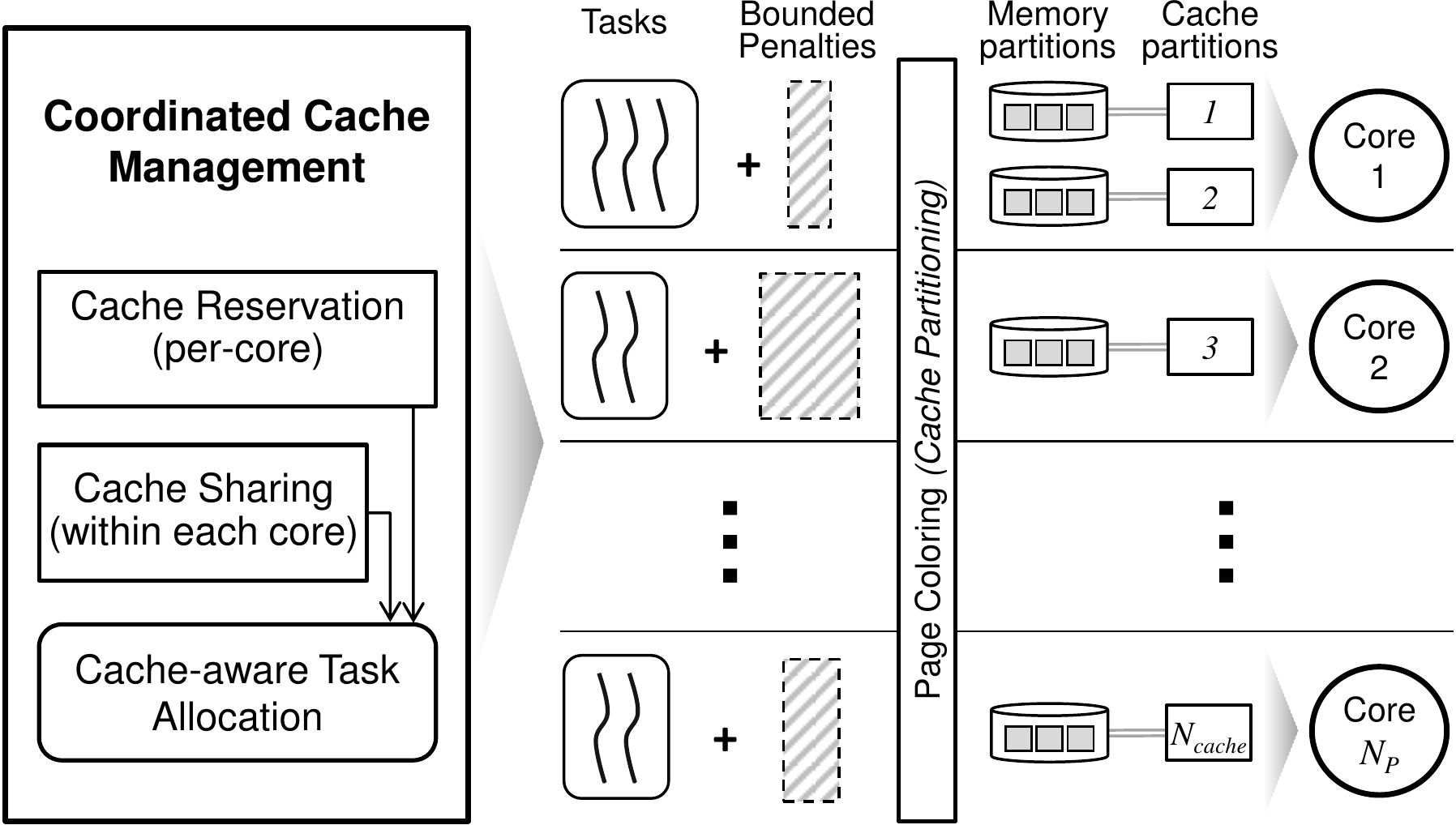}\\
\vspace{-5pt}
  \caption{Overview of the proposed OS-level cache management}\label{fig:CACHE_overview}
\vspace{-5pt}
\end{figure}

In this section, we describe our proposed cache management scheme. \figref{CACHE_overview} shows the overview of our scheme that consists of three components: {\em cache reservation}, {\em cache sharing}, and {\em cache-aware task allocation}. Cache reservation ensures the exclusive use of a portion of the shared cache for each core. Cache sharing enables sharing of cache partitions among tasks within each core. Cache-aware task allocation uses these two components to find efficient cache and task allocation while maintaining feasibility.


\subsection{Cache Reservation}
Due to the inherent difficulties of precisely analyzing inter-core cache interference on a multi-core processor, we reserve a portion of cache partitions for each core to prevent inter-core cache interference. The reserved cache partitions are exclusively used by their owner core, thereby preventing cache contention from other cores. Per-core cache reservation differentiates our scheme from other cache partitioning techniques that allocate exclusive cache partitions to each task. Within each core, cache partitions reserved for the core {\em can} be shared by tasks running on the core. This approach allows the core to execute more tasks than the number of cache partitions allocated to that core. The execution time of a task can potentially be further reduced by providing more cache partitions to the task. Moreover, since the sharing of a cache partition means the sharing of an associated memory partition, it can significantly reduce the waste of cache and memory space caused by the memory co-partitioning problem due to page coloring.

Cache partitions are reserved for a core by allocating associated memory partitions to the core. Each core manages the state of pages in its memory partitions. When a new task is assigned to a core, the task's memory requests are handled by allocating free pages from the core's memory partitions. The appropriate number of cache partitions for each core depends on the tasks running on the core. This cache allocation will be determined by our cache-aware task allocation, discussed in Section~\ref{CACHE_task_allocation}.


\subsection{Cache Sharing: Bounding Intra-core Penalties}
Suppose that a certain number of cache partitions is allocated to a core by cache reservation. Our scheme allows tasks running on the core to share the given partitions, but sharing causes intra-core cache interference. Intra-core cache interference can be further subdivided into two types: 
\begin{enumerate}
  \item {\em Cache warm-up delay}: occurs at the beginning of each period of a task and arises due to the execution of other tasks while the task is inactive.
  \item {\em Cache-related preemption delay}: occurs when a task is preempted by a higher-priority task and is imposed on the preempted task.
\end{enumerate}
Previous work on bounding cache interference on single-core platforms~\cite{Altmeyer_RTSS11, CGLee_IEEE01, Lunniss_RTAS13} assumes that the cache warm-up delay can be taken into account in the WCET of a task by static cache analyses. However, such static cache analysis tools may not be readily available for modern multi-core processors. We therefore consider the cache warm-up delay as an extrinsic factor to a task's WCET. This approach enables measurement-based WCET analysis to estimate the task WCET in isolation from other tasks.\footnote{Appropriate ``error margins'' that are proportional to system criticality can be applied to these measurements, as is done in practice.} For instance, once a task is launched, the task's cache is initially warmed up during the startup phase or the very first execution of the task. If the task runs alone in the system or uses its cache all by itself, subsequent task instances do not experience any cache warm-up delay at run-time \cite{Liedtke_RTAS97}. By considering the cache warm-up delay as an extrinsic factor, the WCET obtained in such an isolated environment can be safely used even when the task's cache is shared. 


We formally define cache warm-up delay and cache-related preemption delay. 
$\omega_{j,i}$ is $\tau_j$'s cache warm-up delay, which is caused by the tasks that (i) have priorities higher than or equal to $\tau_i$ and (ii) share cache partitions with $\tau_j$. 
$\gamma_{j,i}$ is the cache-related preemption delay caused by $\tau_j$ and imposed on the tasks that (i) have priorities lower than $\tau_j$ and higher than or equal to $\tau_i$ and (ii) share cache partitions with $\tau_j$.
Hence, $\omega_{j,i}$ and $\gamma_{j,i}$ are represented as follows:
\vspace{-3pt}
\[
\begin{split}
&\omega_{j,i}=\left|\mathbb{S}_j\cap \bigcup_{\tau_k \in \mathbb{P}(\tau_i) \land \tau_k \ne \tau_j \land \pi_k \ge \pi_i}\mathbb{S}_k\right|\cdot\Delta\\
&\gamma_{j,i}=\left|\mathbb{S}_j\cap \bigcup_{\tau_k\in \mathbb{P}(\tau_i) \land \pi_k < \pi_j \land \pi_k \ge \pi_i}\mathbb{S}_k\right|\cdot\Delta
\end{split}
\]
where $\mathbb{S}_j$ is the set of cache partitions assigned to $\tau_j$, $\Delta$ is the maximum time to refill one cache partition, $\mathbb{P}(\tau_i)$ is the core of $\tau_i$, and $\pi_k$ is the priority of $\tau_k$. Note that, in case of a write-back cache, $\Delta$ should take into account the effect of a {\em dirty} cache line that requires two memory accesses to fetch a new cache line~\cite{Sebek_TR01}.

The utilization of a taskset $\Gamma_j$, which is allocated to a core $j$, with intra-core cache interference penalties, $\omega$ and $\gamma$, can be calculated by extending Liu and Layland's schedulability condition \cite{Liu_Layland} as follows:
\begin{equation} \label{eq:CACHE_utilization}
util(\Gamma_j) = \sum_{\tau_i \in \Gamma_j } \left(\frac{C_i}{T_i}+\frac{\omega_{i,n}}{T_i}+\frac{\gamma_{i,n}}{T_i}\right)
\end{equation}
where $n$ is the index of the lowest-priority task in $\Gamma_j$. It is based on the Basumallick and Nilsen's technique \cite{Basumallick_94}, but we explicitly consider cache warm-up delay $\omega$.  

The iterative response time test \cite{Joseph_J86} can be extended as follows to incorporate the two types of intra-core cache interference:
\vspace{-4pt}
\begin{equation} \label{eq:CACHE_response_time_test}
\begin{split}
W_i^{k+1}&=C_i+\omega_{i,n}+
\sum_{\tau_h \in \mathbb{P}(\tau_i)\land \pi_h>\pi_i}\left\lceil\frac{W_i^{k}}{T_h}\right\rceil C_h +\\
&\sum_{\tau_h \in \mathbb{P}(\tau_i)\land \pi_h>\pi_i}\!\!\left\{\omega_{h,n}\!+\!\left(\left\lceil\frac{W_i^{k}}{T_h}\right\rceil\!-\! 1\right)\omega_{h,i}\right\}\!+\!\!\sum_{\tau_h \in \mathbb{P}(\tau_i)\land \pi_h>\pi_i}\!\!\left\lceil\frac{W_i^{k}}{T_h}\right\rceil\!\gamma_{h,i}
\end{split}
\end{equation}
where $W_i^{k}$ is the worst-case response time of $\tau_i$ at the $k^{th}$ iteration, and $n$ is the index of the lowest-priority task on $\tau_i$'s core. The test terminates when $W_i^{k+1} = W_i^{k}$. Task $\tau_i$ is schedulable if its response time is before its deadline: $W_i^k \le D_i$. 
We represent the amount of $\omega$ and $\gamma$ delays caused by the execution of a higher priority task $\tau_j$ within the worst-case response time $W_i^k$ in the second and the third summing terms of \eqref{eq:CACHE_response_time_test}. Note that the first execution of a higher priority task $\tau_h$ within $W_i^k$ causes a cache warm-up delay of $\omega_{h,n}$, but the subsequent executions of $\tau_h$ cause only $\omega_{h,i}$ because tasks with lower priorities than $\tau_i$ are not scheduled while $\tau_i$ is executing. 

\begin{figure}[t]
\centering
\vspace{-8pt}
  \includegraphics[width=0.7\textwidth]{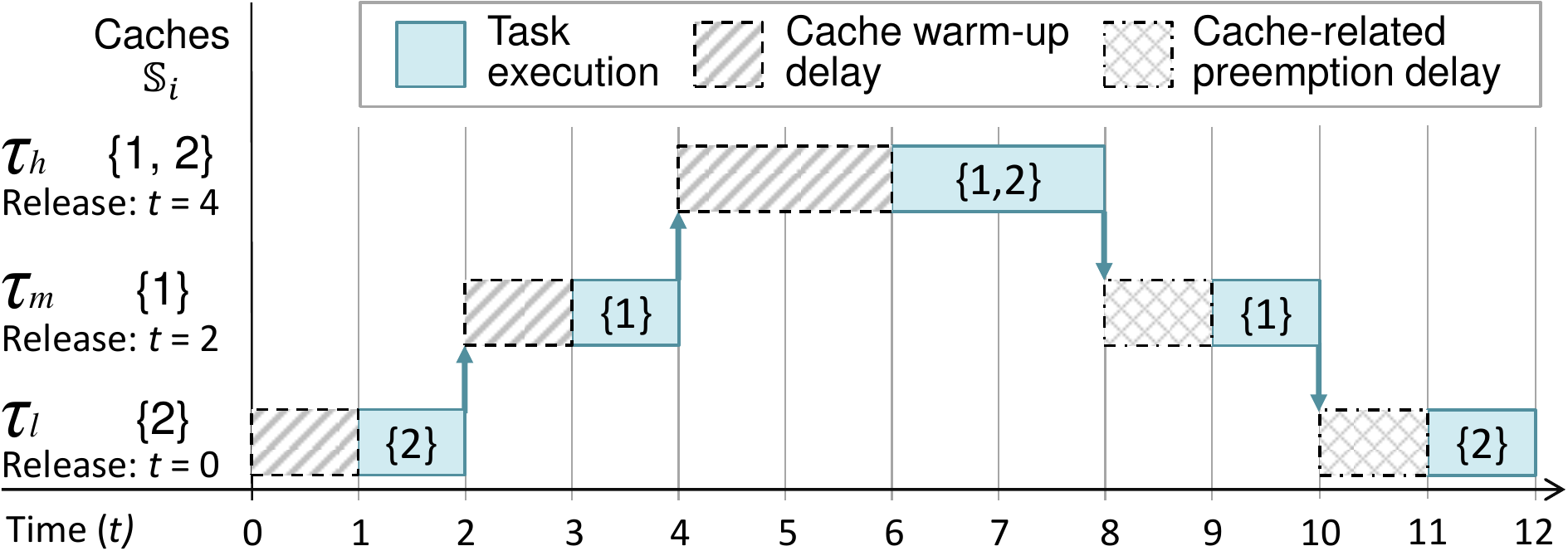}\\
\vspace{-9pt}
  \caption[Tasks sharing cache partitions with cache interference penalties]{Three tasks sharing cache partitions with intra-core cache interference penalties}\label{fig:CACHE_response_time_fig}
\end{figure}

\figref{CACHE_response_time_fig} shows an example taskset $\{\tau_h, \tau_m, \tau_l\}$ sharing a set of cache partitions $\{1, 2\}$. In this taskset, $\tau_h$ is a high-priority task, $\tau_m$ is a medium-priority task, and $\tau_l$ is a low-priority task. Assume that the cache partitions are assigned to the tasks as follows: $\mathbb{S}_h$ is $\{1,2\}$, $\mathbb{S}_m$ is $\{1\}$, and $\mathbb{S}_l$ is $\{2\}$. All tasks have the same execution time $C_i=2$ and the same periods and deadlines $T_i=D_i=12$. The cache partition refill time $\Delta$ is 1 in this example. When $\tau_h$ starts its execution, it needs to refill its two cache partitions. $\tau_m$ has one cache warm-up delay and one cache-related preemption delay due to $\tau_h$. $\tau_l$ also has one cache warm-up delay and one cache-related preemption delay. 

It is worth noting that Eqs.~\eqref{eq:CACHE_utilization} and \eqref{eq:CACHE_response_time_test} are independent of the specific cache analysis used. If a precise cache analysis tool is available for a target multi-core processor's shared cache, the cache partition refilling time $\Delta$ can be more tightly estimated.

\subsection{Cache Sharing: How to Share Cache Partitions}
We now describe how cache partitions are allocated to tasks within a core such that schedulability is preserved and memory requirements are guaranteed despite sharing the partitions. 
There are two conditions for a cache allocation to be feasible. The first condition is the response time test given by Eq.~\eqref{eq:CACHE_response_time_test}. The factors affecting a task's response time are as follows: (i) cache-related task execution time $C_i(k)$, (ii) cache partition refill time $\Delta$, (iii) the number of other tasks sharing the task's cache partitions, and (iv) the periods of the tasks sharing the cache partitions. Factors (i) and (ii) are explicitly used to calculate the response time. If factor (iii) increases or factor (iv) is relatively short, the response time may be lengthened due to cache penalties caused by frequent preemptions. 

\begin{figure}[t]
\centering
\vspace{-5pt}
  \includegraphics[width=0.7\textwidth]{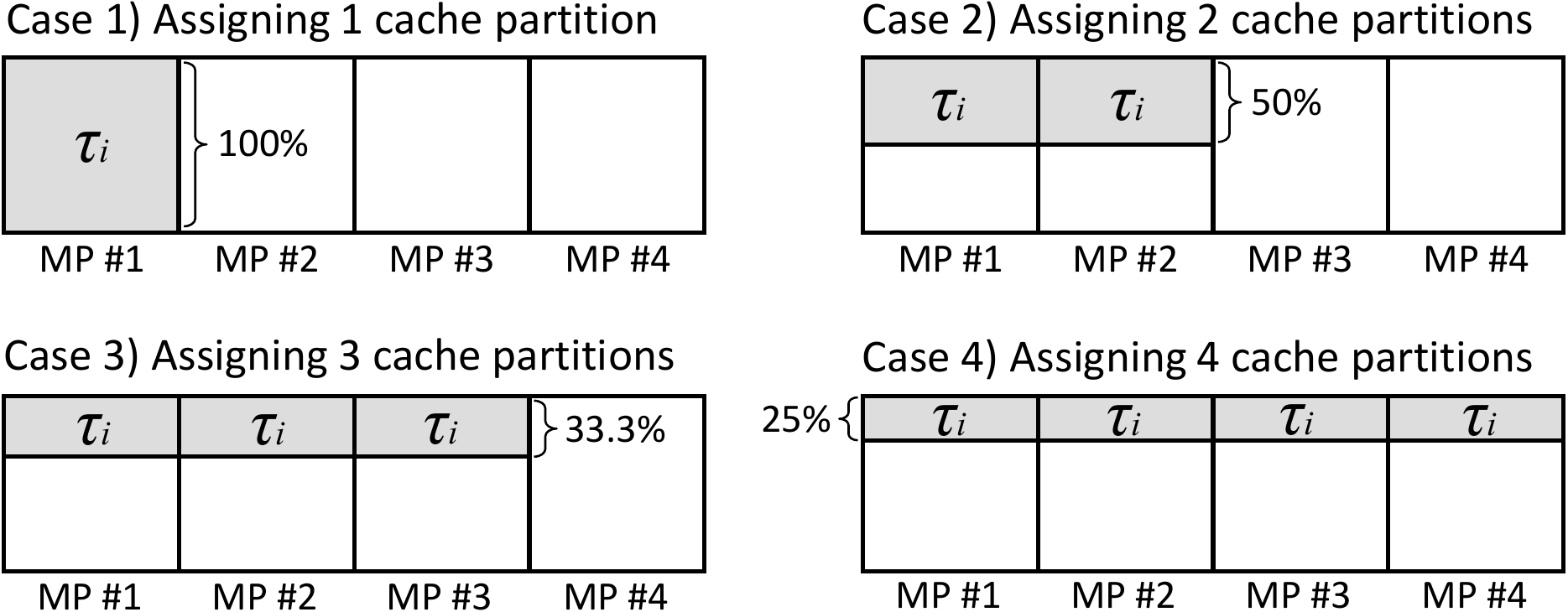}\\
\vspace{-8pt}
  \caption{Page allocations for different cache allocation scenarios}\label{fig:CACHE_mem_sharing_1}
\vspace{-5pt}
\end{figure}

The second condition is related to the task memory requirements. Before defining this condition, we show in \figref{CACHE_mem_sharing_1} an example of page allocations for different cache allocation cases. In each case, there are four memory partitions and one task $\tau_i$. Each memory partition is depicted as a square and the shaded area represents the memory space allocated to $\tau_i$. The task $\tau_i$'s memory requirement $M_i$ is equal to the size of one memory partition. If we assign only one cache partition to $\tau_i$, all pages for $\tau_i$ are allocated from one memory partition (Case~1 in \figref{CACHE_mem_sharing_1}). If we assign more than one cache partition to $\tau_i$, our scheme allocates pages to $\tau_i$ from the corresponding memory partitions in round-robin order.\footnote{If a page is deallocated from $\tau_i$, the deallocated page is used ahead of never-allocated free pages to service $\tau_i$'s next page request. This enables multiple memory partitions to be allocated at the same rate without explicit enforcement such as in memory reservation \cite{Eswaran2005}.} Thus, the same amount of pages from each of the corresponding memory partitions is allocated to $\tau_i$ at its maximum memory usage (Cases~2, 3, and 4 in \figref{CACHE_mem_sharing_1}). The reason behind this approach is to render the page allocation deterministic, which is required for each task's cache access behavior to be consistent. For instance, if pages are allocated randomly, a task may have different cache performance when it re-launches.

A cache partition can be shared among tasks by sharing a memory partition. We present a necessary and sufficient condition for cache sharing to meet the task memory requirements under our page allocation approach. For each cache partition $\rho$, the following condition must be satisfied:
\begin{equation} \label{eq:CACHE_memory_condition}
\sum_{\forall\tau_{i}:\,\mathbb{S}_i\ni \rho} \frac{M_i}{|\mathbb{S}_i|}\le M_{total}/N_{cache}
\end{equation}
where $M_i$ is the size of the memory requirement of $\tau_i$, $|\mathbb{S}_i|$ is the number of cache partitions assigned to $\tau_i$, $M_{total}$ is the entire memory size, and $N_{cache}$ is the number of cache partitions. Hence. $M_{total}/N_{cache}$ represents the size of a single memory partition. $M_i\over|\mathbb{S}_i|$ represents $\tau_i$'s per-memory-partition memory usage. This condition means that the sum of the per-memory-partition usage of the tasks sharing the cache partition $\rho$ should not exceed the size of one memory partition. If this condition is not satisfied, tasks may experience memory pressure or swapping. 

\begin{algorithm}[t]
\caption[MinCacheAlloc($\Gamma_j$, $N'_{cache}$): finds a feasible cache allocation with the minimum CPU utilization]{MinCacheAlloc($\Gamma_j$, $N'_{cache}$)}
\label{CACHE_algo_cache_allocation}
{\scriptsize
\begin{algorithmic}[1]
\REQUIRE $\Gamma_j$: a taskset assigned to the core $j$, $N'_{cache}$: the number of available cache partitions
\ENSURE $\varphi_{min}$: a cache allocation with the minimum CPU utilization ($\varphi_{min} =\varnothing$, if no allocation is feasible), $minUtil$: the CPU utilization of $\Gamma_j$ with $\varphi_{min}$
\STATE $\varphi_{min} \gets \varnothing$
\STATE $minUtil \gets 1$
\STATE $\Phi \gets$ a set of candidate allocations of $N'_{cache}$ to $\Gamma_j$
\FOR {each allocation $\varphi_i$ in $\Phi$}
\STATE Apply $\varphi_i$ to $\Gamma_j$
\IF {$\Gamma_j$ satisfies both Eq. \eqref{eq:CACHE_response_time_test} and Eq. \eqref{eq:CACHE_memory_condition}}
	\STATE $currentUtil \gets$ CPU utilization from Eq. \eqref{eq:CACHE_utilization}
	\IF {$minUtil \ge currentUtil$}
		\STATE $\varphi_{min} \gets \varphi_i$
		\STATE $minUtil \gets currentUtil$
	\ENDIF
\ENDIF
\ENDFOR
\RETURN $\{\varphi_{min}, minUtil\}$
\end{algorithmic}
}
\end{algorithm}

\begin{algorithm}[t]
\caption[FindBestFit($\tau_i$, $N_P$, $A_\Gamma$, $A_{cache}$): finds the best-fit core for a given task in the presence of cache interference]{FindBestFit($\tau_i$, $N_P$, $A_\Gamma$, $A_{cache}$)}
\label{CACHE_algo_best_fit}
{\scriptsize
\begin{algorithmic}[1]
\REQUIRE $\tau_i$: a task to be allocated, $N_P$: the number of cores, $A_\Gamma$: an array of a taskset allocated to each core, $A_{cache}$: an array of the number of cache partitions assigned to each core
\ENSURE $cid$: the best-fit core's index ($cid = 0$, if no core can schedule $\tau_i$)
\STATE $space \gets 1$
\STATE $cid \gets 0$
\FOR {$j \gets 1$ \textbf{to} $N_P$}
	\STATE $\{\varphi, util\} \gets$ MinCacheAlloc($\tau_i \cup A_\Gamma[j]$, $A_{cache}[j]$)
	\IF {$\varphi \ne \varnothing$ and $space \ge 1 - util$} 
		\STATE $space \gets 1 - util$
		\STATE $cid \gets j$
	\ENDIF
\ENDFOR
\RETURN $cid$
\end{algorithmic}
}
\end{algorithm}

Algorithm~\ref{CACHE_algo_cache_allocation} shows a procedure for finding a feasible cache allocation with the minimum CPU utilization. It first creates a set of candidate cache allocations to be examined, which are combinations of given cache partitions for a given taskset. Then, it checks the feasibility of each candidate allocation by using Eqs.~\eqref{eq:CACHE_response_time_test} and \eqref{eq:CACHE_memory_condition}. Many methods can be used to generate the candidate cache allocations, such as exhaustive search and heuristics. We use an exhaustive search in the evaluation of this chapter. A heuristic approach will be introduced in Section~\ref{VCACHE_allocating_cache_colors_to_tasks}.

\subsection{Cache-Aware Task Allocation}
  \label{CACHE_task_allocation}

Cache-aware task allocation is an algorithm to allocate tasks and cache partitions to cores while exploiting the benefits of cache reservation and cache sharing. The objective of our algorithm is to reduce the number of cache partitions required to schedule a given taskset on a given number of cores, because remaining cache partitions can be used for many purposes, such as for non-real-time tasks or for saving the CPU utilization. 

Under our scheme, tasks may share cache partitions when they are assigned to the same core. This means that, to take advantage of cache sharing, it is desired to pack tasks into the same core as much as possible. Hence, our algorithm is based on the best-fit decreasing bin-packing algorithm that results in load concentration. For cache allocation, our algorithm gradually assigns cache partitions to cores while allocating tasks to cores by using cache reservation and cache sharing. 

We first explain Algorithm~\ref{CACHE_algo_best_fit} that finds the best-fit core in our task allocation algorithm. Once the task to be allocated is given, Algorithm~\ref{CACHE_algo_best_fit} checks whether the task is schedulable on each core and estimates the total utilization of each core with the task. Then, it selects the core where the task fits best.

\begin{algorithm}[t]
\caption[CacheAwareTaskAlloc($\Gamma$, $N_P$, $N_{cache}$): allocates tasks and cache partitions to cores]{CacheAwareTaskAlloc($\Gamma$, $N_P$, $N_{cache}$)}
\label{CACHE_algo_task_allocation}
{\scriptsize
\begin{algorithmic}[1]
\REQUIRE $\Gamma$: a taskset to be allocated, $N_P$: the number of cores, $N_{cache}$: the number of available cache partitions
\ENSURE True/False: the schedulability of $\Gamma$, $A_\Gamma$: an array of a taskset allocated to each core, $A_{cache}$: an array of the number of cache partitions assigned to each core, $N_{P'}$: the number of remaining cache partitions
\STATE Sort tasks in $\Gamma$ in decreasing order of their average utilization
\STATE Initialize elements of $A_\Gamma$ to $\varnothing$ and $A_{cache}$ to 0 
\FOR {each task $\tau_i$ in $\Gamma$}
	\STATE $cid \gets$ FindBestFit($\tau_i$, $N_P$, $A_\Gamma$, $A_{cache}$)
	\IF {$cid > 0$}
		\STATE /* Found the core for $\tau_i$ */
		\STATE Insert $\tau_i$ to $A_\Gamma[cid]$
		\STATE Mark $\tau_i$ schedulable
		\STATE \textbf{continue}
	\ENDIF
	\STATE /* Try with $k$ more partitions */
	\FOR {$k \gets 1$ \textbf{to} $N_{cache}$ }
		\FOR {$j \gets 1$ \textbf{to} $N_P$}
			\STATE $A_{tmp}[j] \gets A_{cache}[j] + k$
		\ENDFOR
		\STATE $cid \gets$ FindBestFit($\tau_i$, $N_P$, $A_\Gamma$, $A_{tmp}$)
		\IF {$cid > 0$}
			\STATE Insert $\tau_i$ to $A_\Gamma[cid]$
			\STATE Mark $\tau_i$ schedulable
			\STATE $N_{cache} \gets N_{cache} - k$ /* Assign $k$ to the core */
			\STATE $A_{cache}[cid] \gets A_{cache}[cid] + k$
			\STATE \textbf{break}
		\ENDIF
	\ENDFOR
\ENDFOR
\IF {all tasks schedulable}
	\RETURN $\{$True$, A_\Gamma, A_{cache}, N_{cache}\}$
\ELSE
	\RETURN $\{$False$, A_\Gamma, A_{cache}, N_{cache}\}$
\ENDIF
\end{algorithmic}
}
\end{algorithm}

Our cache-aware task allocation algorithm is given in Algorithm~\ref{CACHE_algo_task_allocation}. Before allocating tasks, it sorts tasks in decreasing order of their average utilization, i.e. $(\sum_{k=1}^{N_{cache}} C_i(k)/T_i)/N_{cache}$. The number of cache partitions for each core is set to zero. Then, the algorithm initiates task allocation. If a task to be allocated is not schedulable on any core and the number of remaining cache partitions is not zero, the algorithm increases the number of each core's cache partitions by 1 and finds the best-fit core again, until the cache partition increment per core exceeds $N_{cache}$. When the algorithm finds the best-fit core, only the best-fit core maintains its increased number of cache partitions and other cores return to their previous number of cache partitions.

The algorithm returns the number of remaining cache partitions along with the task allocation and cache assignment. Here, we employ a simple solution to save the CPU utilization with the remaining cache partitions: assigning each remaining cache partition to a core which will obtain the greatest saving in utilization when an additional cache partition is given to it. We use this approach in our experiments when we measure the CPU utilization with a specified number of cache partitions.

\subsection{Tasks with Shared Memory Regions}
\label{shared_mem}
Like previous work on cache-aware response time analysis \cite{Altmeyer_RTSS11,CGLee_IEEE01} and software cache partitioning schemes \cite{Liedtke_RTAS97,Lin_HPCA08,Zhang_EuroSys09}, we have so far assumed that tasks do not use share memory regions. 
However, recent operating systems widely use shared pages, not only for inter-process communication and shared libraries, but also the kernel's copy-on-write technique and file caches~\cite{Kim_RTCSA12}. 
Suppose that two tasks share a memory region and they are allocated to different cores. 
Then, the tasks may experience inter-core cache interference because the shared memory region causes the sharing of cache partitions among those tasks. 

We suggest one simple but effective strategy for this problem. Tasks that share their memory regions are bundled together and each task bundle is allocated together as a single task into a core. Hence, the tasks in the same bundle will not cause inter-core cache interference to each other as well as to the other tasks on other cores. This strategy can be integrated into our cache-aware task allocation and be performed before allocating tasks with no shared memory regions. 

If a task bundle cannot be allocated to a single core and the shared memory regions are read/writable data regions, we can assign exclusive cache partitions to each of the shared data regions. Since the shared data regions are typically protected by mutually-exclusive locks (mutexes) to avoid race conditions, only one task in the bundle accesses each data region, which prevents inter-core cache interference among tasks in the bundle. Of course, other tasks will not experience any cache interference from the use of the shared data regions because those regions are assigned exclusive cache partitions.
If the shared memory regions are read-only regions and not protected by mutexes, such as shared libraries and file caches, the system may duplicate the pages of the memory regions for each core and assign exclusive cache partitions to the duplicated regions to avoid inter-core cache interference.

\section{Evaluation}
\label{CACHE_evaluation}
In this section, we evaluate our proposed cache management scheme. We first describe the implementation of our scheme and then show the experimental results of cache reservation, cache sharing, and cache-aware task allocation.

\subsection{Implementation}
We have implemented our scheme in Linux/RK, based on the Linux 2.6.38.8 kernel. To easily implement page coloring, we have used the memory reservation mechanism \cite{Eswaran2005, Kim_RTCSA12} of Linux/RK. Memory reservation maintains a global page pool to manage unallocated physical pages. In this page pool, we categorize pages into memory partitions with their color indices. 
When a real-time taskset is given, our scheme assigns a core index and color indices to each task. Then, a memory reservation is created for each task from the page pool, using the task's memory demand and assigned color indices, and each task only uses pages within its memory reservation during execution.

\begin{figure}[t]
\centering
\vspace{-15pt}
  \includegraphics[width=0.6\textwidth]{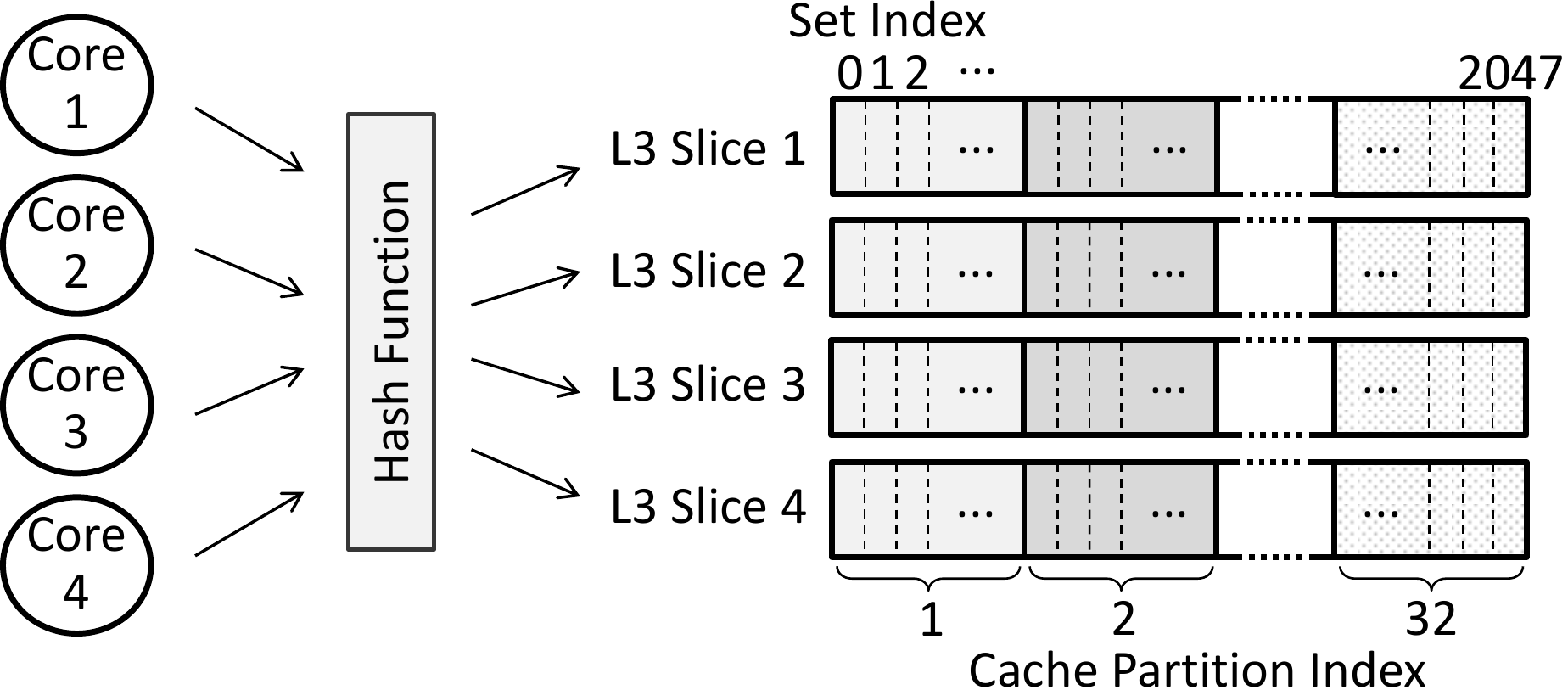}\\
\vspace{-5pt}
  \caption{Page coloring on the Intel i7-2600 L3 cache}\label{fig:CACHE_coloring_implementation}
\vspace{-5pt}
\end{figure}

The target system is equipped with the Intel Core i7-2600 3.4GHz quad-core processor. The system is configured for 4KB page frames and a 1GB memory reservation page pool. The processor has a unified 8MB L3 shared cache that consists of four cache slices. Each cache slice has 2MB and is 16-way set associative with a line size of 64B, thereby having 2048 sets. For the entire L3 cache to be shared among all cores, the processor distributes all physical addresses across the four cache slices by using an on-chip hash function \cite{IntelDevDoc, Intel_HC23}.\footnote{Intel refers to this technique, which is unrelated to cache partitioning, as a {\em Smart Cache}. The details on the hash function are proprietary in nature.} \figref{CACHE_coloring_implementation} shows the implementation of page coloring on this cache configuration. 
Regardless of the hash function, the cache set index for a given physical address is independent from the cache slice index. Hence, with page coloring, we can use 32 cache partitions and each cache partition spans the four cache slices. Page coloring divides the L3 cache into 32 cache partitions of 256KB and the page pool into 32 memory partitions of 32MB. The cache partition refill time $\Delta$ in the target system is 45.3 $\mu$sec,\footnote{The cache partition refill time is the time to fetch from memory to the L3 cache. Thus, it is hardly affected by the fact that the Intel i7's core-to-L3 access time varies from 26 to 31 cycles. Our WCET estimation covers such L3 access variations.} which is an empirically obtained from a cache calibration tool, as given in \cite{CacheCalibrator}.

\subsection{Taskset}
\label{CACHE_exp_taskset}
\tableref{CACHE_task_desc} shows four periodic tasks that we have created for the evaluation.  The task functions are selected from the PARSEC benchmark suite \cite{PARSEC} to create a taskset consisting of cache-sensitive and cache-insensitive tasks. We utilize them as representative components of complex real-time embedded applications such as sensor fusion and computer vision in an autonomous vehicle \cite{Kim_ICCPS13}. Each task has a relative deadline $D_i$ equal to its period $T_i$ and a memory requirement $M_i$. Due to the memory co-partitioning of page coloring, $M_i$ determines the minimum required number of cache partitions $k$ for a task $\tau_i$, given by $\lceil\frac{M_i}{M_{total}/N_{cache}}\rceil \le k$. Task priorities are assigned by the deadline-monotonic scheduling policy.

\begin{figure*}[t]
\centering
  \subfloat[$\tau_1$: \texttt{p\_streamcluster}] {\label{fig:CACHE_wcet_t1}
  \includegraphics[width=0.35\textwidth]{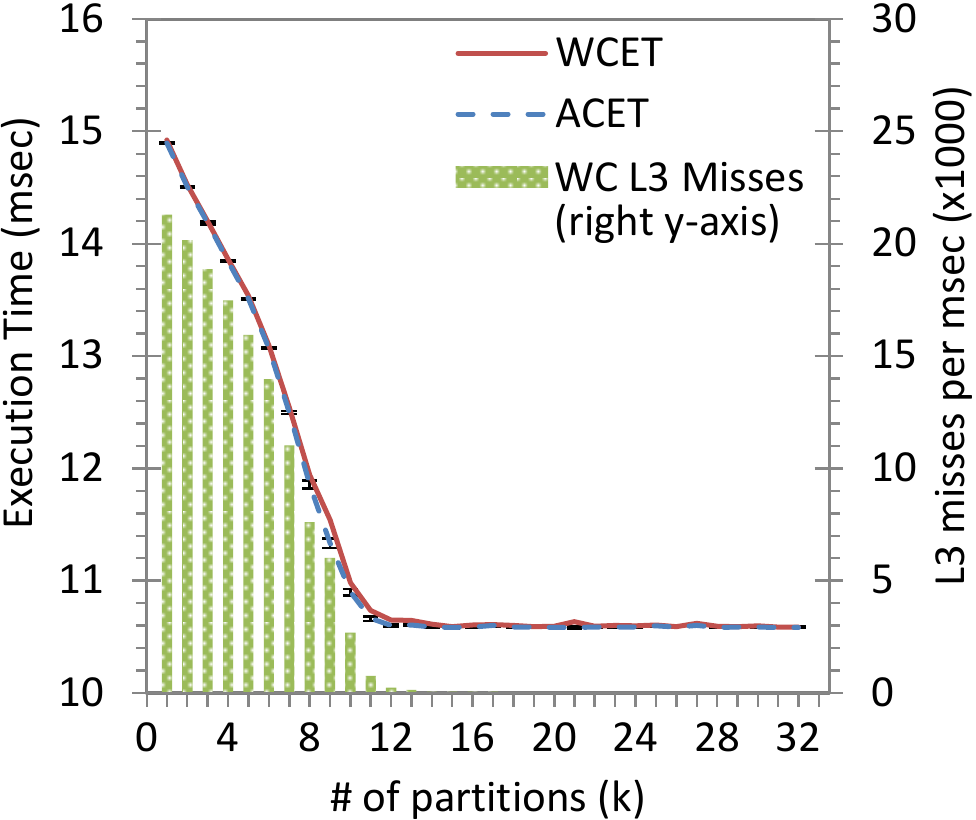}
  }
  \subfloat[$\tau_2$: \texttt{p\_ferret}] {\label{fig:CACHE_wcet_t2}
  \includegraphics[width=0.35\textwidth]{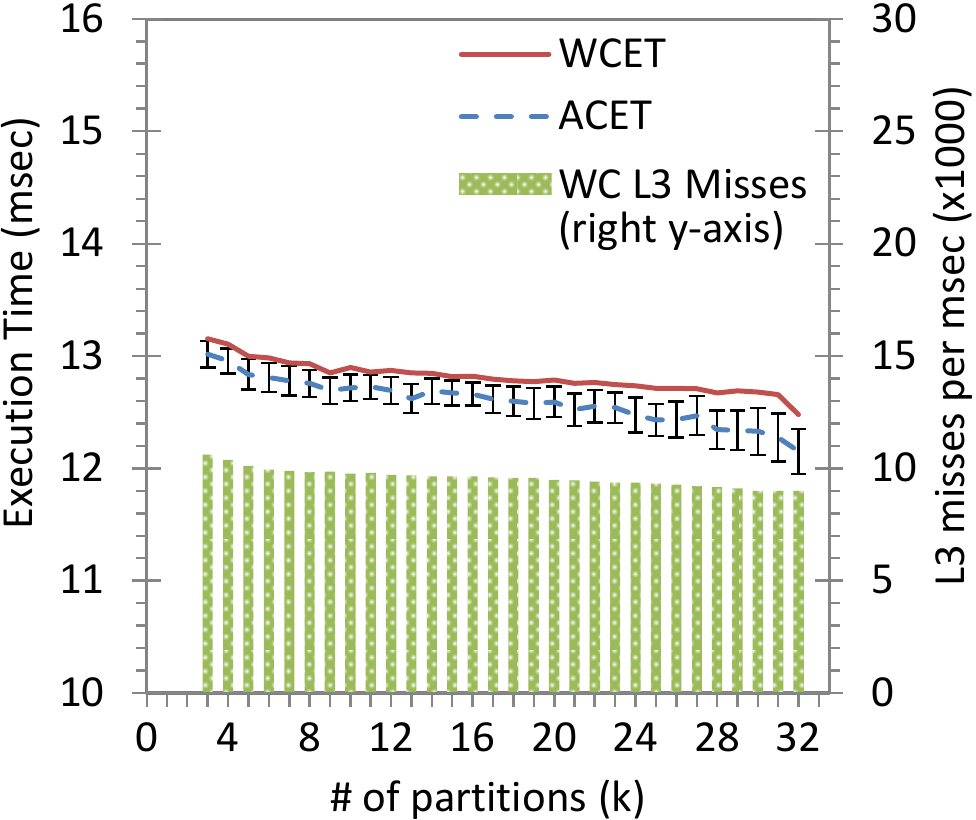}
  }\\
  \subfloat[$\tau_3$: \texttt{p\_canneal}] {\label{fig:CACHE_wcet_t3}
  \includegraphics[width=0.35\textwidth]{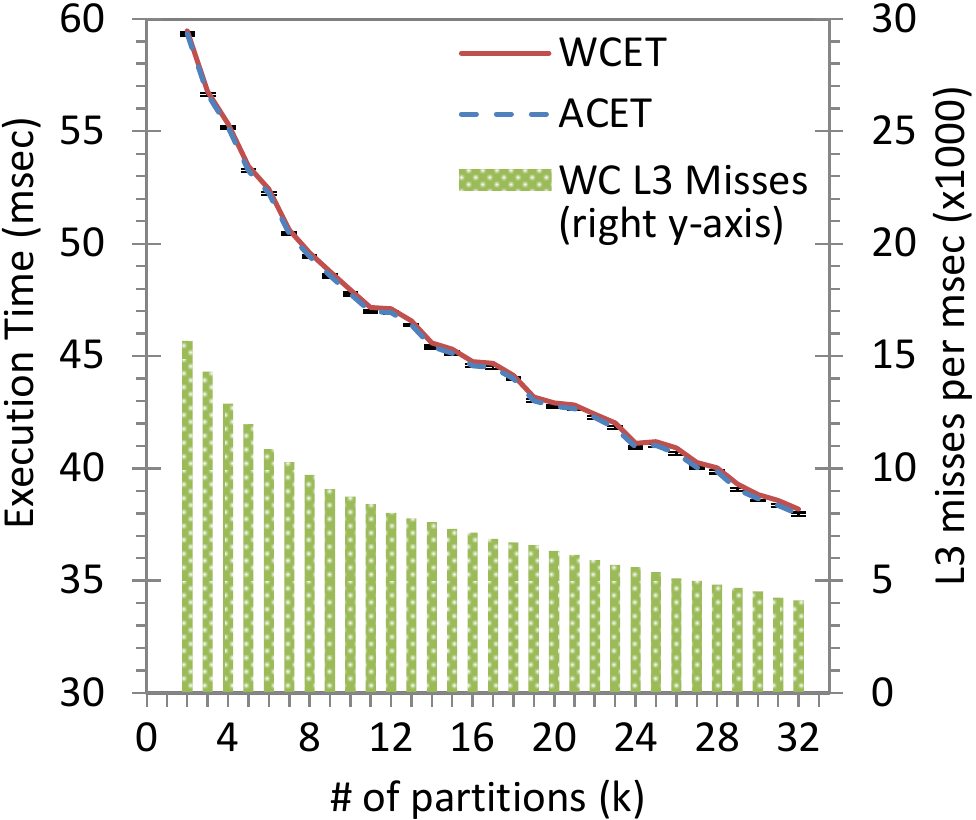}
  }
  \subfloat[$\tau_4$: \texttt{p\_fluidanimate}] {\label{fig:CACHE_wcet_t4}
  \includegraphics[width=0.35\textwidth]{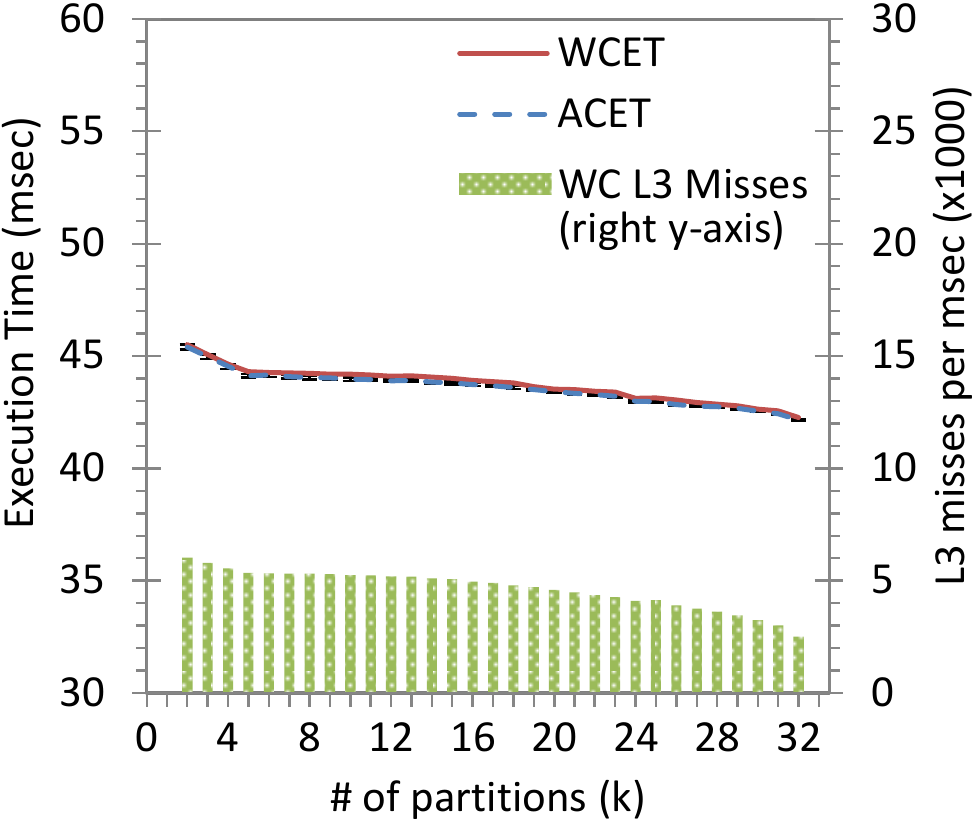}
  }
\caption{Execution time and L3 misses of each task as the number of cache partitions increases when running alone in the system}\label{fig:CACHE_wcet_task}
\end{figure*}

For the WCET analysis, we used the measurement-based approach. To reduce inaccuracies in measurement, we disabled the processor's simultaneous multithreading and dynamic clock frequency scaling. All unrelated system services such as GUI and networking were also disabled during the experiments. We used the processor's hardware performance counters to measure the task execution time and the L3 misses, when each of the tasks were running alone in the system. Then, we chose the maximum observed execution time and the maximum observed L3 misses as the WCET estimate and the worst-case L3 misses, respectively. \figref{CACHE_wcet_task} shows each task's per-period execution time as the number of assigned cache partitions increases. In each sub-figure, the WCET and the average-case execution time (ACET) are plotted as a solid line and a dotted line, respectively. The worst-case L3 misses per period are presented as a bar graph with the scale on the right y-axis. 

The taskset used in our evaluation is a mixture of cache-sensitive and cache-insensitive tasks.
We can confirm this from \figref{CACHE_wcet_task}. $\tau_1$ and $\tau_3$ are cache-sensitive tasks. The $\tau_1$'s WCET $C_1(k)$ drastically decreases as the number of cache partitions $k$ increases, until $k$ exceeds 12. The number of $\tau_1$'s L3 misses also decreases as $k$ increases. $\tau_3$'s WCET $C_3(k)$ continuously decreases as $k$ increases. In terms of utilization, the difference between the maximum and the minimum utilization of $\tau_1$ is $(C_1(32) - C_1(1))/T_1 = 10.82 \%$. The utilization difference of $\tau_3$ is $11.83 \%$. On the other hand, $\tau_2$ and $\tau_4$ are cache-insensitive. The utilization differences of $\tau_2$ and $\tau_4$ are merely $0.56 \%$ and $0.54 \%$, respectively.  

\begin{table}[t]
\centering
{
\vspace{-15pt}
\footnotesize
\caption[Task parameters for cache management experiments]{Taskset information}\label{tab:CACHE_task_desc}
\vspace{-5pt}
\begin{tabularx}{.95\textwidth}{@{\hspace{0.5em}}c @{\hspace{1.2em}}c @{\hspace{1.2em}}c @{\hspace{1.2em}}c @{\hspace{1.2em}}c @{\hspace{1.2em}}X@{\hspace{0.5em}}} 
\hline
Task		&$T_i$=$D_i$	&$M_i$	&Min.	&Cache	&\multirow{2}{*}{Name and description}\\
$\tau_i$	&(msec)			&(MB)	&$k$	&Sensitive	&\\
\hline
$\tau_1$	&40 	&18	&1	&Yes	&{\tt p\_streamcluster}: computes clustering of data points\\
\hline
$\tau_2$	&120 	&66	&3	&No		&{\tt p\_ferret}: image-based similarity search engine\\
\hline
$\tau_3$	&180 	&52	&2	&Yes	&{\tt p\_canneal}: graph restructuring for low routing cost\\
\hline
$\tau_4$	&600 	&50	&2	&No		&{\tt p\_fluidanimate}: simulates fluid motion for animations\\
\hline
\end{tabularx}
}
\end{table}

\begin{figure*}[]
\centering
  \subfloat[$\tau_1$: \texttt{p\_streamcluster}] {\label{fig:CACHE_cache_rsv_1}
  \includegraphics[width=0.49\textwidth]{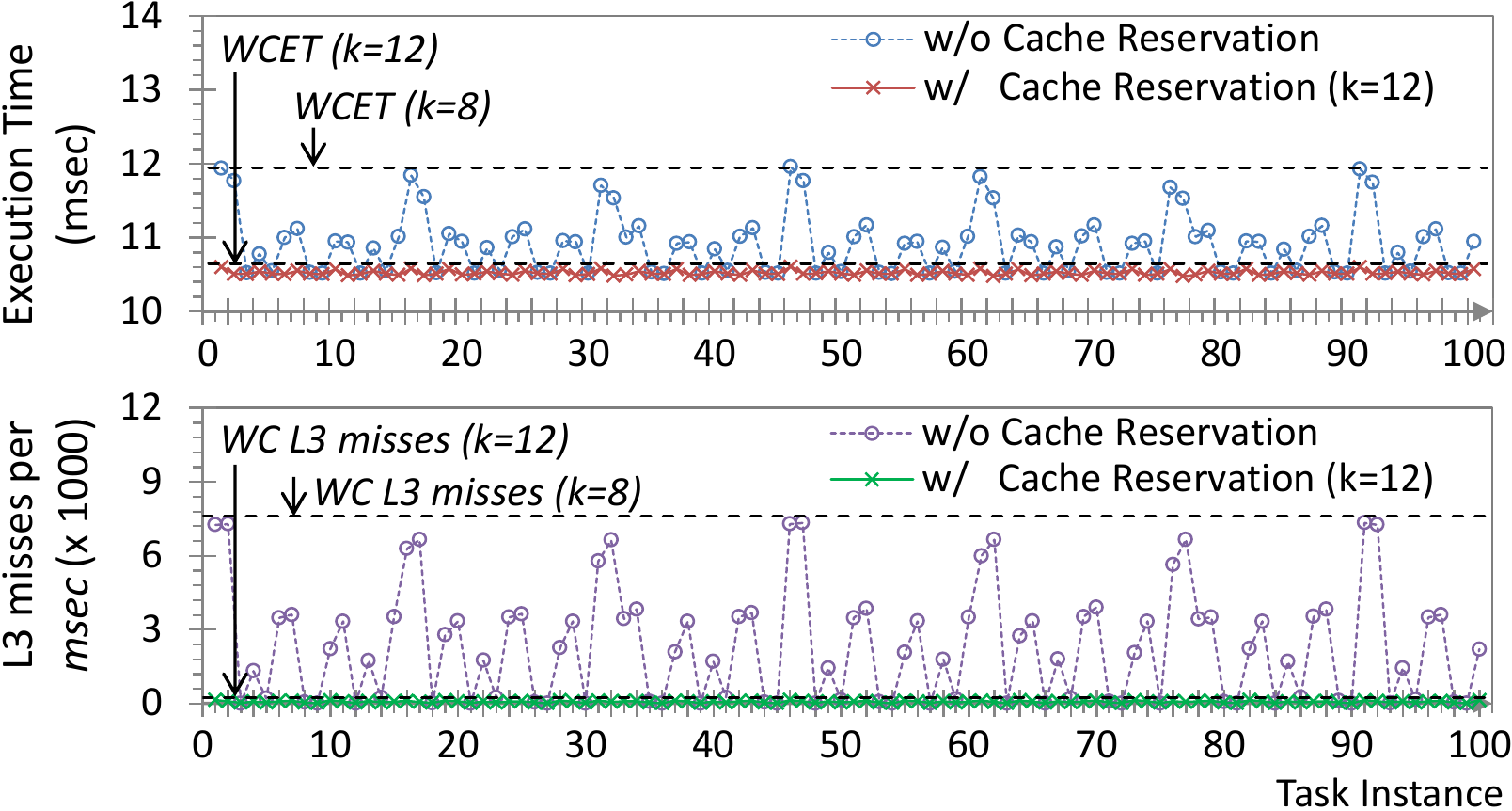}
  }
  \subfloat[$\tau_2$: \texttt{p\_ferret}] {\label{fig:CACHE_cache_rsv_2}
  \includegraphics[width=0.49\textwidth]{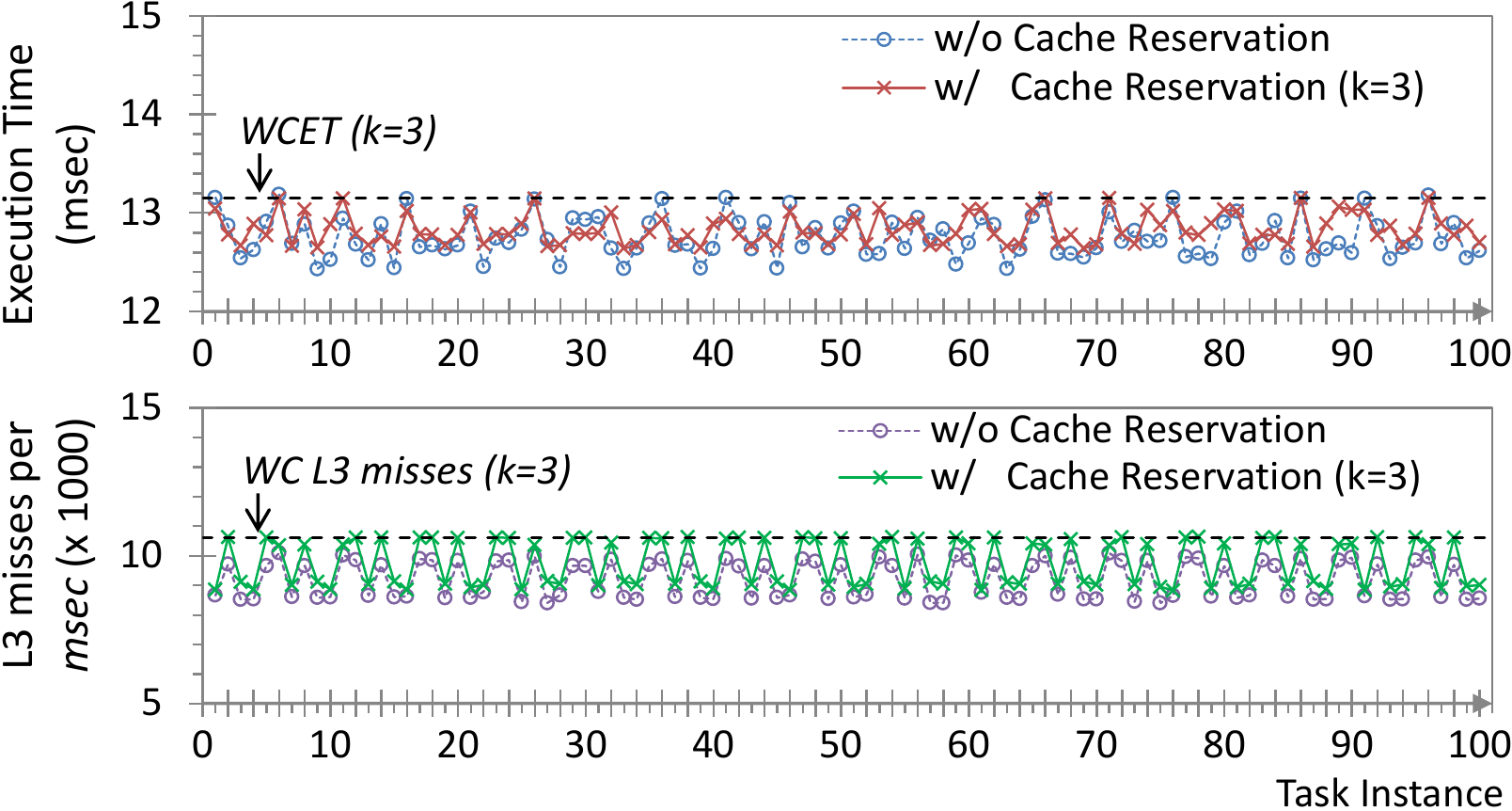}
  }\\
  \subfloat[$\tau_3$: \texttt{p\_canneal}] {\label{fig:CACHE_cache_rsv_3}
  \includegraphics[width=0.49\textwidth]{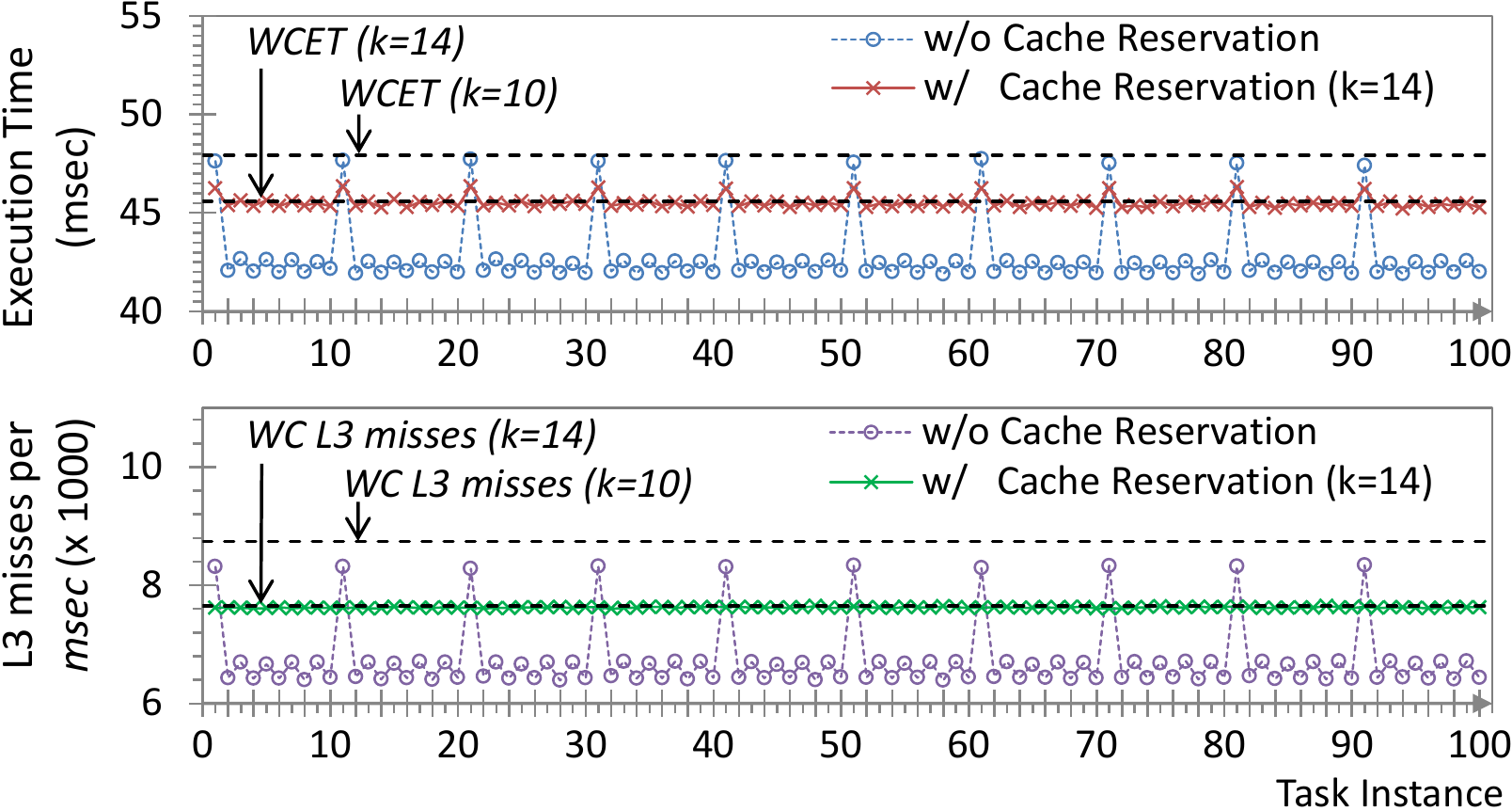}
  }
  \subfloat[$\tau_4$: \texttt{p\_fluidanimate}] {\label{fig:CACHE_cache_rsv_4}
  \includegraphics[width=0.49\textwidth]{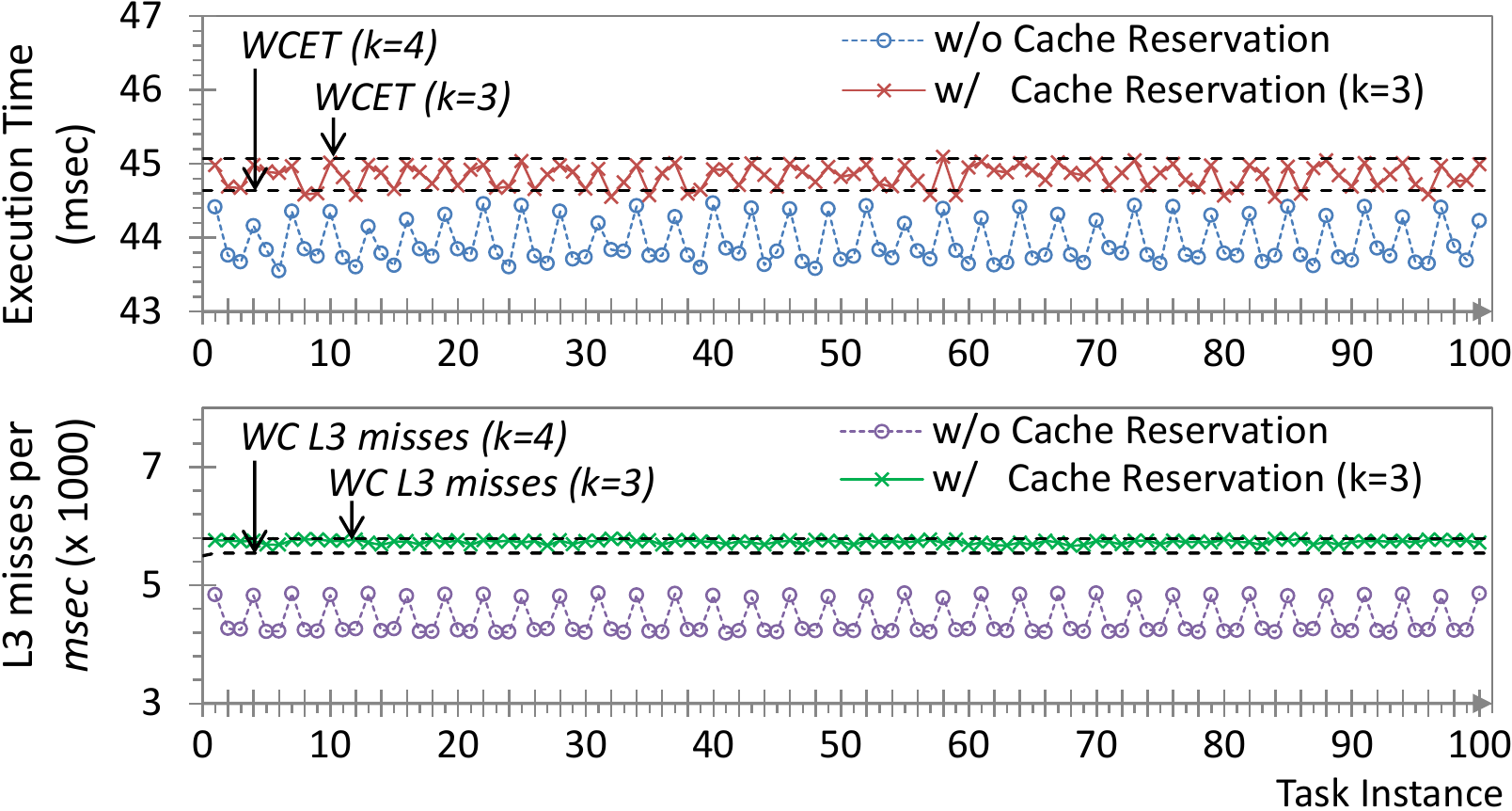}
  }
\caption{Observed execution time and L3 misses of tasks when each task runs simultaneously on different cores}\label{fig:CACHE_cache_rsv_exp}\vspace{-10pt}
\end{figure*}

\subsection{Cache Reservation}
The purpose of this experiment is to verify how effective cache reservation is in avoiding inter-core cache interference. We ran each task on different cores simultaneously, i.e. $\tau_i$ on Core $i$, under two cases: with and without cache reservation. Memory reservation was used in both cases. Without cache reservation, all tasks competitively used the entire cache space. With cache reservation, the number of cache partitions for each core was as follows: 12 partitions for Core 1, 3 for Core 2, 14 for Core 3, and 3 for Core 4. These numbers are determined to reduce the total CPU utilization by the observation of \figref{CACHE_wcet_task}. The cache partitions assigned to each core were solely used by the task on that core. 

\figref{CACHE_cache_rsv_exp} presents the observed execution time and the L3 misses of four tasks with and without cache reservation, when they ran simultaneously on different cores. In each sub-figure, the upper graph shows the execution time of each task instance and the lower graph shows the number of L3 misses for each instance. The x-axis on each graph indicates the instance numbers of a task. Tasks are released at the same instance using \texttt{hrtimers} in Linux.

The execution times of all tasks without cache reservation vary significantly compared to the execution times with cache reservation. Without cache reservation, tasks compete for the L3 cache and higher worst-case L3 misses are encountered. The correlation between execution time and L3 misses is clearly shown in \figref{CACHE_cache_rsv_1} and \figref{CACHE_cache_rsv_3}. The average execution time of tasks without cache reservation may not be much higher. However, the absence of cache reservation contributes to poor timing predictability. The longest execution time of $\tau_1$ without cache reservation is close to its WCET with 8 dedicated cache partitions ($C_1(8)$), that of $\tau_2$ is close to $C_2(3)$, that of $\tau_3$ is close to $C_3(10)$, and that of $\tau_4$ is close to $C_4(4)$. Note that, without cache reservation, the longest execution times cannot be obtained before profiling the whole taskset. Hence, the profiling may need to be re-conducted whenever a single parameter of the taskset changes. In addition, without cache reservation, the cache is not effectively utilized. The total number of cache partitions for the above longest execution times is $8+3+10+4=25$. This means that 7 partitions are wasted in terms of WCET. 

With cache reservation, the execution times of $\tau_1$, $\tau_2$, and $\tau_4$ do not exceed their WCETs that are estimated in isolation from other tasks. $\tau_3$ also does not exceed its WCET except at the beginning of each hyper-period of 1800 msec. $\tau_3$ exceeds its WCET by less than $2\%$ once in a hyper-period. However, this is not caused by inter-core cache interference. As shown in \figref{CACHE_cache_rsv_3}, the L3 misses of $\tau_3$ instances are always lower than its worst-case L3 misses even at the beginning of each hyper-period, meaning that cache reservation successfully avoids inter-core cache interference. Since all task instances start their execution concurrently at the beginning of each hyper-period, we strongly suspect that the observed execution time slightly greater than the WCET is caused by other shared memory resources in the system, such as a memory controller and memory buses. We will address this issue in Chapter~\ref{chapter_bounding_and_reducing_memory_interference}.

\subsection{Cache Sharing}

\begin{table}[t]
\centering
{
\footnotesize
\caption[Cache-to-task allocation for cache management experiments]{Cache allocation to tasks with cache sharing}\label{tab:CACHE_cache_alloc_tasks}
\begin{tabularx}{.7\textwidth}{c c c >{\centering\arraybackslash}m{1.8cm} c}
\hline
\multirow{2}{*}{$\tau_i$}	&Allocated cache  				&\multicolumn{1}{c}{WCET}
	&\multicolumn{2}{c}{Worst-Case Response-Time}\\\cline{4-5}
				&partitions $\mathbb{S}_i$						&\multicolumn{1}{c}{(msec)}
	&NoCInt				&CInt\\
\hline
\T\B$\tau_1$			&{\scriptsize \{1, 2, 3, 4, 5, 6, 7, 8\}}	&$C_1(8)$ = 11.94
	&11.94			&12.30\\
\hline
\T\B$\tau_2$			&{\scriptsize \{1, 2, 3\}}\hspace{47px}		&$C_2(3)$ = 13.15
	&25.09			&25.72\\
\hline
\T\B$\tau_3$			&{\scriptsize \{1, 2, 3, 4, 5, 6, 7, 8\}}	&$C_3(8)$ = 49.58
	&98.55			&101.36\\
\hline
\T\B$\tau_4$			&\hspace{29px}{\scriptsize \{4, 5, 6, 7, 8\}}	&$C_4(5)$ = 44.30
	&179.88			&273.78\\
\hline
\end{tabularx}
}
\end{table}

\begin{figure}[t]
\centering
  \subfloat[$\tau_1$: p\_streamcluster] {\label{fig:CACHE_exp_intra_rt_t1}
  \includegraphics[width=.48\textwidth]{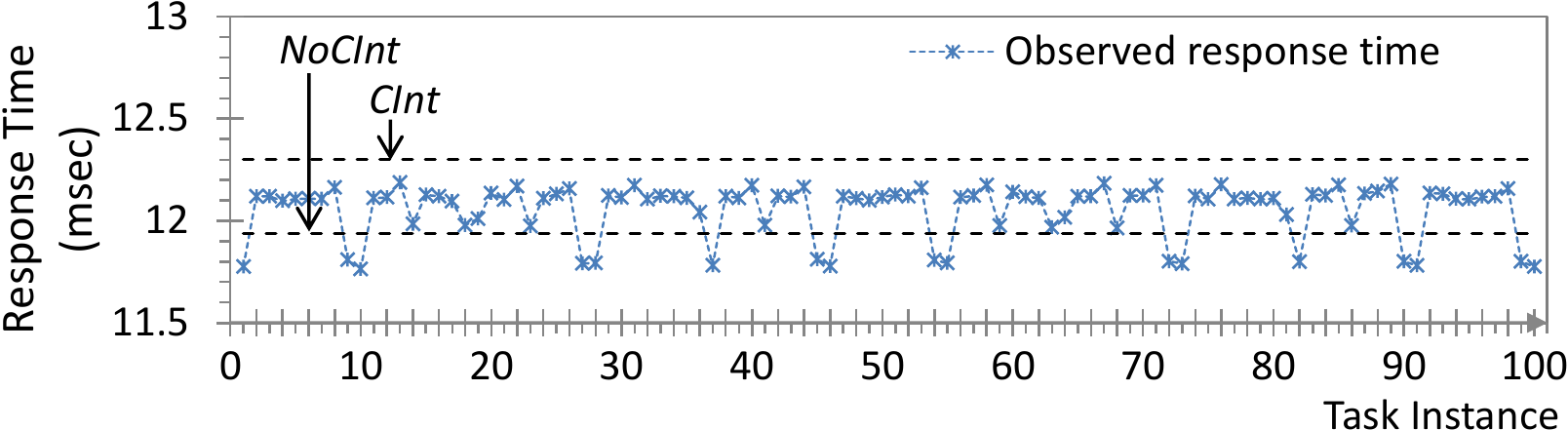}
  }
  \subfloat[$\tau_2$: p\_ferret] {\label{fig:CACHE_exp_intra_rt_t2}
  \includegraphics[width=.48\textwidth]{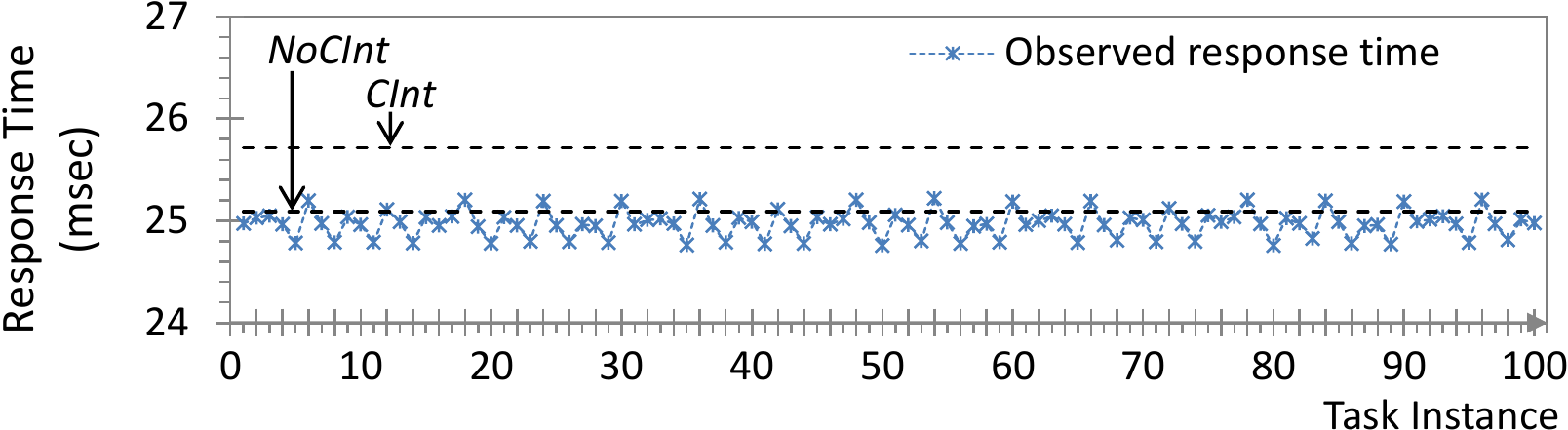}
  }\\
  \subfloat[$\tau_3$: p\_canneal] {\label{fig:CACHE_exp_intra_rt_t3}
  \includegraphics[width=.48\textwidth]{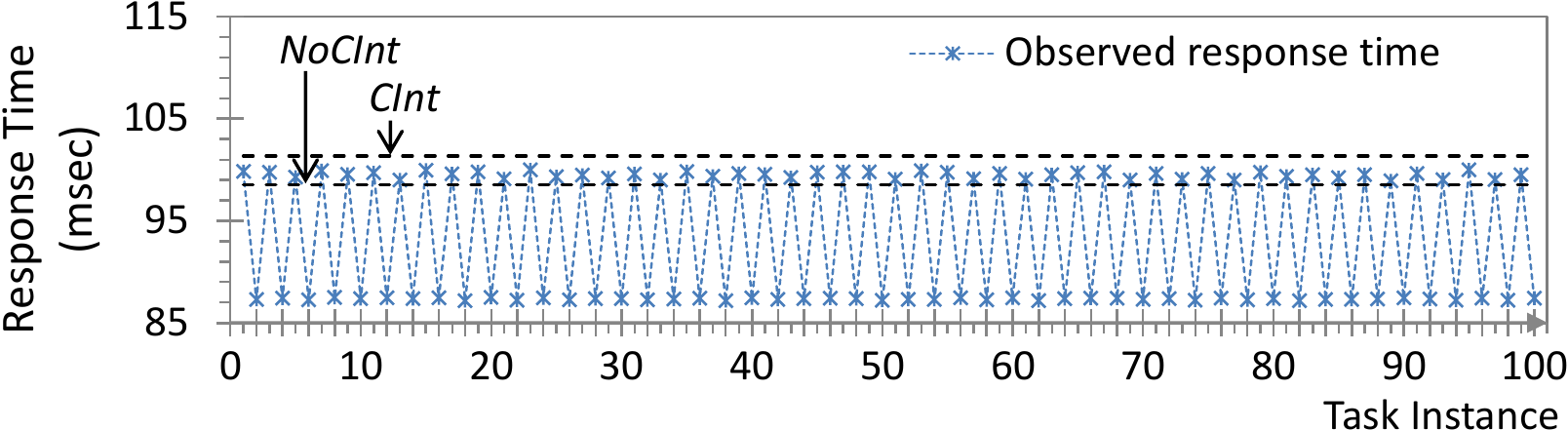}
  }
  \subfloat[$\tau_4$: p\_fluidanimate] {\label{fig:CACHE_exp_intra_rt_t4}
  \includegraphics[width=.48\textwidth]{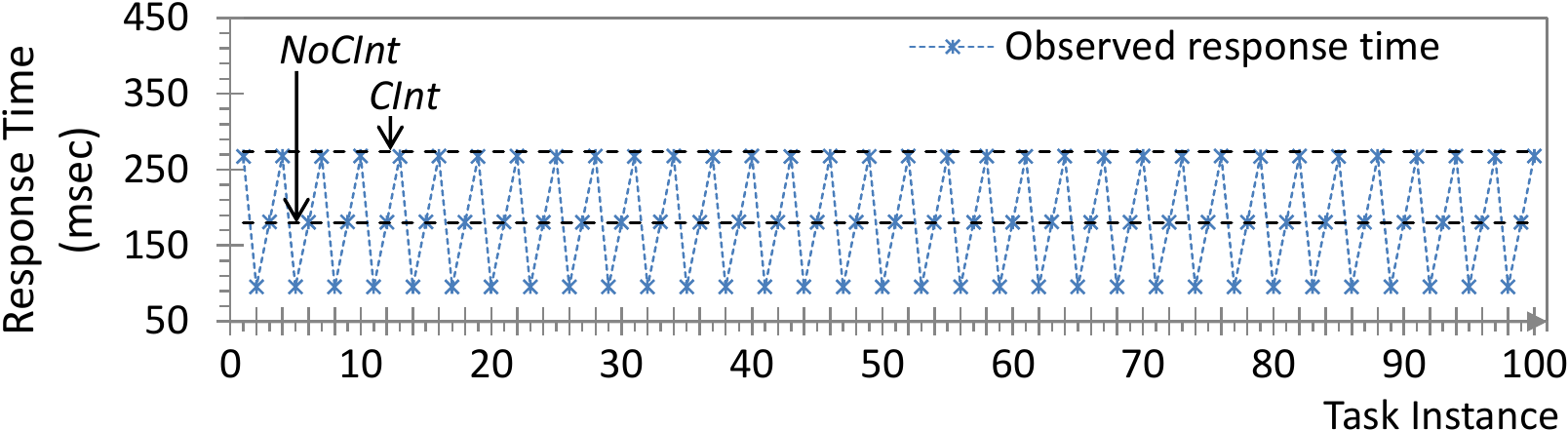}
  }
\caption{Response time of tasks with cache sharing on a single core}\label{fig:CACHE_exp_intra_rt}
\end{figure}

We first evaluate the effectiveness of our proposed equations in predicting the worst-case response time (WCRT) of a task with cache sharing. In this experiment, all tasks run on a single core with 8 cache partitions. \tableref{CACHE_cache_alloc_tasks} shows the cache partition allocations to the tasks by the cache-sharing technique and the predicted WCRT of the tasks. The WCRT is calculated with two methods: ``{\em NoCInt}'' means intra-core cache interference is not taken into account, and ``{\em CInt}'' means the WCRT is calculated by Eq.~\eqref{eq:CACHE_response_time_test}. 

\begin{figure}[t]
\centering
  \includegraphics[width=.7\textwidth]{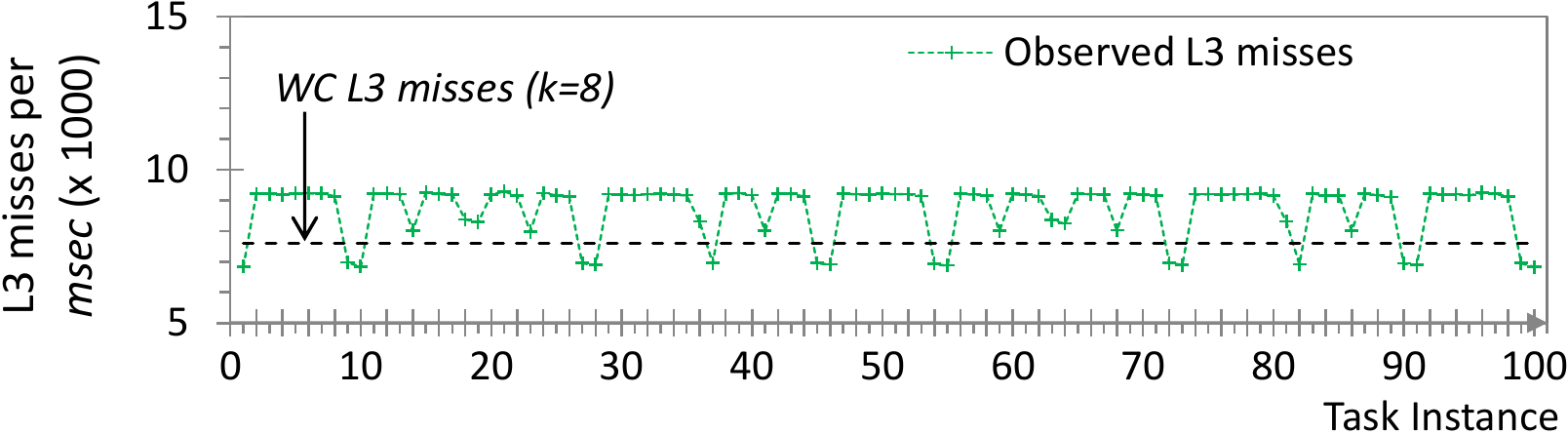}\\
  \caption{L3 misses of task $\tau_1$ with cache sharing}\label{fig:CACHE_exp_intra_l3miss_t1}
\end{figure}

\figref{CACHE_exp_intra_rt} illustrates the observed response time of each task. The WCRT values with {\em NoCInt} and {\em CInt} are depicted as straight lines in each graph. In all tasks, the observed response time exceeds the WCRT with {\em NoCInt}, but does not exceed the WCRT with {\em CInt}. For $\tau_1$, the observed response time greater than the WCRT with {\em NoCInt} is solely caused by the cache warm-up delay, because $\tau_1$ has the highest priority task and does not experience any cache-related preemption delay. \figref{CACHE_exp_intra_l3miss_t1} supports this observation. It shows the observed L3 misses of $\tau_1$'s instances. Since $\tau_1$ shares its cache partitions with other tasks, the observed L3 misses are higher than the worst-case L3 miss value that is estimated when $\tau_1$ does not share cache partitions. The correlation between $\tau_1$'s observed response time and observed L3 misses is also clearly shown. Hence, we can identify that $\tau_1$'s observed response time greater than the WCRT with {\em NoCInt} is caused by increased L3 cache misses, rather than jitter. This result shows the effect of our response time test that explicitly considers the cache warm-up delay. Task $\tau_4$ shows a significant 93.9 msec difference between {\em NoCInt} and {\em CInt}. Since the WCRT with {\em NoCInt} is close to the period of $\tau_3$, timing penalties from intra-core cache interference make the response time exceed the period of $\tau_3$. Then, the next instance of $\tau_3$ preempts $\tau_4$, thereby increasing the response time of $\tau_4$ significantly. 

\begin{figure}[t]
\centering
  \includegraphics[width=.7\textwidth]{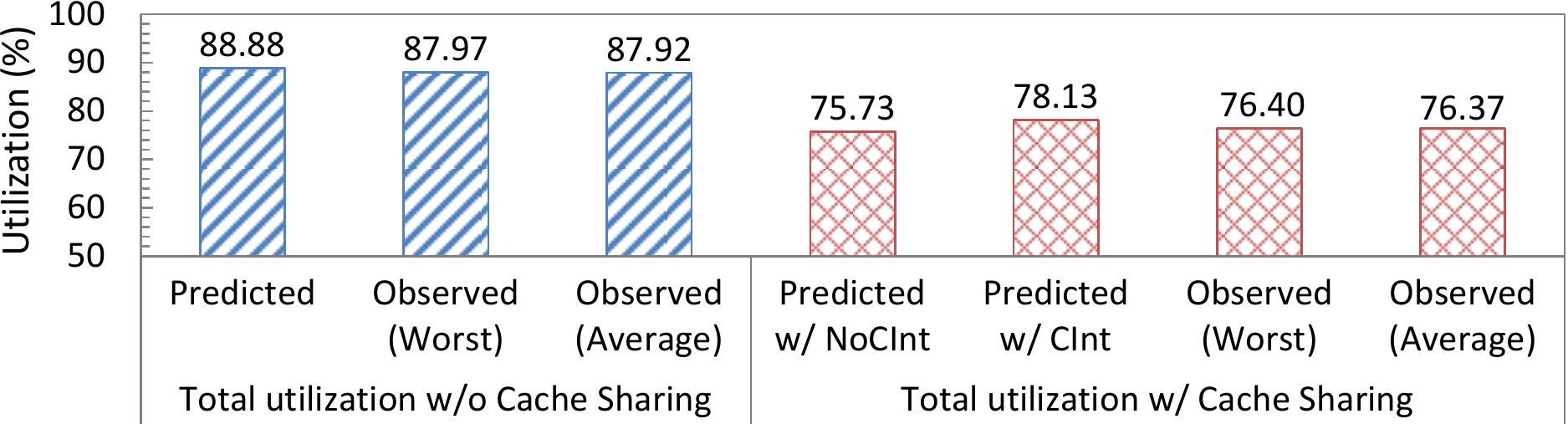}\\
  \caption{Total CPU utilization with and without cache sharing}\label{fig:CACHE_exp_intra_util}
\end{figure}

Secondly, we identify the utilization benefit of the cache-sharing technique by comparing the total CPU utilization with and without cache sharing. Without cache sharing, cache allocations are as follows: $\tau_1$ is assigned 1~partition, $\tau_2$ is assigned 3~partitions, $\tau_3$ is assigned 2~partitions, and $\tau_4$ is assigned 2~partitions. Note that this is the only possible cache allocation without cache sharing because the number of available cache partitions is eight, which is equal to the sum of each task's minimum cache requirement. With cache sharing, the same cache allocations as in the \tableref{CACHE_cache_alloc_tasks} are used. \figref{CACHE_exp_intra_util} depicts the total CPU utilization with and without cache sharing. The left three bars are the predicted and the observed values without cache sharing and the right four bars are the values with cache sharing. The utilization values with cache sharing are almost $10\%$ lower than the values without cache sharing. This result shows that cache sharing is very beneficial for saving the CPU utilization. Furthermore, with cache sharing, both the worst-case and the average-case observed utilization are higher than the predicted utilization with {\em NoCInt} but lower than the predicted value with {\em CInt}. This implies that Eq.~\eqref{eq:CACHE_utilization} provides a safe upper bound on the total utilization with intra-core cache interference. 

\subsection{Cache-Aware Task Allocation}
We now evaluate the effectiveness of our cache-aware task allocation (CATA) algorithm that exploits cache reservation and cache sharing. Note that it is not appropriate to compare CATA against previous approaches such as in \cite{Berna_RTNS12, Paolieri_RTAS11}, since (i) they do not consider the task memory requirements, which is essential to prevent page swapping when page coloring is used, and (ii) they require non-preemptive EDF scheduling due to the lack of intra-core cache interference analysis. Hence, for comparison, we consider the best-fit decreasing (BFD) and the worst-fit decreasing (WFD) bin-packing algorithms. BFD and WFD is each combined with a conventional software cache partitioning approach. Before allocating tasks, BFD and WFD evenly distribute given cache partitions to all $N_P$ cores and sort tasks in decreasing order of task utilization with the number of per-core cache partitions. During task allocation, they do not use cache sharing. 

The system parameters used in this experiment are as follows: the number of tasks $n=\{8, 12, 16\}$, the number of cores $N_P=4$, the number of total cache partitions $N_{cache}=32$, and the size of total system memory $M_{total}=\{1024, 2048\}$ MB. To generate more than the four tasks in \tableref{CACHE_task_desc}, we have duplicated the taskset such that the number of tasks is a multiple of four. 

\begin{figure}[t]
\centering
  \includegraphics[width=0.7\textwidth]{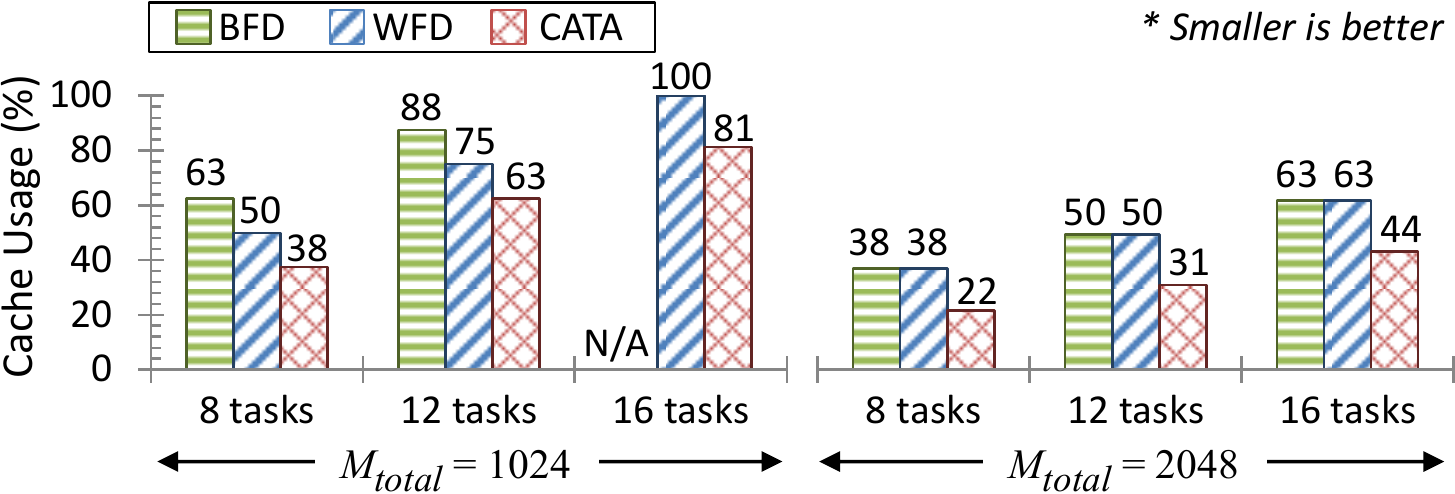}\\
  \caption{Minimum amount of cache required to schedule given tasksets}\label{fig:CACHE_benefit_1_cache}
\end{figure}
\begin{figure}[t]
\centering
  \includegraphics[width=0.7\textwidth]{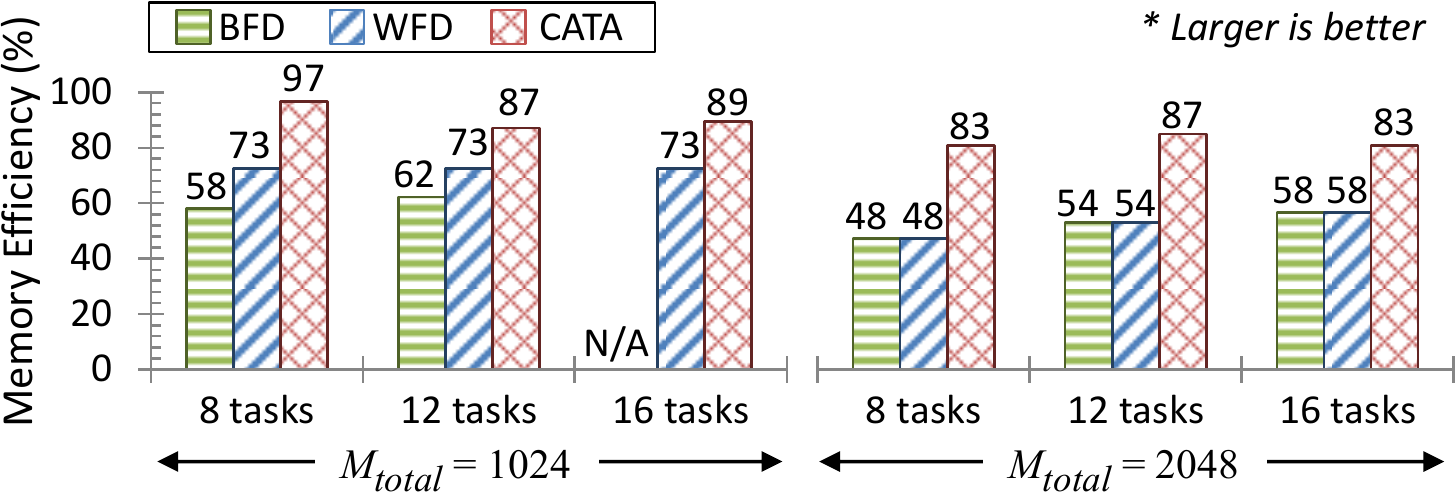}\\
  \caption{Memory space efficiency at minimum cache usage}\label{fig:CACHE_benefit_2_mem}
\end{figure}
\begin{figure}[t]
\centering
  \includegraphics[width=0.7\textwidth]{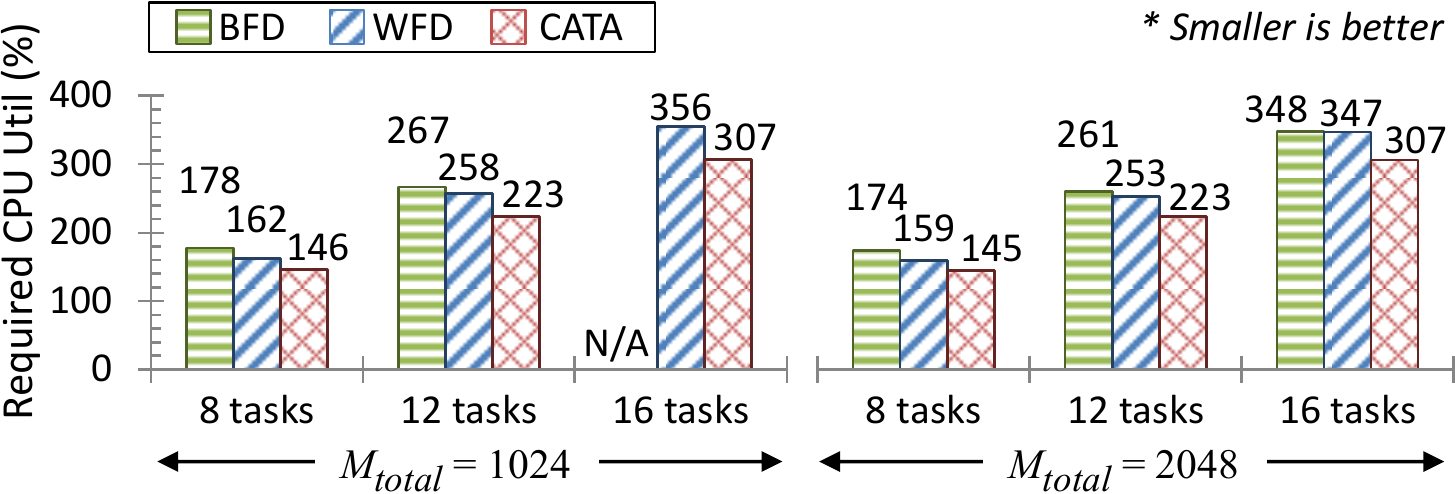}\\
  \caption{Total CPU utilization required to schedule given tasksets}\label{fig:CACHE_benefit_3_cpu}
\end{figure}

We first compare in \figref{CACHE_benefit_1_cache} the minimum number of cache partitions required to schedule a given taskset under BFD, WFD, and CATA. The y-axis represents the cache partition usage as a percentage to $N_{cache}$, for ease of comparison. CATA schedules given tasksets by using 16\% to 25\% and 12\% to 19\% less cache partitions than BFD and WFD, respectively. All algorithms consume more cache partitions when $M_{total}=1024$, compared to when $M_{total}=2048$, due to the task memory requirements. BFD fails to schedule a taskset with 16 tasks when $M_{total}=1024$ but schedules the taskset when $M_{total}=2048$. We next compare the memory space efficiency of the algorithms at their minimum cache partition usage. The memory space efficiency in our context is the ratio of the total memory usage of tasks to the size of allocated memory partitions, computed as $(\sum M_i) / \{(M_{total}/N_{cache}) \times (\text{\# of allocated memory partitions})\}$. \figref{CACHE_benefit_2_mem} shows the memory space efficiency. CATA is 25\% to 39\% and 14\% to 35\% more memory space efficient than BFD and WFD, respectively. Since BFD and WFD suffer from the memory co-partitioning problem, they exhibit poor memory space efficiency. On the other hand, CATA shows 97\% of memory space efficiency when $n=8$ and $M_{total}=1024$, meaning that only 3\% of slack space exists in the allocated memory partitions. Lastly, we compare in \figref{CACHE_benefit_3_cpu} the total accumulated CPU utilization required to schedule given tasksets under BFD, WFD, and CATA when all cache partitions are used. CATA requires 29\% to 44\% and 14\% to 49\% less CPU utilization than BFD and WFD, respectively. The utilization benefit of CATA becomes larger as the number of tasks increases. This is because CATA utilizes cache sharing but BFD and WFD suffer from the availability of a limited number of cache partitions. Based on these results, we therefore conclude that our scheme efficiently allocates cache partitions to tasks and significantly mitigates the memory co-partitioning problem and the availability of a limited number of cache partitions.

\section{Summary}
\label{CACHE_conclusions}
In this chapter, we introduced a coordinated OS-level cache management scheme for a multi-core platform. While providing predictable performance on architectures with shared caches across cores, our scheme addresses the two major challenges of page coloring: the memory co-partitioning problem and the availability of only a limited number of cache partitions. 
Experimental results indicate that our scheme significantly mitigates the negative impact of the memory co-partitioning problem, by yielding as much as 39\% higher memory efficiency than the conventional approaches. Also, experimental results show that our scheme is effective in overcoming the limited number of cache partitions, by consuming up to 25\% fewer cache partitions for satisfying timing constraints compared to the conventional approaches. 
Our scheme can be used not only for developing new multi-core systems but also for migrating existing applications from single-core to multi-core platforms.

\chapter{Bounding and Reducing Memory Interference}
\label{chapter_bounding_and_reducing_memory_interference}

Prior work on addressing memory interference~\cite{Dasari_11, Yun_ECRTS12, Pellizzoni_DATE10, Schliecker_DATE10, Andersson_SIGBED10} has modeled main memory as a {\em black-box} system, where each memory request takes a constant service time and memory requests from different cores are serviced in either Round-Robin (RR) or First-Come First-Serve (FCFS) order. 
However, such an over-simplified memory model used by prior work may produce pessimistic or optimistic estimates on the memory interference delay in a COTS multi-core system.

In this chapter, we propose a {\em white-box} approach for bounding and reducing memory interference.  
By explicitly considering the timing characteristics of major resources in the DRAM system, including the re-ordering effect of FR-FCFS and the rank/bank/bus timing constraints, we obtain an upper bound on the worst-case memory interference delay for a task when it executes in parallel with other tasks. 
Our technique combines two approaches: a {\em request-driven} approach focused on the task's own memory requests and a {\em job-driven} approach focused on interfering memory requests during the task's execution.    
Combining them, our analysis yields a tighter upper bound on the worst-case response time of a task in the presence of memory interference. To reduce the negative impact of memory interference, we use software DRAM bank partitioning~\cite{Liu_PACT12, Suzuki_ICESS13, Yun_RTAS14, Kim_RTAS14, Kim_RTS16, Xie_HPCA14}. We consider both dedicated and shared bank partitions due to the limited availability of DRAM banks, and our analysis results in an upper bound on the interference delay in both cases. In the evaluation section, we show the effectiveness of our analysis on a well-known COTS multi-core platform. 

In addition, we develop a memory interference-aware task allocation algorithm that accommodates memory interference delay during the allocation phase. The key idea of our algorithm is to co-locate memory-intensive tasks on the same core with dedicated DRAM banks. This approach reduces the amount of memory interference among tasks, thereby improving task schedulability. Experimental results indicate that our algorithm yields a significant improvement in task schedulability over existing approaches such as in~\cite{Paolieri_RTAS11}, with as much as 96\% more tasksets being schedulable.

To focus on the memory interference problem, we make the following assumptions in this chapter. First, tasks fit in the memory capacity. In a system with bank partitioning, this assumption can be satisfied by configuring each bank partition to have multiple DRAM banks. Second, each task is assigned private cache partitions, thereby no cache interference among tasks. This assumption will be relaxed in Section~\ref{combining_with_cache_interference_analysis} by combining our cache and memory interference analyses. Third, each task is assumed to have sufficient cache space of its own to store one row of each DRAM bank assigned to it. This assumption is used to bound the re-ordering effect of the memory controller, which will be described in Section~\ref{MEM_request-reiven}. In fact, this is a reasonable assumption in a modern multi-core system which typically has a large LLC.\footnote{For instance, \figref{MEM_address-mapping} shows a physical address mapping to the LLC and the DRAM in the Intel Core-i7 system. For the LLC mapping, the last 6 bits of a physical address are used as a cache line offset, and the next 11 bits are used as a cache set index. For the DRAM mapping, the last 13 bits are used as a column index and the next 4 bits are used as a bank index. In order for a task to store one row in its cache, consecutive $2^{13-6}=128$ cache sets need to be allocated to the task. With page coloring, this is equal to 2 out of 32 cache partitions in the example system.} For brevity, we use the following notation in this chapter:
\begin{itemize}
	\item $bank(p)$: the set of bank partitions assigned to a core $p$
	\item $shared(p,q)$: the intersection of $bank(p)$ and $bank(q)$
	\item $\Gamma_p$: the set of tasks allocated to a core $p$
	
\end{itemize}

The background and related prior work on memory interference have been presented in Section~\ref{MEM_background}. The system model including assumptions and notation for DRAM-based main memory and tasks can be found in Chapter~\ref{system_model}.

The rest of this chapter is organized as follows. Section~\ref{MEM_analysis} presents our memory interference analysis. Section~\ref{MEM_task_allocation} provides our memory interference-aware allocation algorithm. Section~\ref{MEM_evaluation} provides a detailed evaluation. Section~\ref{MEM_conclusions} summarizes this chapter.

\section{Bounding Memory Interference Delay}
\label{MEM_analysis}
The memory interference delay that a task can suffer from other tasks can be bounded by using either of two factors: (i) the number of memory requests generated by the task itself, and (ii) the number of interfering requests generated by other tasks that run in parallel. For instance, if a task $\tau_i$ does not generate any memory requests during its execution, this task will not suffer from any delays regardless of the number of interfering memory requests from other tasks. In this case, the use of factor (i) will give a tighter bound. Conversely, assume that other tasks simultaneously running on different cores do not generate any memory requests. Task $\tau_i$ will not experience any delays because there is no extra contention on the memory system from $\tau_i$'s perspective, so the use of factor (ii) will give a tighter bound in this case. 

In this section, we present our approach for bounding memory interference based on the aforementioned observation. We first analyze the memory interference delay using two different approaches: {\em request-driven} (Sec.~\ref{MEM_request-reiven}) and {\em job-driven} (Sec.~\ref{MEM_job-reiven}). Then by combining them, we present a response-time-based schedulability analysis that tightly bounds the worst-case memory interference delay of a task (Sec.~\ref{MEM_response_time_test}). We also discuss the effect of write batching in memory controllers (Sec.~\ref{MEM_write_batching}), and present how memory interference and cache interference analyses can be combined (Sec.~\ref{combining_with_cache_interference_analysis}).

\smallskip
\noindent\textbf{DRAM Commands:}
Four DRAM commands are considered in our analysis: precharge (PRE), activate (ACT), read (RD) and write (WR). Depending on the current state of the bank, the memory controller generates a sequence of DRAM commands for a single read/write memory request:
\begin{itemize}
	\item {\em Row-hit} request: RD/WR
	\item {\em Row-conflict} request: PRE, ACT and RD/WR
\end{itemize}
Each DRAM command is assumed to have the same priority and arrival time as the corresponding memory request. Note that the auto-precharge commands (RDAP/WRAP) are not generated under the open-row policy. We do not consider the refresh (REF) command because the effect of REF in memory interference delay is rather negligible compared to that of other commands.\footnote{The effect of REF ($E_{R}$) in memory interference delay can be roughly estimated as $E_{R}^{k+1}=\lceil\text{\{(total delay from analysis)}+E_{R}^k\}/t_{REFI}\rceil\cdot t_{RFC}$, where $E_R^0=0$. For the DDR3-1333 with 2Gb density below 85\textcelsius, $t_{RFC}/t_{REFI}$ is $160\text{ns}/7.8\mu\text{s}=0.02$, so the effect of REF results in only about 2\% increase in the total memory interference delay. A more detailed analysis on REF can be found in~\cite{Bhat_ECRTS10}.} The DRAM timing parameters used in this work are summarized in \tableref{MEM_dram_timing_param} and are taken from Micron's datasheet~\cite{Micron_DDR3}.

\begin{table}[t]
	\centering
	{
		\footnotesize
		\caption{DRAM timing parameters}\label{tab:MEM_dram_timing_param}
		\begin{tabular}{l|c|c|c}
			\hline
			Parameters & Symbols & DDR3-1333 & Units\\\hline
			DRAM clock cycle time& $t_{CK}$ & 1.5 & nsec\\
			Precharge latency & $t_{RP}$ & 9 & cycles\\
			Activate latency & $t_{RCD}$ & 9 & cycles\\
			CAS read latency & {\it CL} & 9 & cycles\\
			CAS write latency & {\it WL} & 7 & cycles\\
			Burst Length & {\it BL} & 8 & columns\\
			Write to read delay & $t_{WTR}$ & 5 & cycles\\
			Write recovery time & $t_{WR}$ & 10 & cycles\\
			Activate to activate delay & $t_{RRD}$ & 4 & cycles\\
			Four activate windows & $t_{FAW}$ & 20 & cycles\\
			Activate to precharge delay & $t_{RAS}$ & 24 & cycles \\ 
			Row cycle time & $t_{RC}$ & 33 & cycles \\
			Read to precharge delay & $t_{RTP}$ & 5 & cycles\\
			Refresh to activate delay & $t_{RFC}$ & 160 & nsec\\
			Average refresh interval & $t_{REFI}$ & 7.8 & $\mu$sec\\
			Rank-to-rank switch delay & $t_{RTRS}$ & 2 & cycles\\
			\hline
		\end{tabular}
	}
\end{table}

\subsection{Request-Driven Bounding Approach}
\label{MEM_request-reiven}
The request-driven approach focuses on the number of memory requests generated by a task $\tau_i$ ($H_i$) and the amount of additional delay imposed on each request of $\tau_i$. In other words, it bounds the total interference delay by $H_i \times$(per-request interference delay), where the per-request delay is bounded by using DRAM and processor parameters, not by using task parameters of other tasks.

The interference delay for a memory request generated by a processor core $p$ can be categorized into two types: {\em inter-bank} and {\em intra-bank}. 
If there is one core $q$ that does not share any bank partitions with $p$, the core $q$ only incurs inter-bank memory interference delay to $p$. If there is another core $q'$ that shares bank partitions with $p$, the core $q'$ incurs intra-bank memory interference. We present analyses on the two types of interference delay and calculate the total interference delay based on them.

\begin{figure}[t]
	\centering
	\subfloat{
		\includegraphics[width=0.65\textwidth]{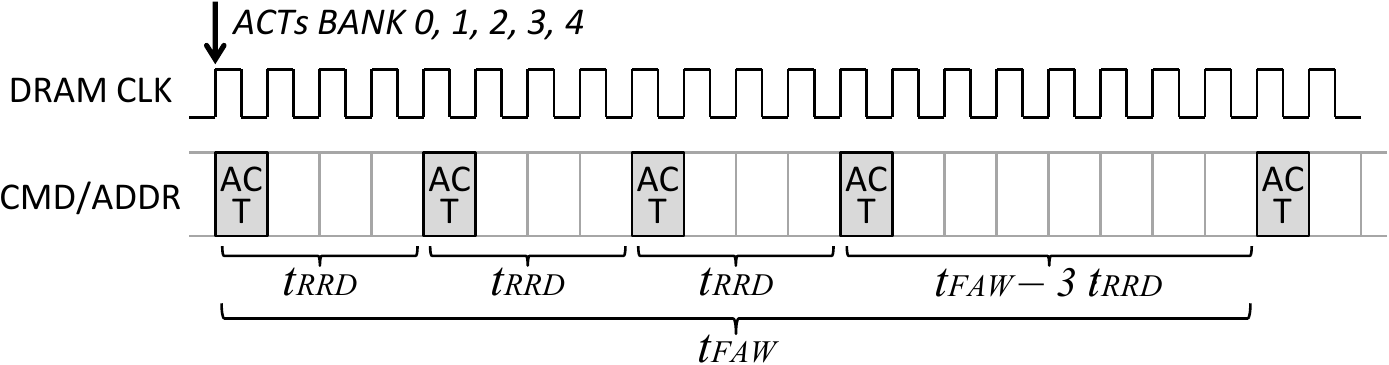}
	}
	\caption{Inter-bank row-activate timing constraints}
	\label{fig:MEM_row-activate}
\end{figure}


\begin{figure}[t]
	\centering
	\subfloat[WR followed by RD, different banks in the same rank] {\label{fig:MEM_wr-rd-delay}
		\includegraphics[width=0.6\textwidth]{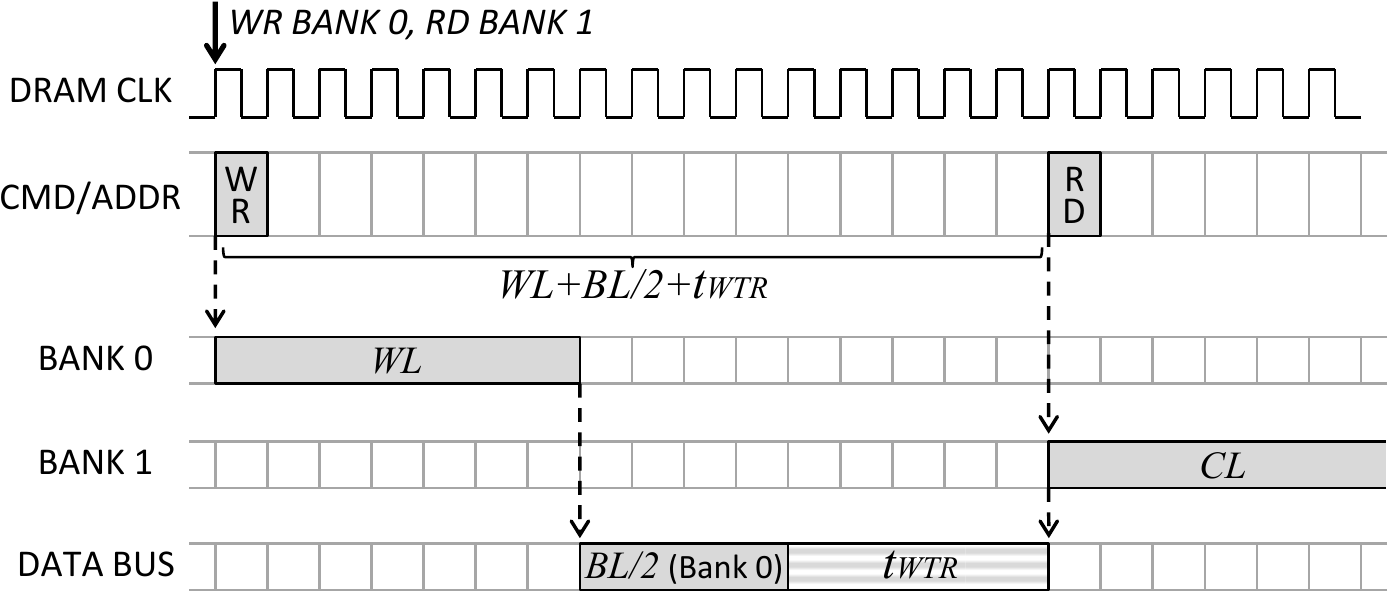}
	}\\
	\subfloat[RD followed by WR, different banks in the same rank] {\label{fig:MEM_rd-wr-delay}
		\includegraphics[width=0.6\textwidth]{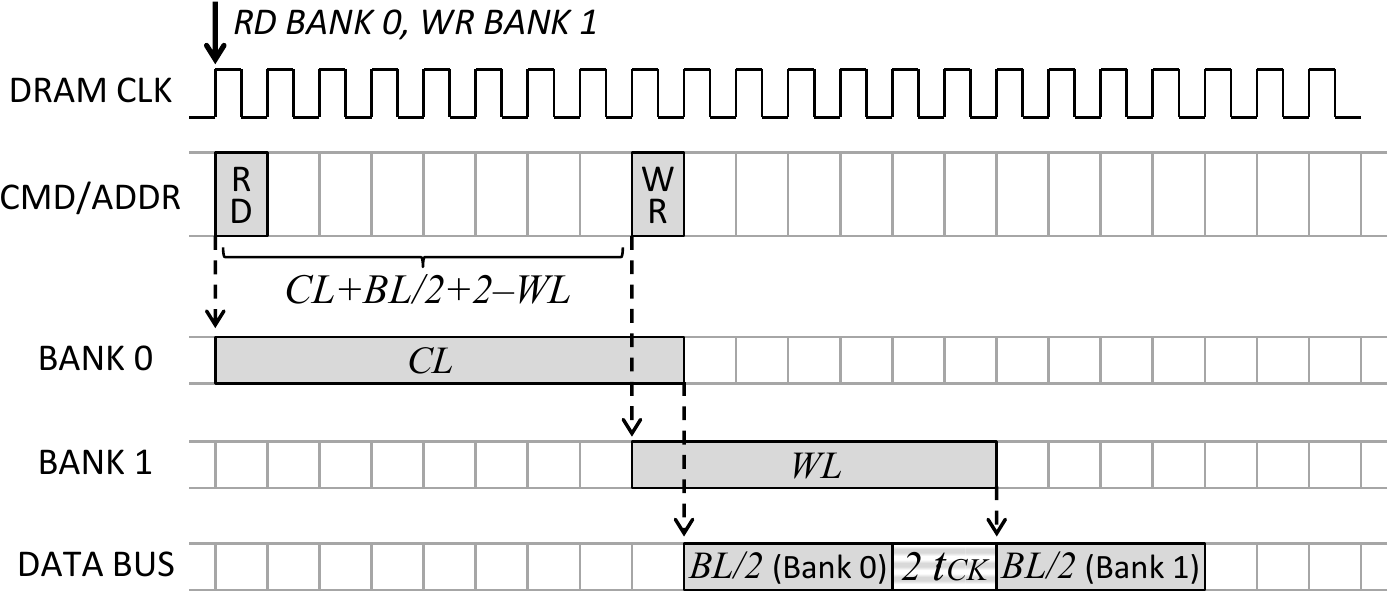}
	}
	\caption{Data bus turn-around delay}\label{fig:MEM_data-bus-turn-around}  
\end{figure}

\begin{figure}[t]
	\centering
	\subfloat[WR followed by RD, different ranks] {\label{fig:MEM_rank-to-rank-delay-wr-rd}
		\includegraphics[width=0.6\textwidth]{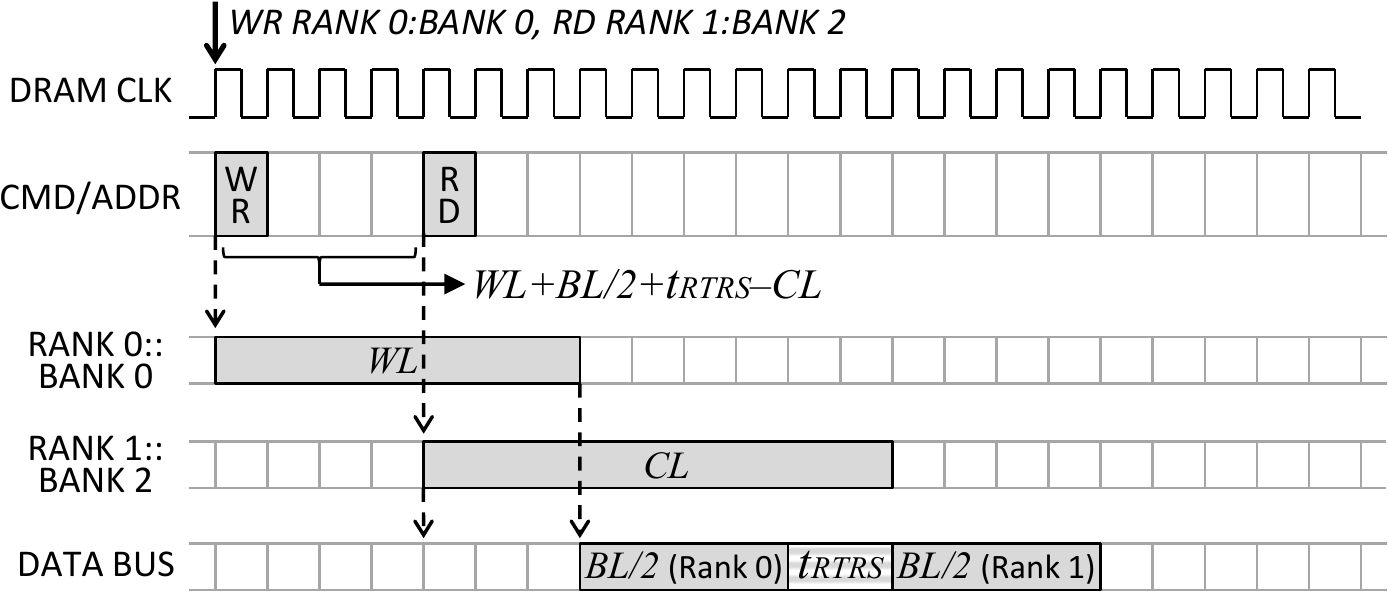}
	}\\
	\subfloat[RD followed by WR, different ranks] {\label{fig:MEM_rank-to-rank-delay-rd-wr}
		\includegraphics[width=0.6\textwidth]{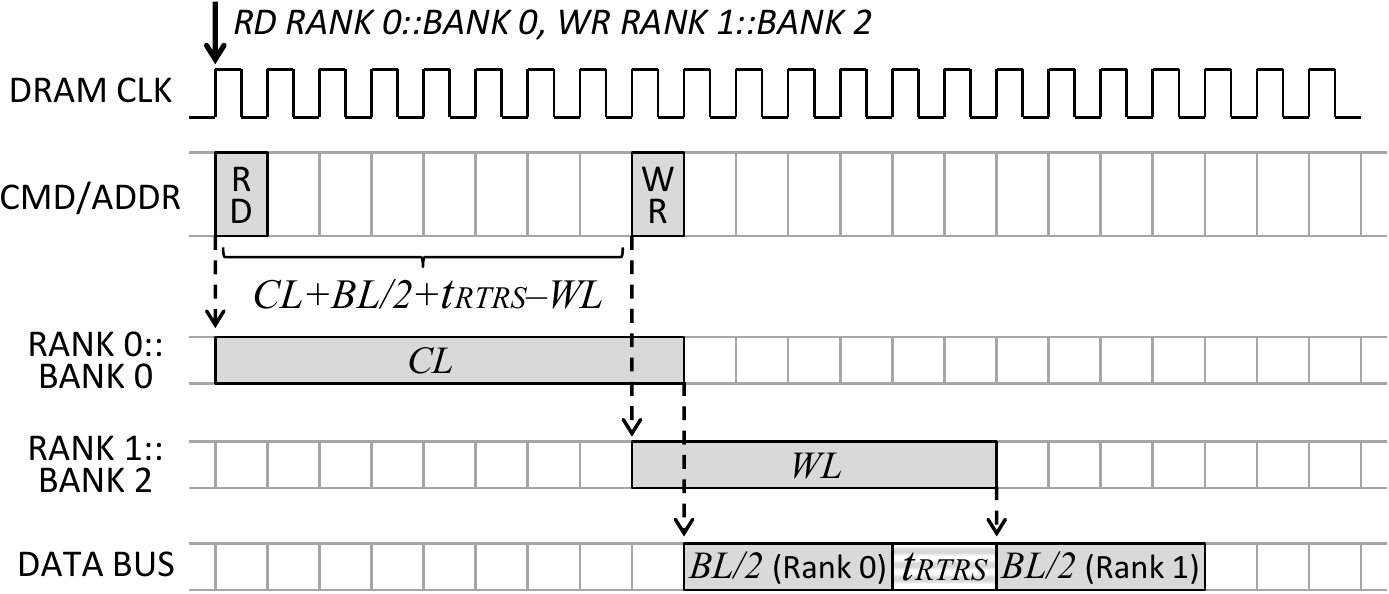}
	}\\
	\subfloat[RD followed by RD, different ranks] {\label{fig:MEM_rank-to-rank-delay-rd-rd}
		\includegraphics[width=0.6\textwidth]{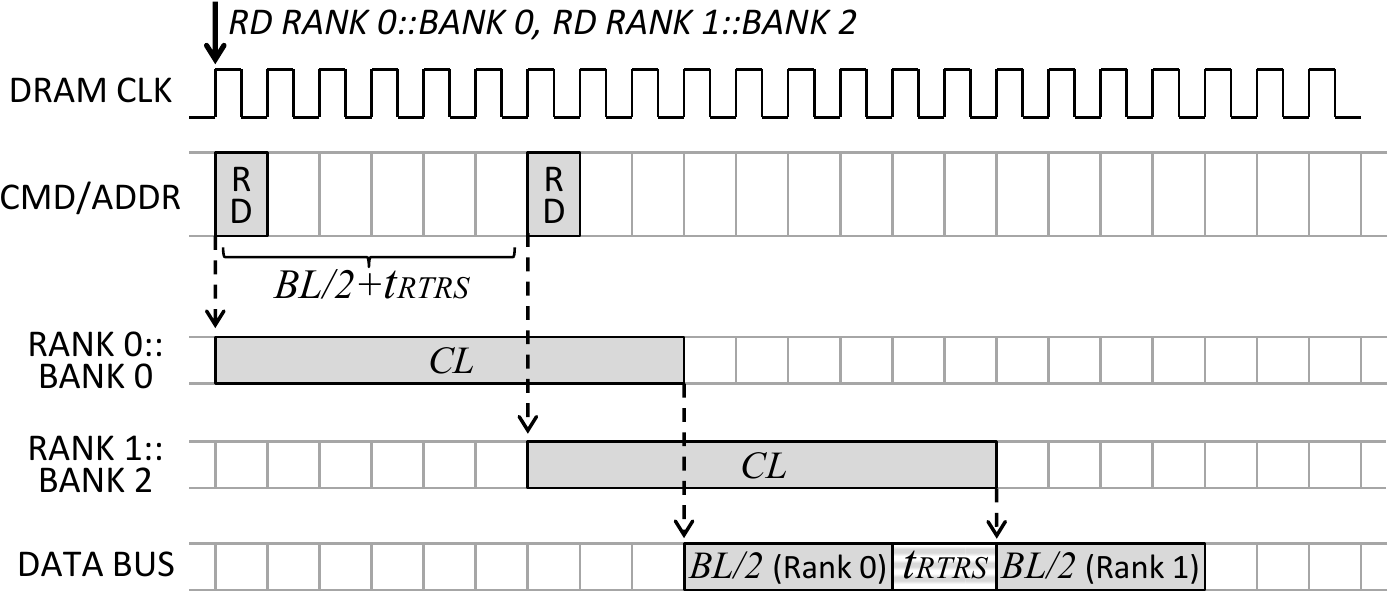}
	}\\
	\caption{Rank-to-rank switch delay}\label{fig:MEM_rank-to-rank-delay}  
\end{figure}

\smallskip
\noindent\textbf{Inter-bank interference delay:} 
Suppose that a core $p$ is assigned dedicated bank partitions. 
When a memory request is generated by one task on $p$, the request is enqueued into the request queue of the corresponding DRAM bank. Then, a sequence of DRAM commands is generated based on the type of the request, i.e., one command (RD/WR) for a row-hit request, and three commands (PRE, ACT, RD/WR) for a row-conflict request. At the bank scheduler, there is no interference delay from other cores because $p$ does not share its banks. In contrast, once a command of the request is sent to the channel scheduler, it can be delayed by the commands from other banks, because the FR-FCFS policy at the channel scheduler issues {\em ready} commands (with respect to the channel timing constraints) in the order of arrival time. 
The amount of delay imposed on each DRAM command is determined by the following factors:
\begin{itemize}
	\item {\it Address/command bus scheduling time}: Each DRAM command takes one DRAM clock cycle on the address/command buses. For a PRE command, as it is not affected by other timing constraints, the delay caused by each of the commands that have arrived earlier is: 
	$$L_{inter}^{PRE}=t_{CK}$$
	\item {\it Inter-bank row-activate timing constraints}: The JEDEC standard \cite{JEDEC_DDR3} specifies that there be a minimum separation time of $t_{RRD}$ between two ACTs to different banks, and no more than four ACTs can be issued during $t_{FAW}$ (\figref{MEM_row-activate}). Thus, in case of an ACT command, the maximum delay from each of the commands that have arrived earlier is:
	$$L^{ACT}_{inter}=\max(t_{RRD}, t_{FAW}-3\cdot t_{RRD})\cdot t_{CK}$$
	
	\item {\it Data bus turn-around and rank-to-rank switch delay}: When a RD/WR command is issued, data is transfered in burst mode on both the rising and falling edges of the DRAM clock signal, resulting in $\text{\it BL}/2$ of delay due to data bus contention. In addition, if a WR/RD command is followed by an RD/WR command to different banks in the same rank, the data flow direction of the data bus needs to be reversed, resulting in {\em data bus turn-around delay}. 
	\figref{MEM_data-bus-turn-around} depicts the data bus contention and bus turn-around delay in two cases. 
	When a WR command is followed by an RD command to different banks in the same rank, RD is delayed by $\text{\it WL}+\text{\it BL}/2+t_{WTR}$ cycles (\figsubref{MEM_wr-rd-delay}). When RD is followed by WR, WR is delayed by $\text{\it CL}+\text{\it BL}/2+2-\text{\it WL}$ cycles (\figsubref{MEM_rd-wr-delay}). If two consecutive WR/RD commands are issued to different ranks, there is {\em rank-to-rank switch delay}, $t_{RTRS}$, between the resulting two data transfers. \figref{MEM_rank-to-rank-delay} shows the rank-to-rank switch delay in three cases. When WR is followed by RD to different ranks, RD is delayed by $\text{\it WL}+\text{\it BL}/2+t_{RTRS}-\text{\it CL}$ cycles (\figsubref{MEM_rank-to-rank-delay-wr-rd}). When RD is followed by WR, WR is delayed by $\text{\it CL}+\text{\it BL}/2+t_{RTRS}-\text{\it WL}$ cycles (\figsubref{MEM_rank-to-rank-delay-rd-wr}). Lastly, when the two commands are of the same type, the latter is delayed by $\text{\it BL}/2+t_{RTRS}$ cycles (\figsubref{MEM_rank-to-rank-delay-rd-wr}). 
	Therefore, for a WR/RD command, the maximum delay from each of the commands that have arrived earlier is given by:
	\[
	\begin{split}
	L^{RW}_{inter}=\max(&\text{\it WL}+\text{\it BL}/2+t_{WTR}, \\
	&\text{\it CL}+\text{\it BL}/2+2-\text{\it WL},\\
	&\text{\it WL}+\text{\it BL}/2+t_{RTRS}-\text{\it CL},\\
	&\text{\it CL}+\text{\it BL}/2+t_{RTRS}-\text{\it WL},\\  
	&\text{\it BL}/2+t_{RTRS})\cdot t_{CK}
	\end{split}
	\]
\end{itemize}

Using these parameters, we derive the inter-bank interference delay imposed on each memory request of a core $p$. Recall that each memory request may consist of up to three DRAM commands: PRE, ACT and RD/WR. Each command of a request can be delayed by all commands that have arrived earlier at other banks. The worst-case delay for $p$'s request occurs when (i) a request of $p$ arrives after the arrival of the requests of all other cores that do not share banks with $p$, and (ii) each previous request causes PRE, ACT and RD/WR commands. Therefore, the worst-case per-request inter-bank interference delay for a core $p$, $RD_p^{inter}$, is given by: 
\begin{equation} \label{eq:MEM_RD_inter}
\begin{split}
{RD}_p^{inter}=\sum_{\substack{\forall q:\,q \ne p \,\wedge\, \\shared(q,p)=\emptyset}} \!\!\!\!\left(L_{inter}^{PRE}+L_{inter}^{ACT}+L_{inter}^{RW}\right)\\
\end{split}
\end{equation}

\begin{figure}[t]
	\centering
	\subfloat{
		\includegraphics[width=1\textwidth]{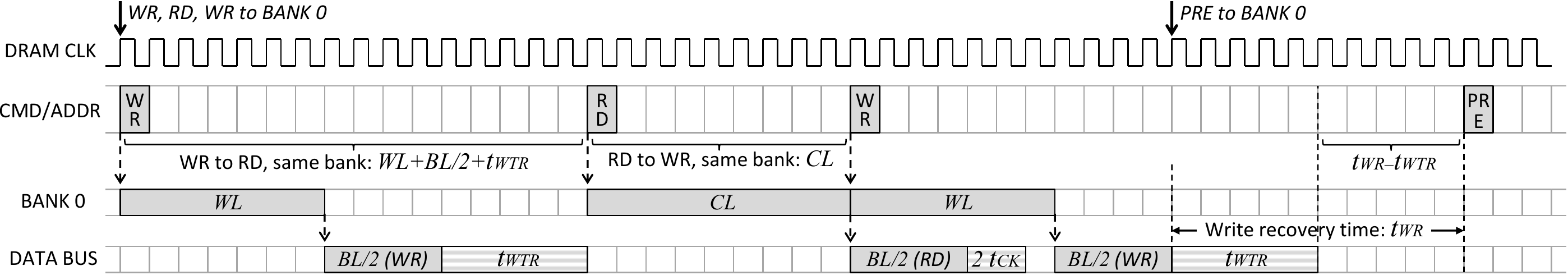}
	}
	\caption{Timing diagram of consecutive row-hit requests}
	\label{fig:MEM_consecutive-row-hit}
\end{figure}

\smallskip
\noindent\textbf{Intra-bank interference delay:}
Memory requests to the same bank are queued into the bank request buffer and their service order is determined by the bank scheduler. A lower-priority request should wait until all higher priority requests are completely serviced by the bank. 
The delay caused by each higher-priority request includes (i) the inter-bank interference delay for the higher priority request, and (ii) the service time of the request within the DRAM bank. The inter-bank interference delay can be calculated by Eq.~\eqref{eq:MEM_RD_inter}. The service time within the DRAM bank depends on the type of the request:
\begin{itemize}
	\item {\it Row-hit service time}: The row-hit request is for a requested column already in the row-buffer. Hence, it can simply read/write its column. In case of read, an RD command takes $\text{\it CL}+\text{\it BL}/2$ for data transfer and may cause 2 cycles of delay to the next request for data bus turn-around time~\cite{JEDEC_DDR3}. Note that the read-to-precharge delay ($t_{RTP}$) does not need to be explicitly considered here because the worst-case delay of an RD command is larger than $t_{RTP}$ in DDR3 SDRAM~\cite{JEDEC_DDR3} (or \tableref{MEM_dram_timing_param} for DDR3-1333), i.e., $t_{RTP} < \text{\it CL}+\text{\it BL}/2 + 2$.
	In case of write, a WR command takes $\text{\it WL}+\text{\it BL}/2$ for data transfer and may cause $\max(t_{WTR},t_{WR})$ of delay to the next request for bus turn-around or write recovery time, depending on the type of the next request. Thus, in the worst case, the service time for one row-hit request is:
	\[
	L_{hit}=\max\{\text{\it CL}+\text{\it BL}/2+2, \text{\it WL}+\text{\it BL}/2+\max(t_{WTR},t_{WR})\}\cdot t_{CK}
	\]
	\item {\it Row-conflict service time}: The row-conflict request should open a row before accessing a column by issuing PRE and ACT commands, which may take up to $t_{RP}$ and $t_{RCD}$ cycles, respectively. Hence, the worst-case service time for one row-conflict request is represented as follows:
	\[
	\begin{split}
	L_{conf}=(t_{RP}+t_{RCD})\cdot t_{CK}+L_{hit}
	\end{split}
	\]
	If the next request is also row-conflict and issues PRE and ACT commands, constraints on the active-to-precharge delay ($t_{RAS}$) and the row-cycle time ($t_{RC}$, a minimum separation between two ACTs in the same bank) should be satisfied. The row-conflict service time $L_{conf}$ satisfies $t_{RAS}$ because $t_{RCD} \cdot t_{CK}+L_{hit}$ is larger than $t_{RAS}\cdot t_{CK}$ in DDR3 SDRAM~\cite{JEDEC_DDR3} (or \tableref{MEM_dram_timing_param} for DDR3-1333). $L_{conf}$ also satisfies $t_{RC}$, because $t_{RC}$ is equal to $t_{RAS}+t_{RP}$ where $t_{RP}$ is time for the PRE command of the next request to be completed.
	\item {\it Consecutive row-hit requests}: If $m$ row-hit requests are present in the memory request buffer, their service time is much smaller than $m\cdot L_{hit}$. Due to the data bus turn-around time, the worst-case service time happens when the requests alternate between read and write, as depicted in \figref{MEM_consecutive-row-hit}. WR followed by RD causes $\text{\it WL}+\text{\it BL}/2+t_{WTR}$ of delay to RD, and RD followed by WR causes \text{\it CL} of delay to WR. As WR-to-RD causes larger delay than RD-to-WR in DDR3 SDRAM~\cite{JEDEC_DDR3, Lee_TR2010}, $m$ row-hits takes $\lceil {m\over 2}\rceil \cdot (\text{\it WL}+\text{\it BL}/2+t_{WTR})+ \lfloor {m \over 2}\rfloor \cdot \text{\it CL}$ cycles. In addition, if a PRE command is the next command to be issued after the $m$ row-hits, it needs to wait an extra $t_{WR}-t_{WTR}$ cycles due to the write recovery time. Therefore, the worst-case service time for $m$ consecutive row-hit requests is:
	\[
	\begin{split}
	L_{conhit}(m)=\{&\left\lceil {m/2}\right\rceil \cdot (\text{\it WL}+\text{\it BL}/2+t_{WTR})+\left\lfloor {m/2}\right\rfloor \cdot \text{\it CL}+(t_{WR}-t_{WTR})\}\cdot t_{CK}
	\end{split}
	\]
\end{itemize}

Under the FR-FCFS policy, the bank scheduler serves row-conflict requests in the order of their arrival times. When row-hit requests arrive at the queue, the bank scheduler {\em re-orders} memory requests such that row-hits are served earlier than older row-conflicts. For each open row, the maximum row-hit requests that can be generated in a system is represented as $N_{cols}/\text{\it BL}$, where $N_{cols}$ is the number of columns in one row. This is due to the fact that, as described in the system model, (i) each task is assumed to have enough cache space to store one row of each bank assigned to it, (ii) the memory request addresses are aligned to the size of {\it BL}, and (iii) tasks do not share memory. Once the tasks that have their data in the currently open row fetch all columns in the open row into their caches, all the subsequent memory accesses to the row will be served at the cache level and no DRAM requests will be generated for those accesses. If one of the tasks accesses a row different from the currently open one, this memory access causes a row-conflict request so that the re-ordering effect no longer occurs. In many systems, as described in~\cite{Mutlu_MICRO07, Moscibroda_2007, Ausavarungnirun_ISCA12}, the re-ordering effect can also be bounded by a hardware threshold $N_{cap}$, which caps the number of re-ordering between requests. Therefore, the maximum number of row-hits that can be prioritized over older row-conflicts is:
\begin{equation} \label{eq:MEM_N_reorder}
N_{reorder}=\min\left({N_{cols}/ \text{\it BL}}, N_{cap}\right)
\end{equation}
The exact value of $N_{cap}$ is not publicly available on many platforms. Even so, we can still obtain a theoretical bound on $N_{reorder}$ by ${N_{cols}/ \text{\it BL}}$, the parameters of which are easily found in the JEDEC standard~\cite{JEDEC_DDR3}. 

We now analyze the intra-bank interference delay for each memory request generated by a processor core $p$. Within a bank request buffer, each request of $p$ can be delayed by both the re-ordering effect and the previous memory requests in the queue. Therefore, the worst-case per-request interference delay for a core $p$ ($RD_p^{intra}$) is calculated as follows:
\begin{equation} \label{eq:MEM_RD_intra}
\begin{split}
&{RD}_p^{intra}=reorder(p)+\!\!\!\!\!\sum_{\substack{\forall q:\,q \ne p \,\wedge\, \\shared(q,p) \ne \emptyset}}\!\!\!\!\!\!\!\left(L_{conf}+ RD_q^{inter}\right)\\
\end{split}
\end{equation}
\vspace{-10pt}
\begin{equation} \label{eq:MEM_reordering}
\begin{split}
reorder(p)=&\left\{ \begin{array}{l}
\scalebox{0.9}{$0 \quad\quad\quad\quad\quad\quad\quad\quad\quad\quad\quad\quad\quad\quad\quad\quad\quad\quad\quad\quad{ \text{if }} {\nexists q :q\ne p\wedge shared(q,p)\ne \emptyset}$}\\
\scalebox{0.9}{$\displaystyle \!L_{conhit}(N_{reorder})+\Bigg(\!N_{reorder}\cdot \!\!\!\!\!\!\!\!\!\!\sum_{\substack{\forall q:\,q\ne p\,\wedge\,\\ shared(q,p)= \emptyset}}\!\!\!\!\!\!\!\!L_{inter}^{RW}\Bigg) + (t_{RP}+t_{RCD})\cdot t_{CK}$}\quad\,\,{ \text{otherwise}}\\
\end{array} \right.\\
\end{split}
\end{equation}
In Eq.~\eqref{eq:MEM_RD_intra}, the summation part calculates the delay from memory requests that can be queued before the arrival of $p$'s request. It considers processor cores that share bank partitions with $p$. Since row-conflict causes a longer delay than row-hit, the worst-case delay from each of the older requests is the sum of the row-conflict service time ($L_{conf}$) and the per-request inter-bank interference delay ($RD_q^{inter}$). The function $reorder(p)$ calculates the delay from the re-ordering effect. As shown in Eq.~\eqref{eq:MEM_reordering}, it gives zero if there is no core sharing bank partitions with $p$. Otherwise, it calculates the re-ordering effect as the sum of the consecutive row-hits' service time ($L_{conhit}(N_{reorder})$) and the inter-bank delay for the row-hits ($N_{reorder} \cdot \sum L_{inter}^{RW}$). In addition, since the memory request of $p$ that was originally row-hit could become row-conflict due to interfering requests from cores sharing bank partitions with $p$, Eq.~\eqref{eq:MEM_reordering} captures delays for additional PRE and ACT commands ($(t_{RP}+t_{RCD})\cdot t_{CK}$).

\smallskip
\noindent\textbf{Total interference delay:} A memory request from a core $p$ experiences both inter-bank and intra-bank interference delay. Hence, the worst-case per-request interference delay for $p$, $RD_p$, is represented as follows:
\begin{equation}
RD_p=RD_p^{inter}+RD_p^{intra}
\end{equation}
Since $RD_p$ is the worst-case delay for each request, the total memory interference delay of $\tau_i$ is upper bounded by $H_i\cdot RD_p$.

\subsection{Job-Driven Bounding Approach}
\label{MEM_job-reiven}
The job-driven approach focuses on how many interfering memory requests are generated during a task's job execution time. In the worst case, every memory request from other cores can delay the execution of a task running on a specific core. Therefore, by capturing the maximum number of requests generated by the other cores during a time interval $t$, the job-driven approach bounds the memory interference delay that can be imposed on tasks running on a specific core in any time interval $t$. 

We define $A_p(t)$, which is the maximum number of memory requests generated by the core $p$ during a time interval $t$ as:
\begin{equation} \label{eq:MEM_eq4}
\begin{split}
A_p(t)=\sum_{\forall \tau_i \in \Gamma_p}\left(\left\lceil{t\over T_i}\right\rceil+1\right)\cdot H_i
\end{split}
\end{equation}
where $\Gamma_p$ is the set of tasks assigned to the core $p$. The ``+1'' term is to capture the carry-in job of each task during a given time interval $t$. Note that this calculation is quite pessimistic, because we do not make assumptions on memory access patterns (e.g. access rate or distribution). It is possible to add this type of assumptions, such as the specific memory access pattern of the tasks~\cite{Dasari_11, Andersson_SIGBED10} or using memory request throttling mechanisms~\cite{Shen_USENIX09, Ebrahimi_ASPLOS10, Yun_ECRTS12}. This helps to calculate a tighter $A_p(t)$, while other equations in our work can be used independent of such additional assumptions.

\smallskip
\noindent\textbf{Inter-bank interference delay:} The worst-case inter-bank interference delay imposed on a core $p$ during a time interval $t$ is represented as follows:
\begin{equation} \label{eq:MEM_JD_inter}
\begin{split}
&{JD}_p^{inter}(t)=
\!\!\!\!\!\!
\sum_{\substack{\forall q:\,q\ne p\,\wedge\,\\ shared(q,p)=\emptyset}}
\!\!\!\!\!\!\!\!
A_q(t)\cdot \left(L_{inter}^{ACT}+L_{inter}^{RW}+L_{inter}^{PRE}\right)
\end{split}
\end{equation}
In this equation, the summation considers processor cores that do not share bank partitions with $p$. The other cores sharing banks with $p$ will be taken into account in Eq.~\eqref{eq:MEM_JD_intra}.
The number of memory requests generated by other cores ($A_q(t)$) is multiplied by the maximum inter-bank interference delay from each of these requests ($L_{inter}^{ACT}+L_{inter}^{RW}+L_{inter}^{PRE}$).  

\smallskip
\noindent\textbf{Intra-bank interference delay:} The worst-case intra-bank interference delay imposed on a core $p$ during $t$ is as follows:\begin{equation} \label{eq:MEM_JD_intra}
\begin{split}
&{JD}_p^{intra}(t)=
\!\!\!\!\!\!\!
\sum_{\substack{\forall q:\,q\ne p\,\wedge\,\\ shared(q,p)\ne \emptyset}}
\!\!\!\!\!\!\!\!\!
\left(A_q(t)\cdot L_{conf}+JD_q^{inter}(t)\right)
\end{split}
\end{equation}
Eq.~\eqref{eq:MEM_JD_intra} considers other cores that share bank partitions with $p$. The number of requests generated by each of these cores during $t$ is calculated as $A_q(t)$.
Since a row-conflict request causes larger delay than a row-hit one, $A_q(t)$ is multiplied by the row-conflict service time $L_{conf}$. In addition, $JD_q^{inter}$ is added because each interfering core $q$ itself may be delayed by inter-bank interference depending on its bank partitions. 
Note that the re-ordering effect of the bank scheduler does not need to be considered here because Eq.~\eqref{eq:MEM_JD_intra} captures the worst case where all the possible memory requests generated by other cores arrived ahead of any request from $p$.

\smallskip
\noindent\textbf{Total interference delay:} The worst-case memory interference delay is the sum of the worst-case inter-bank and intra-bank delays. Therefore, the memory interference delay for a core $p$ during a time interval $t$, $JD_p(t)$, is upper bounded by:\begin{equation} \label{eq:MEM_JD_p}
JD_p(t)=JD_p^{inter}(t)+JD_p^{intra}(t)
\end{equation}
It is worth noting that the job-driven approach will give a tighter bound than the request-driven approach when the number of interfering memory requests from other cores is relatively small compared to the number of the memory requests of the task under analysis. Conversely, in the opposite case, the request-driven approach will give a tighter bound than the job-driven approach. We will compare the results of these two approaches in Section~\ref{MEM_evaluation}.

\subsection{Response-Time Based Schedulability Analysis}
\label{MEM_response_time_test}
We have presented the request-driven and the job-driven approaches to analyze the worst-case memory interference delay. Since each of the two approaches bounds the interference delay by itself, a tighter upper bound can be obtained by taking the smaller result from the two approaches. Based on the analyses of the two approaches, the iterative response time test \cite{Joseph_J86} is extended as follows to incorporate the memory interference delay:
\begin{equation} \label{eq:MEM_response_time}
\begin{split}
W_i^{k+1}&=C_i+\sum_{\tau_h\in \Gamma_p \land \pi_h > \pi_i }\left\lceil{W_i^k\over T_h}\right\rceil\cdot C_h\\
&+ \min \left\{ H_i \cdot  RD_p +      
\sum_{\tau_h\in \Gamma_p \land \pi_h > \pi_i }     \left\lceil{W_i^k\over T_h}\right\rceil \cdot H_h \cdot RD_p,\,\,
JD_p(W_i^k)  \right\}
\end{split}
\end{equation}
where $W_i^{k}$ is the worst-case response time of $\tau_i$ at the $k^{th}$ iteration, $p$ is the core of $\tau_i$, $\Gamma_p$ is the set of tasks assigned to the core $p$, and $\pi_i$ is the priority of $\tau_i$. The test terminates when $W_i^{k+1} = W_i^{k}$. The task $\tau_i$ is schedulable if its response time does not exceed its deadline: $W_i^k \le D_i$. The first and the second terms are the same as the classical response time test. In the third term, the memory interference delay for $\tau_i$ is bounded by using the two approaches. The request-driven approach bounds the delay with the addition of $H_i\cdot RD_p$ and $\sum\lceil{W_i^k\over T_h}\rceil\cdot H_h\cdot RD_p$, which is the total delay imposed on $\tau_i$ and its higher priority tasks. The job-driven approach bounds the delay by $JD_p(W_i^k)$, that captures the total delay incurred during $\tau_i$'s response time.

\subsection{Memory Controllers with Write Batching} 
\label{MEM_write_batching}
Many recent memory controllers handle write requests in batches when the write buffer is close to full so that the bus turn-around delay can be amortized across many requests~\cite{Lee_TR2010,Seshadri_ISCA14}. Although the modeling of memory controllers using write batching is not within the scope of our work, we believe that our analysis could still be used to bound memory interference in systems with such memory controllers. 
If a memory controller uses write batching, the worst-case delay of a single memory operation can be much larger than the one computed by $L_{inter}^{PRE}+L_{inter}^{ACT}+L_{inter}^{RW}$, due to write-buffer draining.\footnote{Note that the write-buffer draining does not completely block read requests until all the write requests are serviced. In a memory controller with write batching, read requests are always exposed to the memory controller, but write requests are exposed to and scheduled by the memory controller only when the write buffer is close to full~\cite{Lee_TR2010}. Hence, even when the write buffer is being drained, a read request can be scheduled if its commands are ready with respect to DRAM timing constraints (e.g., read and write requests to different banks).}
However, this does not restrict the applicability of our theory on such memory controllers. We discuss this from two cases as follows.

%

First, consider a job of task $\tau_i$ and how it experiences interference from a job of task $\tau_j$ where $\tau_j$ is assigned to a different core than $\tau_i$.
If the job of $\tau_i$ starts its execution at a time when the write buffer is fully filled with the memory requests of the job of $\tau_j$, then the job of $\tau_i$ suffers additional interference from at most $w$ memory requests, where $w$ is the size of the write buffer. However, this effect can only happen once per the job of $\tau_i$ and be bounded by a constant value. Afterwards, the total number of interfering memory requests remains the same during the execution of the job of $\tau_i$. In addition, since the use of write batching reduces memory bus turn-around delay, it may even shorten the response time of the job of $\tau_i$. 

Second, consider a job of task $\tau_i$ and how it experiences interference from a job of task $\tau_j$ where $\tau_j$ is assigned to the same core as $\tau_i$.
If the job of $\tau_i$ starts its execution at a time when the write buffer is full with the memory requests of the job of $\tau_j$, all the memory requests in the write buffer need to be serviced first, which can delay the execution of the job of $\tau_i$. 
However, this effect can only happen once per context switch and hence it can be accounted for as a preemption cost.



\subsection{Combining with Cache Interference Analysis}
\label{combining_with_cache_interference_analysis}

We have so far assumed that there is no cache interference among tasks. Now, we relax this assumption by combining our memory interference analysis with our cache interference analysis presented in Chapter~\ref{chapter_coordinated_cache_management}. For simplicity, we assume that cache warm-up delay has been taken into account in the WCET of each task. Then, the analysis for the worst-case response time of a task $\tau_i$ in the presence of cache interference is re-written as follows:
\begin{equation} \label{eq:MEM_cache_response_time_test}
\begin{split}
W_i^{k+1}&=C_i+
\sum_{\tau_h \in \Gamma_p \land \pi_h>\pi_i}\left\lceil\frac{W_i^{k}}{T_h}\right\rceil C_h +
\sum_{\tau_h \in \Gamma_p \land \pi_h>\pi_i}\left\lceil\frac{W_i^{k}}{T_h}\right\rceil\gamma_{h,i}\\
\end{split}
\end{equation}
where $W_i^{k}$ is the worst-case response time of $\tau_i$ at the $k^{th}$ iteration, $p$ is the core where $\tau_i$ is allocated, $\pi_h$ is the priority of $\tau_h$, and $\Gamma_p$ is the set of tasks allocated to the core $p$, and $\gamma_{h,i}$ is the cache-related preemption delay caused by $\tau_h$. 
The term $\gamma_{h,i}$ is calculated as follows:
\begin{equation}
\gamma_{h,i}=\left|\mathbb{S}_h\cap \bigcup_{\tau_j\in \Gamma_p \land \pi_j < \pi_h \land \pi_j \ge \pi_i}\mathbb{S}_j\right|\cdot\Delta
\end{equation}
where $\mathbb{S}_h$ is the set of cache partitions assigned to $\tau_h$, and $\Delta$ is the maximum time needed to reload data in one cache partition. 

In Eq~\eqref{eq:MEM_cache_response_time_test}, the last term captures the total amount of cache interference delay caused by higher-priority tasks. To identify the number of memory requests generated by cache interference, we introduce a new term, $\gamma_{h,i}^*$:
\begin{equation}
	\gamma_{h,i}^*=\left|\mathbb{S}_h\cap \bigcup_{\tau_j\in \Gamma_p \land \pi_j < \pi_h \land \pi_j \ge \pi_i}\mathbb{S}_j\right|\cdot\delta
\end{equation}
where $\delta$ is the number of memory requests needed to reload one cache partition. Note that $\delta$ is determined by the size of a cache partition in the system. In case of a write-back cache, $\delta$ should take into account the effect of a {\em dirty} cache line that requires two memory accesses to fetch a new cache line~\cite{Sebek_TR01}.

Then, we incorporate $\gamma_{h,i}^*$ in the request-driven and job-driven approaches. For the request-driven approach, the total number of memory requests generated by cache interference during the response time of a task $\tau_i$ is given by:
\begin{equation} \label{eq:MEM_memory_requests_from_cache_interference}
	\begin{split}
		H_i^*(W_i)=\sum_{\tau_h \in \Gamma_p \land \pi_h>\pi_i}\left\lceil\frac{W_i}{T_h}\right\rceil\gamma_{h,i}^*
	\end{split}
\end{equation}

For the job-driven approach, the $A_p(t)$ function given in Eq.~\eqref{eq:MEM_eq4}, which captures the maximum number of memory requests generated by the core $p$ during a time interval $t$, is extended as follows to incorporate cache interference:
\begin{equation} \label{eq:MEM_eq4_cache_interference}
\begin{split}
A_p(t)=\sum_{\forall \tau_i \in \Gamma_p}\left(\left\lceil{t\over T_i}\right\rceil+1\right)\cdot (H_i+\gamma_{i,n}^*)
\end{split}
\end{equation}
where $n$ is the index of the lowest-priority task in $\Gamma_p$.

Finally, the response-time based schedulability analysis incorporating both cache and memory interference delay is given as follows:
\begin{equation} \label{eq:MEM_response_time_combined}
	\begin{split}
		W_i^{k+1}&=C_i+\sum_{\tau_h\in \Gamma_p \land \pi_h > \pi_i }\left\lceil{W_i^k\over T_h}\right\rceil\cdot C_h\\
		&+ \min \left\{ H_i \cdot  RD_p +      
		\sum_{\tau_h\in \Gamma_p \land \pi_h > \pi_i }     \left\lceil{W_i^k\over T_h}\right\rceil \cdot H_h \cdot RD_p+H_i^*(W_i^k)\cdot RD_p,\,\,
		JD_p(W_i^k)  \right\}
	\end{split}
\end{equation}
where $W_i^{k}$ is the worst-case response time of $\tau_i$ at the $k^{th}$ iteration. The test terminates when $W_i^{k+1} = W_i^{k}$. The task $\tau_i$ is schedulable if its response time does not exceed its deadline: $W_i^k \le D_i$. Using this equation, we can check task schedulability in the presence of both cache and memory interference.

\section{Reducing Memory Interference via Task Allocation}

\label{MEM_task_allocation}

In this section, we present our memory interference-aware task allocation algorithm to reduce memory interference during the allocation phase. Our algorithm is motivated by the following observations we have made from our analysis given in Section~\ref{MEM_analysis}: (i) memory interference for a task is caused by other tasks running on other cores in parallel, (ii) tasks running on the same core do not interfere with each other, and (iii) the use of bank partitioning reduces the memory interference delay. These observations lead to an efficient task allocation under partitioned scheduling. By co-locating memory-intensive tasks on the same core with dedicated DRAM banks, the amount of memory interference among tasks can be significantly reduced, thereby providing better schedulability.

Our memory interference-aware allocation algorithm (MIAA) is given in Algorithm~\ref{alg:MEM_miaa}. MIAA takes three input parameters: $\Gamma$ is a set of tasks to be allocated, $N_P$ is the number of available processor cores, and $N_{bank}$ is the number of available bank partitions. MIAA returns {\em schedulable}, if every task in $\Gamma$ can meet its deadline, and {\em unschedulable}, if any task misses its deadline.

\begin{algorithm}[t]
	\caption[MIAA($\Gamma, N_P, N_{bank}$): a memory interference-aware task allocation algorithm]{MIAA($\Gamma, N_P, N_{bank}$)}
	\label{alg:MEM_miaa}
	\algsetup{linenosize=\scriptsize}
	\scriptsize
	\begin{algorithmic}[1]
		\REQUIRE $\Gamma$: a taskset to be allocated, $N_P$: the number of processor cores, $N_{bank}$: the number of available bank partitions
		\ENSURE Schedulability of $\Gamma$
		\STATE $\mathbb{G} \leftarrow $ MemoryInterferenceGraph($\Gamma$) \label{line:MEM_build_graph}
		\STATE $\Gamma_{p_1} \leftarrow \emptyset$ 
		\STATE $bank(p_1) \leftarrow $ LeastInterferingBank($N_{bank}, \Pi, \mathbb{G}, \Gamma$)
		\STATE $\Pi \leftarrow \{p_1\}$
		\STATE $\Phi \leftarrow \{\Gamma\}$ \label{line:MEM_initial_bundle}
		\WHILE {$\Phi \neq \emptyset$} 
		\STATE /* Allocates bundles */
		\STATE $\Phi' \leftarrow \Phi$; $\Phi_{rest} \leftarrow \emptyset$ 
		\FORALL {$\varphi_i \in \Phi'$ in descending order of utilization } \label{line:MEM_bundle_allocation_begin}
		\STATE $\Phi \leftarrow \Phi \setminus \{\varphi_i\}$
		\STATE $p_{bestfit} \leftarrow$ BestFit($\varphi_i, \Pi$)
		\IF {$p_{bestfit} \ne invalid$} \label{line:MEM_bundle_allocated}
		\FORALL {$p_j \in \Pi: p_j \ne p_{bestfit} \land \lnot schedulable(p_j)$}
		\STATE $\Phi \leftarrow \Phi \cup \{$RemoveExcess($p_j, \mathbb{G}$)$\}$ \label{line:MEM_call_remove_excess}
		\ENDFOR
		\ELSE \label{line:MEM_bundle_not_allocated}
		\STATE $\Phi_{rest} \leftarrow \Phi_{rest} \cup \{\varphi_i\}$
		\ENDIF
		\ENDFOR \label{line:MEM_bundle_allocation_end}
		\IF {$|\Phi_{rest}|=0$}
		\STATE {\bf continue}
		\ENDIF
		\STATE /* Breaks unallocated bundles */
		\STATE $all\_singletons \leftarrow true$
		\FORALL {$\varphi_i \in \Phi_{rest}$} \label{line:MEM_break_unallocated_bundles}
		\IF {$|\varphi_i|>1$}
		\STATE $all\_singletons \leftarrow false$
		\STATE $p_{emptiest} \leftarrow \argmin\limits_{p_i \in \Pi} (utilization(p_i))$
		\STATE $(\varphi_j,\varphi_k) \leftarrow $ExtractMinCut($\varphi_i, 1-utilization(p_{emptiest}), \mathbb{G}$) \label{line:MEM_call_extract_min_cut}
		\STATE $\Phi \leftarrow \Phi \cup \{\varphi_j,\varphi_k\} $ \label{line:MEM_result_of_extract_min_cut}
		\ELSE
		\STATE $\Phi \leftarrow \Phi \cup \{\varphi_i\}$
		\ENDIF
		\ENDFOR
		\STATE /* Opens a new processor core */
		\IF {$all\_singletons = true$}
		\IF {$|\Pi| = N_P$}
		\RETURN unschedulable
		\ENDIF
		\STATE $\varphi \leftarrow \bigcup\limits_{\varphi_i\in \Phi}\varphi_i$ \label{line:MEM_merge_all_bundles}
		\STATE $\Gamma_{p_{new}} \leftarrow \emptyset$
		\STATE $bank(p_{new}) \leftarrow $LeastInterferingBank($N_{bank},\Pi,\mathbb{G},\varphi$)
		\STATE $\Pi \leftarrow \Pi \cup \{p_{new}\}$ \label{line:MEM_add_new_core}
		\STATE $\Phi \leftarrow \{\varphi\}$
		\ENDIF
		\ENDWHILE
		\RETURN schedulable
	\end{algorithmic}
\end{algorithm}

\begin{algorithm}[t]
	\caption[MemoryInterferenceGraph($\Gamma$): creates a memory interference graph]{MemoryInterferenceGraph($\Gamma$)}
	\label{alg:MEM_memory_interference_graph}
	\algsetup{linenosize=\scriptsize}
	\scriptsize
	\begin{algorithmic}[1]
		\REQUIRE $\Gamma$: a taskset ($\Gamma=\{\tau_1,\tau_2,...,\tau_n\}$)
		\ENSURE $\mathbb{G}$: a graph with tasks as nodes and memory interference-intensity among nodes as edge weights
		\STATE Construct a fully-connected undirected graph $\mathbb{G}$ with tasks in $\Gamma$ as nodes
		\FOR {$i \leftarrow 1$ \TO $n$}
		\FOR {$j \leftarrow i + 1$ \TO $n$}
		\STATE Let two processors, $p_1$ and $p_2$, share the same bank partition
		\STATE $\Gamma_{p_1} \leftarrow \{\tau_i\}$
		\STATE $\Gamma_{p_2} \leftarrow \{\tau_j\}$
		\STATE $W_i \leftarrow$ response time of $\tau_i$ 
		\STATE $W_j \leftarrow$ response time of $\tau_j$ 
		\STATE $weight(\mathbb{G}, \tau_i, \tau_j) \leftarrow (W_i- C_i)/T_i + (W_j-C_j)/T_j$		\label{line:MEM_graph_weight}
		\ENDFOR
		\ENDFOR
		\RETURN $\mathbb{G}$
	\end{algorithmic}
\end{algorithm}

To understand the intensity of memory interference among tasks, MIAA first constructs a memory interference graph~$\mathbb{G}$ (line~\ref{line:MEM_build_graph} of Alg.~\ref{alg:MEM_miaa}). The graph $\mathbb{G}$ is a fully-connected, weighted, undirected graph, where each node represents a task and the weight of an edge between two nodes represents the amount of memory interference that the corresponding two tasks can generate. Algorithm~\ref{alg:MEM_memory_interference_graph} gives the pseudo-code for constructing $\mathbb{G}$. For each pair of two tasks, $\tau_i$ and $\tau_j$, the edge weight between the two tasks is calculated as follows. First, the two tasks are assumed to be assigned to two empty cores that share the same bank partition. Then, the response times of the two tasks, $W_i$ and $W_j$, are calculated by using Eq.~\eqref{eq:MEM_response_time}, assuming that no other tasks are executing in the system. Since each task is the only task allocated to its core, the task response time is equal to the sum of the task WCET and the memory interference delay imposed on the task. Hence, we use $(W_i-C_i)/T_i+(W_j-C_j)/T_j$ as the edge weight between $\tau_i$ and $\tau_j$ ($weight(\mathbb{G}, \tau_i, \tau_j)$), which represents the amount of CPU utilization penalty that may occur due to memory interference among $\tau_i$ and $\tau_j$.

\begin{algorithm}[t]
	\caption[LeastInterferingBank($N_{bank}, \Pi, \mathbb{G}, \varphi$): finds a bank partition that likely leads to the least amount of memory interference to unallocated tasks]{LeastInterferingBank($N_{bank}, \Pi, \mathbb{G}, \varphi$)}
	\label{alg:MEM_least_contended_bank}
	\algsetup{linenosize=\scriptsize}
	\scriptsize
	\begin{algorithmic}[1]
		\REQUIRE $N_{bank}$: the number of bank partitions, $\Pi$: a set of processor cores, $\mathbb{G}$: a memory interference graph, $\varphi$: a set of tasks that have not been allocated to cores yet
		\ENSURE $b$: a bank partition index ($1 \le b \le N_{bank}$)
		\IF {$|\Pi| < N_{bank}$}
		\RETURN $indexof(unused\_bank\_partition())$ \label{line:MEM_unused_bank_partition}
		\ENDIF
		\STATE $p_{min} \leftarrow p_1$
		\STATE $w_{p_{min}} \leftarrow \infty$
		\FORALL {$p_i \in \Pi$}
		\STATE $w_{p_i} \leftarrow \sum_{\tau_j\in \Gamma_{p_i}}\sum_{\tau_k \in \varphi} weight(\mathbb{G},\tau_j, \tau_k)$
		\IF {$w_{p_{min}} > w_{p_i}$}
		\STATE $p_{min} \leftarrow p_i$
		\STATE $w_{p_{min}} \leftarrow w_{p_i}$
		\ENDIF
		\ENDFOR
		\RETURN $bank(p_{min})$
	\end{algorithmic}
\end{algorithm}

After constructing the graph $\mathbb{G}$, MIAA opens one core, $p_1$, by adding it to the core set $\Pi$. It is worth noting that every time a new core is opened (added to $\Pi$), a bank partition is assigned to it by the {\tt LeastInterferingBank()} function given in Algorithm~\ref{alg:MEM_least_contended_bank}.  The purpose of {\tt LeastInterferingBank()} is to find a bank partition that likely leads to the least amount of memory interference to the tasks that have not been allocated yet (input parameter $\varphi$ of Alg.~\ref{alg:MEM_least_contended_bank}). If the number of cores in $\Pi$ does not exceed the number of bank partitions ($N_{bank}$), {\tt LeastInterferingBank()} returns the index of an unused bank partition (line~\ref{line:MEM_unused_bank_partition} of Alg.~\ref{alg:MEM_least_contended_bank}). Otherwise, {\tt LeastInterferingBank()} tries to find the least interfering bank by using $\mathbb{G}$ as follows. For each core $p_i$, it calculates $w_{p_i}$ that is the sum of the weights of all edges between the tasks in $p_i$ and the tasks in $\varphi$. Then, it returns the bank partition index of a core $p_{min}$ with the smallest $w_{p_{min}}$.

\begin{algorithm}[t]
	\caption[BestFit($\varphi, \Pi$): finds the best-fit core for a given task in the presence of memory interference]{BestFit($\varphi, \Pi$)}
	\label{alg:MEM_find_best_fit}
	\algsetup{linenosize=\scriptsize}
	\scriptsize
	\begin{algorithmic}[1]
		\REQUIRE $\varphi$: a task bundle to be allocated, $\Pi$: a set of available processor cores
		\ENSURE $p_i$: the processor core where $\varphi$ is allocated ($p_i=invalid$, if the allocation fails)
		\FORALL {$p_i \in \Pi$ in non-increasing order of utilization}
		\STATE $\Gamma_{p_i} \leftarrow \Gamma_{p_i} \cup \varphi$
		\IF {$schedulable(p_i)$}
		\RETURN $p_i$
		\ENDIF
		\STATE $\Gamma_{p_i} \leftarrow \Gamma_{p_i} \setminus \varphi$
		\ENDFOR
		\RETURN $invalid$
	\end{algorithmic}
\end{algorithm}

The allocation strategy of MIAA is to group memory-intensive tasks into a single bundle and allocate as many tasks in each bundle as possible into the same core. To do so, MIAA first groups all tasks in $\Gamma$ into a single bundle and assign that bundle as an element of the set of bundles to be allocated (line~\ref{line:MEM_initial_bundle} of Alg.~\ref{alg:MEM_miaa}). Then, it allocates all bundles in $\Phi$ based on the best-fit decreasing (BFD) heuristic (from line~\ref{line:MEM_bundle_allocation_begin} to line~\ref{line:MEM_bundle_allocation_end}). Here, we define the utilization of a bundle $\varphi_i$ as $\sum_{\tau_k\in \varphi_k} C_k/T_k$. Bundles are sorted in descending order of utilization and MIAA tries to allocate each bundle to a core by using the {\tt BestFit()} function given in Algorithm~\ref{alg:MEM_find_best_fit}. This algorithm finds the best-fit core that can schedule a given bundle with the least amount of remaining utilization. The utilization of a core $p_i$ is defined as $\sum_{\tau_k \in \Gamma_{p_i}} C_k/T_k$, where $\Gamma_{p_i}$ is the set of tasks allocated to the core $p_i$. If a bundle is allocated (line~\ref{line:MEM_bundle_allocated} of Alg.~\ref{alg:MEM_miaa}), that bundle may introduce additional memory interference to all other cores. Therefore, we need to check if the other cores can still schedule their tasksets. If any core becomes unschedulable due to the just-allocated bundle, MIAA uses the {\tt RemoveExcess()} function to remove enough tasks from the core in order to make it schedulable again, and puts the removed tasks as a new bundle into $\Phi$ (line~\ref{line:MEM_call_remove_excess}). Conversely, if a bundle is not allocated to any core (line~\ref{line:MEM_bundle_not_allocated}), it is put into $\Phi_{rest}$ and will be considered later. 

\begin{algorithm}[t]
	\caption[RemoveExcess($p_i, \mathbb{G}$): removes a task from a given core for schedualbility]{RemoveExcess($p_i, \mathbb{G}$)}
	\label{alg:MEM_remove_excess}
	\algsetup{linenosize=\scriptsize}
	\scriptsize
	\begin{algorithmic}[1]
		\REQUIRE $p_i$: a processor core, $\mathbb{G}$: a memory interference graph
		\ENSURE $\varphi$: a set of tasks removed from $p_i$
		\STATE $\varphi \leftarrow \emptyset$
		\REPEAT 
		\STATE $w_{\tau_{min}} \leftarrow \infty$
		\FORALL {$\tau_j \in \Gamma_{p_i}$}
		\STATE $w_{\tau_i} \leftarrow \sum_{\tau_k\in \Gamma_{p_i}\land \tau_k \ne \tau_j}weight(\mathbb{G}, \tau_j, \tau_k)$
		\IF {$w_{\tau_{min}} > w_{\tau_i}$}
		\STATE $\tau_{min} \leftarrow \tau_j$
		\STATE $w_{\tau_{min}} \leftarrow w_{\tau_i}$
		\ENDIF
		\ENDFOR
		\STATE $\Gamma_{p_i} \leftarrow \Gamma_{p_i} \setminus \{\tau_{min}\}$
		\STATE $\varphi \leftarrow \varphi \cup \{\tau_{min}\}$
		\UNTIL {$schedulable(p_i)$}
		\RETURN $\varphi$
	\end{algorithmic}
\end{algorithm}

We shall explain the {\tt RemoveExcess()} function before moving onto the next phase of MIAA. The pseudo-code of {\tt RemoveExcess()} is given in Algorithm~\ref{alg:MEM_remove_excess}. The goal of this function is to make the core $p_i$ schedulable again while keeping as many memory-intensive tasks as possible. To do so, the function extracts one task at a time from the core with the following two steps. In step one, it calculates the weight $w_{\tau_i}$ for each task $\tau_i$, which is the sum of all edge weights from $\tau_i$ to the other tasks on the same core. In step two, it removes a task $\tau_{min}$ with the smallest $w_{\tau_{min}}$ from the core. These two steps are repeated until the core becomes schedulable. Then, the function groups the removed tasks into a single bundle and returns it. 

\begin{algorithm}[t]
	\caption[ExtractMinCut($\varphi, max\_util, \mathbb{G}$): breaks a task bundle into two sub-bundles in the presence of memory interference]{ExtractMinCut($\varphi, max\_util, \mathbb{G}$)}
	\label{alg:MEM_extract_min_cut}
	\algsetup{linenosize=\scriptsize}
	\scriptsize
	\begin{algorithmic}[1]
		\REQUIRE $\varphi$: a task bundle to be broken, $max\_util$: the maximum utilization allowed for the first sub-bundle, $\mathbb{G}$: a memory interference graph
		\ENSURE $(\varphi', \varphi'')$: a tuple of sub-bundles
		\STATE Find a task $\tau_i \in \varphi$ with the highest utilization
		\STATE $\varphi' \leftarrow \{\tau_i\}$
		\STATE $\varphi'' \leftarrow \varphi \setminus \{\tau_i\}$
		\WHILE {$|\varphi''| > 1$}
		\STATE $w_{\tau_{max}} \leftarrow -1$
		\FORALL {$\tau_i \in \varphi''$}
		\STATE $w \leftarrow \sum_{\tau_j \in \varphi'} weight(\mathbb{G},\tau_i,\tau_j)$
		\IF {$w_{\tau_{max}} < w$}
		\STATE $\tau_{max}\leftarrow \tau_i$
		\STATE $w_{\tau_{max}} \leftarrow w$
		\ENDIF
		\ENDFOR	
		\IF {$utilization(\varphi' \cup \{\tau_{max}\}) \le max\_util$}
		\STATE $\varphi' \leftarrow \varphi' \cup \{\tau_{max}\}$
		\STATE $\varphi'' \leftarrow \varphi'' \setminus \{\tau_{max}\}$
		\ELSE
		\STATE {\bf break}
		\ENDIF
		\ENDWHILE
		\RETURN $(\varphi',\varphi'')$
	\end{algorithmic}
\end{algorithm}

Once the bundle allocation phase is done, MIAA attempts to break unallocated bundles in $\Phi_{rest}$ (line~\ref{line:MEM_break_unallocated_bundles} of Alg.~\ref{alg:MEM_miaa}). If an unallocated bundle $\varphi_i$ contains more than one task, it is broken into two sub-bundles by the {\tt ExtractMinCut()} function such that the utilization of the first sub-bundle does not exceed the remaining utilization of the emptiest core (line~\ref{line:MEM_call_extract_min_cut}). If $\varphi_i$ has only one task in it, $\varphi_i$ is put again into $\Phi$. Algorithm~\ref{alg:MEM_extract_min_cut} gives the pseudo-code of {\tt  ExtractMinCut()}. The primary goal of this function is to break a bundle into two sub-bundles while minimizing memory interference among them. To meet this goal, the function first finds a task with the highest utilization in the input bundle and puts that task into the first sub-bundle $\varphi'$. All the other tasks are put into the second sub-bundle $\varphi''$. Then, the function selects a task in $\varphi''$ with the maximum sum of edge weights to the tasks in $\varphi'$ and moves that task to $\varphi'$. This operation repeats as long as $\varphi''$ has enough tasks and the utilization of $\varphi'$ does not exceed the requested sub-bundle utilization ($max\_util$). When {\tt ExtractMinCut()} returns the two sub-bundles, MIAA puts them into $\Phi$ (line~\ref{line:MEM_result_of_extract_min_cut} of Alg.~\ref{alg:MEM_miaa}). 

If all unallocated bundles are singletons, meaning that none of them can be broken into sub-bundles, and the number of cores used is less than $N_P$, MIAA adds a new core to $\Pi$ (line~\ref{line:MEM_add_new_core}). Since the addition of a new core opens up the possibility of allocating all remaining bundles together to the same core, MIAA merges the remaining bundles into a single bundle (line~\ref{line:MEM_merge_all_bundles}) and puts it into $\Phi$. MIAA then repeats the whole process again until $\Phi$ becomes empty. 

MIAA is based on the BFD heuristic which has $O(n\cdot m)$ complexity, where $n$ is the number of tasks and $m$ is the number of processor cores used. On the one hand, the complexity of MIAA could be better than that of BFD due to the bundled allocation of tasks. On the other hand, the complexity of MIAA could be worse than that of BFD due to {\tt RemoveExcess()} which can undo task allocation. However, MIAA is guaranteed to complete in bounded time. The worst case of {\tt RemoveExcess()} happens when it removes all the previously-allocated tasks from cores. Then, MIAA opens a new core until there is any remaining core. If there is no remaining core, MIAA completes and returns a failure result.

It is worth noting that MIAA allocates at most one bank partition to each core, assuming that one bank partition is sufficient to meet the memory requirements of any set of tasks that may be allocated to a single core. This assumption can be satisfied by configuring one bank partition to have multiple DRAM banks, as discussed in Section~\ref{MEM_background_bank_partitioning}. However, we believe that explicitly modeling each task's memory requirement can help in providing better schedulability, which remains as our future work. 

\section{Evaluation}
\label{MEM_evaluation}

In this section, we first compare the memory interference delay observed in a real platform with the one predicted by our analysis.
Then, we evaluate our memory interference-aware allocation algorithm.

\subsection{Memory Interference in a Real Platform}
\subsubsection{Experimental Setup}
The target platform is equipped with an Intel Core i7-2600 quad-core processor running at 3.4 GHz. The on-chip memory controller of the processor supports dual memory channels, but by installing a single DIMM, only one channel is activated in accordance with our system model.\footnote{This is why the DRAM address mapping in \figref{MEM_address-mapping} does not have a bit for channel selection.} The platform uses a single DDR3-1333 DIMM that consists of 2 ranks and 8 banks per each rank. The timing parameters of the DIMM are shown in \tableref{MEM_dram_timing_param}.

We used the latest version of Linux/RK~\cite{LinuxRK, ResourceKernel} for software cache and bank partitioning~\cite{Kim_ECRTS13, Suzuki_ICESS13}.\footnote{Linux/RK is available at \url{https://rtml.ece.cmu.edu/redmine/projects/rk}.} Cache partitioning divides the shared L3 cache of the processor into 32 partitions, and bank partitioning divides the DRAM banks into 16 partitions (1 DRAM bank per partition). For the measurement tool, we used the Linux/RK profiler~\cite{Kim_RTSSWork12} that records execution times and memory accesses (last-level cache misses) using hardware performance counters. 
In addition, we used the memory reservation mechanism of Linux/RK~\cite{Eswaran2005, Kim_RTCSA12} to protect each application against unexpected page swap-outs.
To reduce measurement inaccuracies and improve predictability, we disabled the stride-based and adjacent cache-line prefetchers, simultaneous multithreading, and dynamic clock frequency scaling of the processor. All unrelated system services such as GUI and networking were also disabled. 

It is worth noting that some of our assumptions do not hold in the target platform. First, the processor of the target platform is not fully timing-compositional, in that it can generate multiple outstanding memory requests and hide memory access delay by out-of-order execution. Second, the memory controller of the target platform uses write batching, and there may be other discordances between the memory controller and our system model because detailed information on the memory controller is not open to the public. However, we have chosen the target platform because (i) it is equipped with DDR3 SDRAM which is our main focus in this work, and (ii) it can run an OS that provides the software cache and DRAM bank partitioning features needed for our experiments. We will explore how the aforementioned differences between the target platform and our system model influences our experimental results. 

\subsubsection{Results with Synthetic Tasks}
\label{MEM_results_with_synthetic_tasks}
Our focus here is on analyzing the memory interference of the two types of synthetic tasks. At first, we use the synthetic {\em latency} task~\cite{Yun_RTAS14}, which traverses a randomly ordered linked list. Due to the data dependency of pointer-chasing operations in linked-list traversals, the {\em latency} task generates only one outstanding memory request at a time, nullifying the effect of multiple outstanding memory requests in the target platform. We configure the working set size of the {\em latency} task to be four times of the L3 cache in order to minimize cache hits. In addition, we configure the {\em latency} task to generate only read requests in order to avoid the write-batching effect of the memory controller. 
We execute multiple instances of the {\em latency} task to generate interfering memory requests and to measure delay caused by them. Each instance is allocated to a different core, and assigned 4 cache partitions and 1 bank partition. We evaluate two cases where the instances share and do not share bank partitions. 

\begin{figure}[!ht]
	\centering
	\subfloat[Private bank] {\label{fig:MEM_latency-vs-latency-bankp}
		\hspace{-4pt}\includegraphics[width=0.32\textwidth]{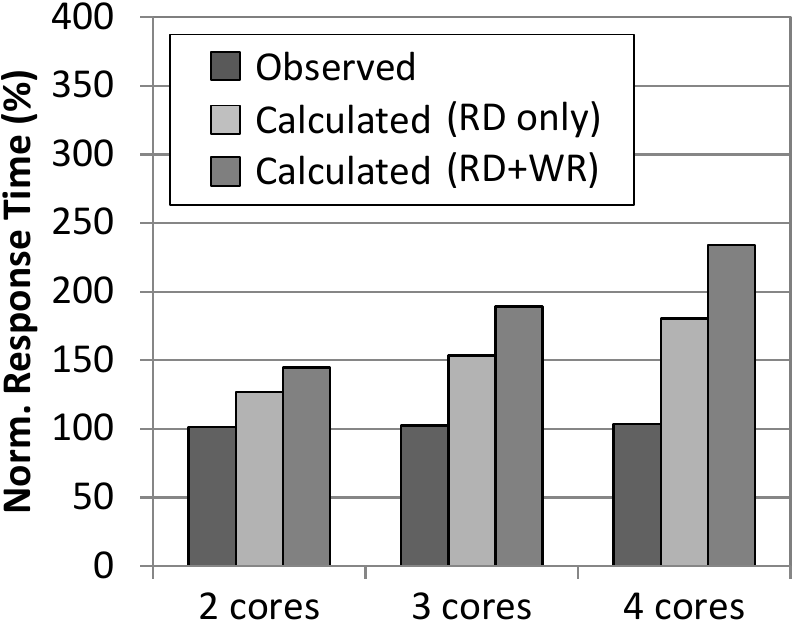}
	}\hspace{30pt}
	\subfloat[Shared bank] {\label{fig:MEM_latency-vs-latency-nobankp}
		\hspace{-4pt}\includegraphics[width=0.32\textwidth]{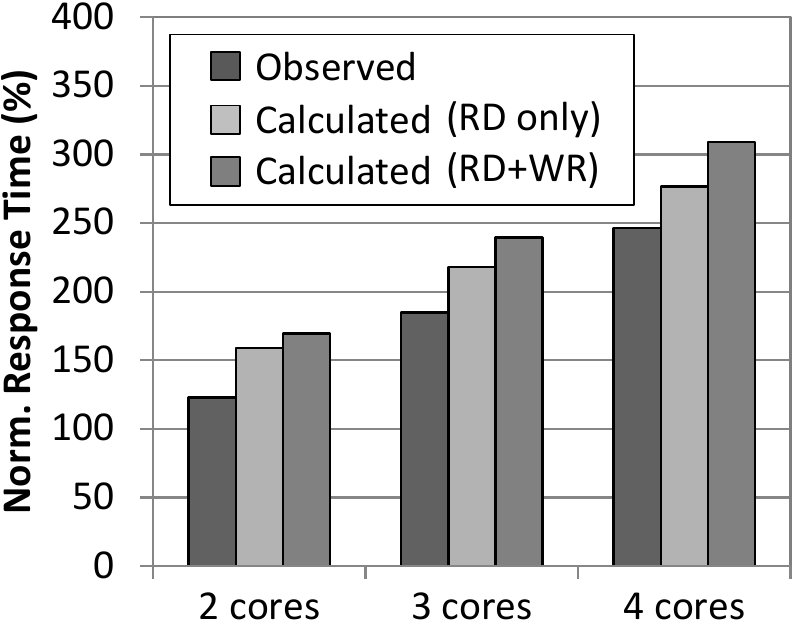}
	}\\
	\caption{Response times of a synthetic task that generates one outstanding read memory request at a time}\label{fig:MEM_latency-vs-latency}
\end{figure}

\figref{MEM_latency-vs-latency} compares the maximum observed response times of one instance of the {\em latency} task with the calculated response times from our analysis, while the other instances are running in parallel. Since the {\em latency} task generates only read requests, we present results from a variation of our analysis, {\em ``Calculated (RD only)''}, which considers only read requests in $L_{inter}^{RW}$ and $L_{hit}$, in addition to {\em ``Calculated (RD+WR)''}, which is our original analysis considering both read and write requests.
The x-axis of each subgraph denotes the total number of cores used, e.g., ``2 cores'' means that two instances run on two different cores and the other cores are left idle. The y-axis shows the response time of the instance under analysis, normalized to the case when it runs alone in the system. Since each instance is allocated alone to each core, the response time increase is equal to the amount of memory interference suffered from other cores. The difference between the observed and calculated values represents the pessimism embedded in our analysis. \figsubref{MEM_latency-vs-latency-bankp} shows the response times when each instance has a private bank partition. We observed a very small increase in response times even when all four cores were used. This is because (i) each instance of the {\em latency} task does not experience intra-bank interference due to its private bank partition, and (ii) each instance generates a relatively small number of memory requests, so each memory request is likely serviced before the arrival of requests from other cores. However, our analysis pessimistically assumes that each memory request may always be delayed by the memory requests of all other cores. In addition, the executions of DRAM commands at different banks can be overlapped as long as DRAM timing constraints are not violated, but our analysis does not consider such an overlapping effect.

\figsubref{MEM_latency-vs-latency-nobankp} depicts the response times when all cores share the same bank partition. We set the re-ordering window size $N_{reorder}$ to zero in our analysis, because the {\em latency} task accesses a randomly ordered linked list and has very low row-buffer locality, thereby hardly generating row-hit requests. As can be seen in this figure, the results from both of our analyses bound the observed response times. The pessimism of our analysis in the shared bank case is not as significant as the one in the private bank case. This is due to the fact that the use of a single shared bank serializes the executions of DRAM commands from multiple cores, making their executions close to the worst-case considered by our analysis.

\begin{figure}[!ht]
	\centering
	\subfloat[Private bank] {\label{fig:MEM_stream-vs-stream-bankp}
		\hspace{-4pt}\includegraphics[width=0.32\textwidth]{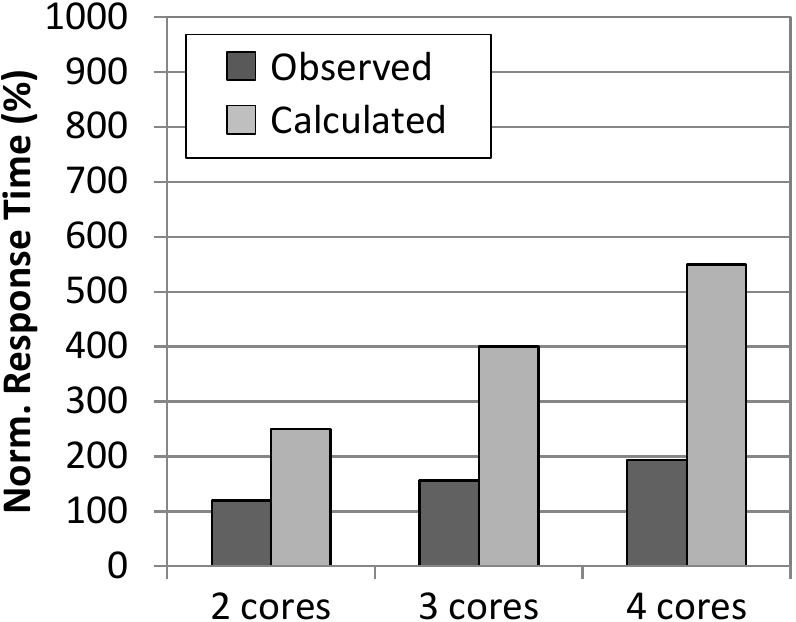}
	}\hspace{30pt}
	\subfloat[Shared bank] {\label{fig:MEM_stream-vs-stream-nobankp}
		\hspace{-4pt}\includegraphics[width=0.32\textwidth]{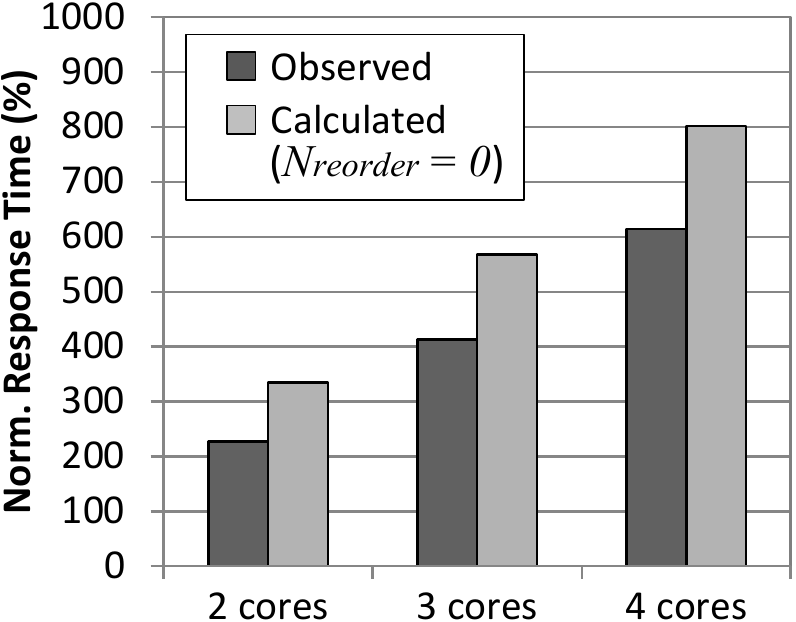}
	}\\
	\caption{Response times of a synthetic task that generates multiple outstanding read and write memory requests at a time} \label{fig:MEM_stream-vs-stream}
\end{figure}

Next, we use a synthetic {\em memory-intensive} task which has the opposite characteristics of the {\em latency} task. The memory-intensive task is a modified version of the {\em stream} benchmark~\cite{STREAM}. The memory-intensive task generates a combination of read and write requests with very high row-buffer locality and little computation. In addition, it can generate multiple outstanding memory requests in the target platform due to the lack of data dependency. Therefore, by using the memory-intensive task, we can identify the effects of the differences between the target platform and our analysis. Similar to the {\em latency} task experiments, we execute multiple instances of the memory-intensive task, with each assigned 4 cache partitions and 1 bank partition, and compare private and shared bank cases.

\figref{MEM_stream-vs-stream} compares the response times of one instance of the memory-intensive task, while the other instances are running in parallel. Since the memory-intensive task generates both read and write requests, we do not consider our read-only analysis used in \figref{MEM_latency-vs-latency}. \figsubref{MEM_stream-vs-stream-bankp} shows the response times with a private bank partition per core. Since the memory-intensive task generates a larger number of memory requests than the {\em latency} task, the observed response times of the memory-intensive task is longer than the ones of the {\em latency} task. Interestingly, although the memory-intensive task might generate multiple outstanding memory requests at a time, our analysis could bound memory interference delay. This is because the extra penalty caused by multiple outstanding memory requests can be compensated by various latency-hiding effects in the target platform. First, an increase in the memory access latency can be hidden by the out-of-order execution of the target processor. Second, the memory controller handles the write requests in batches, which can reduce the processor stall time. However, in order to precisely analyze memory interference in a platform like ours, both the extra penalty caused by multiple outstanding memory requests and the latency-hiding effects from out-of-order execution and write batching should be accounted for by analysis, which remains as future work.

\figsubref{MEM_stream-vs-stream-nobankp} illustrates the response times when all cores share the same bank partition. Since the memory-intensive task has very high row-buffer locality, we expected that a large re-ordering window size $N_{reorder}$ would be needed for our analysis to bound the re-ordering effect.  However, as shown in this figure, our analysis could bound memory interference even when we set $N_{reorder}$ to zero. We suspect that the re-ordering effect on the memory-intensive task is canceled out by the memory latency-hiding techniques of the target platform.

\subsubsection{Results with PARSEC Benchmarks}

We now analyze the memory interference delay of the PARSEC benchmarks~\cite{PARSEC}, which are closer to the memory access patterns of real applications compared to the synthetic tasks used in Section~\ref{MEM_results_with_synthetic_tasks}. A total of eleven PARSEC benchmarks are used in this experiment. Two PARSEC benchmarks, {\em dedup} and {\em facesim}, are excluded from the experiment due to their frequent disk accesses for data files. In order to compare the impact of different amounts of interfering memory requests, we use the two types of synthetic tasks, {\em memory-intensive} and {\em memory-non-intensive}. Each PARSEC benchmark is assigned to Core 1 and the synthetic tasks are assigned to the other cores (Core 2, 3, 4) to generate interfering memory requests. To meet the memory size requirement of the benchmarks, each benchmark is assigned 20 private cache partitions.\footnote{Software cache partitioning simultaneously partitions the entire physical memory space into the number of cache partitions. Therefore the spatial memory requirement of a task determines the minimum number of cache partitions for that task~\cite{Kim_ECRTS13}.} The synthetic tasks are each assigned 4 private cache partitions. Each of the benchmarks and the synthetic tasks is assigned 1 bank partition, and we evaluate two cases where tasks share or do not share bank partitions.
The memory-intensive task is the one used in Section~\ref{MEM_results_with_synthetic_tasks}.  When running in isolation, the memory-intensive task generates up to 40K DRAM requests per msec (combination of read and write). Since it has very high row-buffer locality with little computations, ``40K requests per msec'' is likely close to the maximum possible value that a single core can generate with a single bank partition in the target system. The memory-non-intensive task has a similar structure to the {\em stream} benchmark~\cite{STREAM}, but it has multiple non-memory operations between memory operations, thereby generating much fewer DRAM requests. When running alone in the system, the memory-non-intensive task generates up to 1K DRAM requests per msec. 

\begin{figure}[!ht]
	\centering
	\subfloat[One memory-intensive task on Core 2] {\label{fig:MEM_mem-intensive-bankp-2cores}
		\hspace{-4pt}\includegraphics[width=0.7\textwidth]{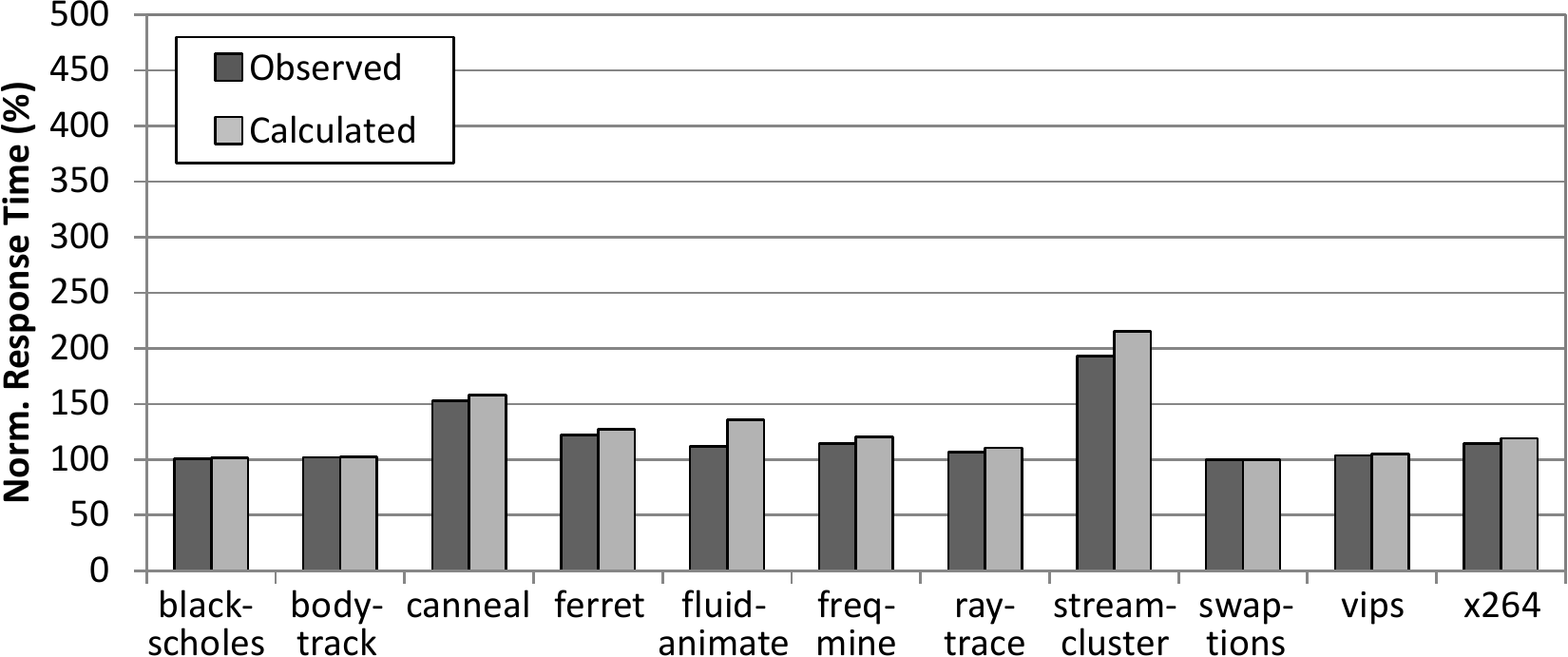}
	}\\
	\subfloat[Two memory-intensive tasks on Core 2 and 3] {\label{fig:MEM_mem-intensive-bankp-3cores}
		\hspace{-4pt}\includegraphics[width=0.7\textwidth]{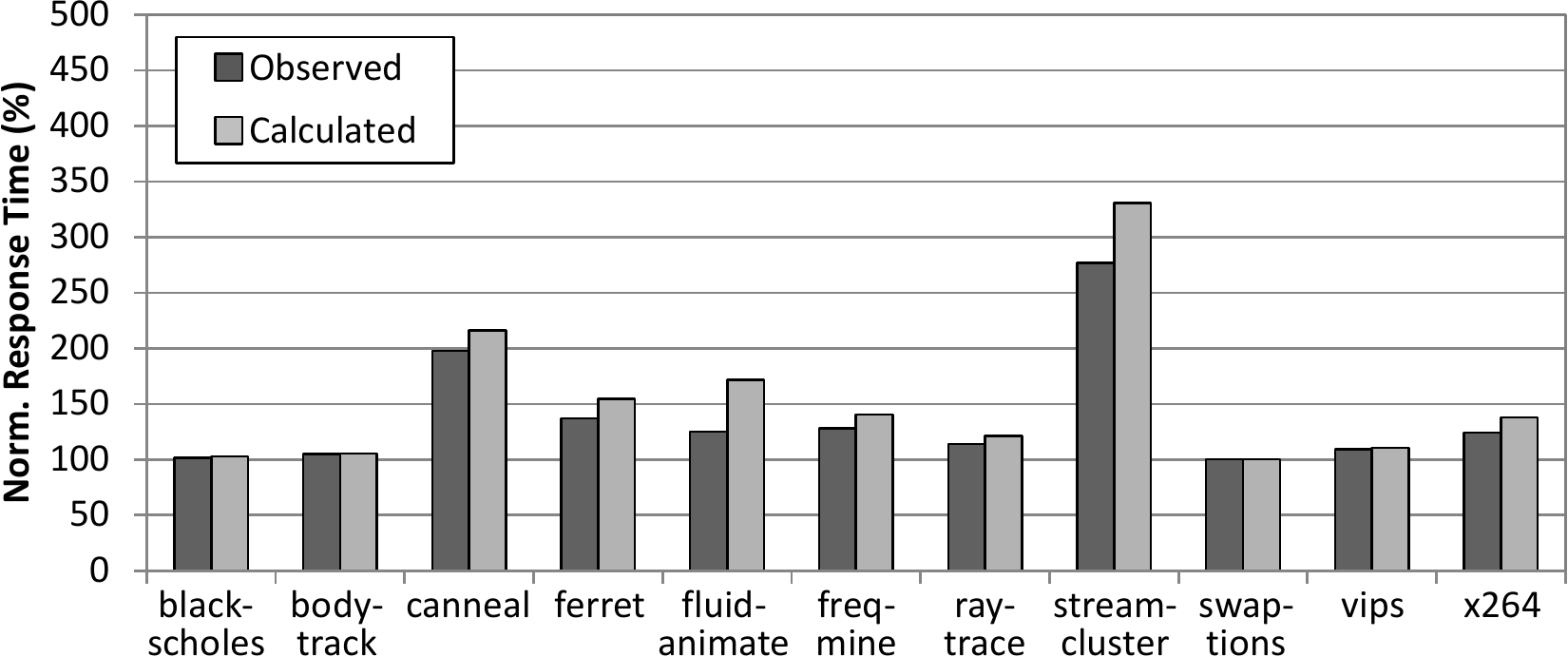}
	}\\
	\subfloat[Three memory-intensive tasks on Core 2, 3 and 4] {\label{fig:MEM_mem-intensive-bankp-4cores}
		\hspace{-4pt}\includegraphics[width=0.7\textwidth]{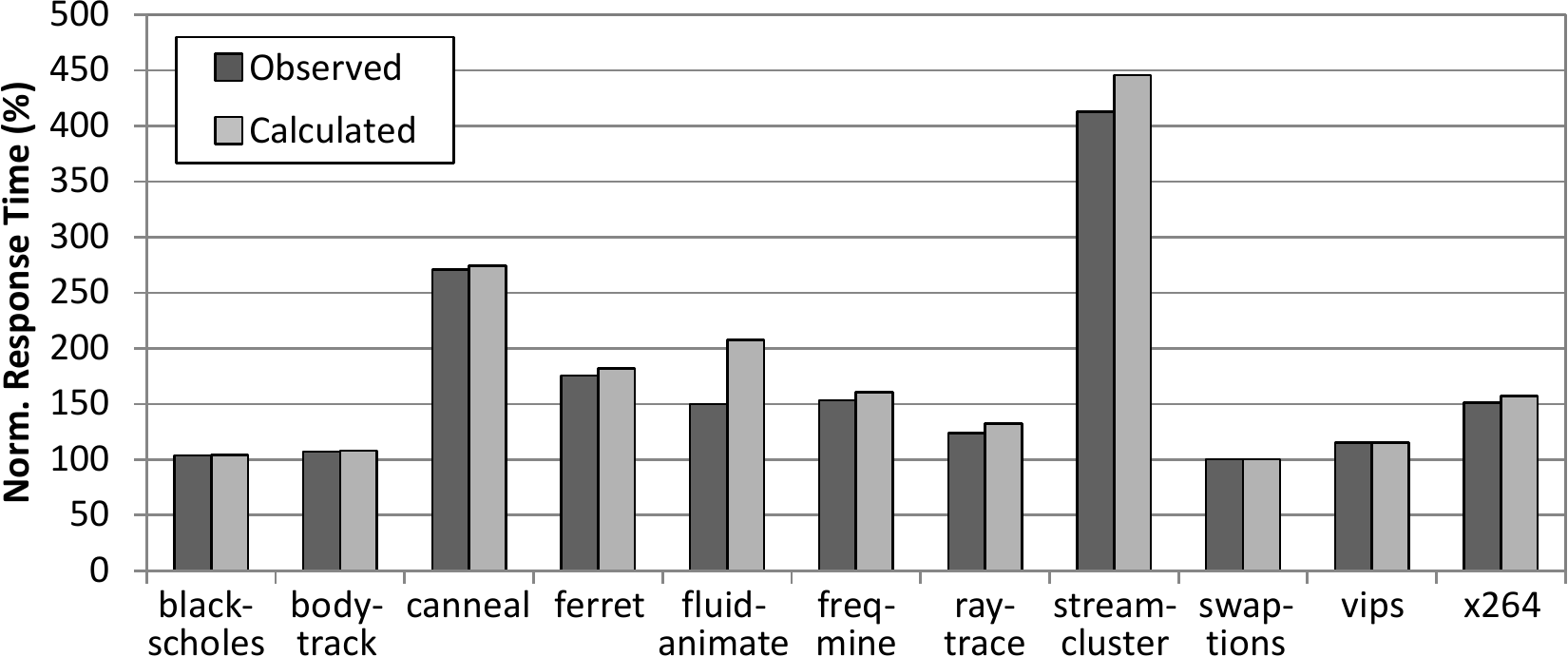}
	}
	\caption{Response times of benchmarks with a private bank partition when memory-intensive tasks run in parallel}\label{fig:MEM_mem-intensive-bankp}
\end{figure}

\begin{figure}[!ht]
	\centering
	\subfloat[One memory-intensive task on Core 2] {\label{fig:MEM_mem-intensive-nobankp-2cores}
		\hspace{-4pt}\includegraphics[width=0.7\textwidth]{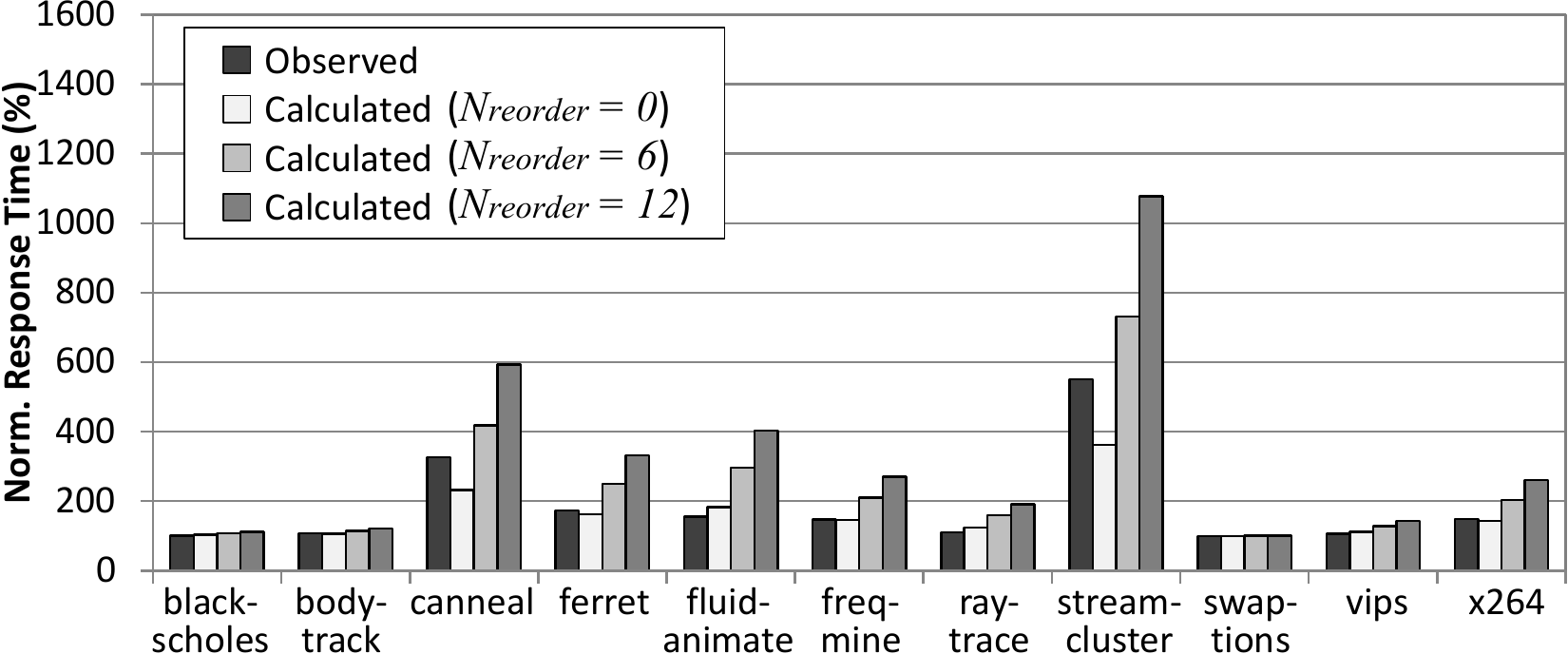}
	}\\
	\subfloat[Two memory-intensive tasks on Core 2 and 3] {\label{fig:MEM_mem-intensive-nobankp-3cores}
		\hspace{-4pt}\includegraphics[width=0.7\textwidth]{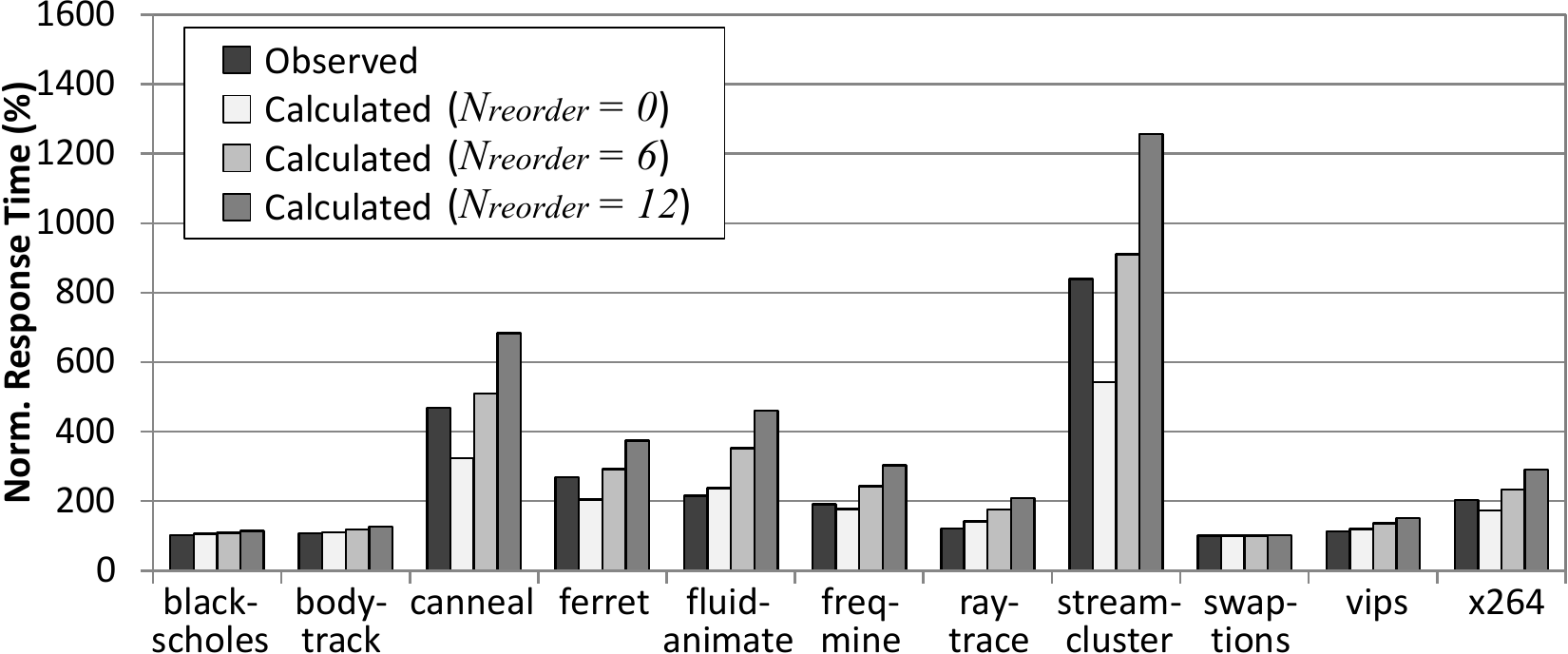}
	}\\
	\subfloat[Three memory-intensive tasks on Core 2, 3 and 4] {\label{fig:MEM_mem-intensive-nobankp-4cores}
		\hspace{-4pt}\includegraphics[width=0.7\textwidth]{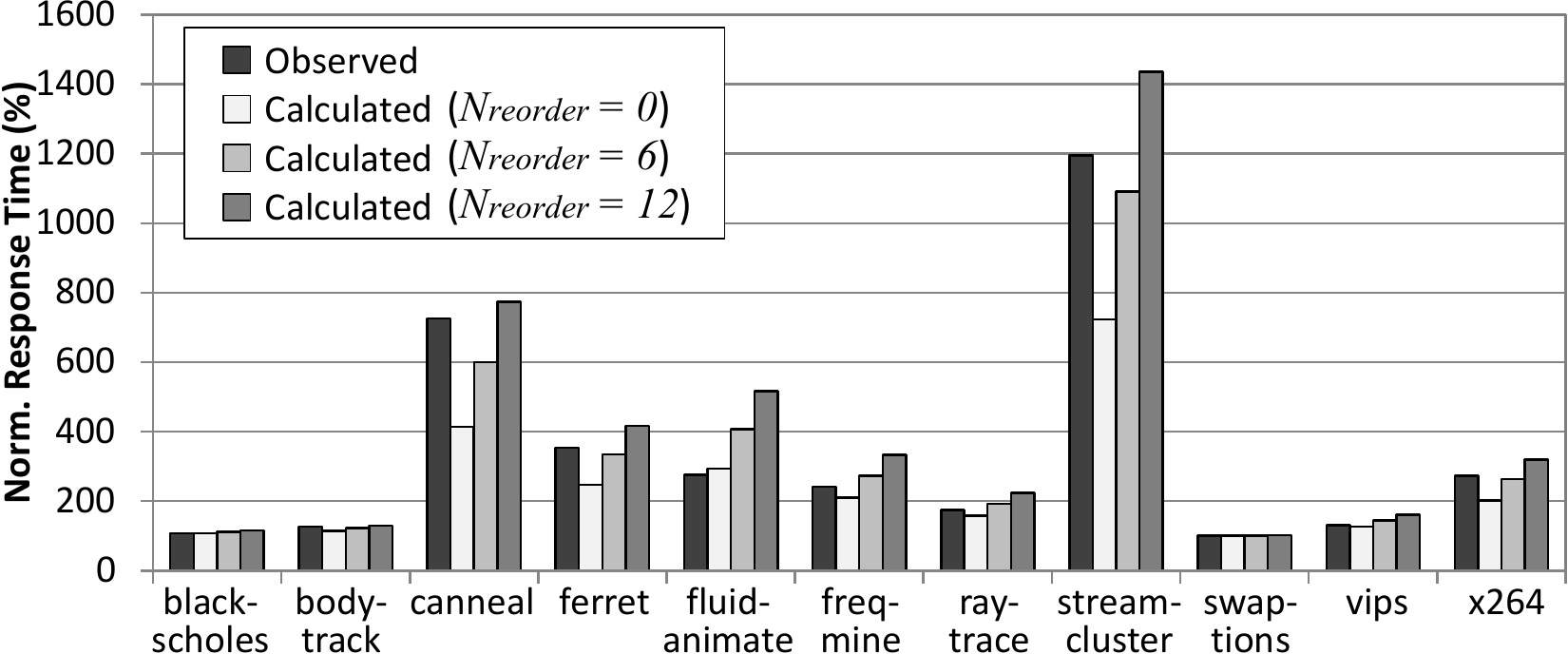}
	}
	\caption{Response times of benchmarks with a shared bank partition when memory-intensive tasks run in parallel}\label{fig:MEM_mem-intensive-nobankp}
\end{figure}

\begin{figure}[t]
	\centering
	\subfloat[Private bank partition] {\label{fig:MEM_mem-non-intensive-bankp}
		\hspace{-4pt}\includegraphics[width=0.7\textwidth]{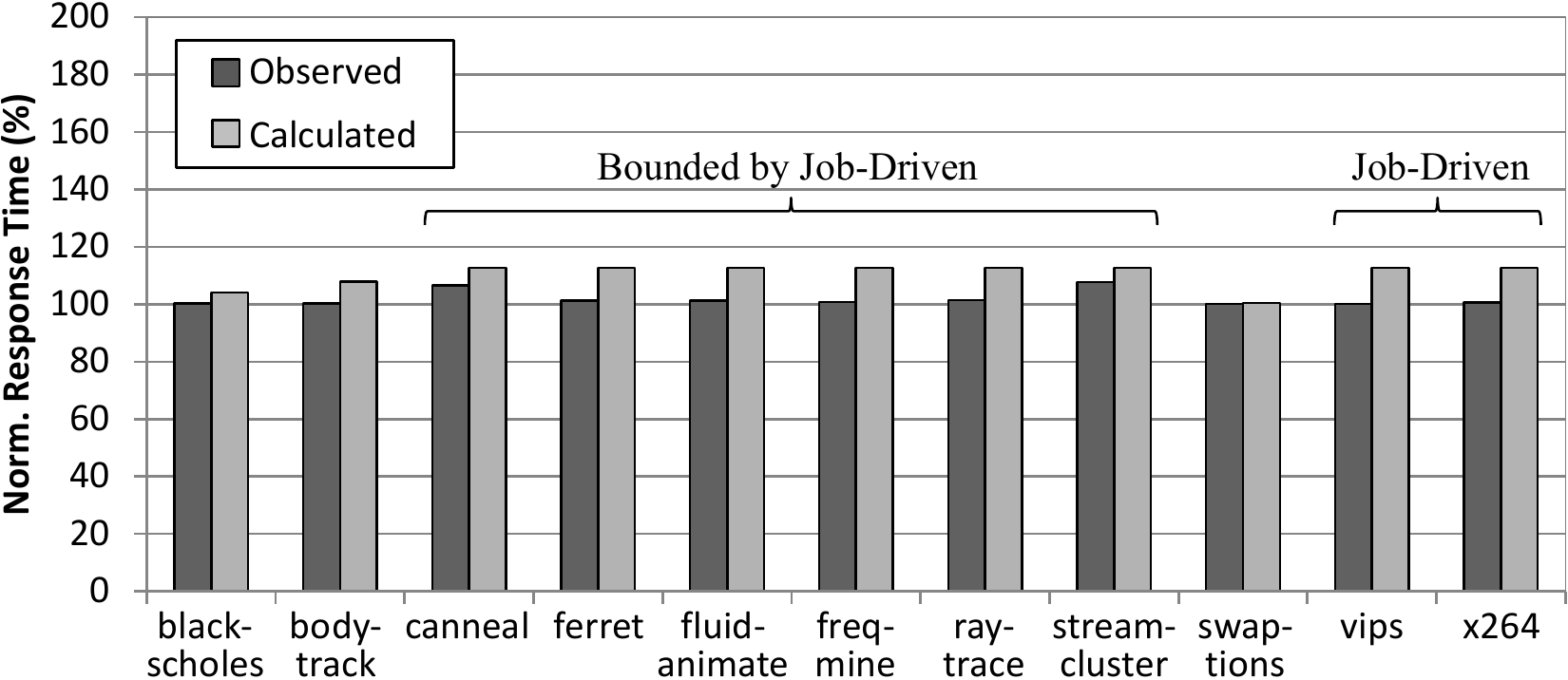}
	}\\
	\subfloat[Shared bank partition] {\label{fig:MEM_mem-non-intensive-nobankp}
		\hspace{-4pt}\includegraphics[width=0.7\textwidth]{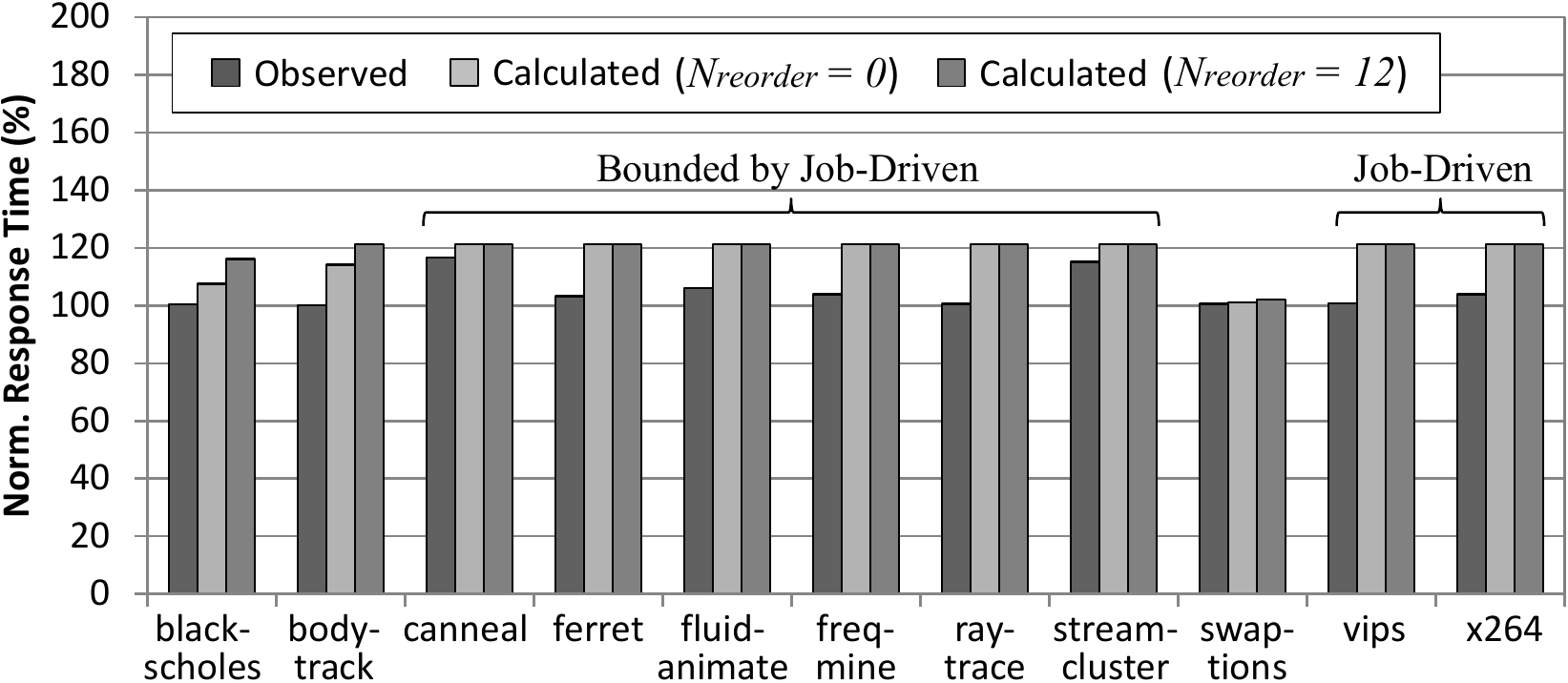}
	}
	\caption{Response times of benchmarks with three memory-non-intensive tasks}\label{fig:MEM_mem-non-intensive-result}
\end{figure} 

We first evaluate the response times of benchmarks with memory-intensive tasks. \figref{MEM_mem-intensive-bankp} and \figref{MEM_mem-intensive-nobankp} compare the maximum observed response times with the calculated response times from our analysis, when memory-intensive tasks are running in parallel. The x-axis of each subgraph denotes the benchmark names, and the y-axis shows the response time of each benchmark normalized to the case when it runs alone in the system. 
\figref{MEM_mem-intensive-bankp} shows the response times with a private bank partition per core. We observed up to 4.1x of response time increase with three memory-intensive tasks in the target system ({\em streamcluster} in \figsubref{MEM_mem-intensive-bankp-4cores}). Our analysis could bound memory interference delay in all cases. The worst over-estimation is found in {\em fluidanimate}. We suspect that this over-estimation comes from the varying memory access patten of the benchmark, because our analysis considers the worst-case memory access scenario. 
Recall that our analysis bounds memory interference based on two approaches: request-driven and job-driven. In this experiment, as the memory-intensive tasks generate an enormous number of memory requests, the response times of all benchmarks are bounded by the request-driven approach. When only the job-driven approach is used, the results are unrealistically pessimistic ($>$10000x; not shown in the figure for simplicity). Thus, these experimental results show the advantage of the request-driven approach.

\figref{MEM_mem-intensive-nobankp} illustrates the response times when all cores share the same bank partition. With bank sharing, we observed up to 12x of response time increase in the target platform. Our analysis requires the re-ordering window size $N_{reorder}$ to calculate the response time when a bank partition is shared. However, we cannot obtain the precise $N_{reorder}$ value because the $N_{cap}$ value of the target platform is not publicly available. Although the $N_{cap}$ value is crucial to reduce the pessimism in our analysis, $N_{reorder}$ can still be bounded without the knowledge of the $N_{cap}$ value, as given in Eq.~\eqref{eq:MEM_N_reorder}. The DRAM used in this platform has $N_{cols}$ of 1024 and $BL$ of 8, so the $N_{reorder}$ value does not exceed 128. In this figure, for purposes of comparison, we present the results from our analysis when $N_{reorder}$ is set to 0, 6, and 12. If we disregard the re-ordering effect of FR-FCFS in this platform ($N_{reorder}=0$), the analysis generates overly optimistic values. In case of {\em streamcluster} with three memory-intensive tasks (\figsubref{MEM_mem-intensive-nobankp-4cores}), the analysis that does not account for the re-ordering effect results in only about half of the observed one. When $N_{reorder}=12$, our analysis can find bounds in all cases. However, this does not necessarily mean that the re-ordering window size of the memory controller is 12. As we have discussed in Section~\ref{MEM_results_with_synthetic_tasks}, multiple outstanding memory requests and various latency hiding techniques cancel their effects on each other in the target platform. Hence, the exact size re-ordering window size of the memory controller can be either greater or smaller than 12. 

We next evaluate the response times with memory-non-intensive tasks. \figsubref{MEM_mem-non-intensive-bankp} and \figsubref{MEM_mem-non-intensive-nobankp} depict the response times of benchmarks with a private and a shared bank partition, respectively, when three memory-non-intensive tasks run in parallel. In contrast to the memory-intensive case, the smallest upper-bounds on the response times are mostly obtained by the job-driven approach due to the low number of interfering memory requests. The experimental results show that our analysis can closely estimate memory interference delay under scenarios with both high and low memory contention.

\subsection{Memory Interference-Aware Task Allocation}

In this subsection, we evaluate the effectiveness of our memory interference-aware allocation (MIAA) algorithm. To do this, we use randomly-generated tasksets and capture the percentage of schedulable tasksets as the metric.

\begin{table}[t]
	\centering
	{
		\footnotesize
		\caption[Base parameters for memory interference experiments]{Base parameters for allocation algorithm experiments}\label{tab:MEM_taskset_param}
		\begin{tabular}{l|c}
			\hline
			Parameters & Values\\\hline
			Number of processor cores ($N_P$) & 8\\
			Number of bank partitions ($N_{bank}$) & 8\\
			Number of tasks to be allocated & 20\\
			Task period ($T_i$) & uniform from [100, 200] msec\\
			Task utilization ($U_i$) & uniform from [0.1, 0.3] \\
			Task WCET ($C_i$) & $U_i\cdot T_i$\\
			Task deadline ($D_i$) & equal to $T_i$ \\
			Ratio of memory-intensive tasks to memory-non-intensive tasks & 5:5\\
			$H_i$ for memory-intensive task & uniform from [10000, 100000]\\
			$H_i$ for memory-non-intensive task & uniform from [100, 1000]\\
			\hline
		\end{tabular}
	}
\end{table}

\subsubsection{Experimental Setup} 
The base parameters we use for experiments are summarized in Table~\ref{tab:MEM_taskset_param}. Once a taskset is generated, the priorities of tasks are assigned by the Rate Monotonic Scheduling (RMS) policy~\cite{Liu_Layland}. The same DRAM parameters as in \tableref{MEM_dram_timing_param} are used, and the re-ordering window size of 12 is used ($N_{reorder}=12$). 

We consider the following six schemes for performance comparison: (i) the best-fit decreasing algorithm (BFDnB), (ii) BFD with bank partitioning (BFDwB), (iii) the first-fit decreasing algorithm (FFDnB), (iv) FFD with bank partitioning (FFDwB), (v) the IA$^3$ algorithm proposed in \cite{Paolieri_RTAS11} (IA3nB), and (vi) IA$^3$ with bank partitioning (IA3wB). The BFD and FFD algorithms are traditional bin-packing heuristics, and IA$^3$ is a recent  interference-aware task allocation algorithm based on FFD. As none of these algorithms is originally designed to consider bank partitioning, all cores share all available bank partitions under BFDnB, FFDnB and IA3nB. Conversely, under BFDwB, FFDwB and IA3wB, bank partitions are assigned to cores in round-robin order so that each core can have a dedicated bank partition. In all these algorithms, we use our response-time test given in Eq.~\eqref{eq:MEM_response_time} to check if a task to be allocated can fit into a candidate core.

IA$^3$ requires each task to have a set of WCET values to represent memory interference as part of the task's WCET. Specifically, IA$^3$ assumes that the worst-case memory interference is affected only by the number of cores used and is not affected by the memory access characteristics of other tasks running in parallel. Hence, under IA3nB and IA3wB, we calculate each task's WCET value as $C_i'=C_i+RD\cdot H_i$, and use $C_i'$ instead of $C_i$ when the FFD module of IA$^3$ sorts tasks in descending order of utilization. We have observed that the use of $C_i'$ with a conventional response-time test~\cite{Joseph_J86} to check if a task to be allocated can fit into a candidate core yields worse performance than the use of $C_i$ with our response-time test. Hence, we use only $C_i'$ when sorting tasks in utilization. In addition, IA$^3$ is designed to allocate cache partitions to cores as well, but we do not consider this feature in our experiments.

\begin{figure}[t]
	\centering
	\subfloat{
		\includegraphics[width=0.7\textwidth]{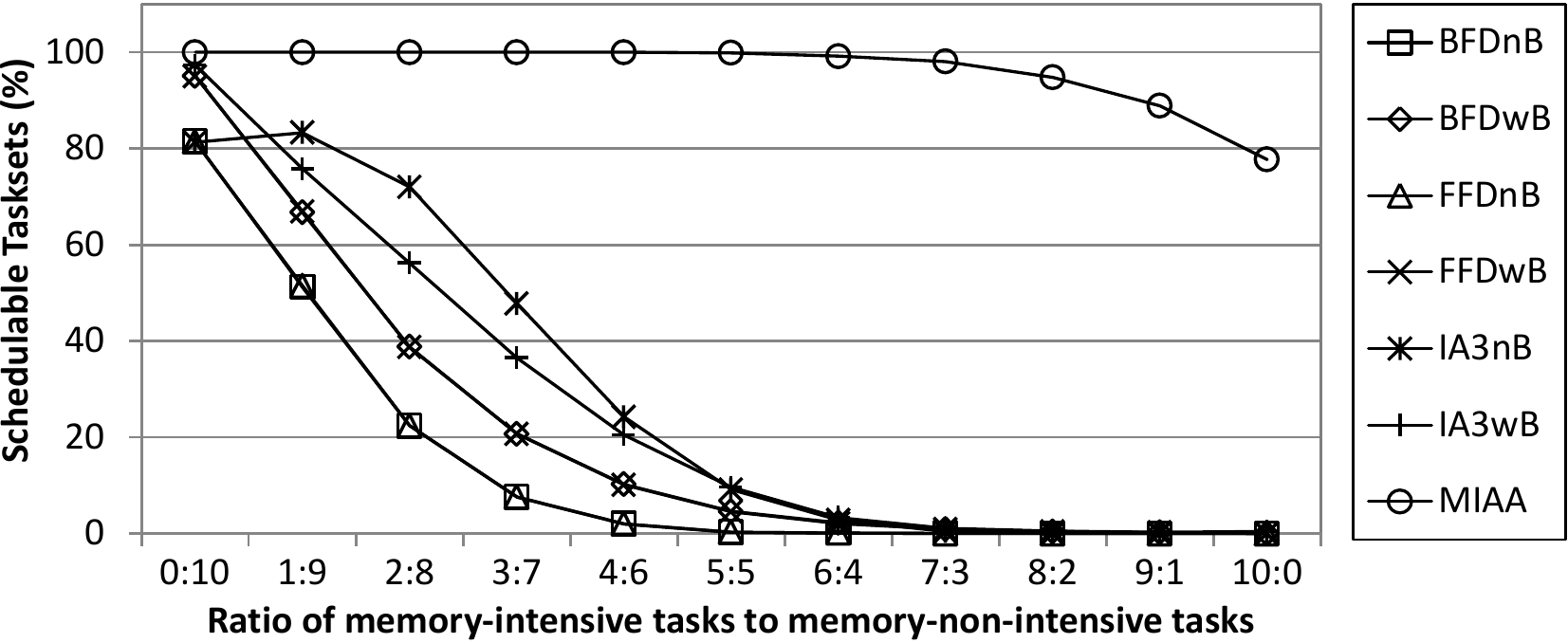}
	}
	\caption{Taskset schedulability as the ratio of memory-intensive tasks increases}
	\label{fig:MEM_task-alloc-memory-intensity}
\end{figure}

\begin{figure}[t]
	\centering
	\subfloat{
		\includegraphics[width=0.7\textwidth]{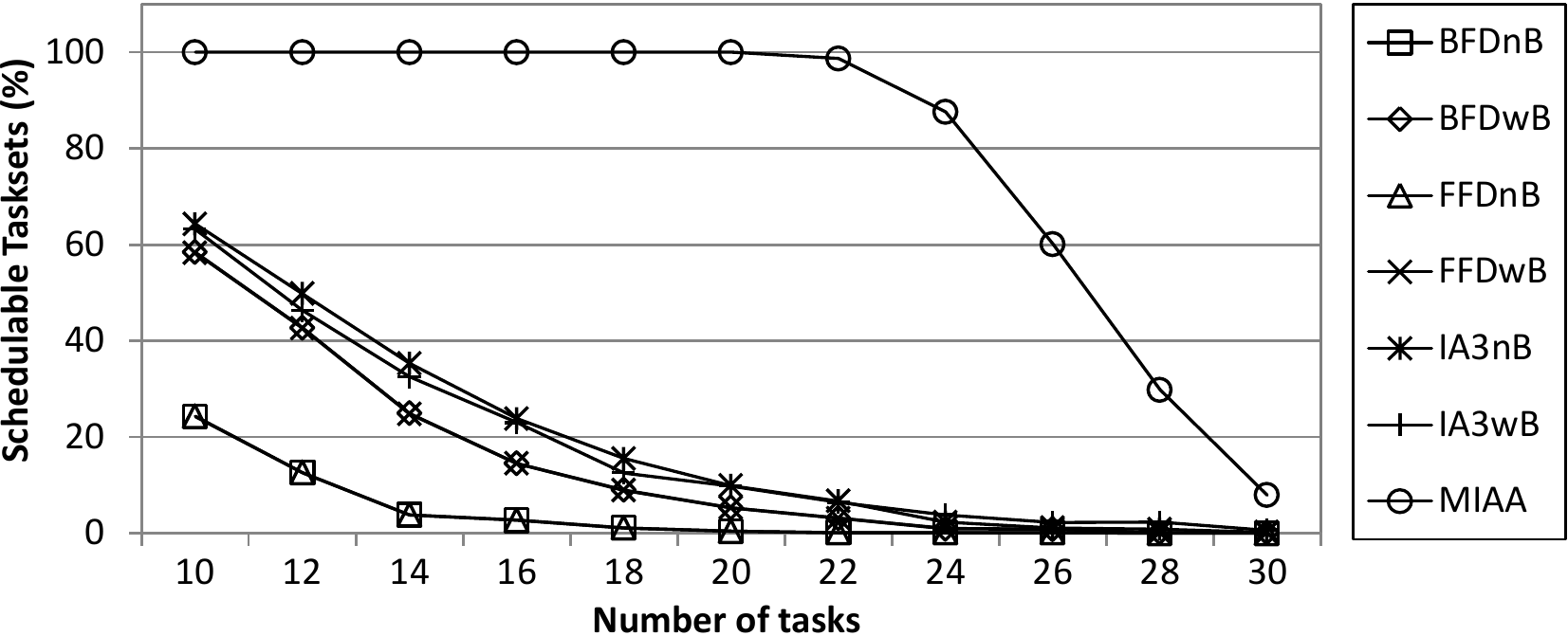}
	}
	\caption{Taskset schedulability as the number of tasks increases}
	\label{fig:MEM_task-alloc-number-of-tasks}
\end{figure}

\begin{figure}[t]
	\centering
	\subfloat{
		\includegraphics[width=0.7\textwidth]{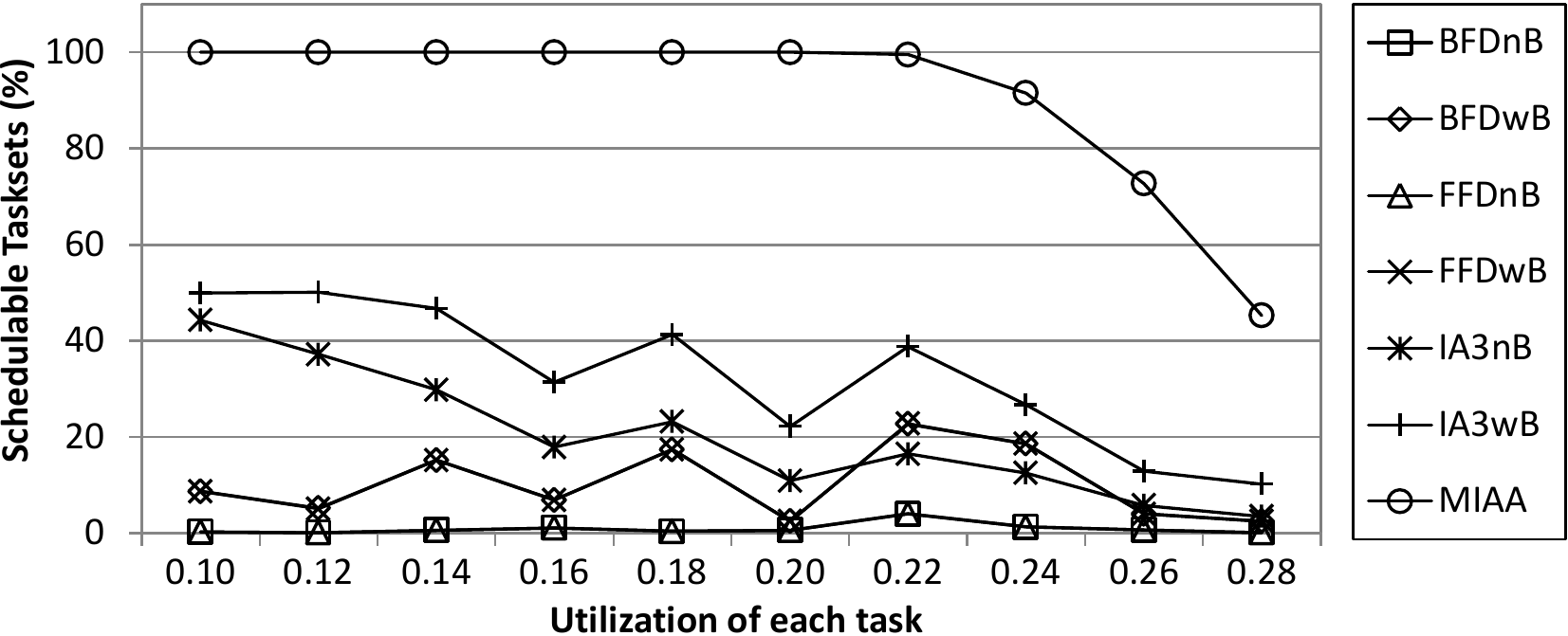}
	}
	\caption{Taskset schedulability as the utilization of each task increases}
	\label{fig:MEM_task-alloc-utilization-of-task}
\end{figure}

\begin{figure}[t]
	\centering
	\subfloat{
		\includegraphics[width=0.7\textwidth]{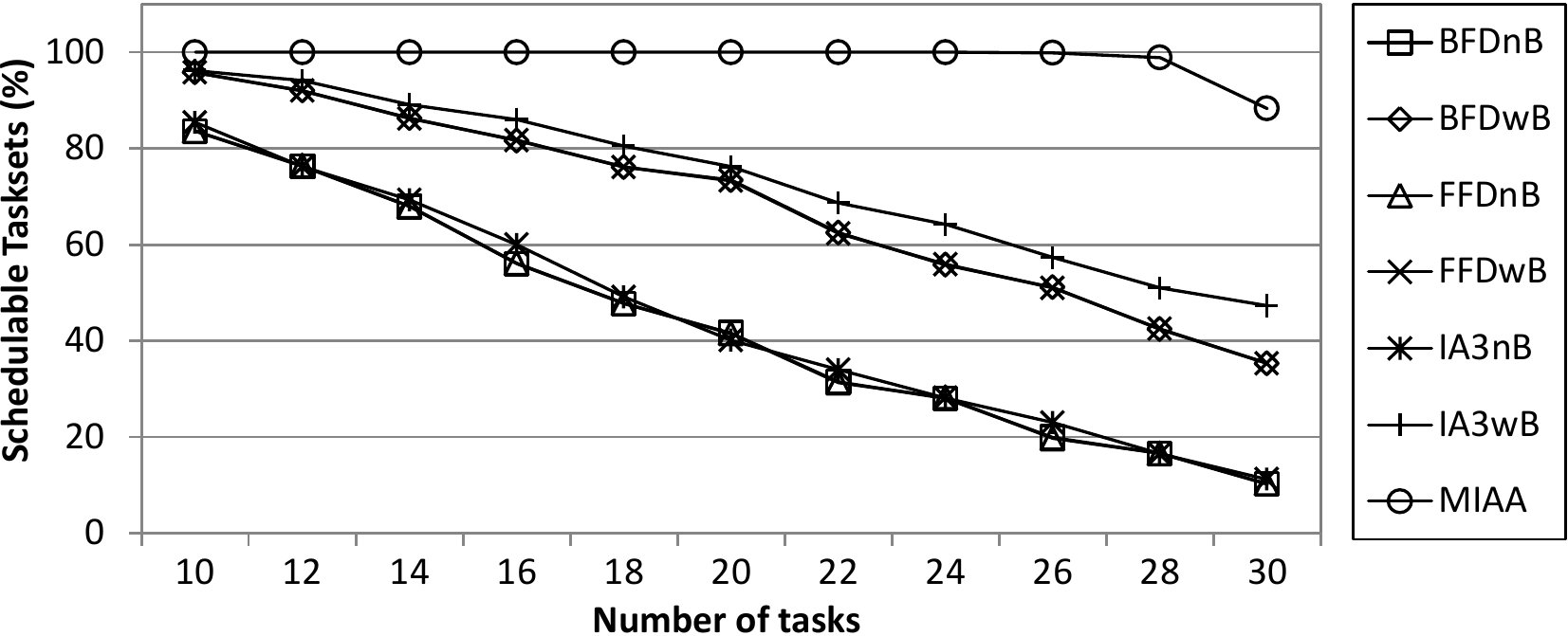}
	}
	\caption{Taskset schedulability when tasks have medium memory intensity}
	\label{fig:MEM_task-alloc-number-of-tasks-medium-intensity}
\end{figure}

\begin{figure}[t]
	\centering
	\subfloat{
		\includegraphics[width=0.7\textwidth]{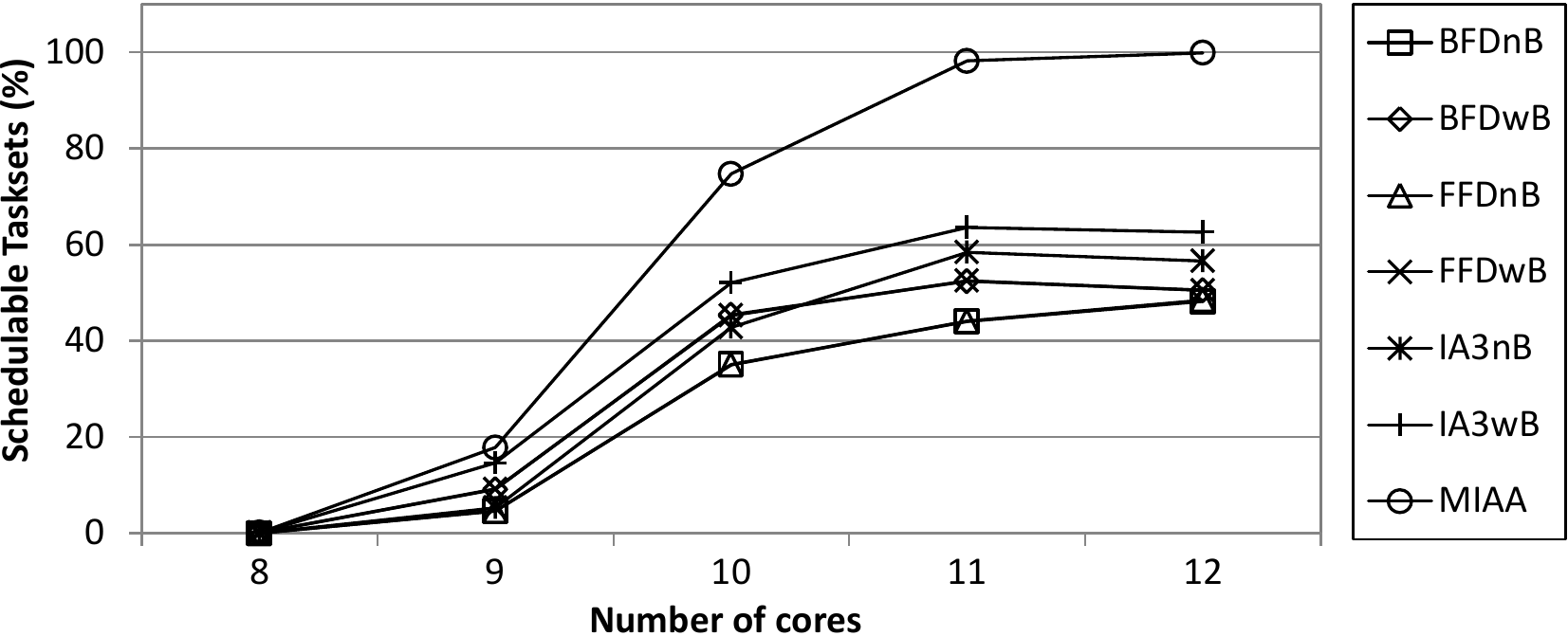}
	}
	\caption{Taskset schedulability as the number of cores increases}
	\label{fig:MEM_task-alloc-number-of-cores}
\end{figure}

\subsubsection{Results}
We explore four main factors that affect taskset schedulability in the presence of memory interference: (i) the ratio of memory-intensive tasks, (ii) the number of tasks, (iii) the utilization of each task, and (iv) the number of cores allowed to use. We generate 10,000 tasksets for each experimental setting, and record the percentage of tasksets where all the tasks pass the response-time test given in Eq.~\eqref{eq:MEM_response_time}. 

\figref{MEM_task-alloc-memory-intensity} shows the percentage of schedulable tasksets as the ratio of memory-intensive tasks to memory-non-intensive tasks increases. The left-most point on the x-axis represents that all tasks are memory-non-intensive, and the right-most point represents the opposite. The percentage difference between MIAA and the other schemes becomes larger as the ratio of memory-intensive tasks increases. For instance, when the ratio is 7:3, MIAA schedules 98\% of tasksets while the other schemes schedule only less than 2\% of tasksets. This big difference mainly comes from the fact that MIAA tries to co-allocate memory-intensive tasks to the same core to avoid memory interference among them. The schedulability of MIAA also decreases as the ratio of approaches to 10:0. This is a natural trend, because the amount of memory interference increases while the number of cores remains unchanged.

We now explore the trend of increasing the number of tasks and the utilization of each task. \figref{MEM_task-alloc-number-of-tasks} and \figref{MEM_task-alloc-utilization-of-task} depict the results. In \figref{MEM_task-alloc-utilization-of-task}, each point $k$ on the x-axis represents that the utilization of each task ranges $[k-0.01,k+0.01]$. As can be seen, MIAA performs the best, and BFDnB and FFDnB show the worst performance. Especially, all the schemes except MIAA schedule less than 70\% of tasksets even if there are only 10 tasks in each taskset or the utilization of each task ranges only $[0.09,0.11]$. This is due to the fact that, when a new task is allocated, tasks that have been allocated earlier on other cores may become unschedulable due to the memory interference from the new task. As MIAA is the only scheme accommodating such a case, it provides significantly higher schedulability than the other schemes.

In \figref{MEM_task-alloc-number-of-tasks-medium-intensity}, we consider the case where tasks have medium memory intensity. For this purpose, we randomly select the $H_i$ value for each task in the range of $[100,10000]$. The results in this figure show that MIAA also outperforms the other schemes when tasks do not have a bimodal distribution of memory intensity.

Lastly, we compare in \figref{MEM_task-alloc-number-of-cores} the percentage of schedulable tasksets under different schemes when the number of cores is more than the number of bank partitions. In this experiment, the number of tasks per taskset is set to 25, and the $H_i$ value and the utilization $U_i$ of each task are randomly selected from $[100,10000]$ and $[0.2,0.4]$, respectively. The percentage under MIAA increases with the number of cores. Especially, MIAA can schedule 98\% of tasksets with 11 cores. However, the other schemes cannot schedule more than 70\% of tasksets, even with 12 cores. These experimental results show that a task allocation algorithm cannot scale well on multi-core platforms without explicit consideration of memory interference and MIAA yields a significant improvement in task schedulability compared to previous schemes. We expect that the performance of MIAA can be further improved by elaborating the functions used by MIAA, such as {\tt LeastInterferingBank()} and {\tt ExtractMinCut()}. This remains as part of future work.

\section{Summary}
\label{MEM_conclusions}


In this chapter, we presented an analysis for bounding memory interference on a multi-core platform with a COTS DRAM system. Our analysis is based on a realistic memory model, which considers the Joint Electron Device Engineering Council (JEDEC) DDR3 SDRAM standard, the FR-FCFS policy of the memory controller, and shared/private DRAM banks. To provide a tighter upper-bound on the memory interference delay, our analysis uses the combination of the request-driven and job-driven approaches. Experimental results from a real hardware platform show that, although some of our assumptions do not hold in the platform used, our analysis can closely estimate the memory interference delay under workloads with both high and low memory contention. 

We also presented a memory interference-aware task allocation algorithm that accommodates memory interference delay during the task allocation phase. Experimental results indicate that our algorithm yields significant benefits in task schedulability, with as much as 96\% more tasksets being schedulable than previous schemes.

As memory-access-intensive tasks become prevalent in cyber-physical systems, contention in shared main memory should be seriously considered and mitigated. We believe that our analysis and task allocation algorithm can be effectively used for designing predictable systems with multi-core platforms. 
Interesting future directions in this area include: (i) analysis on the effect of hardware prefetchers on memory interference delay, (ii) extensions to a non-timing-compositional architecture that allows out-of-order execution and multiple outstanding cache misses, and (iii) examining the effects of upcoming memory schedulers that serve heterogeneous agents.

\chapter{Predictable Cache Management for Virtualization}
\label{chapter_cache_management_for_virtualization}

\begin{figure}[t]
	\centering
	\VS{-5pt}
	\subfloat{
		\includegraphics[width=0.7\textwidth]{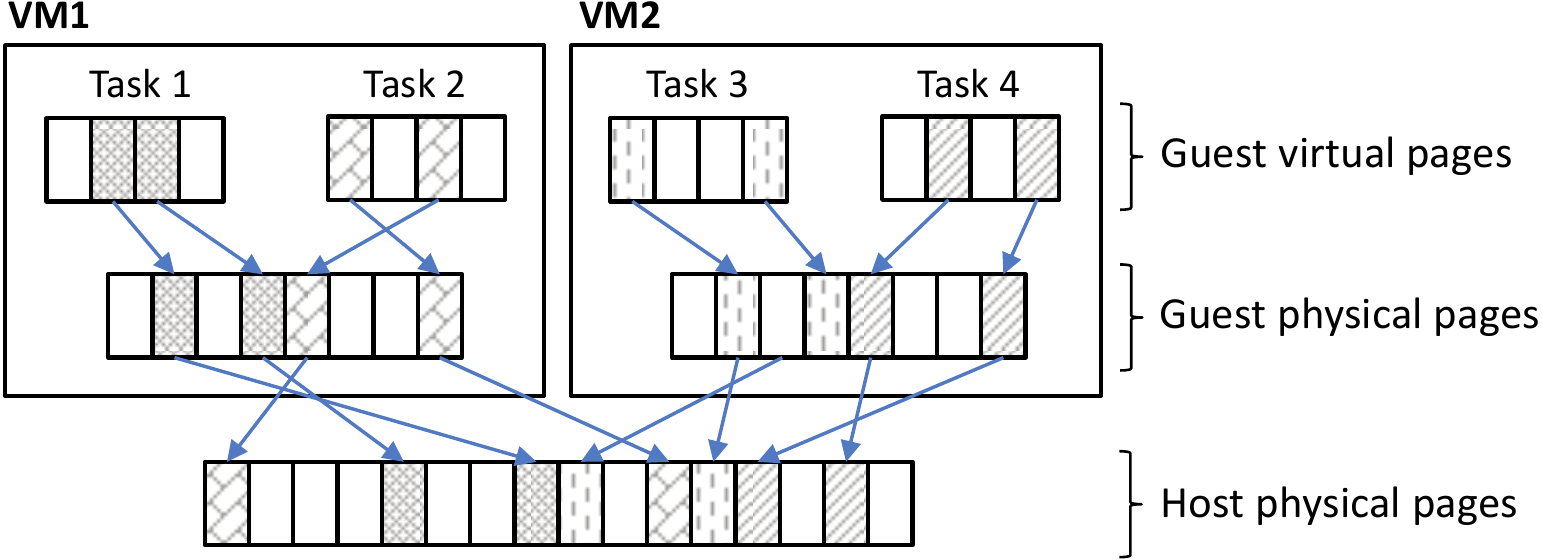}
	}
	\VS{-7pt}
	\caption{Address translation layers in virtualization}
	\VS{-4pt}
	\label{fig:VCACHE_vm_address_space}
\end{figure}

With the growth of processing core counts on recent processors, there is a strong demand for consolidating multiple systems onto a single hardware platform. One of the promising solutions for such consolidation is virtualization. With virtualization, each consolidated system is contained within a virtual machine (VM), which is spatially isolated from other VMs by an additional address translation layer introduced by a hypervisor. \figref{VCACHE_vm_address_space} illustrates the three address layers in modern virtualization platforms, such as Xen~\cite{Xen} and KVM~\cite{KVM}. Guest virtual pages for application tasks within a VM are mapped to {\em guest physical pages} by the guest OS of that VM, and those guest physical pages are mapped to {\em host physical pages} by the hypervisor. 
Using this approach, the hypervisor ensures that any software failure in one VM does not propagate to other VMs.

The additional address layer at the hypervisor, however, makes page coloring and OS-level cache management schemes based on it not to function properly in a VM. Although a guest OS selects guest physical pages for page coloring, those pages may be mapped to host physical pages corresponding to cache sets different from the ones intended by the guest OS, resulting in unpredictable cache allocation. 
Even if page coloring works in a VM, tasks running on other guest OSs that do not support page coloring will suffer from cache interference. Also, cache allocation algorithms developed for native execution environments cannot provide an efficient solution to design VMs in a cache-aware manner and to allocate the host machine's cache to VMs to be consolidated.

In this chapter, we propose a predictable cache management framework for a multi-core virtualization environment. 
To address the problem of cache-to-task allocation in a VM, our framework supports two new hypervisor-level techniques, named vLLC and vColoring. 
vLLC is designed for a VM that runs a guest OS with page coloring support. vLLC provides such a VM with a portion of the host machine's last-level shared cache (LLC) in the form of a {\em virtual LLC}. Then, vLLC enables the guest OS to control the virtual LLC by using its own implementation of page coloring. vColoring, on the other hand, is designed for a VM that runs a guest OS having no page coloring support. vColoring allows the hypervisor to directly assign a portion of the host LLC to a task running in a VM. Hence, with vColoring, we can even control the cache allocation of tasks running on proprietary, closed-source OSs that do not support page coloring. We have implemented prototypes of vLLC and vColoring in the KVM hypervisor running on x86 and ARM multi-core platforms. Experimental results show that vLLC and vColoring are effective in controlling cache allocation to tasks and in addressing cache interference, on both an OS with page coloring (Linux/RK~\cite{Kim_ECRTS13,LinuxRK}) and OSs without page coloring (vanilla Linux and MS Windows Embedded).

In addition, we propose a new cache management scheme as part of our framework. Our scheme determines a cache-to-task allocation that reduces taskset utilization while satisfying timing constraints. Our scheme also designs a VM in a way that the VM's computational demand is captured with respect to the number of cache partitions allocated. Lastly, when VMs are consolidated into the host machine, our scheme finds a cache-to-VM allocation that minimizes the total VM utilization. We use randomly-generated tasksets for the evaluation of our cache management scheme. Experimental results indicate that our scheme yields a significant benefit in VM utilization over other approaches.

The background and related prior work on cache interference were presented in Section~\ref{literature_review_cache_interference}. The system model including assumptions and notation for a last-level cache, tasks and virtual machines can be found in Chapter~\ref{system_model}.

The rest of this chapter is organized as follows. 
Section~\ref{VCACHE_virt_cache_control} presents our vLLC and vColoring techniques. Section~\ref{VCACHE_virt_cache_algo} presents our cache management scheme.
Section~\ref{VCACHE_evaluation} provides detailed evaluation, and Section~\ref{VCACHE_conclusions} summarizes this chapter.

\section{Cache Control in Virtualization}\label{VCACHE_virt_cache_control}

In this section, we provide a brief description on address translation in virtualization. Then, we present our vLLC and vColoring techniques. Both techniques provide a way to allocate cache partitions to individual tasks running in a VM. They do not rely on the page-fault exception of shadow paging or the hardware support of two-dimensional paging. Our techniques differ in their target guest OSs: vLLC is for guest OSs with page coloring (coloring-aware OSs) and vColoring is for guest OSs without page coloring (coloring-unaware OSs). 


\subsection{Address Translation in Virtualization}
\label{VCACHE_address_translation_in_virtualization}

There are three types of addresses in a virtualized environment: guest virtual addresses (GVA), guest physical address (GPA), and host physical address (HPA). Whenever a GVA is accessed, it needs to be translated to the corresponding HPA. Shadow paging and two-dimensional paging are techniques to do such translation in {\em full virtualization} scenarios, where unmodified guest OSs can be used.

\smallskip
\noindent\textbf{Shadow paging:} Under shadow paging, the hypervisor generates shadow page tables where GVAs are directly mapped to HPAs.  Although a guest OS still maintains its own page tables, the memory management unit (MMU) uses the shadow page tables for address translation so that a GVA can be directly translated to its corresponding HPA without having GVA-to-GPA translation. To maintain the validity of contents of the shadow page tables, the hypervisor has to keep track of any change in the guest page tables. 
A well-known approach to doing this is to write-protect the guest page tables, which triggers a page-fault exception to the hypervisor whenever any change is made to the guest page tables.


\smallskip
\noindent\textbf{Two-dimensional paging:} Two-dimensional paging refers to hardware-assisted address translation techniques introduced in recent processors, e.g., AMD Nested Page Tables (NPT), Intel Extended Page Tables (EPT), and ARM Stage-2 Page Tables. Under two-dimensional paging, the MMU can traverse both guest and host page tables. Hence, when a GVA is accessed, the MMU first translates it to a GPA by using the guest page tables and then translates that GPA to an HPA by using the host page tables. Such two-step address translation requires more memory accesses than the direct GVA-to-HPA translation of shadow paging, but it eliminates the overhead of maintaining valid shadow page tables. 

\smallskip
Neither shadow paging nor two-dimensional paging dominates the other in terms of performance~\cite{Wang_ACM11}. It is also currently unknown which technique is preferable for real-time virtualization. Therefore, our goal is to develop cache control techniques that are independent of a specific address translation technique used.

\label{VCACHE_vLLC_scheme}
\begin{figure}[t]
	\VS{-3pt}
	\centering
	\subfloat{
		\includegraphics[width=0.85\textwidth]{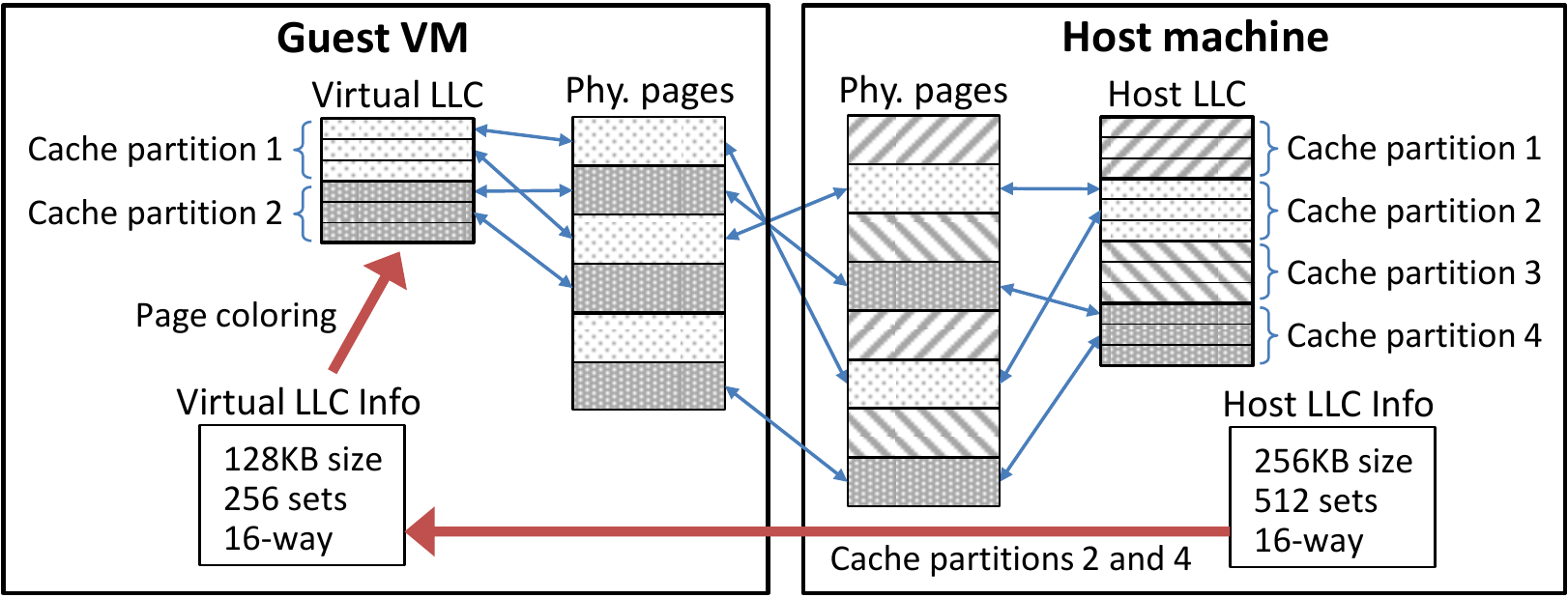}
	}
	\VS{-8pt}
	\caption{vLLC example}
	\label{fig:VCACHE_vLLC_example}
	\VS{-3pt}
\end{figure}

\subsection{vLLC for Coloring-aware Guest OSs}

As previously discussed, page coloring implemented in a guest OS cannot allocate cache partitions to tasks in a VM due to the additional address layer in the hypervisor. vLLC overcomes this limitation. 
The keys to vLLC are (i) to provide a VM with ``virtual LLC (last-level cache)'' information that corresponds to the cache partitions assigned to the VM, and (ii) to map guest physical pages to host physical pages corresponding to the assigned cache partitions. \figref{VCACHE_vLLC_example} illustrates an example of vLLC.
The virtual LLC provided to the VM is different from the actual LLC of the host machine in terms of the size of a cache and the number of cache sets, which are the main factors determining the number of cache partitions. 
In \figref{VCACHE_vLLC_example}, since the hypervisor assigns two cache partitions out of four to the guest VM, the size and the number of cache sets of the virtual LLC are each half of those of the host LLC. Using this virtual LLC, the guest OS can identify that the number of available cache partitions is two. The virtual LLC can be implemented by trapping and emulating cache-related operations, e.g., executions of a {\tt CPUID} instruction on x86 architectures~\cite{IntelDevDoc} and accesses to {\tt CCSIDR} and {\tt CSSERR} registers on an ARM Cortex-A15 architecture~\cite{ARMDevDoc}. 

In addition to the virtual LLC information, vLLC maps guest physical pages (GPPs) to host physical pages (HPPs) such that guest cache partitions are mapped to their corresponding host cache partitions. This can be easily done by the hypervisor because the hypervisor has both the virtual LLC information and the control of the GPP-to-HPP mapping. When a GPP needs to be mapped to an HPP, vLLC in the hypervisor checks the guest cache-partition index of the GPP, finds out the corresponding host cache partition, and maps the GPP to an HPP with that host cache partition. For instance, in \figref{VCACHE_vLLC_example}, cache partitions 2 and 4 of the host machine are represented as cache partitions 1 and 2 in the guest VM, respectively, and GPPs with cache partitions 1 and 2 are mapped to HPPs with cache partitions~2 and 4, respectively. With this approach, a guest OS can allocate cache partitions to tasks. It is worth noting that the GPP-to-HPP mapping happens only once per GPP during the lifetime of a VM. Therefore, once all GPPs used by a task have been populated, vLLC does not cause any runtime overhead to that task. 

There are two constraints in vLLC. First, virtual LLC information should be in accordance with the assumption of page coloring, where the number of cache sets is a power of two. This means that, with vLLC, the number of cache partitions that can be assigned to a VM is restricted to a power of two. Second, it cannot support a guest OS where page coloring is hard-coded (e.g., using fixed cache parameters, instead of checking them when the system boots). If these constraints become a problem, one can disable the page coloring feature of the guest OS and use our vColoring technique.

\subsection{vColoring for Coloring-unaware Guest OSs}
\label{VCACHE_vColoring_scheme}

With vColoring, a VM is assigned two sets of cache partitions, {\em default} and {\em extra}. 
The default set is used whenever a GPP needs to be mapped to an HPP. The hypervisor maps a GPP to an HPP corresponding to one of the cache partitions in the default set.
Hence, by default, all tasks are constrained to use only the default cache partitions. 
The extra set is used for explicit cache allocation requests. When a task running in a VM makes such a request, the hypervisor re-maps all GPPs used by that task to HPPs corresponding to the requested cache partitions in the extra set. 

\begin{figure}[t]
	\VS{-3pt}
	\centering
	\subfloat{
		\includegraphics[width=0.7\textwidth]{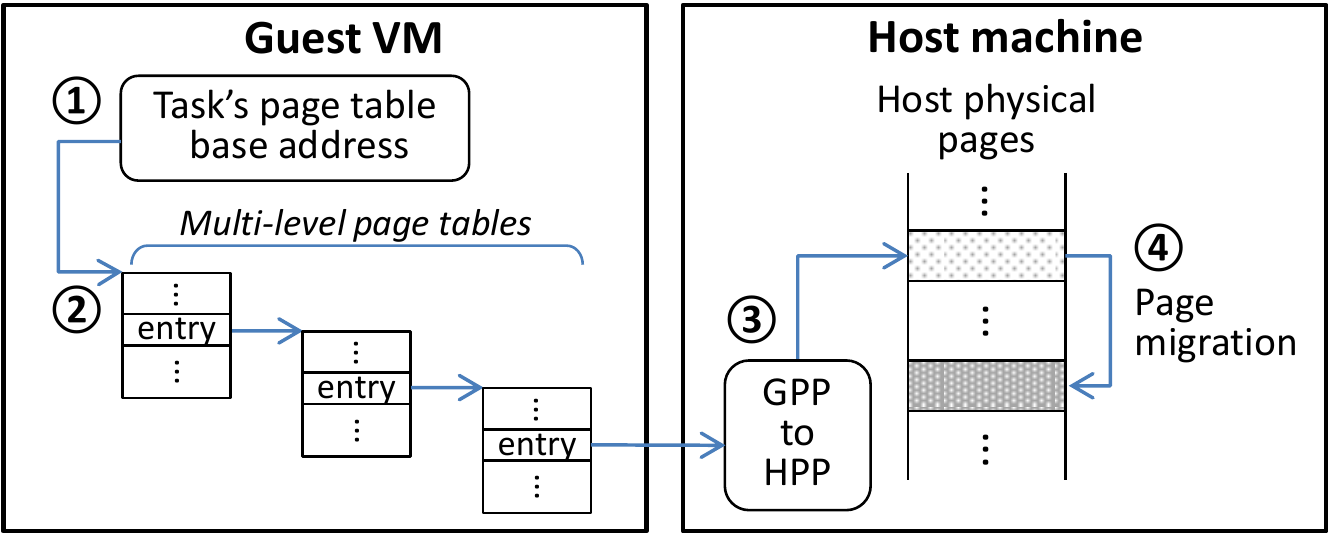}
	}
	\VS{-8pt}
	\caption{Steps for re-mapping GPPs to new HPPs}
	\label{fig:VCACHE_vColoring_page_migration}
	\VS{-2pt}
\end{figure}

\smallskip\noindent\textbf{Re-mapping GPPs to new HPPs:} 
\figref{VCACHE_vColoring_page_migration} shows the detailed steps for re-mapping all the GPPs of a task from the currently-used HPPs to new HPPs for the requested cache partitions. 
The first step is to obtain the task's page table base address (PTBA), which we will explain in detail later. Once the PTBA is obtained, the hypervisor can traverse the task's page tables that are maintained by the guest OS. The second step is to find out {\em present} and {\em user-level accessible} GPPs in the task's page tables. This can be done by checking the information bits of page table entries (PTEs). The third step is to find an HPP mapped to each of the GPPs found in the second step. The fourth step is to migrate each HPP obtained in the third step to a new HPP that corresponds to one of the requested cache partitions. As part of page migration, references to the previous HPP are also updated to the new one.
During all these steps, guest page tables are not changed at all. Therefore, the task can be assigned its requested cache partitions transparent to the guest OS. 
Note that, since the above steps re-map GPPs present at that time, it is desirable to make a cache allocation request at the end of the initialization phase, where a real-time task typically initializes and places all the required data into memory. 

\smallskip\noindent\textbf{PTBA identification:} 
On most processors, the currently-executing task's PTBA is stored in a specific register to facilitate address translation, e.g., a CR3 register in x86 architectures and a Translation Table Base register in ARM architectures. 
We will refer to such a register as a PTB (Page Table Base) register. 
Under shadow paging, the hypervisor traps on write accesses to the PTB register and stores the base address of the corresponding shadow page table into the PTB register. The real PTBA value trapped by the hypervisor is stored in the hypervisor's memory space and used for synchronizing the shadow page table with the guest page table. Under two-dimensional paging, the MMU has two PTB registers, one for a guest PTBA and the other for a host PTBA, and the hypervisor has access to both registers. Therefore, under both address translation techniques, the current task's PTBA can be obtained by the hypervisor. 

\smallskip\noindent\textbf{Cache allocation request:} To make a cache allocation request to the hypervisor, on x86 architectures, a task can use a ``hypercall'' instruction. It can be executed by any user-level task in a VM and results in a world switch to the hypervisor~\cite{IntelDevDoc}. Then, the hypervisor can easily get the task's PTBA because that task is the currently executing one, and the hypervisor can allocate requested cache partitions to the task by following the re-mapping steps explained before. On other architectures, a user-level task is {\em not} allowed to execute a hypercall. Hence, we propose the inclusion of a simple driver that provides a user-level task with an interface to issue a hypercall. Then, the task can make a cache allocation request through the driver interface. Since many recent real-time OSs such as VxWorks~\cite{VxWorks} support implementing device drivers as loadable kernel modules, this approach can be easily used for such OSs without rebuilding the entire kernel image. 

\section{Cache Management Scheme}\label{VCACHE_virt_cache_algo}

In this section, we present our cache management scheme which (i) allocates cache partitions to tasks within a VM while satisfying timing constraints, (ii) designs a VM in a cache-aware manner so that the VM's computational demand is specified w.r.t. the number of cache partitions allocated, and (iii) determines the allocation of cache partitions to a set of VMs to be consolidated.

Cache interference among tasks in a multi-core virtualization environment can be categorized into two types: {\em inter-VCPU} and {\em intra-VCPU}. Inter-VCPU cache interference happens among tasks running on different virtual CPUs (VCPUs). Since those VCPUs can be scheduled on different physical CPU cores (PCPUs) by the hypervisor, tasks on different VCPUs may access the LLC simultaneously. In addition, when a VCPU preempts another VCPU, the cache contents of tasks on the preempted VCPU may be evicted by tasks on the preempting VCPU. Intra-VCPU cache interference happens among tasks running on the same VCPU. Although tasks on the same VCPU cannot access the LLC simultaneously, a task preempting another task may evict the cache contents of the preempted task.

To avoid both inter- and intra-VCPU cache interference, a simple approach would be assigning each task a dedicated set of cache partitions for its own exclusive use. Hence, tasks do not share their assigned cache partitions with others, resulting in no conflicts in the LLC. We will refer to this approach as {\em complete cache partitioning} (CCP). 
However, due to the availability of a limited number of cache partitions, CCP may result in performance degradation. 
Many prior studies in non-virtualized environments~\cite{Busquets_ECRTS97, Bui_RTCSA08, Kim_ECRTS13, Altmeyer_ECRTS14} have shown that sharing of cache partitions among tasks on the same core yields better task schedulability than CCP, and the resulting cache interference can be safely upper-bounded by the notion of {\em cache-related preemption delay} (CRPD). Therefore, our scheme builds on this idea in that (i) cache partitions are not shared among tasks on different VCPUs to prevent inter-VCPU cache interference, and (ii) cache partitions can be shared among tasks on the same core with the cost of intra-VCPU cache interference.

\subsection{Schedulability Analysis}
Before presenting our scheme, we first review VCPU and task schedulability analyses. 
The schedulability of a VCPU $v_i$ can be determined by the following recurrence equation~\cite{Joseph_J86}:\VS{-2pt}
\begin{equation} \label{eq:VCACHE_vcpu_sched_virt_cache}
\begin{split}
W_i^{v,n+1}&=C_i^{v}+\sum_{\substack{v_h\in \mathbb{P}(v_i)\land \pi_h^v>\pi_i^v}} \left\lceil{W_i^{v,n}+J_h^v \over T_h^v}\right\rceil C_h^v\\
\end{split}
\end{equation}
where $W_i^{v,n}$ is the worst-case response time (WCRT) of a VCPU $v_i$ at the $n^{th}$ iteration ($W_i^{v,0} = C_i^{v}$), $\pi_i^v$ is the priority of a VCPU $v_i$, $\mathbb{P}(v_i)$ is the PCPU of $v_i$, and $J_h^v$ is a release jitter ($J_h^v=T_h^v-C_h^v$ for the deferrable server policy and $J_h^v=0$ for the periodic and sporadic server policies~\cite{Bernat_RTSS99}). It terminates when $W_i^{v,n+1}=W_i^{v,n}$, and the VCPU $v_i$ is schedulable if its WCRT does not exceed its period, i.e., $W_i^{v,n}<=T_i^v$. 

The schedulability of task $\tau_j$ running on a VCPU $v_i$ can be determined by:\VS{-2pt}
\begin{equation} \label{eq:VCACHE_task_sched_virt_cache}
\begin{split}
W_j^{n+1}=C_j+
\sum_{\substack{\tau_h\in \mathbb{V}(\tau_j)\\ \land \pi_h>\pi_j}}\left\lceil{W_j^n+J_h \over T_h}\right\rceil (C_h+\gamma_{h,j})+\left\lceil{W_j^n+C_i^v \over T_i^v}\right\rceil(T_i^v-C_i^v)
\end{split}
\end{equation}
where $W_j^n$ is the WCRT of task $\tau_j$ at the $n^{th}$ iteration ($W_j^0=C_j$), $\pi_j$ is the priority of $\tau_j$, $\mathbb{V}(\tau_j)$ is the VCPU of $\tau_j$, $J_h$ is the release jitters of a task $\tau_h$ ($J_h=T_i^v-C_i^v$), and $\gamma_{h,j}$ is the cache-related preemption delay (CRPD) caused by $\tau_h$ and imposed on $\tau_j$. Task $\tau_j$ is schedulable if its WCRT does not exceed its deadline, i.e., $W_j^{n}<=D_j$. 
Note that Eq.~\eqref{eq:VCACHE_task_sched_virt_cache} is based on the task schedulability test under hierarchical scheduling given in~\cite{Saewong_ECRTS02} but extended with the CRPD~\cite{Busquets_ECRTS97, Kim_ECRTS13} to bound intra-VCPU cache interference. For simplicity, we assume that the cache warm-up delay has been taken into account in the WCET of each task. The CRPD $\gamma_{j,i}$ is given by:\VS{-2pt}
\begin{equation} \label{eq:VCACHE_crpd_virt_cache}
\begin{split}
\gamma_{j,i}&=\bigg|\mathbb{S}_j\cap \bigcup_{\tau_k \in \mathbb{V}(\tau_i) \land \pi_k < \pi_j \land \pi_k \ge \pi_i}\mathbb{S}_k\bigg|\cdot\Delta
\end{split}
\end{equation}
where $\mathbb{S}_j$ is the set of cache partitions assigned to $\tau_j$, and $\Delta$ is the maximum time needed to reload data in one cache partitions.\footnote{In case of a write-back cache, $\Delta$ should take into account the effect of a {\em dirty} cache line that requires two memory accesses to fetch a new cache line~\cite{Sebek_TR01}.} 

In the presence of intra-cache VCPU interference, the utilization of a taskset $\Gamma$ allocated to the same VCPU is calculated as follows~\cite{Basumallick_94, Kim_ECRTS13}:\VS{-2pt}
\begin{equation} \label{eq:VCACHE_utilization_virt_cache}
util(\Gamma) = \sum_{\tau_i\in \Gamma} \left(\frac{C_i}{T_i}+\frac{\gamma_{i,n}}{T_i}\right)
\end{equation}
where $n$ is the index of the lowest-priority task in $\Gamma$.

\begin{algorithm}[t]
	\caption[CacheToTaskAlloc($\Gamma, N_{cache}$): finds a cache-to-task allocation]{CacheToTaskAlloc($\Gamma, N_{cache}$)}
	\label{alg:VCACHE_CacheToTaskAlloc}
	\algsetup{linenosize=\scriptsize}
	\scriptsize
	\begin{algorithmic}[1]
		\REQUIRE $\Gamma$: taskset, $N_{cache}$: the number of cache partitions
		\ENSURE Utilization of $\Gamma$ if schedulable, and $\infty$ otherwise
		\IF {$N_{cache} = 0$}
		\RETURN $\infty$
		\ENDIF
		\STATE $cache\_idx \leftarrow 1$
		\FORALL{$\tau_i \in \Gamma$} 
		\STATE /* Find the number of cache partitions for $\tau_i$ */
		\STATE $ S_i \leftarrow \argmin_{1 \le k \le N_{cache}} ({{C_i(k)\over T_i}+{\gamma_{i,n}\over T_i}})$ \label{line:VCACHE_find_number_of_cache_color_for_task}
		\STATE /* Find cache-partition indices for $\tau_i$ */
		\STATE $\mathbb{S}_i \leftarrow \emptyset$
		\FOR {$k \leftarrow 1$ \TO $S_i$} \label{line:VCACHE_find_cache_color_indices_for_task}
		\STATE $\mathbb{S}_i \leftarrow \mathbb{S}_i \cup \{cache\_idx\}$
		\STATE $cache\_idx \leftarrow (cache\_idx + 1)\mod N_{cache}$
		\ENDFOR
		\ENDFOR
		\IF {$schedulable(\Gamma)$}
		\RETURN $util(\Gamma)$
		\ELSE
		\RETURN $\infty$
		\ENDIF
	\end{algorithmic}
\end{algorithm}

\subsection{Allocating Cache Partitions to Tasks}
\label{VCACHE_allocating_cache_colors_to_tasks}

Suppose that we have a set of tasks running on the same VCPU and a set of cache partitions is to be allocated to the tasks. Our goal is to find a cache-to-task allocation that minimizes taskset utilization while satisfying taskset schedulability. When cache sharing is allowed, the problem of cache-to-task allocation is known to be NP-hard~\cite{Bui_RTCSA08}. Hence, we present in Alg.~\ref{alg:VCACHE_CacheToTaskAlloc} a heuristic to solve this problem. It first checks if $N_{cache}$ is non-zero because page coloring requires tasks to be assigned at least one cache partition~\cite{Kim_ECRTS13}. Then, for each task $\tau_i$, it finds the number of cache partitions, $S_i$, that minimizes the sum of the utilization of and CRPD caused by $\tau_i$ (line~\ref{line:VCACHE_find_number_of_cache_color_for_task}). Since cache allocation is not done yet, we approximate $\gamma_{i,n}$ by assuming that all other tasks have been allocated all $N_{cache}$ partitions. Once the number of cache partitions for $\tau_i$ is found, our heuristic finds cache-partitions indices to be allocated (line~\ref{line:VCACHE_find_cache_color_indices_for_task}). It records the index of the next cache partition to be allocated in $cache\_idx$ and begins the allocation starting from $cache\_idx$, with an increment of 1 and a modulo of $N_{cache}$. This approach ensures that the difference in the number of tasks sharing each partition does not exceed 1.

\subsection{Designing a Cache-Aware VM}
\label{VCACHE_designing_cache_aware_vm}

\begin{algorithm}[t]
	\caption[CacheAwareVM($\Gamma, N_{vcpu}, N_{cache}, T^v$): a cache-aware virtual machine designing algorithm]{CacheAwareVM($\Gamma, N_{vcpu}, N_{cache}, T^v$)}
	\label{alg:VCACHE_CacheAwareVM}
	\algsetup{linenosize=\scriptsize}
	\scriptsize
	\begin{algorithmic}[1]
		\REQUIRE $\Gamma$: taskset, $N_{vcpu}$: the number of VCPUs, $N_{cache}$: the number of cache partitions, $T^v$: VCPU period
		\ENSURE Success or Fail
		\STATE $\mathcal{V}\leftarrow \{v_1, v_2, ..., v_{N_vcpu}\}$
		\STATE $\forall v_i \in \mathcal{V}: T_i^v \leftarrow T^v, C_i^v(1,...,N_{cache})\leftarrow T^v, S_i^v \leftarrow 0$ \label{line:VCACHE_init_vcpu_budget}
		\STATE $N_{rem}\leftarrow N_{cache}$ /* Remaining cache partitions */
		\STATE /* Phase 1: Allocate task bundles to VCPUs */
		\STATE $\varphi \leftarrow \Gamma; \Phi \leftarrow \emptyset$\label{line:VCACHE_initial_bundling}
		\WHILE {$util(\varphi)>1$}\label{line:VCACHE_check_initial_bundle_size}
			\STATE $(\varphi', \varphi'') \leftarrow \textrm{BreakBundle}(\varphi, 1,N_{cache})$
			\STATE $\Phi \leftarrow \Phi \cup \{\varphi'\}; \varphi \leftarrow \varphi'$\label{line:VCACHE_save_initial_bundle}
		\ENDWHILE
		\STATE $\Phi \leftarrow \Phi \cup \{\varphi\}$
		\WHILE {$\Phi \ne \emptyset$}
		\STATE /* Allocate bundles */
		\STATE $\Phi_{rest} \leftarrow \emptyset$
		\FORALL {$\varphi_i \in \Phi$ in dec. order of average utilization} \label{line:VCACHE_start_allocate_bundle}
		\STATE $(v_{BF},k) \leftarrow \textrm{BestFitWithCache}(\varphi_i,\mathcal{V},N_{rem})$
		\IF {$v_{BF}\ne invalid$} \label{line:VCACHE_bestfit_found}
		\STATE $\Gamma_{BF}\! \leftarrow\! \Gamma_{BF}\cup\varphi_i; S_{BF}^v\!\leftarrow\! S_{BF}^v\! +\! k; N_{rem}\! \leftarrow\! N_{rem}\! -\! k$
		\ELSE
		\STATE $\Phi_{rest} \leftarrow \Phi_{rest} \cup \{\varphi_i\}$ \label{line:VCACHE_bestfit_not_found}
		\ENDIF
		\ENDFOR \label{line:VCACHE_end_allocate_bundle}
		\STATE /* Break unallocated bundles */
		\STATE $\Phi \leftarrow \emptyset; singletons \leftarrow true$
		\FORALL {$\varphi_i \in \Phi_{rest}$} 
		\IF {$|\varphi_i|>1$} \label{line:VCACHE_break_unallocated_bundles}
		\STATE $singletons\!\leftarrow\!false;\, size\! \leftarrow\! 1\!-\!\min_{v_j \in \mathcal{V}}\!util(\Gamma_j)$
		\STATE $(\varphi', \varphi'') \leftarrow \textrm{BreakBundle}(\varphi_i, size,N_{cache})$
		\STATE $\Phi \leftarrow \Phi \cup \{\varphi', \varphi''\}$
		\ELSE
		\STATE $\Phi \leftarrow \Phi \cup \{\varphi_i\}$
		\ENDIF
		\ENDFOR
		\IF{$singletons = true$} \label{line:VCACHE_all_singletons}
		\RETURN Fail
		\ENDIF		
		\ENDWHILE

		\STATE /* Phase 2: Determine VCPU budget */
		\FORALL {$v_i \in \mathcal{V}$}
		\STATE $C_i^v(0)\leftarrow invalid$
		\FOR {$k \leftarrow 1$ \TO $N_{cache}$}
		\IF {$\textrm{CacheToTaskAlloc}(\Gamma_i, k)\le 1$} \label{line:VCACHE_find_vcpu_budget_with_k_colors}
		\STATE $S_i^v\leftarrow k$; $C_i^v(k) \leftarrow$ Budget $x$ found by binary search
		\ELSE 
		\STATE $C_i^v(k) \leftarrow invalid$
		\ENDIF
		\IF {$C_i(k-1)\ne invalid\land (C_i^v(k-1)<C_i^v(k)\lor C_i^v(k)=invalid$) }\label{line:VCACHE_find_vcpu_budget_use_k_minus_one}
		\STATE $C_i^v(k) \leftarrow C_i^v(k-1)$
		\ENDIF
		\ENDFOR
		\ENDFOR
		\RETURN Success
	\end{algorithmic}
\end{algorithm}

The computational demand of a VM is the aggregate of the demands of all VCPUs in that VM, and it is affected by the allocation of tasks to VCPUs. Especially, when cache-sensitive tasks are allocated together to the same VCPU, the benefit of cache sharing increases, thereby reducing the computational demand. Hence, we propose a cache-aware VM designing algorithm (CAVM) that (i) allocates tasks to VCPUs in a way so as to increase the benefit of cache sharing, and (ii) derives each VCPU's computational demand w.r.t. the number of cache partitions allocated to its taskset. Our algorithm can be used for designing a new VM as well as calculating the computational demand of an existing VM.

Alg.~\ref{alg:VCACHE_CacheAwareVM} presents the pseudo-code of CAVM. It takes four input parameters: $\Gamma$ is a taskset to be allocated, $N_{vcpu}$ is the number of VCPUs in the VM, $N_{cache}$ is the number of available cache partitions, and $T^v$ is the VCPU period that will be assigned to all VCPUs in the VM.\footnote{There are many ways to choose $T^v$. For example, system designers may use a hyperperiod to improve VCPU schedulability, or utilize the findings in \cite{Shin_ACM08} to reduce the overhead of hierarchical scheduling.}
CAVM initializes the budget of each VCPU $v_i$ to be full, i.e., $C_i^v=T^v$, and the number of cache partitions for $v_i$ ($S_i^v$) to zero (line~\ref{line:VCACHE_init_vcpu_budget}).

\begin{algorithm}[t]
	\caption[BestFitWithCache($\varphi, \mathcal{V}, N_{rem}$): finds the best-fit virtual CPU core for a given task bundle]{BestFitWithCache($\varphi, \mathcal{V}, N_{rem}$)}
	\label{alg:VCACHE_BeftFitWithCache}
	\algsetup{linenosize=\scriptsize}
	\scriptsize
	\begin{algorithmic}[1]
		\REQUIRE $\varphi$: a bundle of tasks to be allocated, $\mathcal{V}$: a set of VCPUs, $N_{rem}$: the number of cache partitions
		\ENSURE $(v_i,k)$: a tuple of the best-fit VCPU and the number of additional cache partitions needed 
		\FOR {$k\leftarrow 0$ \TO $N_{rem}$}
		\FORALL {$v_i\!\in\!\mathcal{V}$ in decreasing order of $util(\Gamma_i)$}
		\IF {$\textrm{CacheToTaskAlloc}(\Gamma_i \cup \varphi, S_i^v+k)\le 1$}
		\RETURN $(v_i,k)$
		\ENDIF
		\ENDFOR
		\ENDFOR
		\RETURN $(invalid,-1)$
	\end{algorithmic}
\end{algorithm}

CAVM consists of two phases. The first phase is allocating tasks to VCPUs. 
Our allocation strategy is to group cache-sensitive tasks into a ``bundle'' and allocate as many tasks in the bundle as possible onto the same VCPU. To do so, CAVM first groups all tasks in $\Gamma$ into a single bundle $\varphi$. Then, it checks the utilization of $\varphi$, assuming each task in $\varphi$ uses one dedicated cache partition (line~\ref{line:VCACHE_check_initial_bundle_size}). If it is greater than 1, $\varphi$ is broken into two sub-bundles by \texttt{BreakBundle()} such that the size of the first sub-bundle does not exceed 1. The pseudo-code of \texttt{BreakBundle()} is given in Alg.~\ref{alg:VCACHE_BreakBundle}. To keep as many cache-sensitive tasks as possible in the first sub-bundle, \texttt{BreakBundle()} removes tasks from the first sub-bundle in increasing order of cache sensitivity, which is calculated by $(C_i(1) - C_i(N_{cache}))/T_i$, until the size of the first sub-bundle becomes not to exceed the given size constraint. When \texttt{BreakBundle()} returns, CAVM puts the first sub-bundle into $\Phi$ that is the set of bundles to be allocated (line~\ref{line:VCACHE_save_initial_bundle} of Alg.~\ref{alg:VCACHE_CacheAwareVM}), and continues to check the second bundle if it needs to be broken. As a result, each bundle in $\Phi$ has a utilization not exceeding 1 and is ready to be allocated.

CAVM allocates bundles in $\Phi$ to VCPUs based on the best-fit decreasing (BFD) heuristic (from line~\ref{line:VCACHE_start_allocate_bundle} to line~\ref{line:VCACHE_end_allocate_bundle}). Here, we define the average utilization of a bundle $\varphi_i$ as $\sum_{\tau_j\in\varphi_i}\sum_{k=1}^{N_{cache}}\{(C_j(k)/T_j)/N_{cache}\}$. Bundles are sorted in descending order of average utilization and CAVM tries to allocate each bundle to a VCPU by using \texttt{BestFitWithCache()} given in Alg.~\ref{alg:VCACHE_BeftFitWithCache}. This function finds the best-fit VCPU
that can schedule a given bundle with $k$ additional cache partitions assigned to it, where $k$ starts from 0 to the number of remaining cache partitions ($N_{rem}$). If a best-fit VCPU is found (line~\ref{line:VCACHE_bestfit_found} of Alg.~\ref{alg:VCACHE_CacheAwareVM}), the bundle is allocated to that VCPU, and the number of cache partitions of that VCPU ($S_{BF}^v$) and the number of remaining cache partitions are updated. Otherwise, the bundle is put into $\Phi_{rest}$ (line~\ref{line:VCACHE_bestfit_not_found}). 

\begin{algorithm}[t]
	\caption[BreakBundle($\varphi, size, N_{cache}$): breaks a task bundle into two sub-bundles in the presence of cache interference]{BreakBundle($\varphi, size, N_{cache}$)}
	\label{alg:VCACHE_BreakBundle}
	\algsetup{linenosize=\scriptsize}
	\scriptsize
	\begin{algorithmic}[1]
		\REQUIRE $\varphi$: a bundle to be broken, $size$: the size constraint for the first sub-bundle, $N_{cache}$: the number of partitions
		\ENSURE $(\varphi',\varphi'')$: a tuple of sub-bundles 
		\STATE $\varphi'\leftarrow \varphi; \varphi'' \leftarrow \emptyset$
		\FORALL {$\tau_i\in \varphi$ in increasing order of cache sensitivity}
			\STATE $\varphi' \leftarrow \varphi' \setminus \tau_i$; $\varphi'' \leftarrow \varphi'' \cup \tau_i$
			\STATE /* Get $util(\varphi')$ assuming each task uses one partition */
			\IF {$util(\varphi') \le size$}
				\STATE \textbf{break}
			\ENDIF
		\ENDFOR
		\RETURN $(\varphi', \varphi'')$
	\end{algorithmic}
\end{algorithm}

Then, CAVM attempts to break all unallocated bundles in $\Phi_{rest}$. 
If a bundle in $\Phi_{rest}$ has more than one task (line~\ref{line:VCACHE_break_unallocated_bundles}), it is broken into two sub-bundles by \texttt{BreakBundle()} such that the size of the first sub-bundle does not exceed the remaining capacity of a VCPU having the minimum taskset utilization. The resulting two sub-bundles are put into $\Phi$ so that they can be allocated in the next iteration.
If all unallocated bundles are singletons (line~\ref{line:VCACHE_all_singletons} of Alg.~\ref{alg:VCACHE_CacheAwareVM}), CAVM returns {\em fail} because none of these bundles can be broken into sub-bundles. 

After finishing the first phase of task allocation, each VCPU $v_i$ is allocated its own taskset $\Gamma_i$. The second phase of CAVM determines the budget $C_i^v(k)$ of a VCPU $v_i$ for all possible $k$ values ($1\le k \le N_{cache}$). If $\Gamma_i$ with $k$ cache partitions is schedulable (line~\ref{line:VCACHE_find_vcpu_budget_with_k_colors}), CAVM finds the minimum possible budget $x$ of $v_i$ by using a binary search between 0 and $T_v$, and sets $C_i^v(k)$ to $x$. Otherwise, $C_i^v(k)$ is marked as invalid. Here, it may happen that, due to CRPD, $C_i^v(k-1)$ is smaller than $C_i^v(k)$ or is valid while $C_i^v(k)$ is invalid. In such cases (line~\ref{line:VCACHE_find_vcpu_budget_use_k_minus_one}), CAVM sets $C_i^v(k)$ to $C_i^v(k-1)$ and lets $v_i$ use only $k-1$ cache partitions when $k$ partitions are given. With this, CAVM can find $C_i^v(k)$ values that are monotonically decreasing with $k$.

\subsection{Allocating Host Cache Partitions to VMs}

We now present our cache-to-VM allocation algorithm that determines the number of cache partitions for each VCPU of the VMs to be consolidated, while minimizing the total utilization of those VMs. 
Once cache partitions are allocated, conventional bin-packing heuristics such as BFD can be used to allocate the VCPUs of those VMs to PCPUs.

Let $\rho_{i,k}$ denote the number of cache partitions assigned to $v_i$ when a total of $k$ partitions is provided in the host machine, and $\mathcal{V}$ denote a set of VCPUs of all VMs to be consolidated. Then, the total utilization of VMs with $k$ cache partitions is given by:\VS{-5pt}
\begin{equation} \label{eq:VCACHE_total_vm_util}
\begin{split}
\sum_{v_i\in\mathcal{V}} {C_i^v(\rho_{i,k})\over T_i}
\end{split}
\end{equation}

\VS{-2pt}
To find the minimum total utilization of VMs with $k$ cache colors, $U(k)$, 
we use a dynamic programming approach. Let $x_i$ denote the smallest number of cache partitions that gives a valid budget for $v_i$, i.e., $C_i^v(x_i)\ne invalid$ and $C_i^v(x_i-1)=invalid$, and let $z$ denote the minimum number of cache partitions needed to schedule all VCPUs in $\mathcal{V}$. Then, $z$ is calculated by $z=\sum_{v_i\in\mathcal{V}}x_i$, and $\rho_{i,z}$ is equal to $x_i$ because there is only one valid cache allocation to $v_i$ when $z$ cache partitions are provided. For $k<z$, we represent $U(k)$ as $\infty$ because there is no valid allocation. For $k=z$, $U(k)$ can be computed by Eq.~\eqref{eq:VCACHE_total_vm_util} because $\rho_{i,k}=x_i$. For $k=z+1$, $U(k)$ cannot be computed by Eq.~\eqref{eq:VCACHE_total_vm_util} because $\rho_{i,k}$ is unknown. Instead, we can compute $U(k)$ from $U(z)$. Recall that our CAVM algorithm given in Section~\ref{VCACHE_designing_cache_aware_vm} ensures that $C_i^v(k)$ is monotonically decreasing with $k$. Hence, if any additional cache partition is assigned to $v_i$, a non-negative utilization gain is obtainable. Based on this observation, we can compute $U(k=z+1)$ by $U(z)-\max {C_i^v(\rho_{i,z})-C_i^v(\rho_{i,z}+1)\over T_i^v}$, which subtracts the maximum utilization gain made by one additional cache partition from $U(z)$. We can also find $\rho_{i,z+1}$ by recording the number of cache partitions of $v_i$ that leads to $U(z+1)$. For $k=z+2$, $U(k)$ can be calculated by the minimum between $U(z)-\max {C_i^v(\rho_{i,z})-C_i^v(\rho_{i,z}+2)\over T_i^v}$, which subtracts the maximum gain by two additional partitions from $U(z)$, and $U(z+1)-\max {C_i^v(\rho_{i,z+1})-C_i^v(\rho_{i,z+1}+1)\over T_i^v}$, which subtracts the maximum gain by one additional cache partition from $U(z+1)$. This approach can be extended to all $k>z$, and $U(k)$ is given by the following recurrence:\VS{-2pt}
\begin{equation} \label{eq:VCACHE_total_vm_util_recurrence}
\begin{split}
U(k)=\left\{
\begin{array}{lr}
\vspace{5pt}\infty\,\,\textrm{(unschedulable)} &: k<z\\
\scalebox{0.90}{$\displaystyle\sum_{v_i\in\mathcal{V}} {C_i^v(\rho_{i,k})\over T_i} $} &: k=z\\
\scalebox{0.90}{$\displaystyle \min_{z\le k'< k} \left(U(k')-\max_{v_i\in\mathcal{V}} {{C_i^v(\rho_{i,k'})-C_i^v(\rho_{i,k'}+(k-k'))}\over{T_i^v}}\right)$}&: k>z
\end{array}
\right.
\end{split}
\end{equation}

\VS{-3pt}

\begin{algorithm}[t]
	\caption[CacheToVMAlloc($\mathcal{V}, N_{cache}$): determines the number of cache partitions for the virtual machines to be consolidated into the same hardware platform]{CacheToVMAlloc($\mathcal{V}, N_{cache}$)}
	\label{alg:VCACHE_CacheToVMAlloc}
	\algsetup{linenosize=\scriptsize}
	\scriptsize
	\begin{algorithmic}[1]
		\REQUIRE $\mathcal{V}$: a set of VCPUs of all VMs to be consolidated, $N_{cache}$: the number of available cache partitions
		\ENSURE Success or Fail
		\STATE Find $x_i$ for each VCPU $v_i\in \mathcal{V}$
		\STATE $z \leftarrow \sum_{v_i\in\mathcal{V}} x_i$ 
		\IF {$N_{cache} < z$}
		\RETURN Fail \label{line:VCACHE_CacheToVMAlloc_fail}
		\ENDIF
		\STATE $\forall v_i\in\mathcal{V}: \rho_{i,x}\leftarrow x_i$
		\STATE $U(z) \leftarrow \sum_{v_i\in\mathcal{V}} {{C_i^v(\rho_{i,z})}\over{T_i^v}}$ /* $U(z)$: total utilization */
		\FOR {$k\leftarrow z+1$ \TO $N_{cache}$}
		\STATE $ \displaystyle  U(k)\leftarrow \min_{z\le k'< k} \left(U(k')-\max_{v_i\in\mathcal{V}} {{C_i^v(\rho_{i,k'})-C_i^v(\rho_{i,k'}+(k-k'))}\over{T_i^v}}\right)$ \label{line:VCACHE_CacheToVMAlloc_recurrence}
		\STATE $\forall v_i\!\in\!\mathcal{V}\!: \rho_{i,k}\!\leftarrow$ \# of cache partitions of $v_i$ contributing to $U(k)$ \label{line:VCACHE_CacheToVMAlloc_save_intermediate_cache_alloc}
		\ENDFOR
		\STATE $\forall v_i\in \mathcal{V}: S_i^v\leftarrow \rho_{i,N_{cache}}$\label{line:VCACHE_CacheToVMAlloc_save_final_cache_alloc}
		\RETURN Success
	\end{algorithmic}
\end{algorithm}

Alg.~\ref{alg:VCACHE_CacheToVMAlloc} shows our cache-to-VM allocation algorithm based on the recurrence in Eq.~\eqref{eq:VCACHE_total_vm_util_recurrence}. Our algorithm first finds $z$, and if a given number of cache partitions ($N_{cache}$) is smaller than $z$, it returns fail (line~\ref{line:VCACHE_CacheToVMAlloc_fail}). Otherwise, it computes $U(k)$ iteratively (line~\ref{line:VCACHE_CacheToVMAlloc_recurrence}) and saves $\rho_{i,k}$ that leads to $U(k)$ (line~\ref{line:VCACHE_CacheToVMAlloc_save_intermediate_cache_alloc}). Once the iteration completes, our algorithm sets the number of cache partitions for each VCPU to $\rho_{i,N_{cache}}$ and returns success. The time complexity of our algorithm is $O((N_{cache})^2\cdot|\mathcal{V}|)$.

\section{Evaluation}
\label{VCACHE_evaluation}

This section presents our experimental results on our vLLC, vColoring, and cache management scheme. 
\begin{table}[t]
	\VS{-4pt}
	\centering
	{
		\footnotesize
		\caption{Implementation cost of vLLC and vColoring\VS{-10pt}}\label{tab:VCACHE_vcoloring_implementation_cost}
		\begin{tabular}{l|l|c|c}
			\hline
			\multirow{2}{*}{Name}& \multirow{2}{*}{Items} &\multicolumn{2}{c}{Cost (nsec)}\\\cline{3-4}
			& & x86 & ARM \\\hline
			\multirow{2}{*}{vLLC} & Virtual LLC emulation & 787 & 12212 \\
			& Cache partition check in GPP-to-HPP mapping & 34 & 921 \\\hline
			vColoring& Page migration for GPP re-mapping& 2359& 31864\\\hline
		\end{tabular}
		\VS{-1pt}
	}
\end{table}

\subsection{vLLC and vColoring}

\smallskip\noindent\textbf{Experimental Setup:} 
We have implemented vLLC and vColoring on the KVM hypervisor included in the Linux 3.10.39 kernel.
We chose KVM for its convenience, such as supporting various architectures and providing both shadow paging and two-dimensional paging. However, it is worth noting that our techniques, vLLC and vColoring, can also be implemented in other hypervisors. In our experiments, we use two-dimensional paging because it is the default address translation technique of KVM and shadow paging is not yet supported by KVM for ARM. 

We use x86 and ARM platforms as host machines for our experiments. The x86 platform is equipped with an Intel i7-2600 3.4GHz quad-core processor and 16GB of DDR3 1666MHz memory. The Intel processor has a unified 8MB shared LLC that consists of four 2MB cache slices, providing 32 cache partitions. We disabled hardware prefetcher, simultaneous multithreading, and dynamic clock frequency scaling to reduce measurement inaccuracies. The ARM platform used is an ODROID-XU4 board. It has 2GB of LPDDR3 933MHz memory and a Samsung Exynos 5422 SoC that combines a cluster of four ARM Cortex-A15 cores with a cluster of four Cortex-A7 cores. However, we only use the cluster of Cortex-A15 cores because the performance of the other cluster seems inadequate for our experiments. The LLC shared among four Cortex-A15 cores is 2MB, providing 32 cache partitions. We disabled dynamic clock frequency scaling and configured each core to run at its maximum speed, 2GHz.

Since our focus is on cache interference imposed on tasks in a VM, each platform hosts one VM that has four VCPUs (VCPUs 1-4). Each VCPU is allocated to a different PCPU with 100\% of budget. Hence, there is only one VCPU per PCPU on both the x86 and ARM platforms. The VM is assigned all the 32 cache partitions of the host machine. On the host side, VCPU threads are assigned real-time priorities, which prevents unexpected delays from indispensable system services that could not be disabled. 

Three different guest OSs are used in our experiments: Linux/RK and the vanilla Linux kernel 3.10.39 for x86 and ARM, and MS Windows Embedded 8.1 Industry for x86. Linux/RK is used as a guest OS to evaluate vLLC because it supports page coloring. The vanilla Linux and MS Windows Embedded OSs are used to evaluate vColoring because they both do not support page coloring. Specifically, MS Windows Embedded is chosen to verify that vColoring can be used for proprietary, closed-source guest OSs.

\smallskip\noindent\textbf{Implementation Overhead:} 
\tableref{VCACHE_vcoloring_implementation_cost} shows the computational overhead of vLLC and vColoring, measured with hardware performance counters on the x86 and ARM platforms. vLLC performs the virtual LLC emulation when a guest OS reads the VM's LLC information, which is typically done during the system initialization phase. The GPP-to-HPP mapping occurs only once per GPP, as described in Section~\ref{VCACHE_vLLC_scheme}, and the overhead added by the cache partition check of vLLC in the GPP-to-HPP mapping is less than 5\% of the original mapping time on both platforms. Hence, we consider that the overhead of vLLC is acceptably small. 
vColoring re-maps GPPs when cache partitions are assigned to a task. Since the major overhead of this re-mapping is caused by page migration, we present per-page migration time in \tableref{VCACHE_vcoloring_implementation_cost}.

\begin{figure}[t]
\VS{-18pt}
\centering
\subfloat[x86 platform]{\label{fig:VCACHE_expr_latency_wcet_x86}
\includegraphics[width=0.32\textwidth]{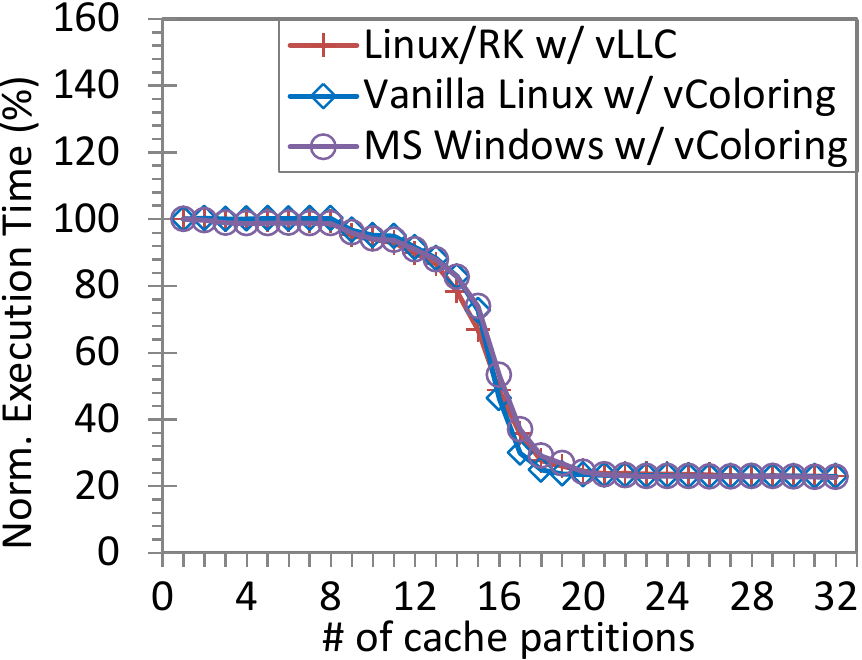}
}\hspace{30pt}
\subfloat[ARM platform]{\label{fig:VCACHE_expr_latency_wcet_arm}
\includegraphics[width=0.32\textwidth]{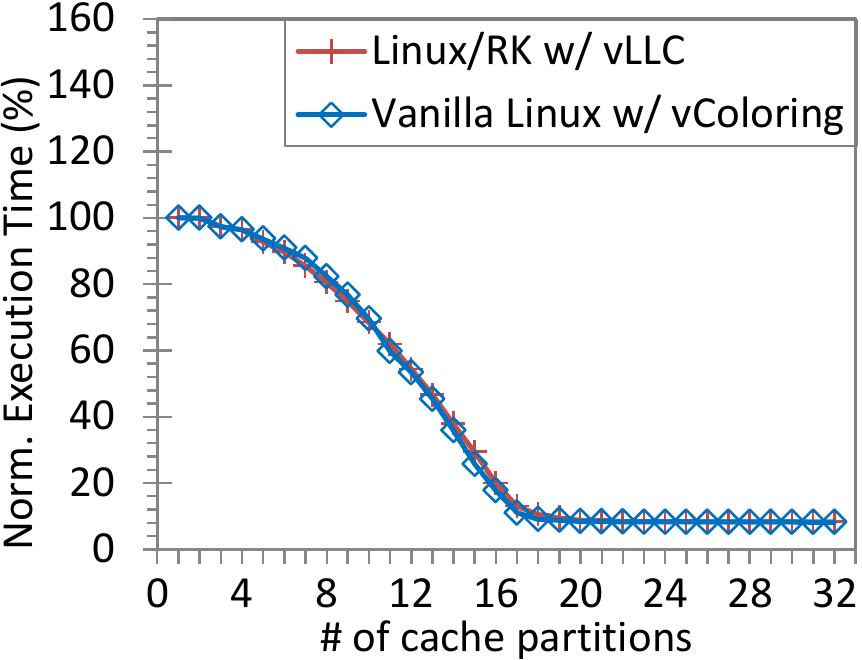}
}
\VS{-9pt}
\caption{Execution times of the {\em latency} task}
\label{fig:VCACHE_expr_latency_wcet}
\VS{-5pt}
\end{figure}

\begin{figure}[t]
	\VS{-3pt}
	\centering
	\subfloat{
		\includegraphics[width=0.7\textwidth]{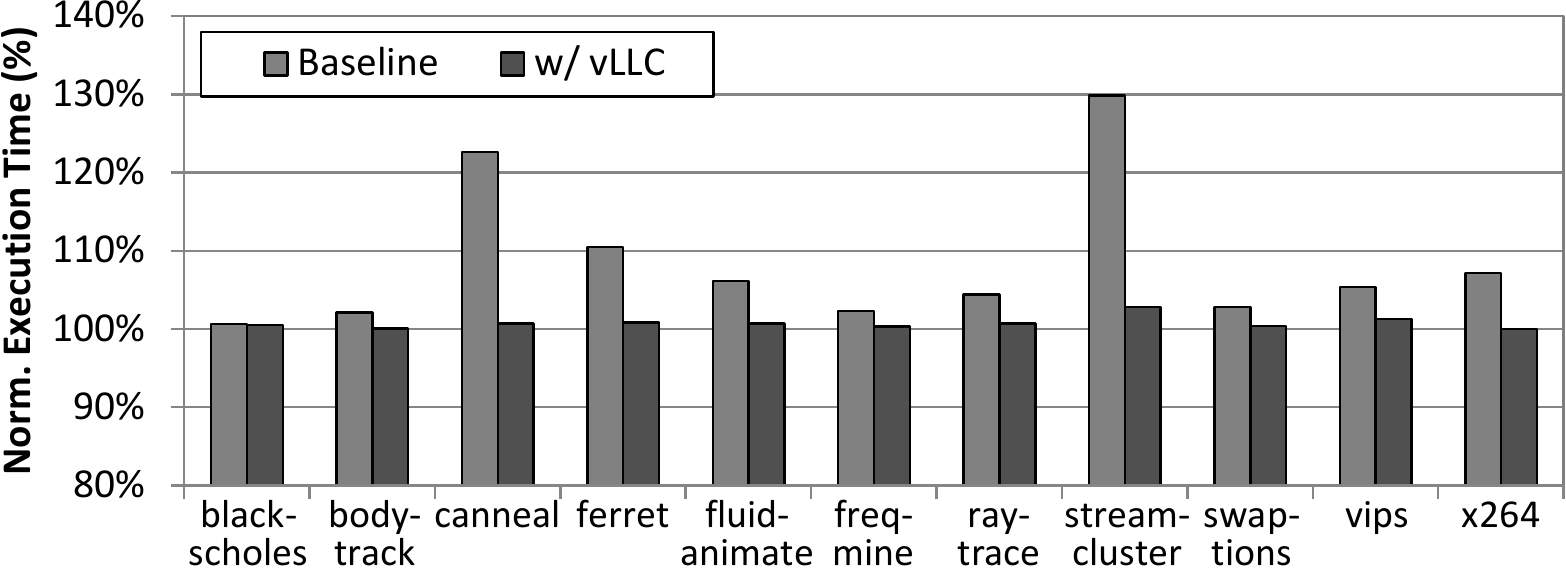}
	}
	\VS{-7pt}
	\caption{Execution times of the PARSEC benchmarks when synthetic tasks run on different VCPUs in parallel}
	\label{fig:VCACHE_expr_parsec_inter_vcpu_interference_x86_vLLC}
\end{figure}

\begin{figure}[t]
	\centering
	\VS{-8pt}
	\subfloat{
		\includegraphics[width=0.7\textwidth]{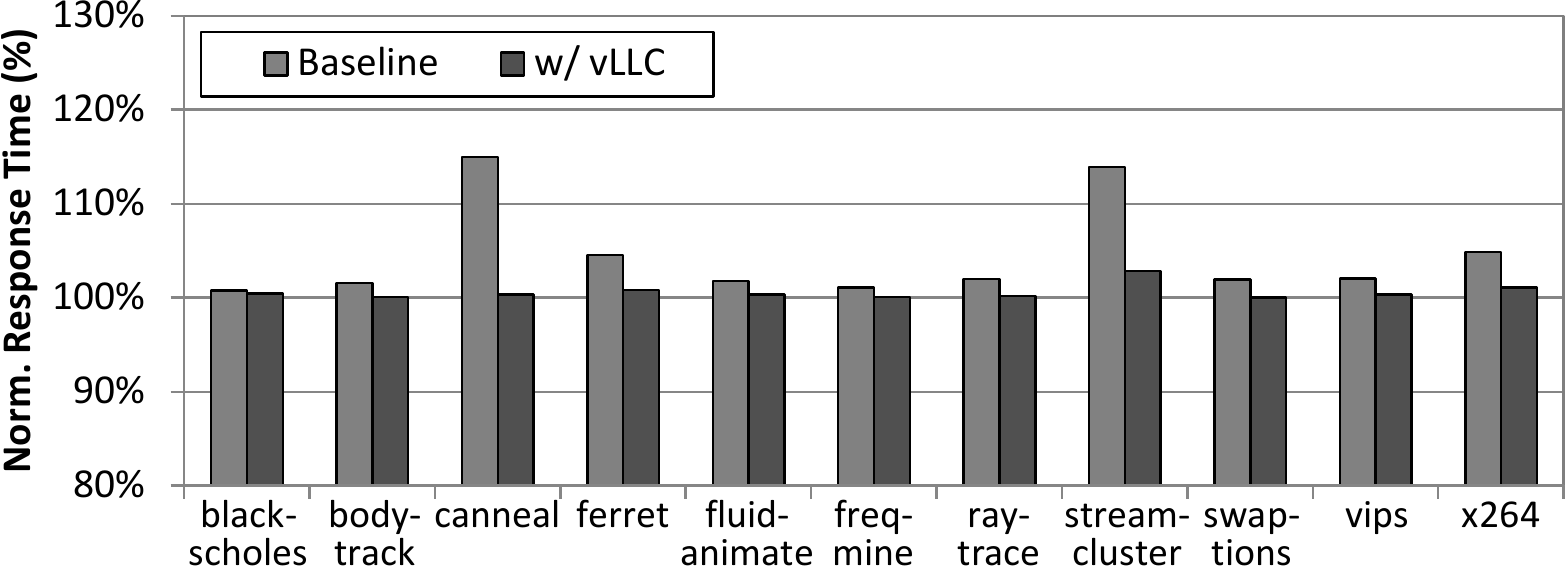}
	}
	\VS{-7pt}
	\caption{Response times of the PARSEC benchmarks when synthetic tasks are scheduled on the same VCPU}
	\label{fig:VCACHE_expr_parsec_intra_vcpu_interference_x86_vLLC}
	\VS{-5pt}
\end{figure}


\smallskip\noindent\textbf{Results with a Synthetic Task:} 
As the first step of our experiments, we check if vLLC and vColoring can correctly assign cache partitions to a task running in a VM. We use the {\em latency} task~\cite{Yun_RTAS14} which traverses a randomly-ordered linked list. The execution time of the {\em latency} task highly depends on the memory access time, due to the data dependency of pointer-chasing operations in linked-list traversals. To make the {\em latency} task cache-sensitive, we configured the working set size of the {\em latency} task to be half of the LLC of each platform, i.e., 4MB on x86 and 1MB on ARM. We compiled this task for both Linux and MS Windows guests on x86.

\figref{VCACHE_expr_latency_wcet} compares the maximum observed execution times of the {\em latency} task when it runs alone in each VM with different numbers of cache partitions assigned to it. The x-axis of each graph denotes the number of cache partitions assigned to the task. The y-axis shows the execution time normalized to the case where the task runs with one cache partition. On both x86 and ARM platforms, the execution time of the task begins to plateau after more than 16 cache partitions are assigned to it. This is because the entire working set of the task can fit into the LLC after that point. On each platform, a very similar execution-time pattern is observed although different guest OSs are used. This shows that both vLLC and vColoring work as expected.

\smallskip\noindent\textbf{Results with PARSEC Benchmarks:} 
We use the PARSEC benchmarks~\cite{PARSEC}, which are closer to the memory access patterns of real applications compared to the synthetic task, {\em latency}. A total of eleven PARSEC benchmarks is used. We have excluded two PARSEC benchmarks, {\em dedup} and {\em facesim}, due to their excessive disk accesses for data files. Since we have shown in the previous subsection that vLLC and vColoring are equivalent in preventing cache interference on x86 and ARM platforms, we use only vLLC on x86 for simplicity.

We first identify the impact of inter-VCPU cache interference on the PARSEC benchmarks. Each benchmark is assigned to VCPU 1 and the three instances of the {\em latency} task are assigned to the other VCPUs to generate interfering cache requests. When vLLC is not used, the benchmark and the three instances share all 32 cache partitions. When vLLC is used, our objective here is to protect the cache behavior of the benchmark from the three instances of {\em latency}. Hence, with vLLC, each benchmark is assigned 31 private cache partitions and the three instances share the remaining 1 partition. 

\figref{VCACHE_expr_parsec_inter_vcpu_interference_x86_vLLC} compares the execution time of each PARSEC benchmark with and without vLLC. The x-axis denotes the benchmark names, and the y-axis shows the execution time of each benchmark normalized to the case when it runs alone in the VM with 32 cache partitions. When vLLC is not used (Baseline), there is up to 30\% of execution time increase. When vLLC is used, only {\em streamcluster} has an execution time increase of 2\% and the other benchmarks have no noticeable difference in their execution times. The reason for the increase in {\em streamcluster}'s execution time is due to the fact that it is assigned a smaller number of cache partitions when vLLC is used, compared to when vLLC is not used.

Next, we explore the impact of intra-VCPU cache interference on the PARSEC benchmarks. Each benchmark and the three instances of {\em latency} are assigned to the same VCPU, and the {\tt SCHED\_RR} policy with a time quantum of 10~msec is used to time-share that VCPU. When vLLC is used, the benchmark is assigned 31 private cache partitions and the three instances share 1 remaining cache partition, just like the inter-VCPU interference experiment.

\figref{VCACHE_expr_parsec_intra_vcpu_interference_x86_vLLC} shows the response time of each benchmark when the three instances of {\em latency} are scheduled on the same VCPU. The response time of a benchmark is normalized to the case when it is scheduled on the same VCPU with three instances of a {\em busyloop} task. {\em busyloop} runs an empty infinite while loop, thereby causing no cache interference. When vLLC is not used, the response time increases by up to 15\%. When vLLC is used, all the benchmarks except {\em streamcluster} have no noticeable difference in their response times. The increase in {\em streamcluster}'s execution time is again because a smaller number of cache partitions is assigned to the benchmark when vLLC is used. To summarize, the results with the PARSEC benchmarks show that both inter- and intra-VCPU cache interference can significantly degrade task performance, and our techniques are effective in allocating cache partitions to tasks running in a VM.

\subsection{Cache Management Scheme} 

In this subsection, we evaluate our real-time cache management scheme for multi-core virtualization. To do this, we use randomly-generated tasksets and capture the total utilization of VMs as the metric.


\begin{table}[t]
	\centering
	\VS{-5pt}
	{
		\footnotesize
		\caption[Base parameters for cache management experiements in virtualization]{Parameters for taskset generation\VS{-10pt}}\label{tab:VCACHE_expr_params}
		\begin{tabular}{p{2cm}|p{5cm}|C{3.5cm}}
			\hline
        \textbf{Type}          & \textbf{Parameters} & \textbf{Values}\VS{-1pt}\\\hline
        \textbf{System} & Number of PCPUs & 4 \\
		& Number of VMs & 2 \\                      
        & Number of VCPUs per VM & 4 \\
        & VCPU replenishment period & 10 msec\\
        & Cache (LLC) size & 2048 KB\\
        & \# of cache partitions ($N_{cache}$)& 32\\
        & Cache hit delay & 26 nsec\\
        & Cache miss delay & 202 nsec\\
        & Cache partition reload time ($\Delta$) & 207 $\mu$sec\\
        \hline
        \textbf{Taskset}           & Total number of tasks & [10, 15]\\
        & Taskset utilization $(U_{taskset})$& 3.0\\
        \hline
        \textbf{WCET}           & Memory accesses per job & [100000, 1000000]\\
        & Neighborhood size & [16, 64]\\
        & Locality & [1.5, 3.0]\\
        & Task memory usage & [8, 40] MB\\
        & *Resulting working-set size & [64 KB, 40 MB]\\
        & *Resulting WCET & [8.47, 202.02] msec\\
			\hline
		\end{tabular}\VS{-2pt}
	}
\end{table}

\begin{figure}[t]
	\VS{-10pt}
	\centering
	\subfloat{
		\includegraphics[width=0.7\textwidth]{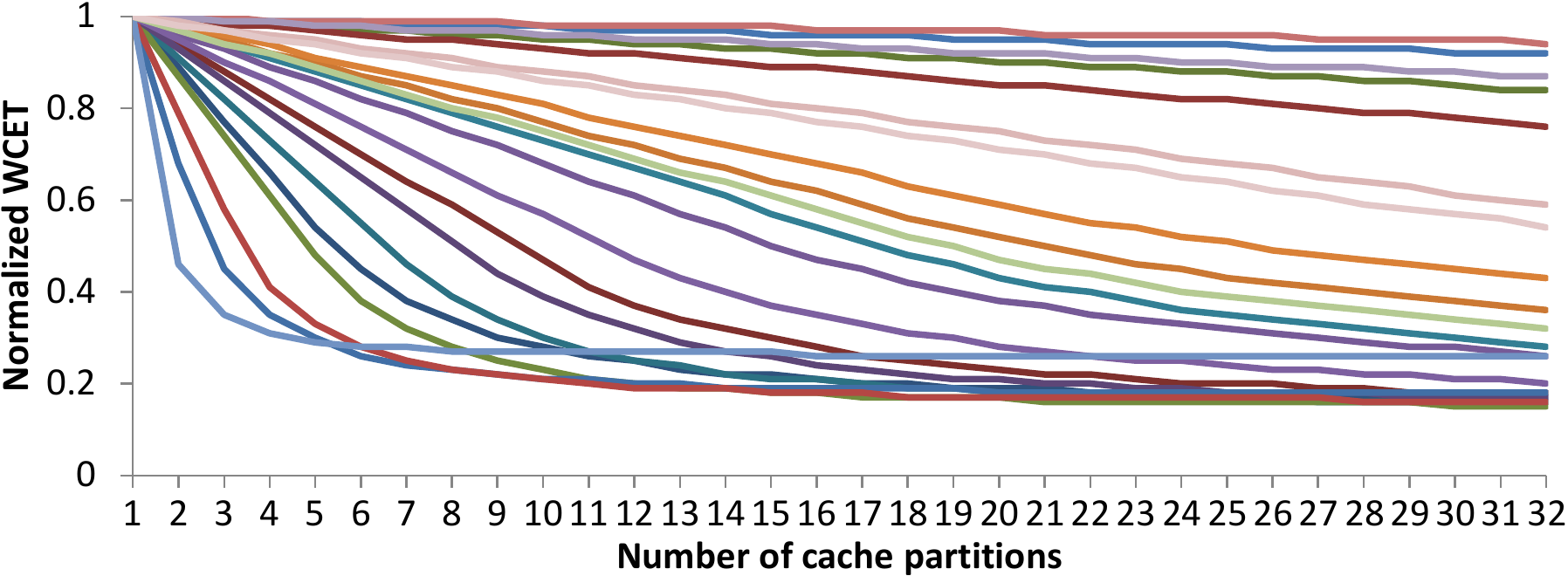}
	}
	\VS{-8pt}
	\caption{Some of WCETs generated for our experiments}
	\label{fig:VCACHE_expr_random_taskset_wcet}
	\VS{-10pt}
\end{figure}

\smallskip\noindent\textbf{Experimental Setup:} 
We generated 10,000 tasksets with the parameters in \tableref{VCACHE_expr_params}. Cache hit/miss delay and cache partition reload time ($\Delta$) were obtained by measurement on our ARM platform. To generate a WCET function ($C_i(k)$) for each task $\tau_i$, we use the method described in \cite{Bui_RTCSA08}. This method first calculates a cache miss rate for given cache size, neighborhood size, locality, and task memory usage, by using the analytical cache behavior model proposed in~\cite{Thiebaut_IEEE92}. It then generates an execution time with the calculated cache miss rate, the timing delay of a cache miss, and the number of memory accesses. With this method, we were able to generate WCETs with different cache sensitivities, as shown in \figref{VCACHE_expr_random_taskset_wcet}. 
Then, the total taskset utilization ($U_{taskset}$) is split into $n$ random-sized pieces, where $n$ is the total number of tasks. The size of each piece represents the utilization of the corresponding task when one cache partition is assigned to it. The period of a task $\tau_i$ is calculated by dividing $C_i(1)$ by its utilization. 
Once a taskset is generated, they are randomly distributed to two VMs, each of which has four VCPUs. Within each VM, the priorities of tasks are assigned by the Rate-Monotonic Scheduling (RMS) policy~\cite{Liu_Layland}. The priorities of VCPUs are arbitrarily assigned since they use the same period. The sporadic server policy is used for VCPU budget replenishment.

\begin{figure}[t]
	\centering
	\subfloat{
		\includegraphics[width=0.7\textwidth]{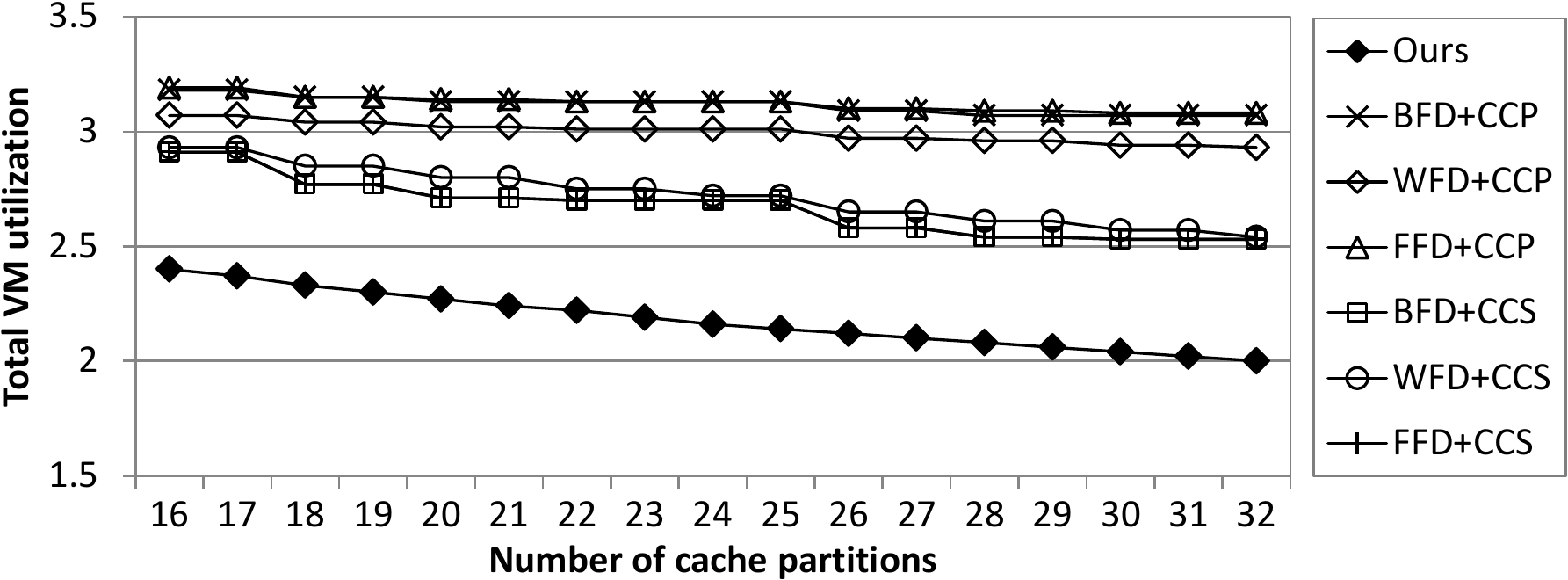}
	}
	\VS{-8pt}
	\caption{VM utilization w.r.t the number of cache partitions}
	\label{fig:VCACHE_expr_vm_cache_alloc_util}
	\VS{-1pt}
\end{figure}

\begin{figure}[t]
	\VS{-10pt}
	\centering
	\subfloat{
		\includegraphics[width=0.7\textwidth]{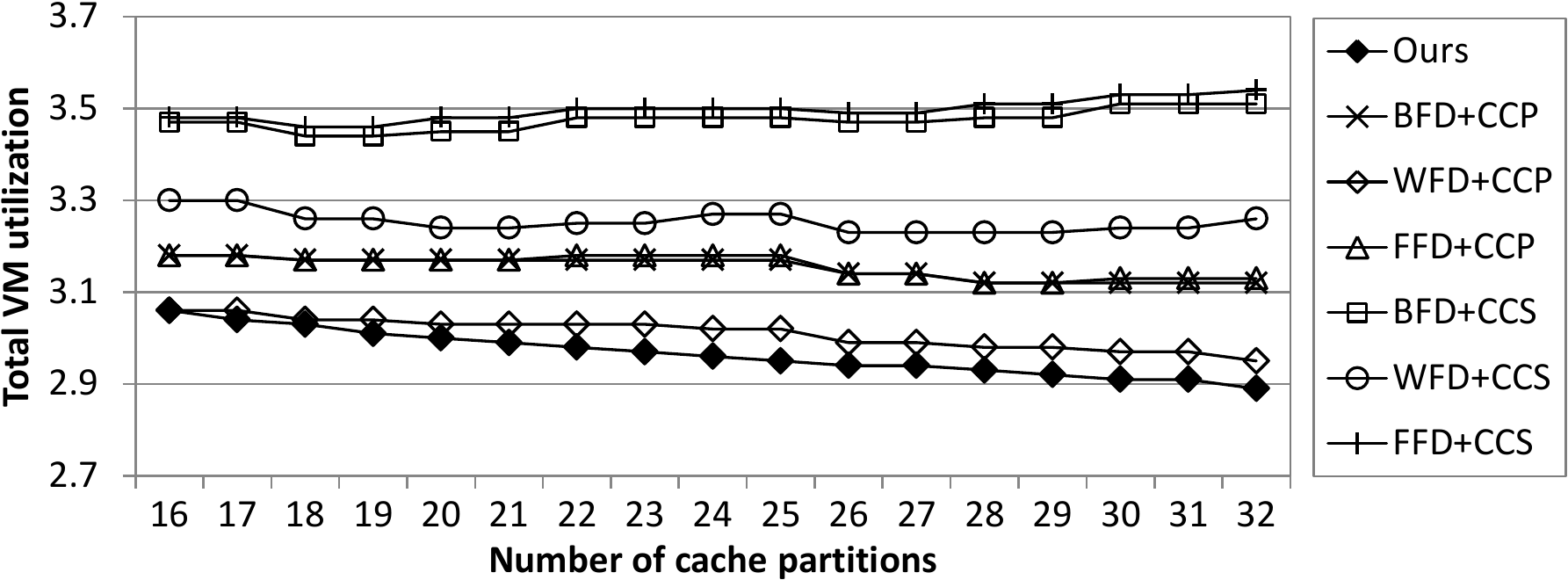}
	}
	\VS{-8pt}
	\caption{VM utilization ($\Delta=10$ msec)}
	\label{fig:VCACHE_expr_vm_cache_alloc_util_delta_10}
	\VS{-1pt}
\end{figure}


\smallskip\noindent\textbf{Results:} 
For comparison with our scheme, we consider variants of the best-fit decreasing (BFD), worst-fit decreasing (WFD), and first-fit decreasing (FFD) heuristics. 
Each heuristic is used for task-to-VCPU allocation within a VM and combined with two different cache-to-task allocation policies: complete cache partitioning (CCP) and complete cache sharing (CCS). CCP allocates private cache partitions to tasks in proportion to their working-set sizes. On the other hand, CCS lets tasks on the same VCPU share all their cache partitions. Hence, we compare our scheme against a total of six approaches: BFD+CCP, WFD+CCP, FFD+CCP, BFD+CCS, WFD+CCS, and FFD+CCS. 
For each approach, $k$ cache partitions, where $1\le k\le N_{cache}$, are evenly distributed to all VCPUs of the two VMs such that the difference in the number of cache partitions of each VCPU does not exceed 1. Tasks are sorted in decreasing order of utilization w.r.t. the number of cache partitions per VCPU. Once task-to-VCPU allocation is done, we determine the budget of each VCPU by the binary search approach used in the Phase~2 of our CAVM algorithm given in Alg.~\ref{alg:VCACHE_CacheToVMAlloc}. Finally, we find the total utilization of VMs by summing up the utilization of all VCPUs.

\figref{VCACHE_expr_vm_cache_alloc_util} shows the total VM utilization as the number of cache partitions increases. Since CCP cannot find a schedulable allocation if the number of partitions is smaller than that of tasks, we compare only the cases where the number of cache partitions is greater than 15. Our scheme outperforms all other approaches, yielding 1.18$\times$ to 1.54$\times$ lower utilization. This is because our scheme allocates cache-sensitive tasks together to the same VCPU to increase the benefit of cache sharing and finds the minimum total VM utilization for a given number of cache partitions. The heuristics with CCS perform better than the ones with CCP. This is because $\Delta$ obtained from our ARM platform is relatively small so that the reduction in task execution time from cache sharing is larger than the resulting CRPD in our experiments.

\figref{VCACHE_expr_vm_cache_alloc_util_delta_10} shows the total VM utilization when $\Delta$ is 10 msec. This experiment is to evaluate our scheme when CRPD is extremely high. Overall, the benefit of using more cache partitions is smaller compared to the previous case. Our scheme outperforms other approaches because it can balance between the utilization gain and CRPD from cache sharing. The heuristics with CCS perform worse than the ones with CCP due to the high CRPD. WFD+CCS is affected less by the high CRPD compared with BFD+CCS and FFD+CCS, because WFD results in less number of tasks per VCPU. Based on these results, we conclude that our scheme allocates cache partitions efficiently in a virtualization environment and yields a significant utilization benefit.

\section{Summary}
\label{VCACHE_conclusions}
In this chapter, we presented our proposed predictable cache management framework for multi-core virtualization. Our framework has vLLC and vColoring, hypervisor-level techniques to enable the cache allocation of individual tasks running in a VM. 
They do not require any hardware feature beyond that available on today's processors.
We have implemented vLLC and vColoring on the KVM hypervisor running on x86 and ARM platforms. Experimental results with three different guest OSs show that both vLLC and vColoring can effectively control the cache allocation of tasks in a VM. vColoring can also be used for DRAM bank partitioning in a virtualized environment, because software-based bank partitioning uses the same approach as page coloring.

Our framework also supports a cache management scheme that determines cache to task allocation, designs a VM in the presence of cache interference, and minimizes the total utilization of VMs to be consolidated into the host machine. Experimental results with randomly-generated tasksets show that our scheme consumes as much as 1.54$\times$ lower CPU utilization for satisfying timing constraints compared with the conventional approaches.
Future work involves addressing temporal interference from main memory in a virtualization environment.

\chapter{Synchronization for Multi-Core Virtual Machines}
\label{chapter_synchronization}

Real-time hierarchical scheduling theory~\cite{Davis_RTSS05,Saewong_ECRTS02,Shin_ECRTS08, Shin_ACM08,Xu_RTSS13} and its implementations~\cite{Kim_RTSS14,Xi_EMSOFT11,Li_VEE14} have established a good foundation for ensuring timing predictability in a virtualized environment. However, the current state of the art still lacks properties required for the sharing of mutually-exclusive resources in virtualization. Specifically, multi-core synchronization mechanisms designed for non-hierarchical scheduling, such as MPCP~\cite{MPCP2,MPCP} and MSRP~\cite{Gai_RTAS03}, can lead to excessive blocking times due to the preemption and budget depletion of VCPUs, as discussed in Section~\ref{SYNC_shared_resource_penalties}. Available solutions in the uni-core hierarchical scheduling context~\cite{Asberg_RTAS13,Behnam_EMSOFT07,Davis_RTSS06} have not yet been extended to multi-core platforms. More importantly, in current virtualization solutions, the hypervisor is unaware of the executions of critical sections of tasks within VCPUs and there is no systematic mechanism to do so. 

In this chapter, we develop a virtualization-aware multi-core priority ceiling protocol (vMPCP) and its framework to address the synchronization issue in a virtualized environment. vMPCP extends the well-known multiprocessor priority ceiling protocol (MPCP) to the multi-core two-level hierarchical scheduling context. vMPCP enables the sharing of resources in a bounded time, within and across VCPUs that could be assigned on different PCPUs. To do so, it uses a para-virtualization approach to expose the executions of critical sections in VCPUs to the hypervisor. Each guest VM can maintain its own priority-numbering scheme and task priorities do not need to be compared across VMs. For the VCPU budget supply and replenishment policy, vMPCP supports both periodic server~\cite{periodic_server} and deferrable server~\cite{deferrable_server} policies. In addition, vMPCP provides an option for VCPUs to overrun their budgets while their tasks are executing critical sections. The effect of the overrun is analyzed and evaluated in detail.

The detailed contributions of our framework are as follows. First, we propose a new synchronization protocol, vMPCP, for multi-core virtualization. We characterize timing penalties caused by critical sections in a virtualized environment and develop a protocol to address such penalties. Second, we analyze the impact of different VCPU budget supply policies, namely periodic and deferrable servers, on synchronization in a multi-core virtualization environment. We also analyze each of the policies with and without VCPU budget overrun. Third, from our analysis and experimental results, we found that the periodic server policy, which has been considered to dominate the deferrable server policy in the literature, does not dominate the deferrable server policy when overrun is used. We also found that the use of overrun does not always yield better results, especially for tasks with relatively long critical sections. Fourth, we have implemented the prototype of vMPCP on the KVM hypervisor running on a multi-core platform. Using this implementation, we identify the effect of vMPCP on a real system by comparing it against a virtualization-unaware synchronization protocol (MPCP).

Since our focus in this chapter is on mutually-exclusive shared resources, we will call them simply ``shared resources''. As described in Section~\ref{background_synchronization}, there are two types of shared resources,  {\em global} and {\em local}, and the critical sections corresponding to those resources are referred to as global critical sections (gcs's) and local critical sections (lcs's), respectively. Each shared resource has a unique index and the function $R(\tau_i,j)$ returns the index of the resource used by the $j$-th critical section of task $\tau_i$. The function $type(\tau_i,j)$ returns $gcs$ or $lcs$, which is the type of the $j$-th critical section of $\tau_i$. In addition, we use $\sigma_i^{gcs}$ and $\sigma_i^{lcs}$ to denote the number of global and local critical section segments of $\tau_i$, respectively. Hence, the total number of critical section segments of $\tau_i$ is $\sigma_i=\sigma_i^{gcs}+\sigma_i^{lcs}$. For brevity, we will also use the following notation in this chapter:
\begin{itemize}
	\item $\mathbb{V}(\tau_i)$: the VCPU where a task $\tau_i$ is allocated 
	\item $\mathbb{P}(v_i)$: the PCPU where a VCPU $v_i$ is allocated
\end{itemize}

The background and related prior work on synchronization were presented in Section~\ref{background_synchronization}. The system model including assumptions and notation for tasks, critical sections and virtual machines can be found in Chapter~\ref{system_model}.

The rest of this chapter is organized as follows. Section~\ref{SYNC_framework} presents the vMPCP framework. Section~\ref{SYNC_analysis} provides the analysis on VCPU and task schedulability under vMPCP. A detailed evaluation is provided in Section~\ref{SYNC_evaluation}. Section~\ref{SYNC_conclusions} summarizes this chapter.

\section{vMPCP Framework}
\label{SYNC_framework}
In this section, we present the virtualization-aware multiprocessor ceiling protocol (vMPCP). We first define vMPCP and explain the optional VCPU budget overrun mechanism for periodic server and deferrable server replenishment policies under vMPCP. Then, we provide the details on the software design to implement vMPCP in the hypervisor.

\begin{figure}[t]
\centering
\subfloat{
\includegraphics[width=0.7\textwidth]{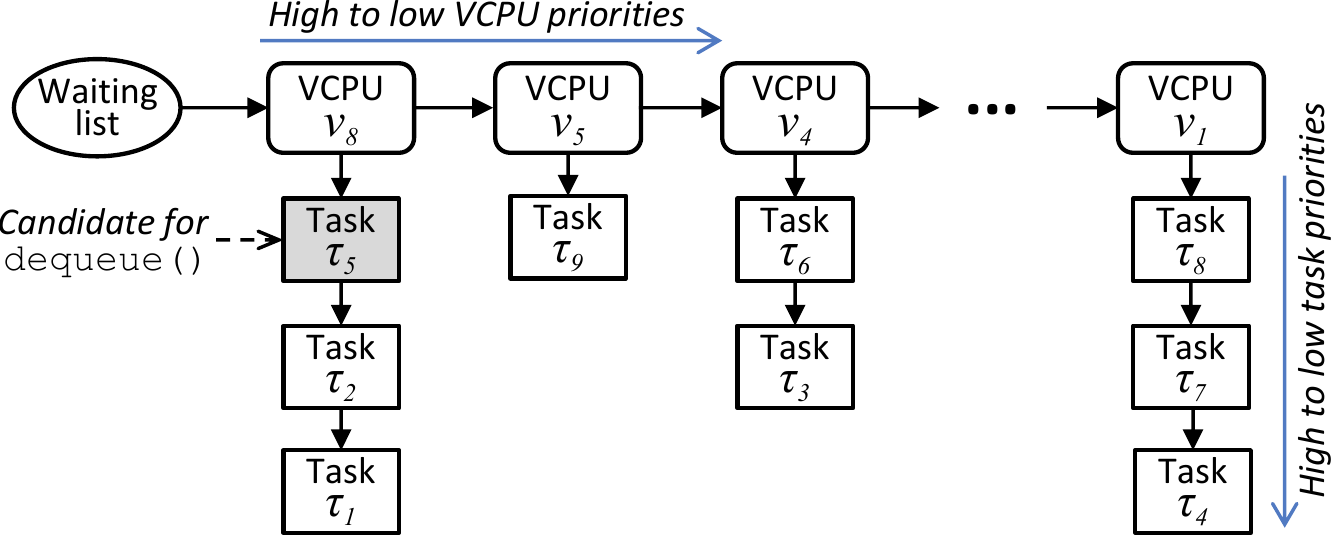}
}
\caption{Two-level priority queue of a global mutex}
\label{fig:SYNC_prio-queue}
\end{figure}

\subsection{Protocol Description}

vMPCP is specifically designed to reduce and bound remote blocking times for accessing global shared resources in a multi-core virtualization environment. To do so, vMPCP uses hierarchical priority ceilings for global critical sections. This approach suppresses both task-level and VCPU-level preemptions while accessing a global resource, thereby reducing the remote blocking times of other tasks waiting on that resource. Global and local resource access rules under vMPCP are defined as follows.


\smallskip
\noindent\textbf{Global shared resources:} vMPCP is based on the multiprocessor priority ceiling protocol (MPCP)~\cite{MPCP2,MPCP}, and extends it to the hierarchical scheduling context. 


\begin{enumerate}
\item Under vMPCP, each mutex protecting a global resource uses a two-level priority queue for its waiting list. \figref{SYNC_prio-queue} shows a logical structure of this two-level priority queue, where the first level is ordered by VCPU priorities and the second level is ordered by task priorities. The key for queue insertion is a pair of VCPU priority and task priority, i.e. $(j,i)$ is a key for a task $\tau_i$ in a VCPU $v_j$. The queue has a \texttt{dequeue} function, which returns the highest priority task of the highest priority VCPU and removes it from the queue.
\item When a task $\tau_i$ requests an access to a global resource $R_k$, the resource $R_k$ can be granted to the task $\tau_i$, if it is not held by another task.
\item While a task $\tau_i$ in a VCPU $v_j$ is holding a resource for its global critical section (gcs), the priority of $\tau_i$ is raised to $\pi_{B,v_j}+\pi_{i}$, where $\pi_{B,v_j}$ is a base task-priority level greater than that of any task in the VCPU $v_j$, and $\pi_{i}$ is the normal priority of $\tau_i$. We refer to $\pi_{B,v_j}+\pi_{i}$ as the {\em task-level priority ceiling} of the gcs of $\tau_i$.
\item While a task $\tau_i$ executes a gcs, the priority of its VCPU $v_j$ is raised to $\pi_{B}^v+\pi_{j}^v$, where $\pi_{B}^v$ is a base VCPU-priority level greater than that of any other VCPUs in the system, and $\pi_{j}^v$ is the normal priority of the VCPU $v_j$. We refer to $\pi_{B}^v+\pi_j^v$ as the {\em VCPU-level priority ceiling} of the gcs of $\tau_i$.
\item When a task $\tau_i$ requests access to a resource $R_k$, the resource $R_k$ cannot be granted to $\tau_i$, if it is already held by another task. In this case, the task $\tau_i$ is inserted to the waiting list (two-level priority queue) of the mutex for $R_k$. 
\item When a global resource $R_k$ is released and the waiting list of the mutex for $R_k$ is not empty, a task dequeued from the head of the queue is granted the resource $R_k$. 
\end{enumerate}

\smallskip
\noindent\textbf{Local shared resources:} vMPCP follows the uniprocessor priority ceiling protocol (PCP)~\cite{PCP} for accessing local resources.\footnote{As an alternative to PCP, the highest locker priority protocol (HLP) can also be used for local resources.}
Unlike the global resource case, a VCPU priority is not affected while its task is accessing a local resource.
\begin{enumerate}
\item Each mutex associated with a local resource $R_k$ is assigned a task-level priority ceiling, which is equal to the highest priority of any task accessing $R_k$. Note that this is valid only within this VCPU.
\item A task $\tau_i$ can access a local resource $R_k$, if the priority of $\tau_i$ is higher than the priority ceilings of any other mutexes currently locked by other tasks in that VCPU.
\item If a task $\tau_i$ is blocked on a local resource by another task that has a lower priority than $\tau_i$, the lower-priority task inherits the priority of $\tau_i$. 
\end{enumerate}

\subsection{VCPU Budget Overrun}

vMPCP provides an option for VCPUs to overrun their budgets when their tasks are in gcs's. This allows tasks to complete their gcs's, even though their VCPU has exhausted its budget. Hence, remote blocking time can be significantly reduced. We present the detailed behavior of the VCPU budget overrun under periodic server and deferrable server policies.


\smallskip
\noindent\textbf{Periodic server with overrun:} The VCPU budget overrun with VCPUs under the periodic server policy works similar to the one presented in~\cite{Davis_RTSS06}. Suppose that a VCPU's budget is exhausted while one of its tasks is in a gcs. If overrun is enabled, the task can continue to execute and finish the gcs. Recall that vMPCP immediately increases the priority of any task executing a gcs to be higher than that of any other normally executing tasks or tasks accessing local resources. Therefore, the amount of overrun time is only affected by the lengths of global critical sections in a VCPU. 

If a VCPU's budget is exhausted while no task of the VCPU is in a gcs, the VCPU suspends until the start of its next replenishment period. Once the VCPU suspends, overrun has no effect. This is to maintain the good property of the periodic server policy, no potential back-to-back interference to lower-priority VCPUs. For instance, consider a task $\tau_i$ waiting for a global resource $R$ that is held by another task on a different physical core. The VCPU of $\tau_i$ is currently suspended due to its budget depletion. If the resource $R$ is released while the VCPU of $\tau_i$ is suspended, the task $\tau_i$ needs to wait until the next replenishment period of its VCPU although overrun is enabled.

\smallskip
\noindent\textbf{Deferrable server with overrun:} Unlike the periodic server policy, VCPUs under the deferrable server policy can overrun more flexibly. Consider a task $\tau_i$ waiting for a global resource $R$ that is held by another task on different physical core. The VCPU of $\tau_i$ has exhausted its regular budget. If the resource $R$ is released, the VCPU of $\tau_i$ is allowed to overrun its budget and the task $\tau_i$ can execute its gcs corresponding to $R$. Once the task $\tau_i$ finishes its gcs, the VCPU of $\tau_i$ suspends again. This difference between periodic server and deferrable server with overrun leads to different values in remote blocking time. We will analyze the details in Section~\ref{SYNC_task_sched}.

\subsection{vMPCP Para-virtualization Interface}

vMPCP increases both the priorities of a task and its VCPU when the task executes a gcs. If a lock corresponding to a global resource is implemented at the hypervisor, e.g., resource sharing among VCPUs from different guest VMs, the hypervisor can manage the priorities of VCPUs appropriately. However, if a lock for a  global resource is implemented within a guest VM image, e.g., resource sharing in a multi-core guest VM hosted on the hypervisor, there is no way for the hypervisor to know if any task of a VCPU of the VM executes a gcs associated with the lock.

To address this issue, vMPCP provides a para-virtualization\footnote{Para-virtualization is a technique involving small modifications to guest operating systems or device drivers to achieve high performance and efficiency.} interface for a VCPU to let the hypervisor know the executions of gcs's in the VCPU. The interface consists of the following two functions:
\begin{itemize}
\item \texttt{vmpcp\_start\_gcs()}: If any task of a VCPU acquires a lock for a global resource, this function is called to let the hypervisor increase the priority of the VCPU by the base VCPU-priority level $\pi_B^v$ of the system. If overrun is enabled, the hypervisor allows the VCPU to continue to execute until \texttt{vmpcp\_finish\_gcs()} is called. The hypervisor may implement an enforcement mechanism for the VCPU not to exceed its pre-determined overrun time that will be given in Sec.~\ref{SYNC_VCPU_sched}.
\item \texttt{vmpcp\_finish\_gcs()}: When there is no global-resource lock held by any task in a VCPU, this function is called to let the hypervisor reduce the priority of the VCPU to its normal priority. Also, if the VCPU's budget is exhausted, the hypervisor suspends the VCPU.
\end{itemize}

\section{vMPCP Schedulability Analysis}
\label{SYNC_analysis}

In this section, we present the schedulability analysis under our proposed vMPCP. 
Our analysis considers each of the periodic server and deferrable server policies with and without VCPU budget overrun. We first analyze the VCPU schedulability on a physical core and the task schedulability on a VCPU. 

\subsection{VCPU Schedulability}
\label{SYNC_VCPU_sched}

vMPCP increases the priority of a VCPU while any task of the VCPU is holding a global resource, which enables a lower-priority VCPU to block a higher-priority VCPU.\footnote{vMPCP does not increase the priority of a VCPU when its task is holding a local resource. Hence, local resources do not affect the VCPU schedulability.} Also, vMPCP results in increased VCPU execution times when overrun is enabled. We now analyze these worst-case effects on VCPU schedulability and derive the VCPU schedulability test under vMPCP.

\smallskip
\noindent\textbf{Blocking from lower-priority VCPUs:} We first focus on the case where the periodic server policy is used. Consider a higher-priority VCPU $v_h$ and a lower-priority VCPU $v_l$, both assigned to the same core. Under the periodic server policy, the higher-priority VCPU $v_h$ never suspends by itself until its budget is exhausted. Hence, the lower-priority VCPU $v_l$ can block $v_h$ only when any global resource that $v_l$'s task has been waiting on is released from another core. The blocking time is equal to the duration of the corresponding gcs (global resource holding time). The worst case happens when all the tasks of $v_l$ have been waiting on global resources and these resources are released from other cores while the higher-priority VCPU $v_h$ is executing. The maximum global resource holding time of $v_l$ is as follows:
\begin{equation} \label{eq:SYNC_ght}
ght(v_l)=\!\!\!\!\sum_{\tau_j\in v_l\land \sigma_j^{gcs}>0}\max_{1\le k \le \sigma_j \land type(\tau_j,k)=gcs}E_{j,k}
\end{equation}
Using Eq.~\eqref{eq:SYNC_ght}, the worst-case blocking time imposed on a VCPU $v_i$ during a time interval $t$ under the periodic server policy is given as follows:
\begin{equation} \label{eq:SYNC_vcpu_block_periodic}
B_i^v(t)=\sum_{v_l\in \mathbb{P}(v_i) \land \pi_l^v<\pi_i^v} ght(v_l)
\end{equation}
where $\pi_i^v$ is the priority of the VCPU $v_i$. Note that the parameter $t$ is used to be consistent with the deferrable server case which will be shown in Eq.~\eqref{eq:SYNC_vcpu_block_deferrable}.

We now consider the case where the deferrable server policy is used. Under this policy, a higher-priority VCPU $v_h$ may suspend itself several times every period. This means that, unlike the periodic server case, the tasks of a lower-priority VCPU $v_l$ may get a chance to request global resources whenever $v_h$ suspends. Hence, each task of the lower-priority VCPU $v_l$ may block the higher-priority VCPU $v_h$ multiple times during $v_h$'s period. The maximum accumulated global resource holding time of the tasks of $v_l$ during a time interval $t$ is given by:
\begin{equation} \label{eq:SYNC_sum_ght}
\begin{split}
sum\_ght(v_l,t)=\sum_{\tau_j\in v_l}\Bigg\{\!\!\left(\!\Big\lceil {t \over T_j}\Big\rceil \!+\!1\right)\cdot\!\!\!\! \sum_{\substack{1\le k \le \sigma_j \land \\ type(\tau_j,k)=gcs}} \!\!\!\!\!\!\!\! E_{j,k}\Bigg\} \\
\end{split}
\end{equation}
Note that the ``+1'' term is to capture the carry-in job of each task during a given time interval $t$. By using Eq.~\eqref{eq:SYNC_sum_ght}, the worst-case blocking time imposed on a VCPU $v_i$ during a time interval $t$ under the deferrable server policy is represented as follows:
\begin{equation} \label{eq:SYNC_vcpu_block_deferrable}
B_i^v(t)=\sum_{v_l\in \mathbb{P}(v_i) \land \pi_l^v<\pi_i^v} sum\_ght(v_l,t)
\end{equation}

\smallskip
\noindent\textbf{Budget overrun time:} If the VCPU budget overrun option is enabled, a VCPU can overrun its budget only when its tasks are executing gcs's. Hence, the maximum time that a VCPU $v_i$ can overrun is bounded by the maximum global resource holding time of that VCPU, which is given in Eq~\eqref{eq:SYNC_ght}. Therefore, the maximum overrun time of a VCPU $v_i$ ($O^v_i$) is equal to $ght(v_i)$ if overrun is enabled, and zero if overrun is not enabled.
 
\smallskip
\noindent\textbf{VCPU schedulability:} The schedulability of a VCPU $v_i$ can be determined by the following recurrence equation:
\begin{equation} \label{eq:SYNC_vcpu_sched}
\begin{split}
W_i^{v,n+1}=&C_i^{v}+O_i^v+B_i^v(W_i^{v,n})\,+\sum_{v_h\in \mathbb{P}(v_i)\land \pi_h^v>\pi_i^v} \left\lceil{W_i^{v,n}+J_h^v \over T_h^v}\right\rceil\cdot (C_h^v+O_h^v)\\
\end{split}
\end{equation}
where $W_i^{v,n}$ is the worst-case response time of $v_i$ at the $n^{th}$ iteration ($W_i^{v,0}=C_i^v+O_i^v$) and $J_h^v$ is a VCPU release jitter ($J_h^v=0$ under the periodic server policy and $J_h^v=T_h^v-C_h^v$ under the deferrable server policy). Eq.~\eqref{eq:SYNC_vcpu_sched} is based on the iterative response time test~\cite{Joseph_J86}. It terminates when $W_i^{v,n+1}=W_i^{v,n}$, and the VCPU $v_i$ is schedulable if its response time does not exceed its period: $W_i^{v,n}<=T_i^v$. In this equation, $O_i^v$ and $O_h^v$ are used to represent the budget overrun of $v_i$ and its higher-priority VCPUs, repectively. The third term represents the blocking time from lower-priority VCPUs during $v_i$'s response time. 

\subsection{Task Schedulability}
\label{SYNC_task_sched}

To determine the schedulability of a task $\tau_i$ under vMPCP, we need to consider the factors discussed in Section~\ref{SYNC_shared_resource_penalties}: (i) local blocking time, (ii) remote blocking time, (iii) back-to-back execution due to remote blocking, (iv) multiple priority inversions, (v) preemptions by higher-priority VCPUs, and (vi) VCPU budget depletion. We take into account factor (iv) when analyzing local blocking time, and factors (v) and (vi) when analyzing remote blocking time. By considering factors (i), (ii) and (iii), we use the following recurrence equation that bounds the worst-case response time of a task $\tau_i$ in a VCPU $v_k$ under vMPCP:
\begin{equation} \label{eq:SYNC_task_sched}
\begin{split}
W_i^{n+1}=&C_i+B_i^{l}+B_i^r+\!\!\!\!\sum_{\tau_h\in \mathbb{V}(\tau_i)\land \pi_h>\pi_i}\!\!\!\bigg\lceil{W_i^n\!+\!J_h\!+\!(W_h\!-\!C_h) \over T_h}\bigg\rceil C_h\\
&+\left\lceil{W_i^n+C_k^v \over T_k^v}\right\rceil(T_k^v-C_k^v)
\end{split}
\end{equation}
where $B_i^l$ is the local blocking time for $\tau_i$, $B_i^r$ is the remote blocking time for $\tau_i$, and $J_h$ is the release jitter of each higher-priority task $\tau_h$ ($J_h=T_k^v-C_k^v$). It terminates when $W_i^{n+1}=W_i^{n}$, and the task $\tau_i$ is schedulable if its response time does not exceed its deadline: $W_i^{n}<=D_i$. Eq.~\eqref{eq:SYNC_task_sched} is based on the response-time test for independent tasks under hierarchical scheduling given in~\cite{Saewong_ECRTS02}. Specifically, the last term of Eq.~\eqref{eq:SYNC_task_sched} is from~\cite{Saewong_ECRTS02}, which captures the execution gap due to the periodic budget supply of the VCPU. The back-to-back execution due to remote blocking from each higher-priority task $\tau_h$ is captured by adding $W_h-C_h$ in the summing term.\footnote{This is a correction made from our previous work~\cite{Kim_RTSS14}. More details on this correction and a suspension-based blocking term in a response-time test can be found in~\cite{Bletsas_TR15}.} 

In the rest of this section, we shall analyze the local and remote blocking times, $B_i^l$ and $B_i^r$. We use $tc_{i,j}$ as the task-level priority ceiling of the $j$-th critical section segment of task $\tau_i$. Similarly, $vc_{i,j}$ is used to represent the VCPU-level priority ceiling of the $j$-th critical section segment of task $\tau_i$. 

\smallskip
\noindent\textbf{Local blocking time:} The local and global critical sections of lower-priority tasks can block the normal execution segment of a higher-priority task $\tau_i$. 
With the local resource access rule of vMPCP based on PCP~\cite{PCP}, only one lower-priority task with a priority ceiling higher than the normal priority of $\tau_i$ can block each normal execution segment of $\tau_i$. Hence, the maximum per-segment blocking time from the local critical sections of lower-priority tasks is given by:
\begin{equation} \label{eq:SYNC_b_llcs}
\begin{split}
B_i^{l\_lcs}=\max_{\substack{\tau_l\in \mathbb{V}(\tau_i) \land \pi_l<\pi_i \\ \land \sigma_l^{lcs}>0}} \bigg(\max_{\substack{1\le u \le \sigma_l \land type(\tau_l,u)=lcs \\\land tc_{l,u}>i}} E_{l,u}\bigg)
\end{split}
\end{equation}

Unlike lcs's, the gcs's of each lower-priority task can block the normal execution segment of $\tau_i$. The maximum per-segment blocking time from the gcs's of lower-priority tasks is given by:
\begin{equation} \label{eq:SYNC_b_lgcs}
\begin{split}
B_i^{l\_gcs}=\sum_{\substack{\tau_l\in \mathbb{V}(\tau_i) \land \pi_l<\pi_i \land \sigma_l^{gcs}>0}} \bigg(\max_{\substack{1\le u \le \sigma_l \land type(\tau_l,u)=gcs}} E_{l,u}\bigg)
\end{split}
\end{equation}

The total local blocking time from both the local and global critical sections of lower-priority tasks is given by:
\begin{equation} \label{eq:SYNC_b_l}
\begin{split}
B_i^{l}=(B_i^{l\_lcs}+B_i^{l\_gcs})\cdot (\sigma_i^{gcs}+1)
\end{split}
\end{equation}
Here, the reason for multiplying by $\sigma_i^{gcs}+1$ is that, before a task $\tau_i$ executes or whenever $\tau_i$ self-suspends due to a global resource, lower-priority tasks may issue requests for local or global resources.

\begin{figure}[t]
\centering
\subfloat{
\includegraphics[width=0.7\textwidth]{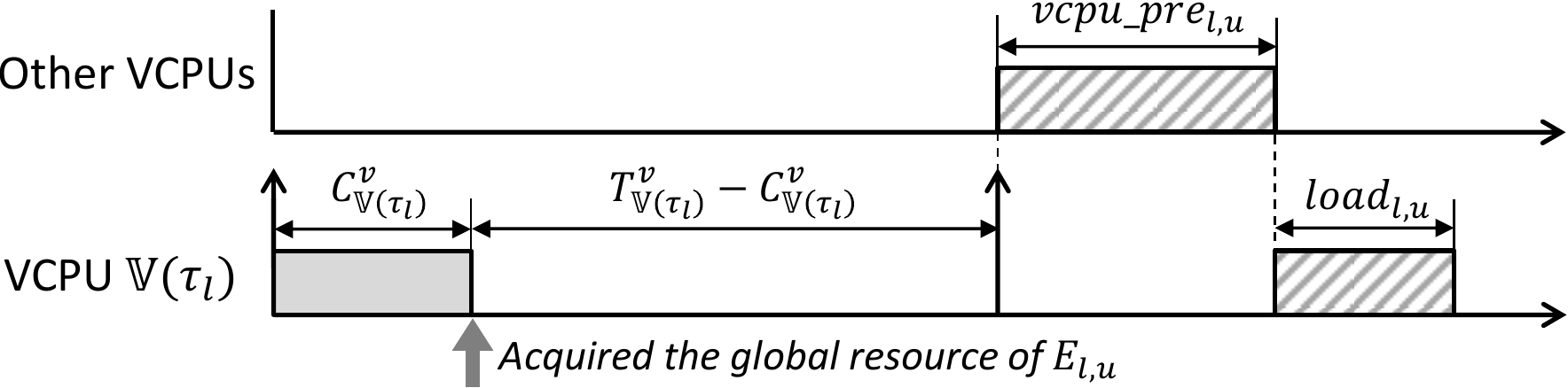}
}
\caption{Periodic server with overrun}
\label{fig:SYNC_periodic_server_with_overrun}
\end{figure}

\smallskip
\noindent\textbf{Remote blocking time:} The remote blocking time $B_i^r$ of a task $\tau_i$ is given by:
\begin{equation} \label{eq:SYNC_b_remote}
\begin{split}
B_i^r=\sum_{1\le j \le \sigma_i \land type(\tau_i,j)=gcs}B_{i,j}^r
\end{split}
\end{equation}
where $B_{i,j}^r$ is the remote blocking time for $\tau_i$ in acquiring the global resource associated with the $j$-th critical section of $\tau_i$. Note that $B_{i,j}^r = 0$ if the $j$-th critical section of $\tau_i$ is a lcs.

The term $B_{i,j}^r$ is bounded by the following recurrence equation:
\begin{equation} \label{eq:SYNC_b_remote_sub}
\begin{split}
B_{i,j}^{r,n+1}=&\max_{\substack{\mathbb{V}(\tau_l) \in lpvcpus(\mathbb{V}(\tau_i)) \\ \land R(\tau_l,u)=R(\tau_i,j)}}W_{l,u}^{gcs} + \sum_{\substack{\mathbb{V}(\tau_h) \in hpvcpus(\mathbb{V}(\tau_i)) \\ \land R(\tau_h,u)=R(\tau_i,j)}}\left(\left\lceil{B_{i,j}^{r,n}\over T_h}\right\rceil+1\right)\cdot W_{h,u}^{gcs}
\end{split}
\end{equation}
where $B_{i,j}^{r,0}=\max_{\substack{\mathbb{V}(\tau_l) \in lpvcpus(\mathbb{V}(\tau_i)) \land R(\tau_l,u)=R(\tau_i,j)}}W_{l,u}^{gcs}$ (the first term of the equation), $lpvcpus(\mathbb{V}(\tau_i))$ is the set of lower-priority VCPUs than the VCPU of $\tau_i$ in the system, $hpvcpus(\mathbb{V}(\tau_i))$ is the set of higher-priority VCPUs than the VCPU of $\tau_i$, and $W_{l,u}^{gcs}$ represents the worst-case response time of the execution $E_{l,u}$ of a gcs after acquiring the corresponding global resource. The first term of Eq.~\eqref{eq:SYNC_b_remote_sub} captures the time for a task in a lower-priority VCPU to finish its gcs. The second term represents the time for tasks in higher-priority VCPUs to execute their gcs's. 

\begin{figure}[t]
\centering
\subfloat{
\includegraphics[width=0.7\textwidth]{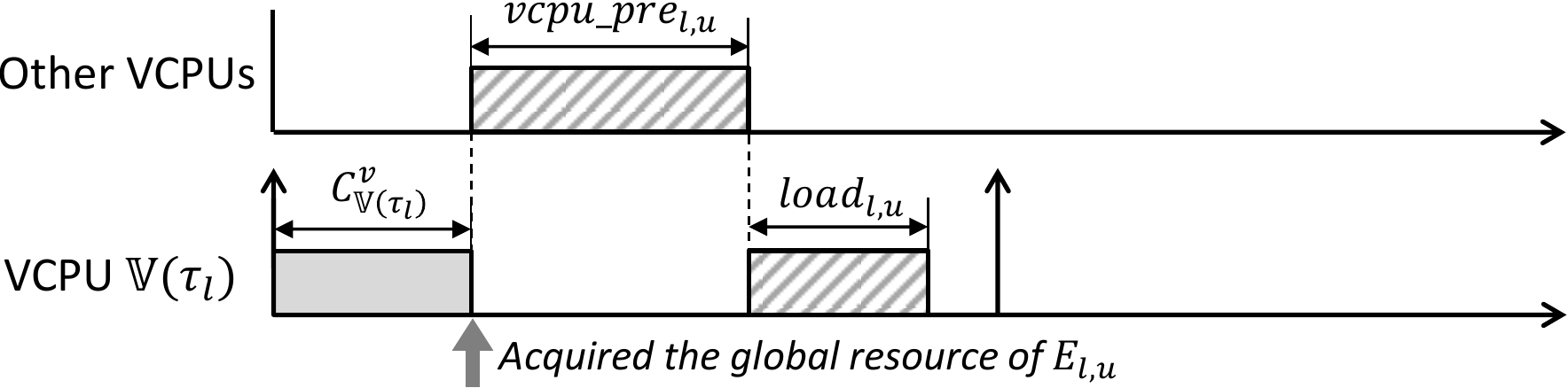}
}
\caption{Deferrable server with overrun}
\label{fig:SYNC_deferrable_server_with_overrun}
\end{figure}

We now analyze $W_{l,u}^{gcs}$, the amount of which depends on which VCPU policy is used and whether overrun is used. We first define two terms, $load_{l,u}$ and $vcpu\_prm_{l,u}$, as follows:
\begin{equation} \label{eq:SYNC_load}
\begin{split}
load_{l,u}=E_{l,u}+\sum_{\tau_x\in \mathbb{V}(\tau_l)} \max_{1\le y\le \sigma_x \land tc_{x,y}>tc_{l,u}}E_{x,y}
\end{split}
\end{equation}
\begin{equation} \label{eq:SYNC_vcpu_prm}
\begin{split}
vcpu\_prm_{l,u}=\!\sum_{\substack{v_z\in \mathbb{P}(\mathbb{V}(\tau_l)) \land v_z \ne \mathbb{V}(\tau_l) }}\,\,\sum_{\tau_x \in v_z} \,\,\max_{\substack{1 \le y \le \sigma_x \land  vc_{x,y}>vc_{l,u}}} \!E_{x,y}
\end{split}
\end{equation}
The term $load_{l,u}$ bounds the maximum VCPU budget required to execute the critical section $E_{l,u}$. It captures the execution time of $E_{l,u}$ and the execution times of gcs's with higher task-level priority ceilings in the same VCPU. Since every gcs has a higher priority than any normal execution segment, we only need to consider one global critical section per task. The term $vcpu\_prm_{l,u}$ bounds the VCPU-level preemptions while $E_{l,u}$ executes. The VCPU of $E_{l,u}$ can only be preempted by other VCPUs that have tasks being executing gcs's with higher VCPU-level priority ceilings. Note that $vcpu\_prm_{l,u}$ increases the response time of $E_{l,u}$ ($W_{l,u}^{gcs}$), but does not consume the budget of $E_{l,u}$'s VCPU.

\begin{itemize}
\item {\it Periodic server with overrun}: The worst-case response time of the execution $E_{l,u}$ of a gcs happens when the corresponding resource is acquired right after its VCPU is suspended. In this case, the execution is delayed until the start of its VCPU's next replenishment period, and this waiting time is up to $T_{\mathbb{V}(\tau_l)}^v-C_{\mathbb{V}(\tau_l)}^v$, as shown in \figref{SYNC_periodic_server_with_overrun}. Once the next period of the VCPU starts, the VCPU can execute and finish $E_{l,u}$ within this period due to overrun. Therefore, $W_{l,u}^{gcs}$ under the periodic server policy with overrun is given by:
\begin{equation} \label{eq:SYNC_w_gcs_periodic}
\begin{split}
W_{l,u}^{gcs}=T_{\mathbb{V}(\tau_l)}^v-C_{\mathbb{V}(\tau_l)}^v+load_{l,u}+vcpu\_prm_{l,u}
\end{split}
\end{equation}

\item {\it Deferrable server with overrun}: In this case, $E_{l,u}$ can be executed without the need to wait until the VCPU's next replenishment period (\figref{SYNC_deferrable_server_with_overrun}). Therefore, $W_{l,u}^{gcs}$ under the deferrable server policy with overrun is given by:
\begin{equation} \label{eq:SYNC_w_gcs_deferrable}
\begin{split}
W_{l,u}^{gcs}=load_{l,u}+vcpu\_prm_{l,u}
\end{split}
\end{equation}

\begin{figure}[t]
\centering
\subfloat{
\includegraphics[width=0.7\textwidth]{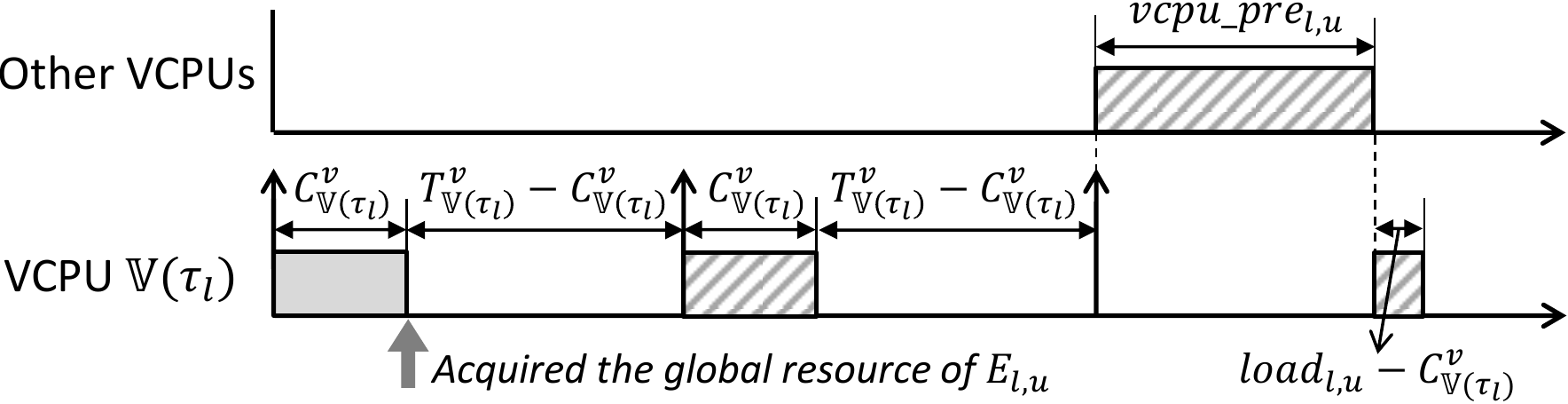}
}
\caption{Periodic/deferrable server without overrun}
\label{fig:SYNC_servers_without_overrun}
\end{figure}

\item {\it Periodic/deferrable server without overrun}: When overrun is not used, the execution of $load_{l,u}$ may span over multiple of its VCPU periods (\figref{SYNC_servers_without_overrun}). The total execution gap is bounded by $\lceil{load_{l,u} \over C_{\mathbb{V}(\tau_l)}^v}\rceil(T_{\mathbb{V}(\tau_l)}^v-C_{\mathbb{V}(\tau_l)}^v)$. Therefore, $W_{l,u}^{gcs}$ under the periodic or deferrable server policy without overrun is given by:
\begin{equation} \label{eq:SYNC_w_gcs_no_overrun}
\begin{split}
W_{l,u}^{gcs}=&\left\lceil{load_{l,u} \over C_{\mathbb{V}(\tau_l)}^v}\right\rceil(T_{\mathbb{V}(\tau_l)}^v-C_{\mathbb{V}(\tau_l)}^v)+load_{l,u}+vcpu\_prm_{l,u}
\end{split}
\end{equation}
Note that, if the amount of $load_{l,u}$ is smaller than the per-period execution budget of the VCPU ($C_{\mathbb{V}(\tau_l)}^v$), Eq.~\eqref{eq:SYNC_w_gcs_no_overrun} becomes equal to Eq.~\eqref{eq:SYNC_w_gcs_periodic}.

\end{itemize}

\section{Evaluation}
\label{SYNC_evaluation}

This section presents our experimental evaluation on vMPCP. 
To our knowledge, vMPCP is the first virtualization-aware multi-core synchronization protocol and there is no schedulability test for existing protocols in the multi-core virtualization environment. We first empirically investigate the performance characteristics of vMPCP in terms of task schedulability, and then compare vMPCP against a virtualization-unaware protocol (MPCP) in terms of response times on a real hardware platform.

\subsection{Comparison of Different Configurations}
\label{SYNC_expr_simulation}

The purpose of this experiment is to explore the impact of different uses of vMPCP on task schedulability. To do this, we use randomly-generated tasksets and capture the percentage of schedulable tasksets as the metric. 

\smallskip
\noindent\textbf{Experimental Setup:} The base parameters we use for experiments are summarized in \tableref{SYNC_expr_params}. As the main interest of our work is in the timing penalties caused by global resources, local resources are not considered. For each experimental setting, we first generate the defined numbers of physical CPU cores in the system, VCPUs for each core, and tasks for each VCPU. Task periods are randomly selected within the defined min/max task period range. 
On each VCPU, the VCPU task utilization is split into $k$ random-sized pieces, where $k$ is the number of tasks in the VCPU. The size of each piece represents the utilization of the corresponding task. 
Then, the WCET of each task is calculated by dividing its utilization by its period. The priorities of tasks and VCPUs are assigned by the Rate-Monotonic Scheduling (RMS) policy~\cite{Liu_Layland} (ties are broken arbitrarily). Once the task information is generated, we determine a VCPU budget value that is used for all VCPUs in the system. Starting from a value equal to the VCPU period, we decrease the VCPU budget by 10~$\mu$secs until all VCPUs pass the VCPU schedulability test given in Eq.~\eqref{eq:SYNC_vcpu_sched}.\footnote{As the minimum time unit in \tableref{SYNC_expr_params} is 10~$\mu$sec, the step size of 10~$\mu$sec is fine-grained enough to find the VCPU budget values in this experiment.} We generate 10,000 tasksets for each experimental setting, and record the percentage of tasksets where all the tasks pass the task schedulability test given in Eq.~\eqref{eq:SYNC_task_sched}.

\begin{table}[t]
\centering
{
\footnotesize
\caption[Base parameters for vMPCP synchronization experiments]{Base parameters for experiments\VS{-5pt}}\label{tab:SYNC_expr_params}
\begin{tabular}{l|c||l|c}
\hline
Parameters & Values& Parameters & Values\VS{-1pt}\\\hline
\# of physical cores & 8 & \# of VCPUs per core & 2\\
\# of tasks per VCPU & 3 & Period of a VCPU & 5 msec\\
Min. task period & 100 msec & Max. task period & 500 msec\\
Per-VCPU task util & 15\% &\# of gcs's per task & 1 \\
\# of lockers per mutex & 2 & Size of a gcs & 10 $\mu$sec\\
\hline
\end{tabular}
}
\end{table}

\smallskip
\noindent\textbf{Results:} We consider the following four uses of vMPCP: periodic server with overrun (PSwO), deferrable server with overrun (DSwO), periodic server with no overrun (PSnO), and deferrable server with no overrun (DSnO). The main factors affecting task schedulability under vMPCP are: (i) the size of a gcs, (ii) the number of lockers per mutex, (iii) the number of gcs's per task, (iv) the VCPU period, and (v) the utilization of tasks in each VCPU. By exploring these factors, we identify the characteristics of the four schemes of vMPCP.

\begin{figure}[t]
\centering
\subfloat{
\includegraphics[width=0.7\textwidth]{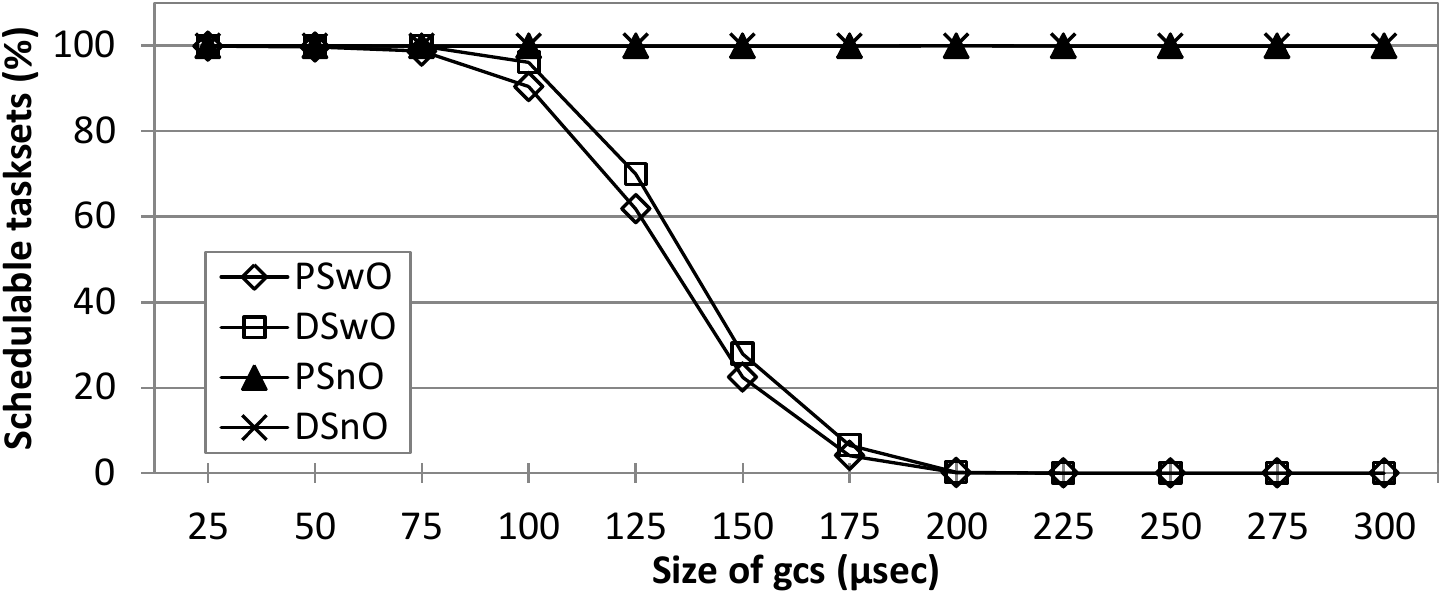}
}
\caption{Taskset schedulability as the size of a gcs increases}
\label{fig:SYNC_expr_size_gcs}
\end{figure}

\figref{SYNC_expr_size_gcs} shows the percentage of schedulable tasksets as the size of a gcs increases. The schemes with no overrun, PSnO and DSnO, are almost unaffected by the size of a gcs. Conversely, the schedulability under the schemes with overrun, PSwO and DSwO, decreases as the size of a gcs increases. This is due to the fact that, without overrun, more VCPU budget can be used for the executions of normal execution segments of tasks. DSwO performs better than PSwO because DSwO results in a shorter response time of the execution of a gcs, as given in Eq.~\eqref{eq:SYNC_w_gcs_deferrable}.

\begin{figure}[t]
\centering
\subfloat{
\includegraphics[width=0.7\textwidth]{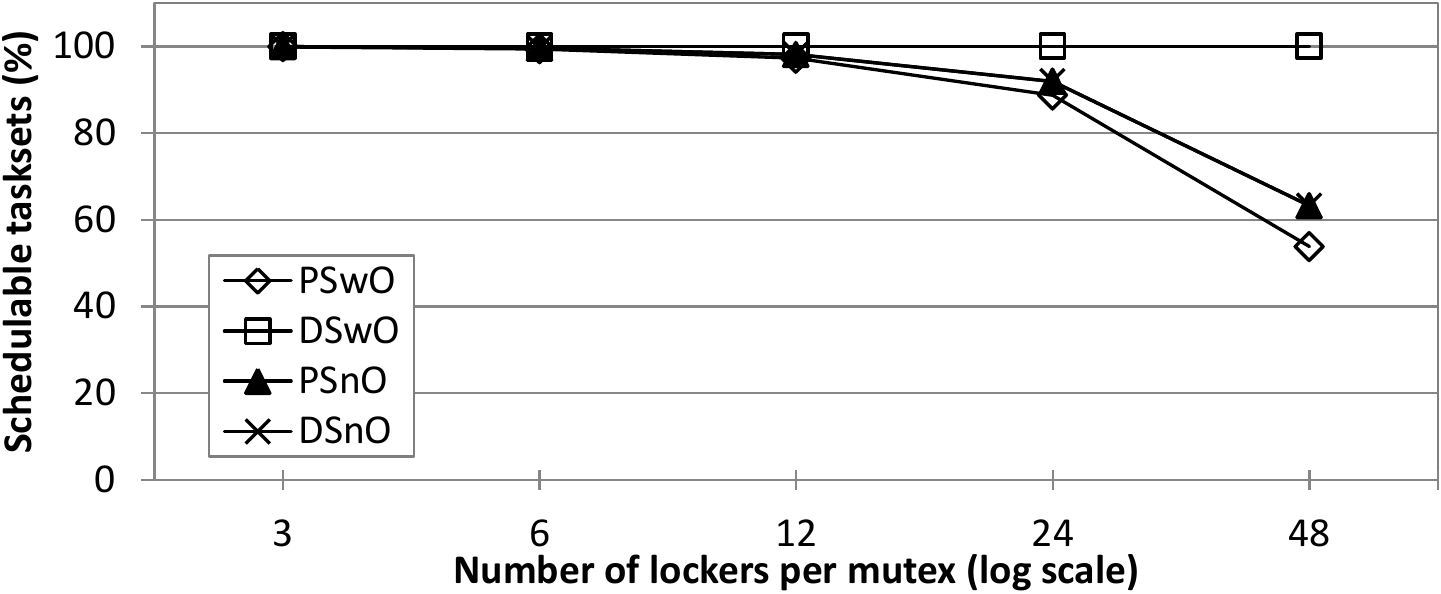}
}
\caption{Taskset schedulability as the number of lockers per mutex increases}
\label{fig:SYNC_expr_n_lockers}
\end{figure}

\figref{SYNC_expr_n_lockers} shows the percentage of schedulable tasksets as the number of lockers per mutex increases. Points on the x-axis represent all possible values for the number of lockers per mutex in our experimental setting. The performance degradation happens only when the number of lockers per muxex is very high ($> 12$). This is because vMPCP uses a two-level priority queue as the waiting list for a mutex. Hence, higher priority tasks or tasks in higher-priority VCPUs do not need to wait until all the lower-priority tasks or tasks in lower-priority VCPUs finish their gcs's. 

\begin{figure}[t]
\centering
\subfloat{
\includegraphics[width=0.7\textwidth]{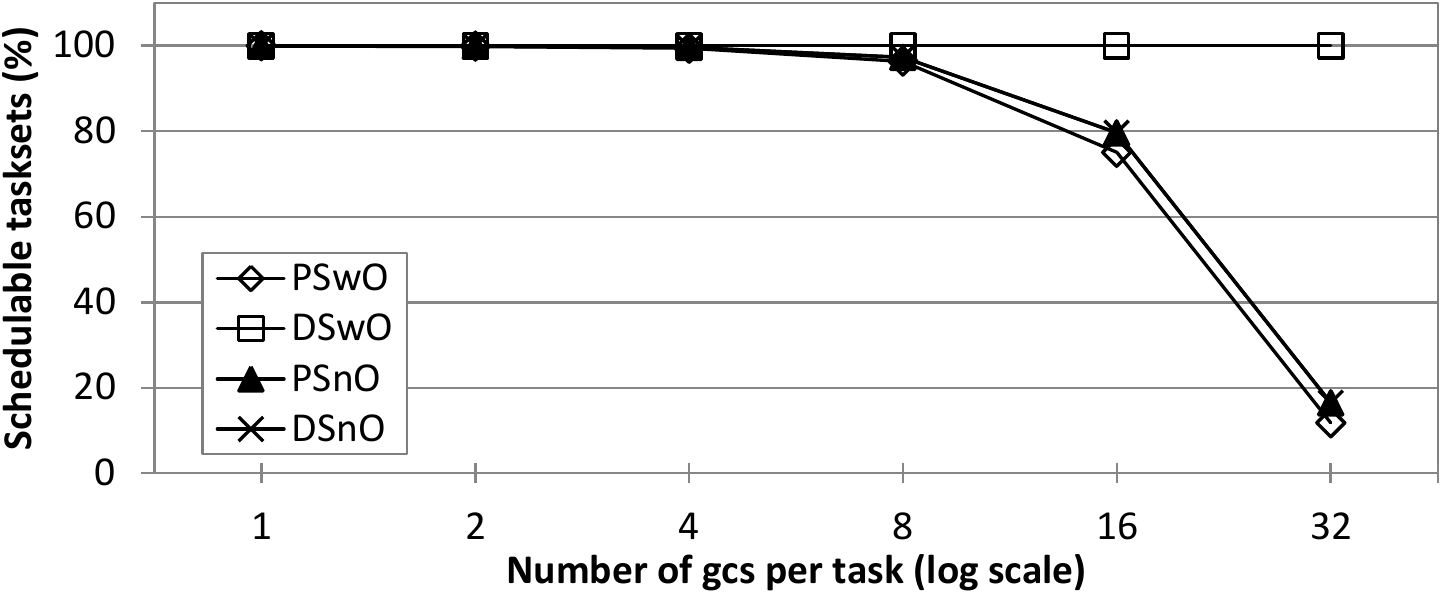}
}
\caption{Taskset schedulability as the number of gcs's per task increases}
\label{fig:SYNC_expr_gcs_per_task}
\end{figure}

\figref{SYNC_expr_gcs_per_task} shows the percentage of schedulable tasksets as the number of gcs's per task increases. The performance difference between DSwO and the other three schemes becomes larger as the number of gcs's per task increases. Even if the number of gcs's per task reaches 32, DSwO does not show any noticeable performance degradation due to its short gcs response time.

\begin{figure}[t]
\centering
\subfloat{
\includegraphics[width=0.7\textwidth]{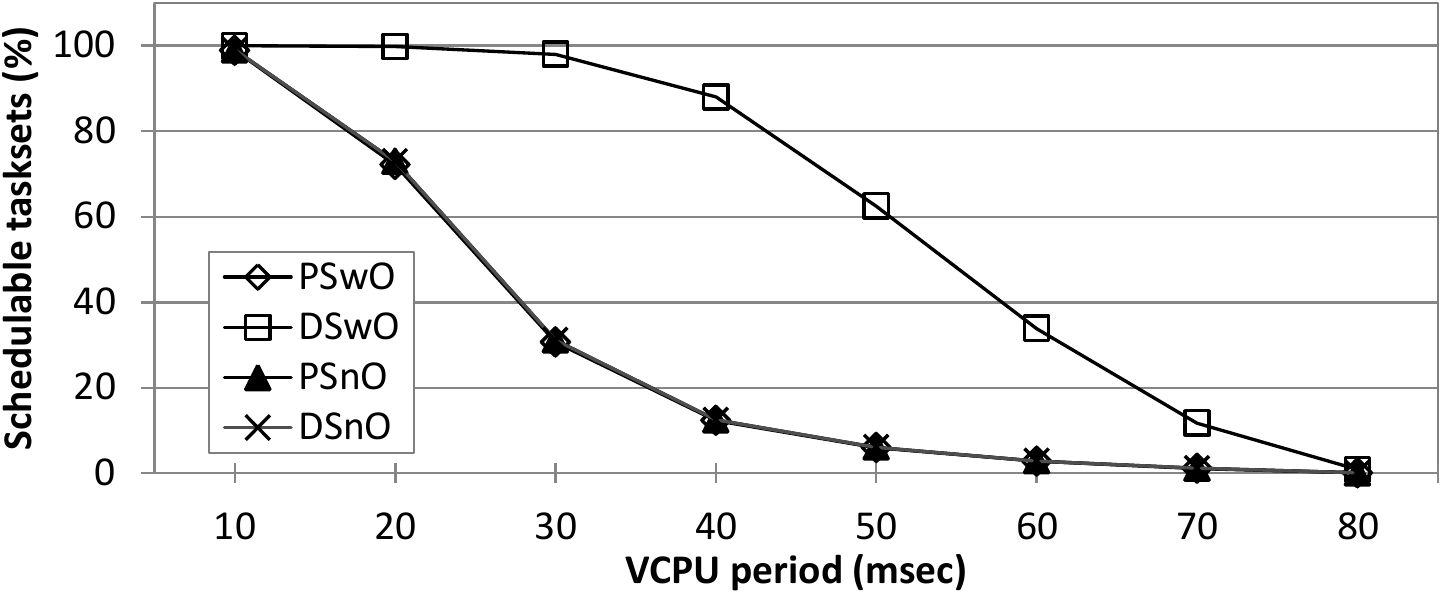}
}
\caption{Taskset schedulability as the VCPU period increases}
\label{fig:SYNC_expr_vcpu_period}
\end{figure}

\figref{SYNC_expr_vcpu_period} shows the percentage of schedulable tasksets as the VCPU period increases. DSwO performs much better than the other three schemes. Especially, when the VCPU period is 40~msec, the difference in the percentage of schedulable tasksets between DSwO and the other schemes is about 80\%. This big difference is due to the fact that PSwO, PSnO and DSnO are sensitive to the VCPU period when accessing global resources, as given by Eq.~\eqref{eq:SYNC_w_gcs_periodic} and Eq.~\eqref{eq:SYNC_w_gcs_no_overrun}. 

\begin{figure}[t]
\centering
\subfloat{
\includegraphics[width=0.7\textwidth]{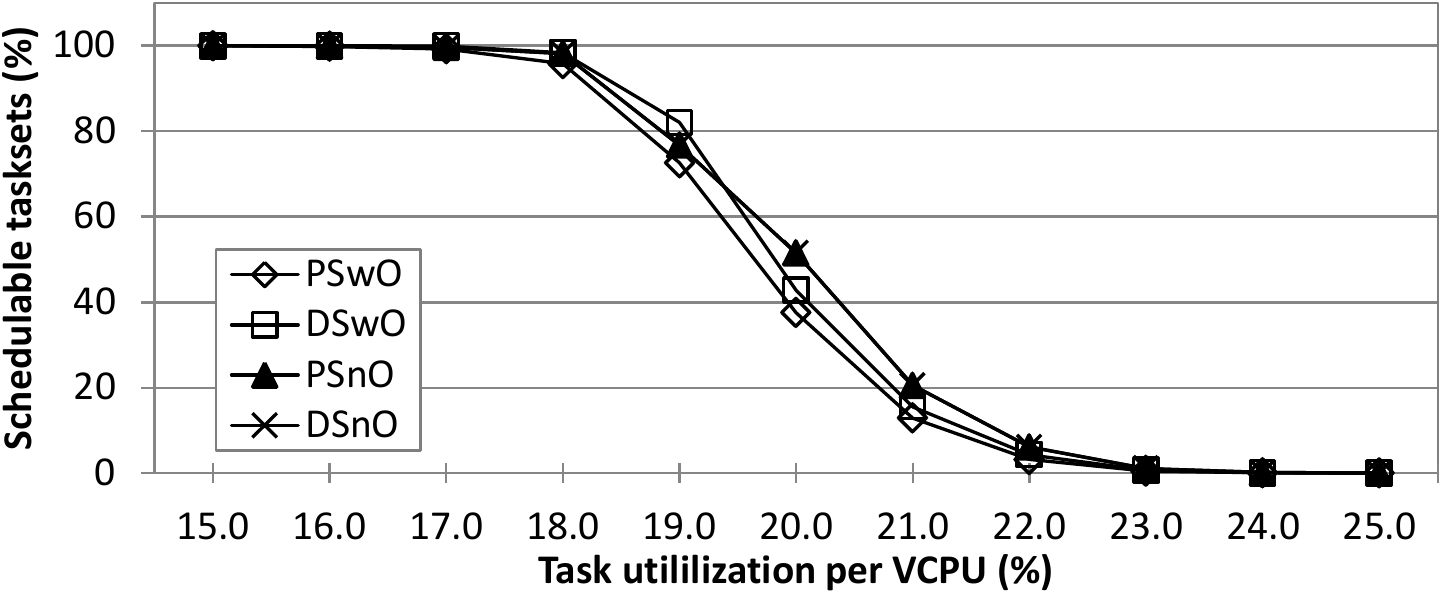}
}
\caption{Taskset schedulability as the task utilization per VCPU increases}
\label{fig:SYNC_expr_taskset_util}
\end{figure}

Lastly, \figref{SYNC_expr_taskset_util} shows the percentage of schedulable tasksets as the utilization of tasks per VCPU increases. For all schemes, the percentage decreases when the per-VCPU utilization is greater than 17.0\%. Interestingly, when the utilization is 19.0\%, DSwO performs better than PSnO and DSnO, but when the utilization is 20.0\%, the result is the opposite.  

In summary, we observe from the results that there is no single scheme that can dominate the others. DSwO generally performs better than PSwO, PSnO and DSnO, due to its short gcs response time. In some cases, PSnO and DSnO outperform DSwO by allowing more VCPU budgets for the normal execution segments of tasks. PSwO gives the worst performance in our experiments. This is because PSwO allows less VCPU budget for normal execution segments than PSnO and DSnO, and gives longer gcs response time than DSwO.

\subsection{Case Study: vMPCP on KVM Hypervisor}
\label{SYNC_expr_case_study}

We now present a case study demonstrating the benefit of vMPCP by using our implementation on the KVM hypervisor.

\smallskip
\noindent\textbf{Implementation:} We have implemented vMPCP on the KVM (Kernel-based Virtual Machine) hypervisor~\cite{KVM} of the latest version of Linux/RK~\cite{LinuxRK, ResourceKernel}.\footnote{Linux/RK is available at \url{https://rtml.ece.cmu.edu/redmine/projects/rk}.} The host machine runs on Linux/RK, and uses KVM to execute guest VMs that also run on Linux/RK. Our implementation supports the deferrable server policy and an optional overrun mechanism. The vMPCP mutex data structures and APIs (e.g., open, lock, unlock) are implemented as part of the Linux/RK kernel module. Specifically, the vMPCP mutexes are classified into {\em intra-VM} and {\em inter-VM} mutexes based on the memory spaces their corresponding global resources belong to. The intra-VM mutexes are for resources shared within a guest VM and use the \texttt{vmpcp\_start\_gcs()} and \texttt{vmpcp\_finish\_gcs()} hypercalls internally. The inter-VM mutexes are for resources shared among guest VMs and the hypervisor. They are implemented by using the per-VCPU {\tt virtqueue} interface of {\tt virtio}~\cite{virtio} for hypervisor-VM communication.


\tableref{SYNC_impl_cost} lists the implementation costs of vMPCP APIs on the KVM hypervisor. The target system used is equipped with an Intel Core i7-2600 quad-core processor running at 3.4~GHz and 8GBytes of RAM. To reduce measurement inaccuracies, we have disabled the simultaneous multithreading and dynamic clock frequency scaling of the processor. The {\tt open} and {\tt destroy} APIs take longer times for intra-VM mutexes than for inter-VM mutexes. This is mainly due to the performance difference between a VM and the hypervisor in memory allocation and deallocation for mutex data structures. The costs of {\tt lock}, {\tt trylock} and {\tt unlock} APIs are similar for both intra- and inter-VM mutexes. The major factor contributing to the lock/unlock costs is the ``world switch'' between a VM and the hypervisor. Since the intra-VM mutexes cause the {\tt vmpcp\_start\_gcs} and {\tt vmpcp\_finish\_gcs} hypercalls, the world switch happens for intra-VM mutexes as well.


\begin{table}[t]
\centering
{
\footnotesize
\caption{Implementation cost of vMPCP on the KVM hypervisor\VS{-6pt}}\label{tab:SYNC_impl_cost}
\begin{tabular}{c|l|c|c}
\hline
Types & \multicolumn{1}{c|}{Mutex APIs} & Avg ($\mu$sec) & Max ($\mu$sec)\VS{-1pt}\\\hline
\multirow{8}{*}{Intra-VM} & {\tt open} (create new mutex) & 4.16 & 7.14\\
& {\tt open} (existing mutex) & 1.87 & 3.64 \\
& {\tt destroy}	&	1.83 &	3.50 \\
& {\tt lock}	&	3.51 &	5.69 \\
& {\tt trylock}	&	2.75 &	5.15 \\
& {\tt unlock}	&	2.26 &	2.68\\
& {\tt *vmpcp\_start\_gcs}	&	2.05 &	2.88\\
& {\tt *vmpcp\_finish\_gcs}	&	1.40 &	1.60\\
\hline
\multirow{6}{*}{Inter-VM} & {\tt open} (create new mutex) & 1.79 & 3.48\\
& {\tt open} (existing mutex) & 1.76 & 3.35 \\
& {\tt destroy}	&	1.49 &	1.78 \\
& {\tt lock}	&	3.09 &	5.31 \\
& {\tt trylock}	&	2.80 &	5.29 \\
& {\tt unlock}	&	1.93 &	2.57\\
\hline
\end{tabular}
}
\end{table}

\smallskip
\noindent\textbf{Case Study:} In this case study, we compare the response times of tasks sharing a global resource under vMPCP and those under a virtualization-unaware multi-core synchronization protocol, MPCP. The target system hosts two guest VMs, each of which has four VCPUs (VM1: $\{v_1, v_3,v_5,v_7\}$, VM2: $\{v_2,v_4,v_6,v_8\}$). All VCPUs have the same budget and period: $v_i=(3,10)$, units in msec. The VCPUs are ordered in increasing order of priorities, i.e., $i<j \implies \pi_i^v < \pi_j^v$. Hence, $v_8$ is the highest-priority VCPU. The release offset of each VCPU is zero. The target machine has four processing cores, Core 1, 2, 3 and 4. Each core is assigned two VCPUs: $\text{Core 1}=\{v_1, v_2\}$, $\text{Core 2}=\{v_3, v_4\}$, $\text{Core 3}=\{v_5, v_6\}$, $\text{Core 4}=\{v_7, v_8\}$. For a taskset, we use eight synthetic tasks, each of which has one gcs. There is one global resource shared among all these tasks. Each task is assigned to a VCPU with the same index number, e.g., $\tau_5 \in v_5$. All tasks except $\tau_2$ have the same timing parameters: $\tau_i=((2, 1, 2), 200)$, where $i\ne 2$, units in msec. Task $\tau_2$ has a slightly longer gcs: $\tau_2=((2, 1.1, 2), 200)$. Each task $\tau_i$ also has a release offset of $i - 1$~msec, e.g., $\tau_5$ is released at $t=4$~msec. We used Linux/RK to set the periods, release offsets, and real-time priorities of VCPUs and tasks. In accordance with our system model, tasks and VCPUs with higher indices are assigned higher priorities. 

\begin{figure}[t]
\centering
\subfloat[MPCP] {\label{fig:SYNC_case_mpcp}
\includegraphics[width=.7\textwidth]{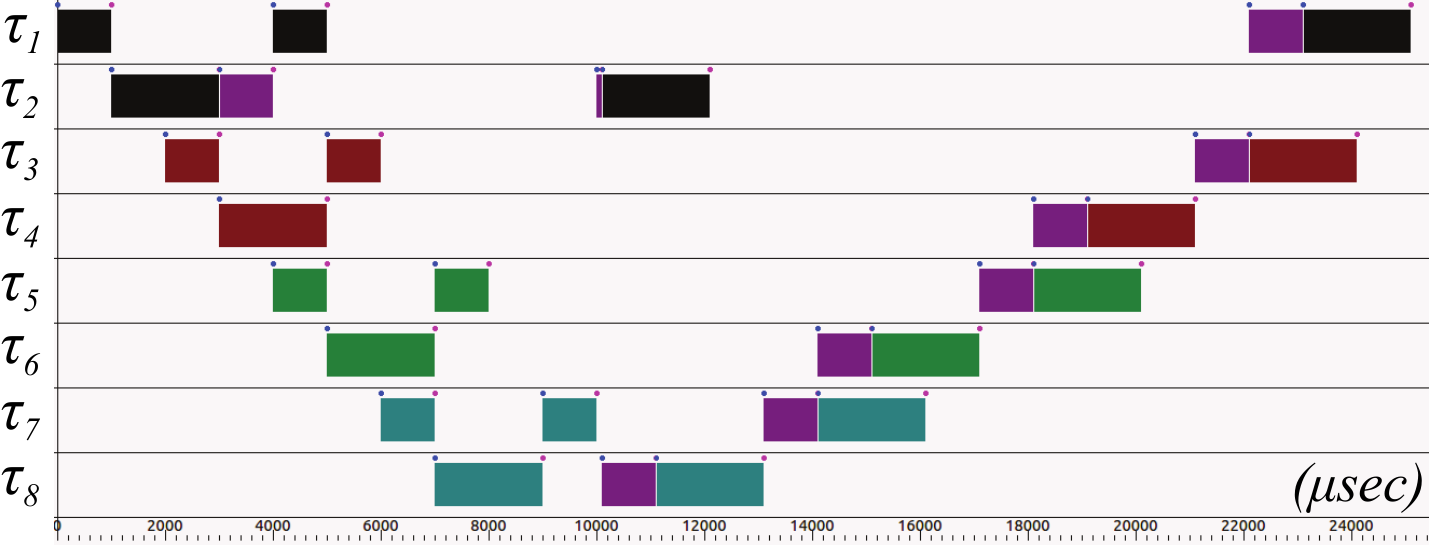}
}\\
\vspace{-5pt}
\subfloat[vMPCP+DSnO (Deferrable server w/ no overrun)] {\label{fig:SYNC_case_vmpcp}
\includegraphics[width=.7\textwidth]{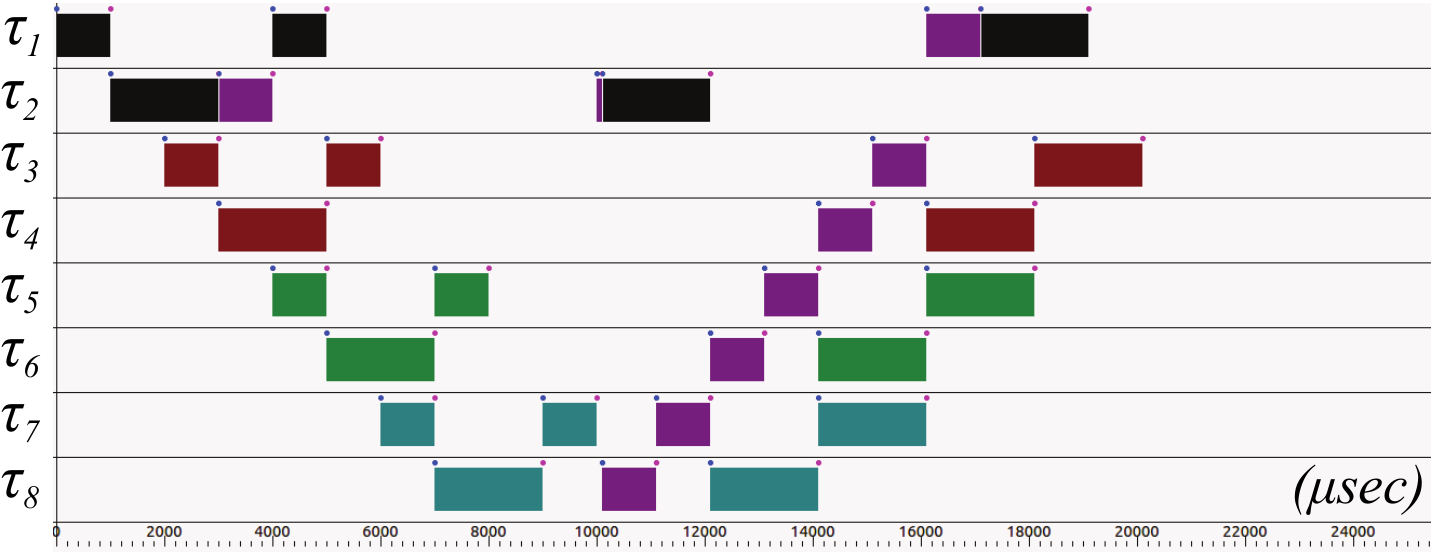}
}\\
\vspace{-5pt}
\subfloat[vMPCP+DSwO (Deferrable server w/ overrun)] {\label{fig:SYNC_case_vmpcp_overrun}
\includegraphics[width=.7\textwidth]{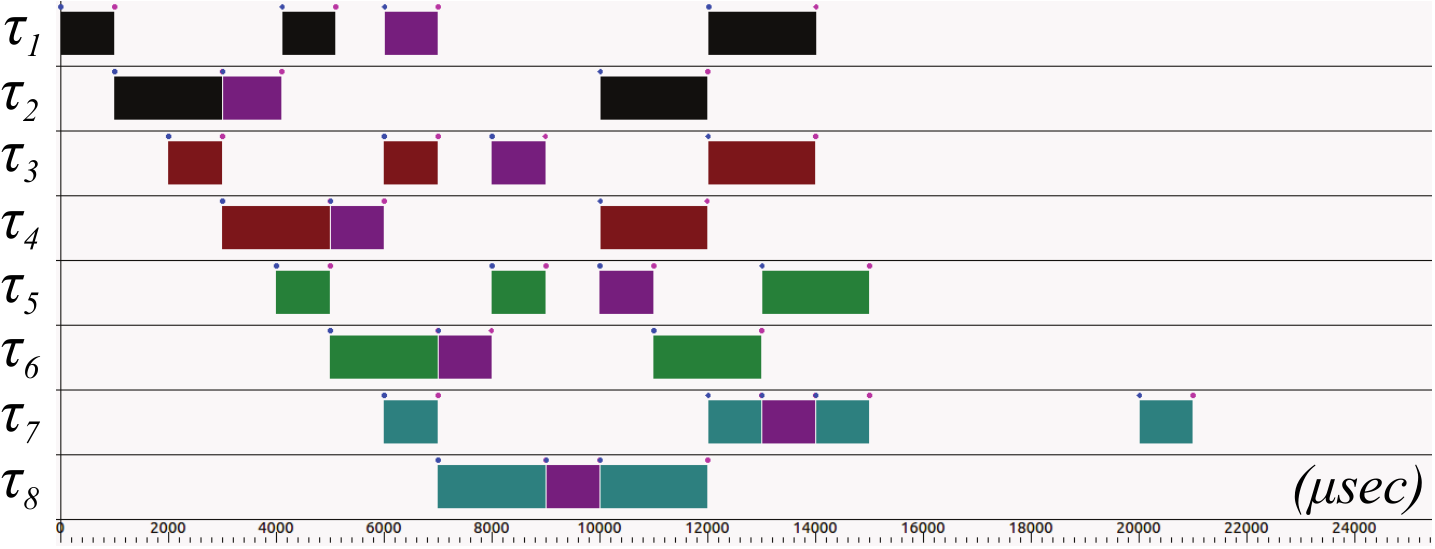}
}
\caption{Task execution timelines under MPCP, vMPCP+DSnO and vMPCP+DSwO}
\label{fig:SYNC_execution_timeline}
\end{figure}

\figref{SYNC_execution_timeline} shows the execution timelines of tasks captured under MPCP, vMPCP with deferrable server and no overrun (vMPCP+DSnO), and vMPCP with deferrable server and overrun enabled (vMPCP+DSwO). As can be seen, the response times of tasks are much shorter under vMPCP+DSnO and vMPCP+DSwO, compared to those under MPCP (7.5\% of resposne time decrease on average under vMPCP+DSnO, and 29.1\% under vMPCP+DSwO). The shared resource is first held by $\tau_2$ at $t=3$, but under MPCP and vMPCP+DSnO, it cannot release the resource due to its VCPU's budget depletion. Hence, the resource is held by $\tau_2$ until the start of its VCPU's next replenish period. Conversely, under vMPCP+DSwO, $\tau_2$ can finish its gcs and release the resource. This allows other tasks to access the resource within the first VCPU period, thereby significantly reducing the response times of tasks. In case of task $\tau_8$, it acquires the resource at $t=10$ under both MPCP and vMPCP+DSnO. Here, the difference happens when $\tau_8$ finishes its gcs. Under MPCP, $\tau_8$ continues to execute because its VCPU has the highest priority on that core. This causes a delay to task $\tau_7$, which is the highest-priority task among the tasks waiting on the resource, to enter its gcs. However, under vMPCP+DSnO, $\tau_7$ starts its gcs right after the resource is released by $\tau_8$. This slightly lengthens the response time of $\tau_8$, but allows other tasks to access the resource much faster. Under vMPCP+DSwO, the response times of all tasks except $\tau_7$ are shorter than those under the other two schemes. The increase in $\tau_7$'s response time is due to the back-to-back execution of the VCPU of $\tau_8$, the amount of which is bounded by our analysis. The case study results show that vMPCP is effective in reducing the response times of tasks accessing shared resources in a multi-core virtualization environment. 


\section{Summary}
\label{SYNC_conclusions}

In this chapter, we developed a novel synchronization framework, vMPCP, to provide bounded blocking time on accessing shared resources in a multi-core virtualization environment. vMPCP reduces the major inefficiencies caused by shared resources, by exposing the executions of global critical sections to the hypervisor. We presented the schedulability analysis under vMPCP, with the periodic and deferrable server policies with and without the budget overrun mechanism. 
From our analysis and experimental results, we made two important findings: (i) the deferrable server outperforms the periodic server when overrun is used, and (ii) the use of overrun does not always yield better schedulability, especially for tasks with long critical sections. We implemented vMPCP on the KVM hypervisor and demonstrated the effect of vMPCP in reducing task response times by an average of 29\% in our case study.
Interesting future directions that can build on our work include the extension of our schedulability analysis to the compositional framework~\cite{Shin_RTSS03, Shin_ACM08}, and the implementation and evaluation of vMPCP on other hypervisors, such as L4/Fiasco~\cite{L4/Fiasco}.

\chapter{Responsive and Enforced Interrupt Handling}
\label{chapter_interrupt_handling}
This chapter describes our proposed interrupt handling scheme for multi-core virtualization, called vINT. 
vINT provides a pseudo-VCPU abstraction to explicitly account for and enforce the CPU usage of virtual interrupt handling. With a pseudo-VCPU, vINT enables tasks within a VCPU to meet their deadlines without suffering from virtual interrupt storms. The use of the pseudo-VCPU abstraction also allows assigning a separate budget and priority to just the interrupt handler and interrupt-triggered tasks of a guest VM. This makes a virtual interrupt be handled although the budget of its original VCPU has been depleted. In addition, virtual interrupt handling is no longer dominated by the budget, replenishment period and priority of its original VCPU, thereby significantly reducing interrupt handling time in a virtualized environment.
vINT does not require making any change to the guest OS code. Hence, it can be easily applicable to {\em full virtualization} scenarios hosting unmodified, proprietary guest OSs. 

We analyze interrupt handling time in a virtualized environment with and without vINT. We also provide analyses on the schedulability of VCPUs and tasks in the presence of physical and virtual interrupts. 
Our experimental results indicate that vINT achieves timely interrupt handling while providing as good task schedulability as when it is not used. 
We have implemented a prototype of vINT on the KVM hypervisor (chosen for convenience). Our case study using this implementation shows the benefits of vINT in providing responsive interrupt handling times and protecting tasks against virtual interrupt storms. \tableref{INTR_comparison_with_previous_work} gives a brief comparison of vINT with closely related prior work.


The background and related prior work on interrupt handling were  discussed in Section~\ref{background_interrupts}. The system model including assumptions and notation for tasks and virtual machines can be found in Chapter~\ref{system_model}.

The rest of this chapter is organized as follows. Section~\ref{INTR_interrupt_handling_time} gives a detailed description on interrupt handling in a virtualization environment and defines interrupt handling time. Section~\ref{INTR_vint_scheme} presents our proposed vINT scheme. Section~\ref{INTR_vint_analysis} shows our analyses on interrupt handling time, and VCPU and task schedulability. Section~\ref{INTR_evaluation} provides detailed evaluation, and Section~\ref{INTR_conclusions} summarizes this chapter.

\begin{table}[t]
	\centering
	\VS{-5pt}
	{
		\scriptsize
		\caption[Comparison with prior work on interrupt handling]{Comparison with previous work\VS{-5pt}}\label{tab:INTR_comparison_with_previous_work}
		\VS{-2pt}
		\begin{tabular}{C{1cm}||C{1.8cm}|C{1.8cm}|C{1.8cm}|C{1.8cm}|C{1.8cm}|C{1.8cm}}
			\hline
			& Priority 		& VCPU  		& Bounded	& Enforced 	& Task 		& Unmodified\\
			Schemes						& based			& temporal 		& interrupt & interrupt	& schedulability 	& guest	OS	\\
			& scheduling		& isolation		& handling	& handling	& analysis  & support 		\\
			\hline
			\cite{Beckert_DAC14}		&				&				& X			& X			& 			& X			\\
			\cite{Kiszka_RTLWS09}		& X				& X				&			&			&			&			\\
			\cite{Lackorzynski_EMSOFT12}& X				& X				&			&			&			&			\\
			\cite{Ma_JISE13}			& X				& X				&			&			&			&			\\
			\hline
			vINT 					& X				& X				& X			& X			& X			& X			\\
			\hline
		\end{tabular}\VS{-10pt}
	}
\end{table}

\section{Interrupt Handling Time in Virtualization}
\label{INTR_interrupt_handling_time}

\begin{figure}[t]
\centering
\VS{-8pt}
\subfloat[Interrupt handling on the same PCPU]{\label{fig:INTR_baseline_interrupt_handling_same_core}
\includegraphics[width=0.7\textwidth]{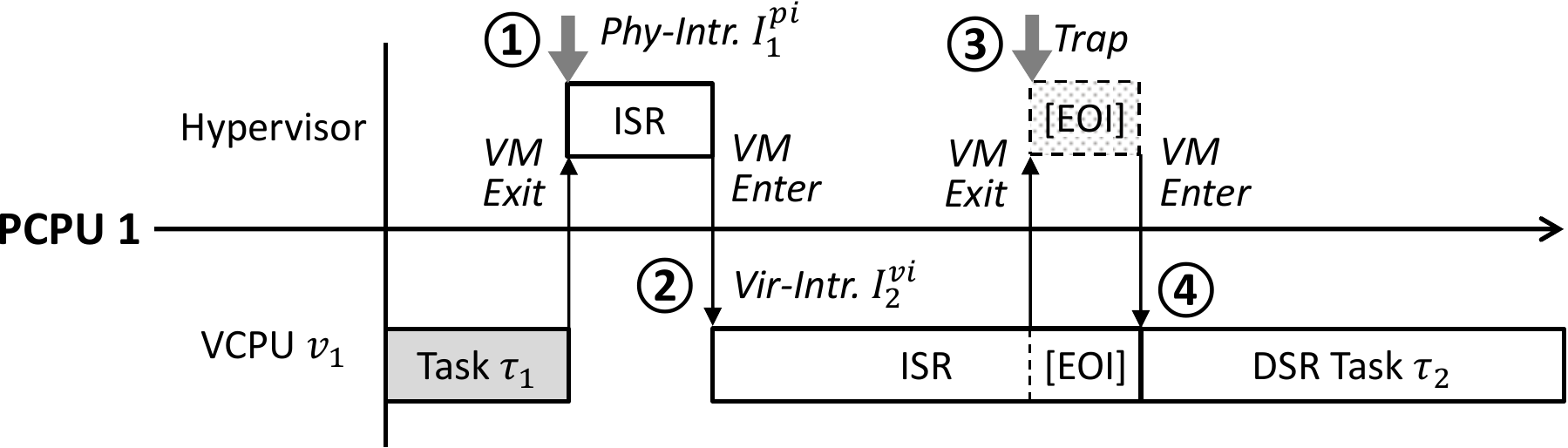}
}\\
\VS{-8pt}
\subfloat[Interrupt handling across two PCPUs]{\label{fig:INTR_baseline_interrupt_handling_diff_core}
\includegraphics[width=0.7\textwidth]{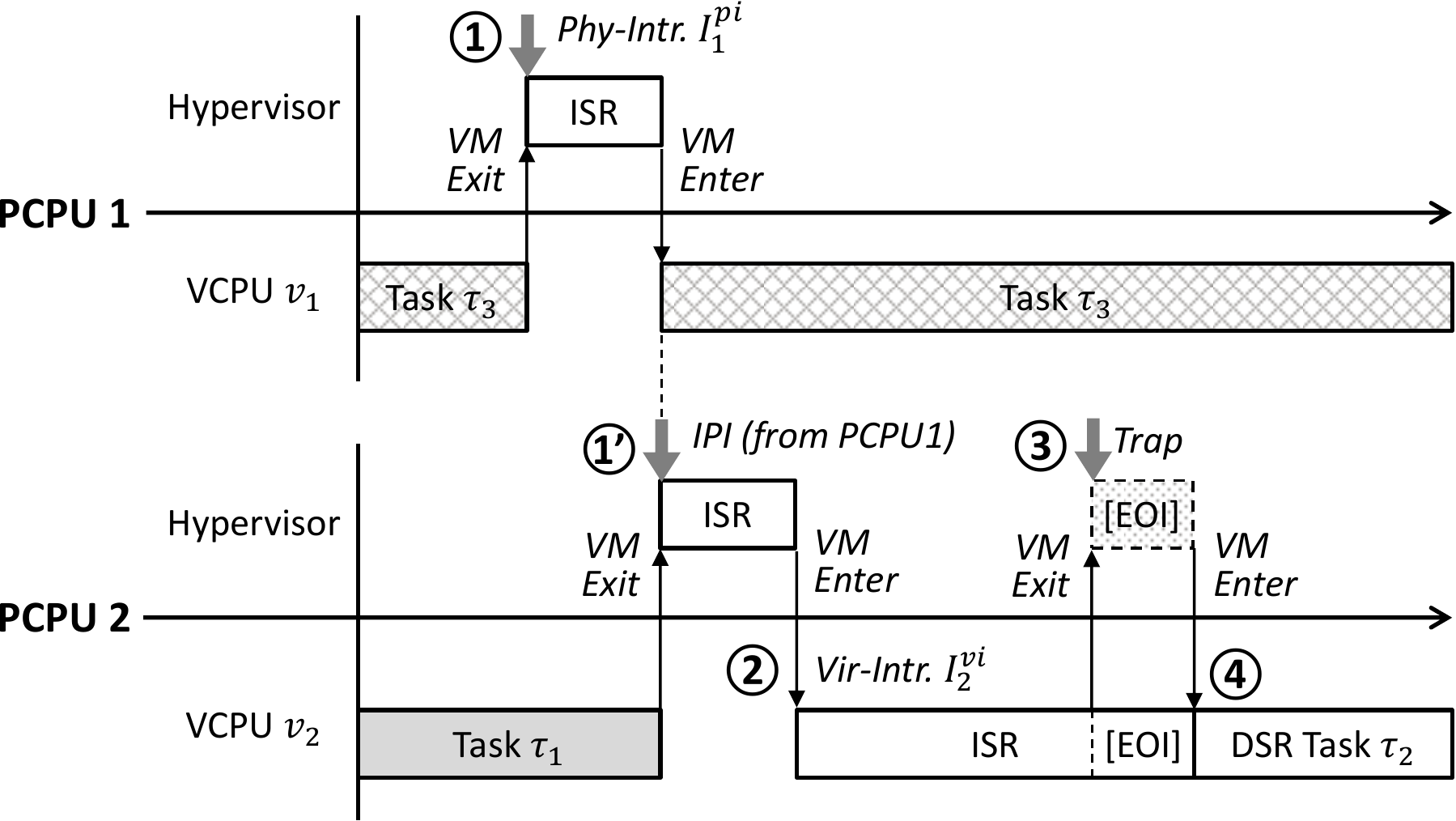}
}
\VS{-5pt}
\caption{Interrupt handling in virtualization}
\VS{-9pt}
\label{fig:INTR_baseline_interrupt_handling}
\end{figure}

We consider two types of interrupts: {\em physical} and {\em virtual}. A physical interrupt $I_i^{pi}$ is a signal issued from a hardware device to a PCPU.
Each physical interrupt is assumed to be statically pinned to one PCPU, which can be easily done in software with the support of a programmable interrupt controller (PIC). 
When a PCPU receives a physical interrupt, the currently executing VCPU on that PCPU is halted and the corresponding ISR of the hypervisor is executed (Step \textcircled{\footnotesize 1} in \figref{INTR_baseline_interrupt_handling}). The CPU time usage of the ISR of a physical interrupt is accounted for as the hypervisor's usage, not as the halted VCPU's usage. Each physical interrupt has a unique priority $\pi_i^{pi}$ determined by the PIC. 
The ISR of a lower-priority physical interrupt can be preempted by that of a higher-priority physical interrupt. The ISRs of physical interrupts are not preemptible by VCPUs. Therefore, a VCPU preempted by an ISR can only resume its execution when all ISRs have been completed. 
A physical interrupt $I_i^{pi}$ is represented as follows:
$$I_i^{pi}:=(C_i^{pi},T_i^{pi})$$
where,
\begin{itemize}
  \item $C_{i}^{pi}$: the WCET of the ISR of $I_i^{pi}$ 
  \item $T_i^{pi}$: the minimum inter-arrival time\footnote{Similarly to prior work~\cite{Brandenburg_JSA11, Leyva_ACM12}, the minimal inter-arrival time of an interrupt refers to a value expected or identified at design time. An interrupt unexpectedly arriving faster than that value may cause an interrupt storm at runtime.} of $I_i^{pi}$
\end{itemize}
The response time of a physical interrupt (or physical interrupt handling time) is the time from the arrival of the physical interrupt signal to the completion of the corresponding ISR. 

A virtual interrupt $I_i^{vi}$ is a software signal from the hypervisor to a guest VM, issued upon the completion of the ISR of a physical interrupt.\footnote{There might be some cases where virtual interrupts are generated as a result of polling at the hypervisor. Considering such a mixed use of interrupts and polling in real-time virtualization remains as future work.} 
Each virtual interrupt is assumed to be statically pinned to one VCPU of a VM. If a target VCPU is located on a PCPU different from that of a physical ISR, e.g., a physical interrupt shared among multiple VCPUs, the delivery of a virtual interrupt from a physical ISR to the VCPU causes an inter-processor interrupt (IPI) that is an additional physical interrupt to notify a state change to the VCPU running on a different PCPU (Steps \textcircled{\footnotesize 1'} and \textcircled{\footnotesize 2} in \figsubref{INTR_baseline_interrupt_handling_diff_core}). Otherwise, a virtual interrupt is immediately delivered to the corresponding VCPU (Step \textcircled{\footnotesize 2} in \figsubref{INTR_baseline_interrupt_handling_same_core}).
When a VCPU receives a virtual interrupt, the currently executing task in that VCPU is halted and the corresponding ISR of the guest OS is executed. 
Each virtual interrupt has a unique priority $\pi_i^{vi}$ given by the emulated PIC. 
Within each VCPU, the ISRs of lower-priority virtual interrupts can be preempted by those of higher-priority virtual interrupts, and virtual ISRs are not preemptible by tasks. As in most CPU architectures, each ISR executes an End-Of-Interrupt (EOI) instruction at the end to notify the completion of the ISR to the PIC. As the EOI is a privileged instruction called by the guest OS, it is trapped and emulated by the hypervisor while consuming the budget of the corresponding VCPU (Step \textcircled{\footnotesize 3} in \figref{INTR_baseline_interrupt_handling}). A virtual interrupt is {\em pending} if it has been injected to the corresponding VCPU but its ISR has not yet been completed.
A virtual interrupt $I_i^{vi}$ is represented as follows:
$$I_i^{vi}:=(C_i^{vi},T_i^{vi})$$
where,
\begin{itemize}
  \item $C_{i}^{vi}$: the WCET of the ISR of $I_i^{vi}$ 
  \item $T_i^{vi}$: the minimum inter-arrival time of $I_i^{vi}$
\end{itemize}

We consider a {\em split interrupt handling} model for guest OSs due to its the wide acceptance in both real-time and non-real-time OSs.
Under split interrupt handling, the ISR of a virtual interrupt performs the minimum amount of work and activates zero or more tasks to execute a deferred service routine (DSR) in the task context (Step \textcircled{\footnotesize 4} in \figref{INTR_baseline_interrupt_handling}). Hence, the priorities of DSRs can be easily configured in contrast to ISRs, and the majority of interrupt handling can be done with desired priorities.
We use $\mathbb{D}(I_i^{vi})$ to denote the set of DSR tasks triggered by the ISR of a virtual interrupt $I_i^{vi}$.
The minimum inter-arrival time of any task in $\mathbb{D}(I_i^{vi})$ is therefore equal to or greater than $T_i^{vi}$. 
The response time of a virtual interrupt (or virtual interrupt handling time) is the time from the arrival of the virtual interrupt to the completion of the corresponding ISR and DSR. Lastly, we denote the sum of the WCETs of the ISR and DSR of a virtual interrupt $I_i^{vi}$ as:
$$\mathbb{C}_i^{vi}=C_i^{vi}+\sum_{\tau_j\in \mathbb{D}(I_i^{vi})}C_j$$





\begin{definition}
An \underbar{interrupt-triggered execution flow} in a virtualized environment is the sequence of executions from the arrival of a physical interrupt to the completion of the ISR and DSR of the corresponding virtual interrupt.
\end{definition}

\begin{definition}
The \underbar{total interrupt handling time} is the amount of time to complete the corresponding interrupt-triggered execution flow.
\end{definition}

\begin{definition} 
An interrupt-triggered execution flow is \underbar{serviceable}, if its total interrupt handling time does not exceed the minimum inter-arrival times of the corresponding physical and virtual interrupts.
\end{definition}


\section{vINT Scheme}
\label{INTR_vint_scheme}


The problems with virtual interrupt handling described in Section~\ref{INTR_interrupt_handling_problems} are caused by the fact that a virtual interrupt is handled by the same VCPU as the one used by other regular tasks. Motivated by this, we propose vINT that can conceptually split virtual interrupt handling from the VCPU of regular tasks in an analyzable way, without modifying the guest OS code. vINT can be selectively used for a subset of virtual interrupts that cannot be serviced within their minimal inter-arrival times by default, or have a possibility of causing virtual interrupt storms. For convenience of explanation, we assume that all virtual interrupts are managed by vINT in Section~\ref{INTR_pseudo_vcpu_abstraction} and \ref{INTR_pseudo_vcpu_realization}. In Section~\ref{INTR_selective_use_of_vint} we relax this assumption. 



\subsection{Pseudo-VCPU Abstraction}
\label{INTR_pseudo_vcpu_abstraction}
vINT uses a {\em pseudo-VCPU} abstraction to represent the resource requirement of the ISR and DSR of a virtual interrupt as a separate VCPU to the hypervisor. The pseudo-VCPU differs from its original VCPU in that it does not have an execution context. In other words, the use of the pseudo-VCPU introduces no additional processing core visible to the guest VM, which is typically a high demand to host legacy guest OSs that may support only uniprocessors. 

Each virtual interrupt can be exclusively associated with one pseudo-VCPU that is located on the same PCPU as its original VCPU. A pseudo-VCPU $v_p$ is described by the same types of parameters as a regular VCPU: $C_p^v$ and $T_p^v$. 
The replenishment period of a pseudo-VCPU $v_p$ is equal to or greater than the minimum inter-arrival time of the associated virtual interrupt $I_i^{vi}$, i.e., $T_p^v \ge T_i^{vi}$. The budget $C_p^v$ of a pseudo-VCPU $v_p$ associated with a virtual interrupt $I_i^{vi}$ is assigned as follows:
\begin{equation} \label{eq:INTR_pseudo_vcpu_budget}
C_p^{v}=\left\lceil{T_p^v\over{T_i^{vi}}} \right\rceil\mathbb{C}_i^{vi}
\end{equation}
It is worth noting that, once a virtual interrupt is assigned its pseudo-VCPU, the budget of its original VCPU can be reduced because the virtual interrupt will be handled by using the budget of the pseudo-VCPU. 


\smallskip
\noindent\textbf{Prioritization of pseudo-VCPUs:}
One of our goals is to provide responsive interrupt handling time, which is challenging due to the VCPU-level preemption while handling a virtual interrupt. To achieve this goal, vINT prioritizes pseudo-VCPUs over regular VCPUs. 
The priority of a pseudo-VCPU $v_p$ associated with a virtual interrupt $I_i^{vi}$ is assigned a priority of $\pi_B^v+(\pi_o^v-1)\cdot L_o+\pi_{\mathbb{D}}(I_i^{vi})$, where $\pi_B^v$ is a base VCPU-priority level greater than that of any regular VCPU on the same PCPU, $\pi_o^v$ is the priority of the original VCPU of $I_i^{vi}$, $L_o$ is the number of priority levels for all DSR tasks in the original VCPU, and $\pi_\mathbb{D}(I_i^{vi})$ is the priority difference between the highest-priority DSR task of $I_i^{vi}$ and the highest-priority DSR task among all DSR tasks in the original VCPU. With this approach, the pseudo-VCPU $v_p$ is not preempted by any regular VCPU, and the relative priority ordering of DSRs within the same original VCPU are preserved.

\subsection{Pseudo-VCPU Realization}
\label{INTR_pseudo_vcpu_realization}

As a pseudo-VCPU does not have an execution context, in its realization, the actual execution of the ISR and DSR of a virtual interrupt still happens within the execution context of their original VCPU. We now explain how vINT handles a virtual interrupt as if it was handled in its pseudo-VCPU. 


\smallskip
\noindent\textbf{DSR task priority adjustment:} Since pseudo-VCPUs are assigned higher priorities than regular VCPUs, the executions of DSRs should not be preempted by regular tasks in the realization. vINT therefore statically adjusts the priority of each DSR task $\tau_j$ to $\pi_{B,v_o}+\pi_j$, where $\pi_{B,v_o}$ is a base task-priority level greater than any regular task in the task $\tau_j$'s original VCPU $v_o$, and $\pi_j$ is the original priority of $\tau_j$. Note that this priority adjustment is not needed if the priorities of DSR tasks are already higher than those of regular tasks in the original VCPU. In addition, since even closed-source, proprietary OSs provide an interface to configure task priorities, the priority adjustment does not violate the requirement of full virtualization.

\smallskip
\noindent\textbf{Virtual interrupt injection:} 
vINT maintains a counter for each pseudo-VCPU to indicate the number of virtual interrupts that can be handled by the pseudo-VCPU at that moment. The maximum possible value of the counter for a pseudo-VCPU $v_p$ associated with a virtual interrupt $I_i^{vi}$ is given by $\lceil {T_p^v / T_i^{vi}}\rceil$. When a virtual interrupt is generated, vINT checks the counter value of the corresponding pseudo-VCPU. If the counter is greater than zero, the counter is decremented by one and the virtual interrupt is injected into its original VCPU. Otherwise, the injection of the virtual interrupt is delayed until the counter becomes greater than zero. The replenishment rule of the counter is similar to that of the VCPU budget. Under the deferrable server policy, the counter is fully replenished at the start of every replenishment period of the pseudo-VCPU. Under the sporadic server policy, the counter is replenished by one at the time when the budget is replenished.

\smallskip
\noindent\textbf{Virtual interrupt handling:} 
We first consider a non-nested interrupt handling scenario. When a virtual interrupt $I_i^{vi}$ is injected, the original VCPU $v_o$ should handle the ISR and DSR of $I_i^{vi}$ by using the priority and budget of the corresponding pseudo-VCPU $v_p$. Hence, vINT immediately raises the priority of $v_o$ to that of $v_p$, and let $v_o$ use the budget of $v_p$ for the amount of $\mathbb{C}_i^{vi}$.
As the DSR tasks of $I_i^{vi}$ have higher priorities than regular tasks, they are guaranteed to be executed as soon as the corresponding ISR finishes. When the VCPU $v_o$ has consumed $\mathbb{C}_i^{vi}$ units of the budget of $v_p$, vINT restores the priority of $v_o$ and lets $v_o$ use its own budget afterwards. The ISR and DSR of $I_i^{vi}$ may be finished earlier than $\mathbb{C}_i^{vi}$ and regular tasks may be executed while their VCPU is still using the budget and priority of the pseudo-VCPU of $I_i^{vi}$. 
However, this does not change the worst-case interference that can be imposed on other VCPUs.

\begin{figure}[t]
\centering
\VS{-7pt}
\subfloat{
\includegraphics[width=0.7\textwidth]{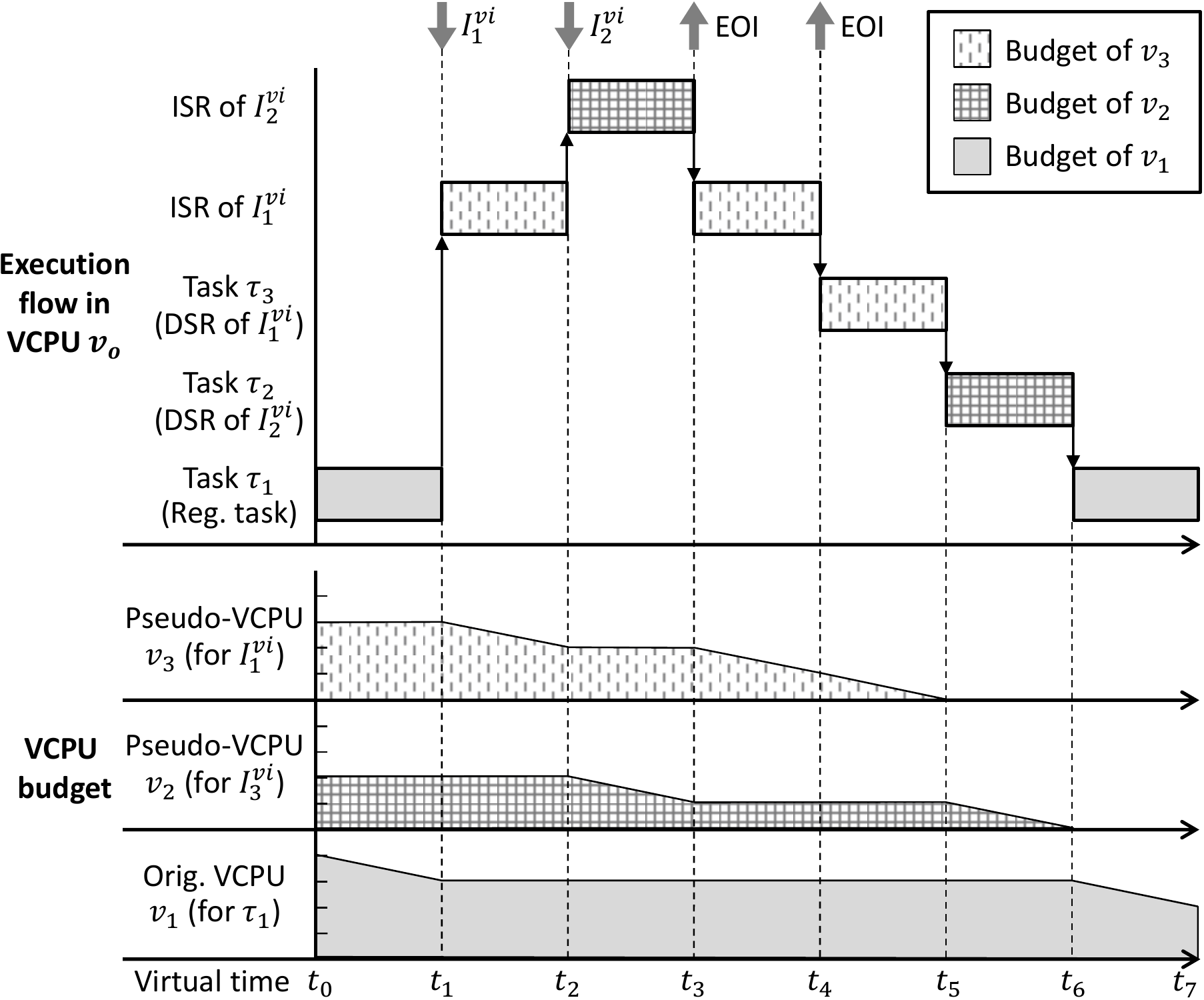}
}
\VS{-5pt}
\caption{vINT nested interrupt handling}  
\VS{-9pt}
\label{fig:INTR_vint_interrupt_handling}
\end{figure}

We next consider a nested interrupt handling scenario. vINT exploits the following two factors to support nested interrupt handling with pseudo-VCPUs: (i) the hypervisor is aware of the set of all pending virtual interrupts in each VCPU, and (ii) the hypervisor traps an EOI instruction called at the end of each virtual ISR. When a new virtual interrupt is injected into a VCPU $v_o$, vINT lets $v_o$ use the budget and priority of the pseudo-VCPU that is associated with the highest-priority virtual interrupt among all pending interrupts. This is because the VCPU executes the ISR of the highest-priority pending interrupt first. When the hypervisor catches an EOI from $v_o$, vINT checks if there is another pending interrupt. If so, vINT lets $v_o$ use the budget and priority of the pseudo-VCPU of the higher-priority pending interrupt, and repeats this until there is no pending interrupt. If there is no pending interrupt, vINT now lets $v_o$ use the budget and priority of the highest-priority pseudo-VCPU, the budget of which has not yet been used for the amount of $\mathbb{C}_i^{vi}$ by $v_o$ to handle the injected interrupt $I_i^{vi}$.
As the relative priorities of pseudo-VCPU follow those of DSR tasks, this approach makes the sequence of the pseudo-VCPU usage correspond to that of the DSR task executions. \figref{INTR_vint_interrupt_handling} shows an example of nested interrupt handling with vINT. In this figure, the x-axis represents the passage of virtual time so the activities of the hypervisor and other VCPUs are omitted. In this figure, tasks, VCPUs, and interrupts are ordered in increasing order of priorities, e.g., $I_2^{vi}$ has higher priority than $I_1^{vi}$.

\subsection{Selective Use of vINT}
\label{INTR_selective_use_of_vint}

We now relax our assumption that all virtual interrupts are managed by vINT. If a virtual interrupt is not managed by vINT, it is not associated with a pseudo-VCPU. The priorities of its DSR tasks remain unchanged. However, the presence of such an unmanaged virtual interrupt affects the pseudo-VCPU budgets of virtual interrupts managed by vINT. Consider a virtual interrupt $I_i^{vi}$ associated with a pseudo-VCPU $v_p$. If there is any virtual interrupt not managed by vINT in the original VCPU of $I_i^{vi}$, the budget $C_p^v$ of $v_p$ is assigned by:
\begin{equation} \label{eq:INTR_pseudo_vcpu_budget_enhanced}
C_p^{v}=\left\lceil{T_p^v\over{T_i^{vi}}} \right\rceil\left(\mathbb{C}_i^{vi}+\sum_{\substack{I_j^{vi} \in \mathbb{V}(I_i^{vi}) \land  pseudo(I_j^{vi})=\emptyset}} \left\lceil{T_i^{vi}\over{T_j^{vi}}} \right\rceil C_j^{vi}\right)
\end{equation}
where, $\mathbb{V}(I_i^{vi})$ is the original VCPU of $I_i^{vi}$, and $pseudo(I_j^{vi})$ is a function returning the pseudo-VCPU of $I_j^{vi}$ if exists, and $\emptyset$ otherwise. 
The second term in the parenthesis of Eq.~\eqref{eq:INTR_pseudo_vcpu_budget_enhanced} is an extra budget for the executions of the ISRs of virtual interrupts not managed by vINT. Since those ISRs may block the handling of $I_i^{vi}$ in the realization, the extra budget allows the ISRs to be executed with the budget and priority of $I_i^{vi}$'s pseudo-VCPU. Therefore, when an instance of $I_i^{vi}$ is injected into its original VCPU $v_o$, vINT lets $v_o$ use the budget and priority of the pseudo-VCPU $v_p$ for the sum of the terms in the parenthesis of Eq.~\eqref{eq:INTR_pseudo_vcpu_budget_enhanced}, instead of only $\mathbb{C}_i^{vi}$.

\section{vINT Timing Analysis}
\label{INTR_vint_analysis}

In this section, we first analyze VCPU and task schedulability in the presence of physical and virtual interrupts. Then, we analyze interrupt handling time with and without vINT. 
For convenience, we use the following notation in this section:
\begin{itemize}
  \item $\mathbb{P}(v_i)$ and $\mathbb{P}(I_j^{pi})$: PCPUs for a VCPU $v_i$ and for a physical interrupt $I_j^{pi}$, respectively
  \item $\mathbb{V}(\tau_i)$ and $\mathbb{V}(I_j^{vi})$: Original VCPUs for task $\tau_i$ and for a virtual interrupt $I_j^{vi}$, respectively
  \item $pseudo(\tau_i)$ and $pseudo(I_j^{vi})$: Pseudo VCPUs for task $\tau_i$ and for a virtual interrupt $I_j^{vi}$, respectively, if exist; $\emptyset$ otherwise.
\end{itemize}


\subsection{VCPU and Task Schedulability}
The schedulability of a VCPU $v_i$ can be determined by the following recurrence equation: 
\begin{equation} \label{eq:INTR_vcpu_sched}
\begin{split}
W_i^{v,n+1}&=C_i^{v}+\sum_{I_u^{pi}\in \mathbb{P}(v_i)} \left\lceil{W_i^{v,n} \over T_u^{pi}}\right\rceil C_u^{pi}+\sum_{v_h\in \mathbb{P}(v_i)\land \pi_h^v>\pi_i^v} \left\lceil{W_i^{v,n}+J_h^v \over T_h^v}\right\rceil C_h^v\\
\end{split}
\end{equation}
where, $W_i^{v,n}$ is the worst-case response time (WCRT) of a VCPU $v_i$ at the $n^{th}$ iteration ($W_i^{v,0} = C_i^{v}$), $\pi_i^v$ is the priority of a VCPU $v_i$, and $J_h^v$ is a release jitter ($J_h^v=T_h^v-C_h^v$ for the deferrable server policy and $J_h^v=0$ for the sporadic server policy). Eq.~\eqref{eq:INTR_vcpu_sched} is based on the iterative response time test in~\cite{Joseph_J86}. It terminates when $W_i^{v,n+1}=W_i^{v,n}$, and the VCPU $v_i$ is schedulable if its WCRT does not exceed its period, i.e., $W_i^{v,n}<=T_i^v$. In this equation, the second term represents the interference from the ISRs of physical interrupts during the execution of $v_i$. 

For task schedulability, we need to consider virtual interrupts. If a virtual interrupt is managed by vINT, regular tasks do not experience any direct interference from that virtual interrupt because it is handled by using the budget of its pseudo-VCPU. On the other hand, if a virtual interrupt is not managed by vINT, it may be handled by the budget of the same VCPU as regular tasks. Hence, we can extend the task response-time test under hierarchical scheduling given in~\cite{Saewong_ECRTS02} as follows to check the schedulability of a regular task $\tau_i$ in a VCPU $v_k$:
\begin{equation} \label{eq:INTR_task_sched_with_and_without_vint}
\begin{split}
W_i^{n+1}=&C_i+
\sum_{\substack{\tau_h\in \mathbb{V}(\tau_i)\land \pi_h>\pi_i\\ \land pseudo(\tau_i)=\emptyset}}\!\left\lceil{W_i^n\!+\!J_h \over T_h}\right\rceil C_h\;\;\;\;\;\;\;\;\;\;\;\;\;\;\;\;\\
&+\left\lceil{W_i^n\!+C_k^v \over T_k^v}\right\rceil(T_k^v-C_k^v)
+\sum_{\substack{I_u^{vi}\in \mathbb{V}(\tau_i)\land \\ pseudo(I_u^{vi})=\emptyset}} \left\lceil{W_i^n\!+J_u^{vi} \over T_u^{vi}} \right\rceil C_u^{vi} 
\end{split}
\end{equation}
where, $W_i^n$ is the WCRT of task $\tau_i$ at the $n^{th}$ iteration ($W_i^0=C_i$), 
$\pi_i$ is the priority of $\tau_i$, and $J_h$ and $J_u^{vi}$ are the release jitters of a task $\tau_h$ and a virtual interrupt $I_u^{vi}$, respectively ($J_h=J_u^{vi}=T_k^v-C_k^v$). It terminates when $W_i^{n+1}=W_i^{n}$, and the task $\tau_i$ is schedulable if its WCRT does not exceed its deadline, i.e., $W_i^{n}<=D_i$. Note that the schedulability result for a task from Eq.~\eqref{eq:INTR_task_sched_with_and_without_vint} is valid only if the task's VCPU passes the VCPU schedulability test given in Eq.~\eqref{eq:INTR_vcpu_sched}. The last summing term of Eq.~\eqref{eq:INTR_task_sched_with_and_without_vint} captures the interference from the ISRs of virtual interrupts that are not managed by vINT. In addition, since Eq.~\eqref{eq:INTR_task_sched_with_and_without_vint} conservatively assumes that the budget of the task's VCPU is available at the latest time possible within each period ($T_k^v-C_k^v$), the interference from physical interrupts does not need to be considered in Eq.~\eqref{eq:INTR_task_sched_with_and_without_vint}.

\subsection{Interrupt Handling Time}
The total interrupt handling time can be bounded by the sum of (i) the WCRT of the ISR of a physical interrupt, (ii) the WCRT of the ISR of a physical IPI if the target VCPU is on a different PCPU, and (iii) the WCRT of the ISR and DSR of the corresponding virtual interrupt. For factors (i) and (ii), the WCRT of the ISR of a physical interrupt $I_i^{pi}$ is bounded by the following recurrence equation:
\begin{equation} \label{eq:INTR_physical_interrupt_resp}
\begin{split}
W_i^{pi,n+1}=C_i^{pi}+\sum_{I_h^{pi}\in \mathbb{P}(I_i^{pi}) \land \pi_h^{pi}>\pi_i^{pi}} \left\lceil{W_i^{pi,n} \over T_h^{pi}}\right\rceil C_h^{pi}
\end{split}
\end{equation}
where, $W_i^{pi,n}$ is the WCRT of a physical interrupt $I_i^{pi}$ at the $n^{th}$ iteration ($W_i^{pi,0}=C_i^{pi}$), and $\pi_i^{pi}$ is the priority of $I_i^{pi}$. 


We now consider the last factor. When vINT is used, as shown in \figref{INTR_vint_interrupt_handling}, the ISR and DSR of a virtual interrupt may be blocked by the ISRs of virtual interrupts that are associated with lower-priority pseudo-VCPUs and executed in the execution context of the same original VCPU. The virtual interrupt may also be blocked by other virtual interrupts that are not managed by vINT. For a virtual interrupt $I_j^{vi}$ associated with a pseudo-VCPU $v_p$, the maximum blocking time from such virtual interrupts during a time interval $t$ is given by:
\begin{equation} \label{eq:INTR_blocking_from_lower_prio_pseudo_vcpus}
\begin{split}
B_{p,j}(t)=\!\!\sum_{\substack{I_u^{vi}\in \mathbb{V}(I_j^{vi}) \land (pseudo(I_u^{vi})=\emptyset \\\lor \pi_{pseudo(I_u^{vi})}^p<\pi_j^p)}}\left\lceil{t \over T_u^{vi}}\right\rceil C_u^{vi}
\end{split}
\end{equation}
where, $\pi_{pseudo(I_u^{vi})}^p$ is the priority of $I_u^{vi}$'s pseudo-VCPU. In addition, the worst case happens when all physical interrupts on the same PCPU arrive with their minimum inter-arrival times and all higher-priority VCPUs fully consume their budgets. The WCRT of a virtual interrupt $I_j^{vi}$ associated with a pseudo-VCPU $v_p$ is therefore bounded by:
\begin{equation} \label{eq:INTR_virtual_interrupt_resp_with_vint}
\begin{split}
W_j^{vi,n+1}&\!\!=\!
\mathbb{C}_j^{vi}+B_{p,j}(W_j^{vi,n})\!+\!\!\sum_{I_u^{pi}\in \mathbb{P}(v_p)}\!\! \left\lceil{W_j^{vi,n} \over T_u^{pi}}\right\rceil C_u^{pi}+\sum_{v_h\in \mathbb{P}(v_p)\land \pi_h^v> \pi_p^v} \left\lceil{W_j^{vi,n}+J_h^v \over T_h^v}\right\rceil C_h^v\\
\end{split}
\end{equation}
where, $W_j^{vi,n}$ is the WCRT of a virtual interrupt $I_j^{vi}$ ($W_j^{vi,0}=\mathbb{C}_j^{vi}$). 
Note that Eq.~\eqref{eq:INTR_virtual_interrupt_resp_with_vint} is similar to the VCPU schedulability test given in Eq.~\eqref{eq:INTR_vcpu_sched}, except the blocking term. This is because the pseudo-VCPU of a virtual interrupt is guaranteed to have enough budget to handle one instance of a virtual interrupt, and there is no other task interfering with the execution of the ISR and DSR of the virtual interrupt in the pseudo-VCPU. 

When vINT is not used, the response time of a virtual interrupt should be captured by considering the executions of other tasks within the same VCPU. Therefore, the WCRT of a virtual interrupt $I_j^{vi}$ in a VCPU $v_k$ is bounded by:
\begin{equation} \label{eq:INTR_virtual_interrupt_resp_without_vint}
\begin{split}
W_j^{vi,n+1}=&\mathbb{C}_j^{vi}+
\sum_{\substack{\tau_h\in \mathbb{V}(I_j^{vi})\land \pi_h>\check{\pi}_\mathbb{D}\\ \land pseudo(\tau_h)=\emptyset}}\!\left\lceil{W_j^{vi,n}\!+\!J_h \over T_h}\right\rceil C_h\;\;\;\;\;\;\;\;\;\;\\
&+\left\lceil{W_j^{vi,n}\!\!+\!C_k^v \over T_k^v}\right\rceil\!(T_k^v\!-\!C_k^v)
+\sum_{\substack{I_{u}^{vi}\in \mathbb{V}(I_{j}^{vi}) \land u\ne j \\ pseudo(I_u^{vi})=\emptyset}} \left\lceil{W_{j}^{vi,n}\!\!+\!J_{u}^{vi} \over T_{u}^{vi}} \right\rceil C_{u}^{vi} 
\end{split}
\end{equation}
where, $\check{\pi}_\mathbb{D}$ is the priority of the lowest-priority task in $\mathbb{D}(I_j^{vi})$. Note that this equation is similar to Eq.~\eqref{eq:INTR_task_sched_with_and_without_vint} which captures the WCRT of a task. 

\section{Evaluation}
\label{INTR_evaluation}

In this section, we first empirically investigate the performance characteristics and benefits of vINT, and then show its effects on a real hardware platform.

\begin{table}[t]
\centering
\VS{-5pt}
{
\footnotesize
\caption[Base parameters for vINT interrupt handling experiments]{Base parameters for our experiments\VS{-5pt}}\label{tab:INTR_expr_params}
\VS{-2pt}
\begin{tabular}{p{8cm}|C{3cm}}
\hline
Parameters & Values\VS{-1pt}\\\hline
Number of PCPUs & 4 \\
Number of VCPUs per PCPU & 3 \\
Number of physical interrupts per PCPU& 6\\
Number of virtual interrupts per VCPU & 2\\
VCPU replenishment period & 10 msec\\
Minimum inter-arrival time of a physical interrupt & $[5, 10]$ msec\\
Minimum inter-arrival time of a regular task & $[100, 500]$ msec\\
WCET of ISR of a physical/virtual interrupt & $[5, 10]$ $\mu$sec\\
WCET of DSR of a virtual interrupt & $[10, 50]$ $\mu$sec\\
Number of regular tasks per VCPU & 3 \\
Number of DSR tasks per VCPU & 2 \\
Task set utilization per VCPU& 10 \%\\
\hline
\end{tabular}\VS{-10pt}
}
\end{table}

\subsection{Experimental Setup}

We consider the following schemes in our experiments: deferrable server without vINT (DSbase), sporadic server without vINT (SSbase), deferrable server with vINT (DSvINT), and sporadic server with vINT (SSvINT). We use randomly-generated task sets and interrupt sets to compare these schemes on how many task sets could be schedulable and how many interrupt sets could be serviced on a timely basis.

Since, in practice, vINT can be selectively applied to a subset of virtual interrupts that cannot be serviced within their virtual interrupt times by the baseline scheme, our experiments only focus on interrupts with short inter-arrival times. \tableref{INTR_expr_params} lists the base parameters we use for our experiments. 
For each experimental setting, we first generate PCPUs, VCPUs, physical interrupts, and tasks and virtual interrupts for each VCPU based on the defined parameters. 
Each virtual interrupt is exclusively associated with one physical interrupt in a random manner, and the minimum inter-arrival time of each virtual interrupt is set equal to that of its associated physical interrupt. 
For each VCPU, the task set utilization per VCPU is split into $k$ random-sized pieces, where $k$ is the number of tasks per VCPU. The size of each piece becomes the utilization of the corresponding task, and the WCET of each task is calculated by dividing its utilization by its minimum inter-arrival time. For DSvINT and SSvINT, we create a pseudo-VCPU for each virtual interrupt with a period equal to the minimum inter-arrival time of the corresponding virtual interrupt and with a budget determined by Eq.~\eqref{eq:INTR_pseudo_vcpu_budget_enhanced}. VCPUs and tasks are assigned unique priorities by using the Rate-Monotonic Scheduling (RMS) policy~\cite{Liu_Layland}, with an arbitrary tie-breaking rule. The priorities of physical and virtual interrupts are assigned randomly. Once this is done, we finally determine the VCPU budget value for each scheme. Starting from a value equal to the VCPU period, the VCPU budget for each scheme is decreased by 1~$\mu$sec until all VCPUs pass the VCPU schedulability test given in Eq.~\eqref{eq:INTR_vcpu_sched}.\footnote{Considering the time-unit granularity used in \tableref{INTR_expr_params}, the step size of 1~$\mu$sec is fine-grained enough to find the maximum-possible VCPU budget for each scheme in our experiments.}

We generate 10,000 task sets and 10,000 interrupt sets for each experimental setting. The metrics used are: (i) the percentage of schedulable task sets where all tasks pass the schedulability test given in Eq.~\eqref{eq:INTR_task_sched_with_and_without_vint}, and (ii) the percentage of serviceable interrupt sets where all interrupt-triggered execution flows are serviceable, checked by Eqs.~\eqref{eq:INTR_physical_interrupt_resp}, \eqref{eq:INTR_virtual_interrupt_resp_with_vint} and \eqref{eq:INTR_virtual_interrupt_resp_without_vint}.

\subsection{Results}

We explore three main factors that affect task schedulability and interrupt serviceability in a virtualized environment: (i) the minimum inter-arrival time of interrupts, (ii) the VCPU period, and (iii) the WCET of interrupt handlers.

\begin{figure}[t]
\centering
\VS{-9pt}
\subfloat{
\includegraphics[width=0.7\textwidth]{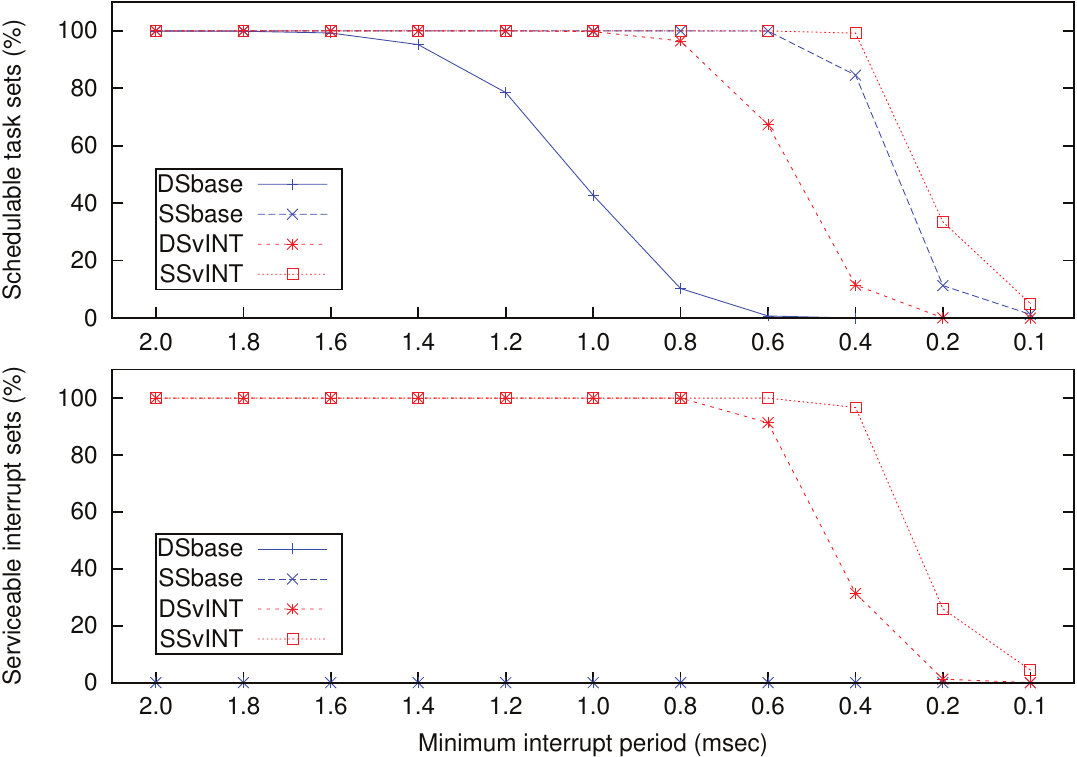}
}
\VS{-7pt}
\caption{Results with short interrupt inter-arrival time}
\VS{-9pt}
\label{fig:INTR_intr_period_short}
\end{figure}

\begin{figure}[t]
\centering
\VS{-9pt}
\subfloat{
\includegraphics[width=0.7\textwidth]{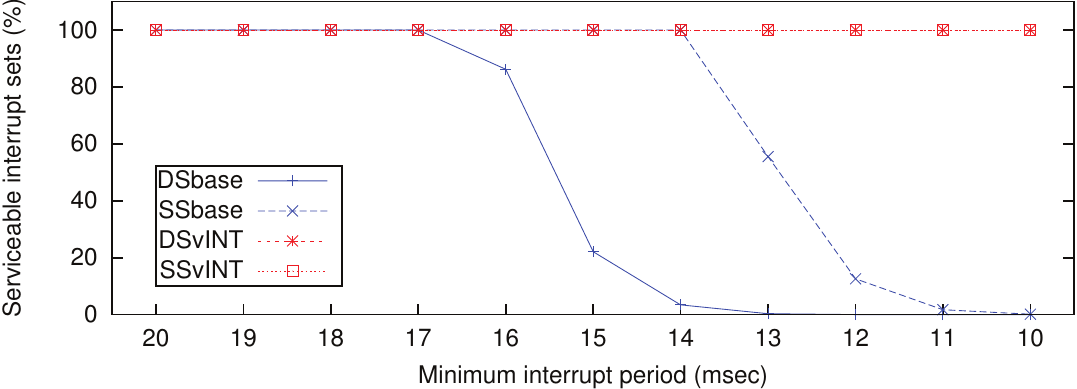}
}
\VS{-7pt}
\caption{Results with long interrupt inter-arrival time}
\VS{-5pt}
\label{fig:INTR_intr_period_long}
\end{figure}

\begin{figure}[t]
\centering
\VS{-10pt}
\subfloat{
\includegraphics[width=0.7\textwidth]{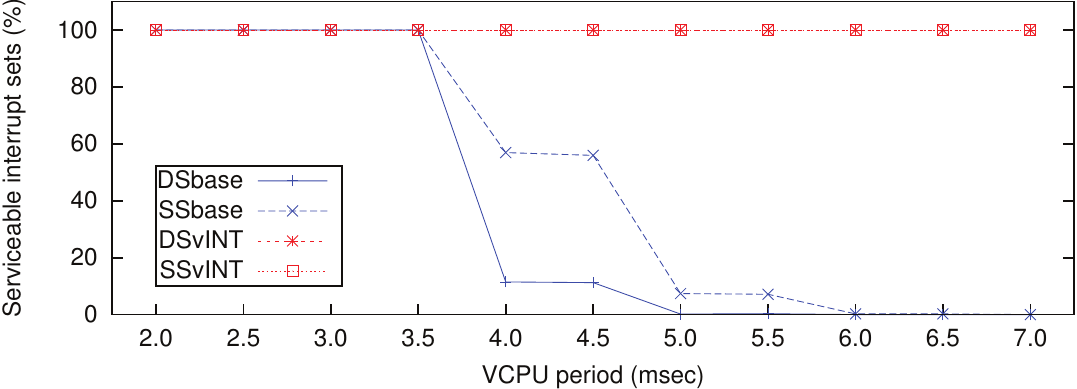}
}
\VS{-7pt}
\caption{Results with the change of VCPU period}
\VS{-9pt}
\label{fig:INTR_vcpu_period}
\end{figure}

\smallskip
\noindent\textbf{Minimum inter-arrival time of interrupts:} \figref{INTR_intr_period_short} shows the percentages of schedulable task sets and serviceable interrupt sets as the minimum inter-arrival time of interrupts decreases. Each point $k$ on the x-axis represents that the minimum inter-arrival time of each interrupt ranges $[k,k+0.5]$~msec. In general, the sporadic server policy (SS) performs better than the deferrable server policy (DS). This is because SS has zero release jitter and allows assigning larger budget values to VCPUs than DS. vINT has benefits in both task scheduling and interrupt handling. DSvINT and SSvINT schedule more task sets than DSbase and SSbase, respectively. Especially, when the range is $[0.6,1.1]$~msec, DSvINT schedules 67\% more task sets than DSbase. The benefit is more significant in interrupt handling. While the schemes without vINT service 0\% of interrupt sets in all cases, the schemes with vINT service more than 99\% of interrupt sets until the range reaches $[0.8, 1.3]$~msec.

When vINT is not used, only the interrupts with slightly longer inter-arrival times can be serviced. \figref{INTR_intr_period_long} depicts the results. In this figure, each point $k$ on the x-axis represents the minimum inter-arrival time of each interrupt in the range of $[k,k+5]$~msec. As all the schemes schedule 100\% of task sets in all cases, we only display the percentage of serviceable interrupt sets in this figure. When the range reaches $[13,18]$~msec, DSbase services less than 1\% of interrupt sets. SSbase performs better than DSbase, but services less than 2\% of interrupt sets when the range becomes $[11,16]$~msec.

\begin{figure}[t]
\centering
\VS{-10pt}
\subfloat{
\includegraphics[width=0.7\textwidth]{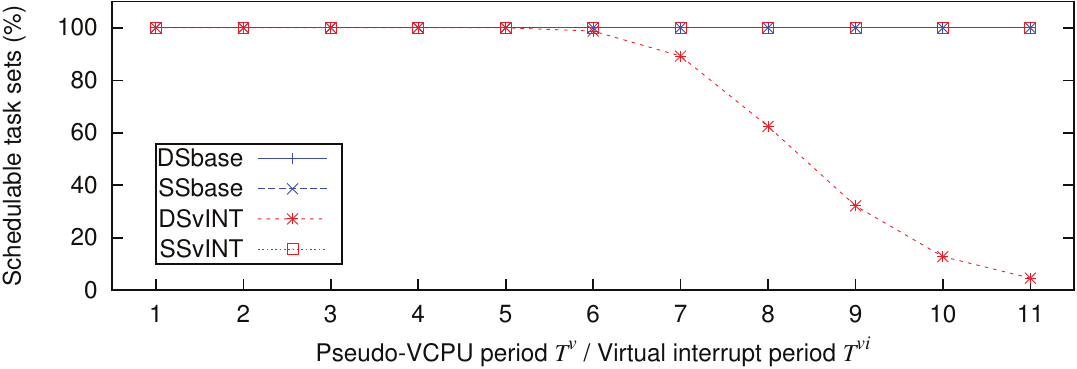}
}
\VS{-6pt}
\caption{Results with the change of pseudo-VCPU period}
\VS{-9pt}
\label{fig:INTR_pseudo_vcpu_period}
\end{figure}

\smallskip
\noindent\textbf{VCPU periods:} Since the interrupt handling time is largely affected by the VCPU period when vINT is not used, we compare in \figref{INTR_vcpu_period} the percentage of serviceable interrupt sets as the VCPU period increases. All the schemes could schedule 100\% of task sets with all VCPU period values depicted in this figure. The schemes with vINT also show 100\% of serviceable interrupt sets in all cases. However, without vINT, the percentage drops significantly when the VCPU period is longer than 3.5~msec.

We have also evaluated the impact of the pseudo-VCPU period. \figref{INTR_pseudo_vcpu_period} shows the percentage of schedulable task sets as the pseudo-VCPU period increases. Each number shown on the x-axis of this figure represents the ratio of the pseudo-VCPU period to the minimum inter-arrival time of interrupts. Hence, a larger value on the x-axis means a longer pseudo-VCPU period. 
As the pseudo-VCPU period increases, task schedulability under DSvINT decreases. This is because DS has a release jitter equal to $T^v-C^v$. Since vINT assigns higher priorities to pseudo-VCPUs, the larger jitter values of pseudo-VCPUs under DS effectively reduce the amount of budget assigned to regular VCPUs. In contrast, SS shows no performance degradation because it has zero release jitter.

\begin{figure}[t]
\centering
\VS{-5pt}
\subfloat{
\includegraphics[width=0.7\textwidth]{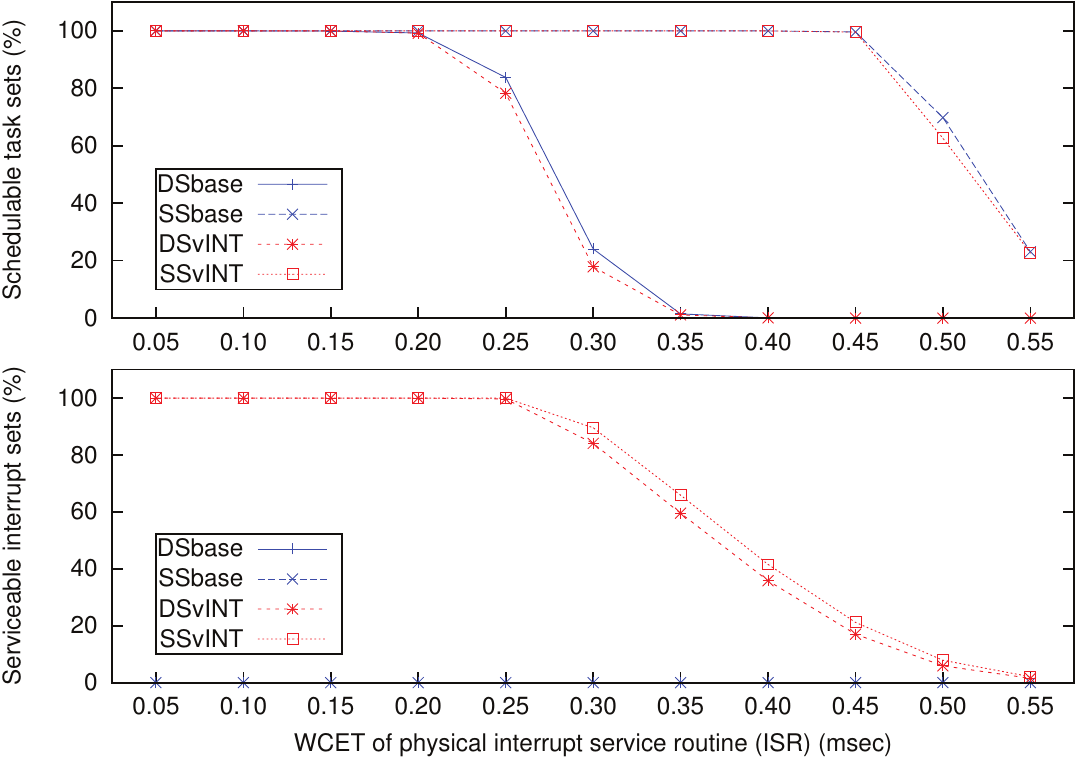}
}
\VS{-6pt}
\caption{Results with the change of physical ISR length}
\VS{-9pt}
\label{fig:INTR_pisr_wcet}
\end{figure}

\begin{figure}[t]
\centering
\VS{-5pt}
\subfloat{
\includegraphics[width=0.7\textwidth]{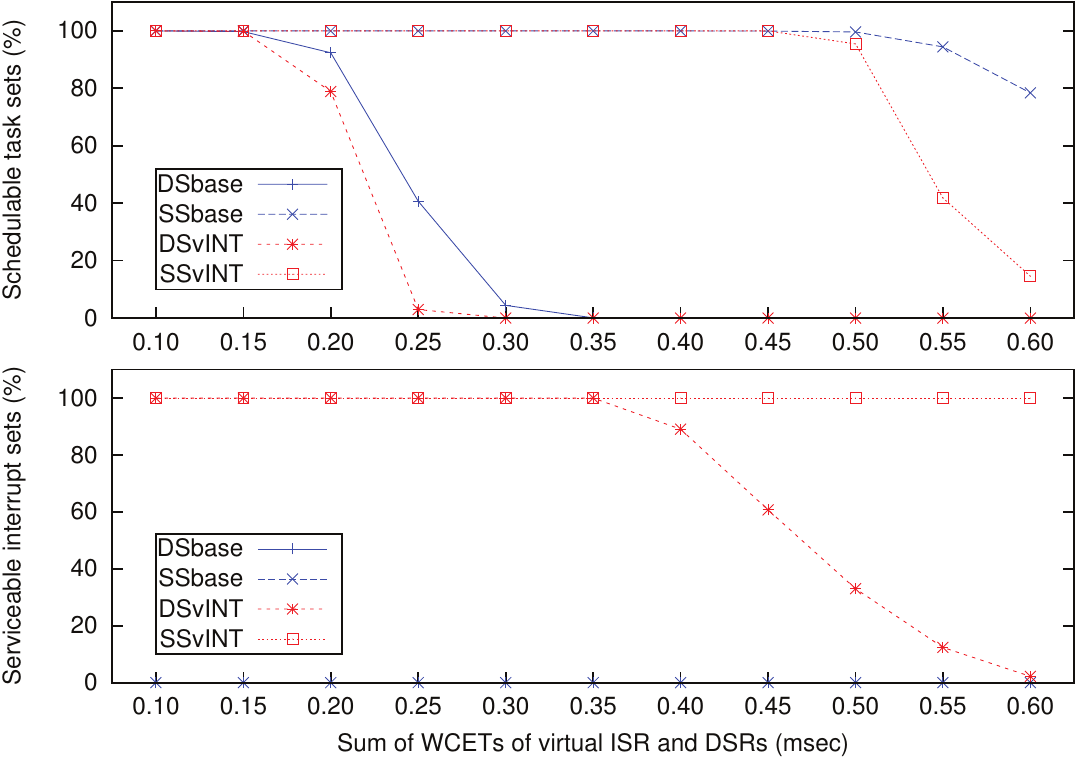}
}
\VS{-6pt}
\caption{Results with the change of virtual ISR and DSR length}
\VS{-5pt}
\label{fig:INTR_visrdsr_wcet}
\end{figure}

\smallskip
\noindent\textbf{WCET of interrupt handlers:} We now evaluate the impact of the length of interrupt handlers. \figref{INTR_pisr_wcet} and \figref{INTR_visrdsr_wcet} show the results when the WCET of a physical ISR, and the sum of the WCETs of virtual ISR and DSR change, respectively. As the WCET increases, both the percentages of schedulable task sets and serviceable interrupt sets decrease. In case of increasing the WCETs of virtual ISRs and DSRs, the schemes with vINT show lower performance in task schedulability than the schemes without vINT, but provide significantly higher performance in interrupt handling. This is mainly due to the fact that vINT creates pseudo-VCPUs and prioritizes them over regular VCPUs in order to reduce interrupt handling time. 

In summary, vINT achieves timely interrupt handling while providing as good task schedulability as when it is not used in most cases.
The benefit of vINT multiplies if the inter-arrival time of interrupts is short. Especially, when the minimum inter-arrival time of interrupts is much shorter than the period of VCPUs, the system with vINT outperforms the system without vINT in both task scheduling and interrupt servicing.

\subsection{Case Study: vINT on KVM Hypervisor}

We present a case study demonstrating the effects of vINT by using our implementation on the KVM hypervisor. We chose KVM because it is open-source software and widely used in real-time virtualization studies~\cite{Cucinotta_COMPSAC09,Kiszka_RTLWS09,Ma_JISE13}. Also, it is useful to observe the overall performance impact of vINT, which can be applied to commercial real-time hypervisors.

\smallskip
\noindent\textbf{Implementation:} We have implemented a prototype version of vINT on the KVM hypervisor~\cite{KVM} of the latest version of Linux/RK~\cite{LinuxRK, ResourceKernel}.\footnote{Linux/RK is available at \url{https://rtml.ece.cmu.edu/redmine/projects/rk}.} 
The KVM of Linux/RK allows the host machine to run multiple guest VMs with the deferrable server policy as the VCPU budget replenishment policy. We use an unmodified Linux kernel v3.10.39 as a guest OS. 

We have applied vINT to the pass-through PCI device management of KVM. Note that PCI pass-through devices do not involve QEMU in interrupt handling. Hence, once a PCI device is assigned to a guest VM in pass-through mode, all physical interrupts generated by the device are handled by the interrupt handler of KVM, and then resulting virtual interrupts are delivered to the corresponding guest VM, without any intervention from QEMU.


\tableref{INTR_vint_impl_cost} lists the implementation costs of vINT. The target system used is equipped with an Intel Core i7-2600 3.4~GHz quad-core processor and a TP-Link PCI Gigabit NIC using a RTL8169 controller. To reduce measurement inaccuracies, we have disabled the simultaneous multithreading and dynamic clock frequency scaling features of the processor. 


\begin{table}[t]
\centering
{
\footnotesize
\caption{Implementation cost of vINT on the KVM hypervisor\VS{-6pt}}\label{tab:INTR_vint_impl_cost}
\VS{-2pt}
\begin{tabular}{l|c|c}
\hline
\multicolumn{1}{c|}{Primitives} & Avg ($\mu$sec) & Max ($\mu$sec)\VS{-1pt}\\\hline
Switching btw. pseudo and reg. VCPUs & 0.703 & 1.192 \\
Pseudo-VCPU budget accounting&	0.341 &	1.265 \\
Pseudo-VCPU budget replenishment&	0.621 & 3.045  \\
\hline
\end{tabular}
}
\VS{-5pt}
\end{table}

\begin{figure*}[t]
\centering
\VS{-22pt}
\subfloat[Idle]{\label{fig:INTR_case_study_netperf_latency_idle}
\includegraphics[width=0.35\textwidth]{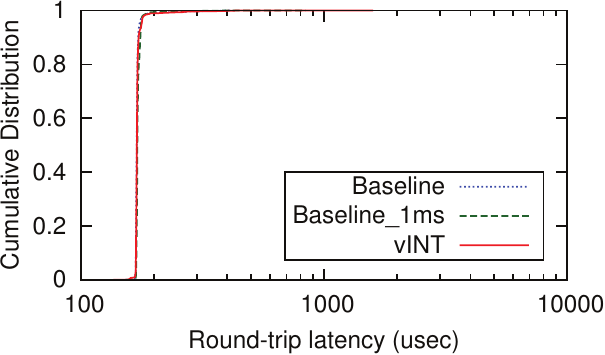}
}\hspace{-5pt}
\subfloat[MPlayer only]{\label{fig:INTR_case_study_netperf_latency_mplayer}
\includegraphics[width=0.35\textwidth]{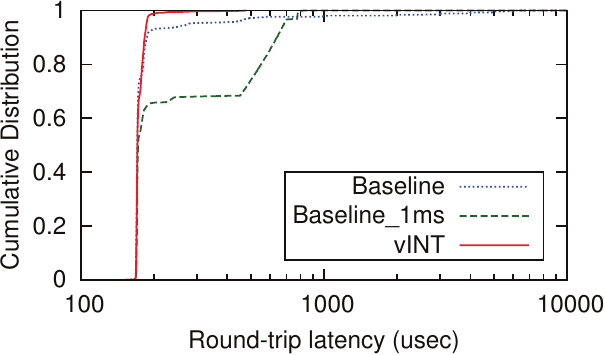}
}\hspace{-5pt}
\subfloat[Both MPlayer and busyloop]{\label{fig:INTR_case_study_netperf_latency_busyloop}
\includegraphics[width=0.35\textwidth]{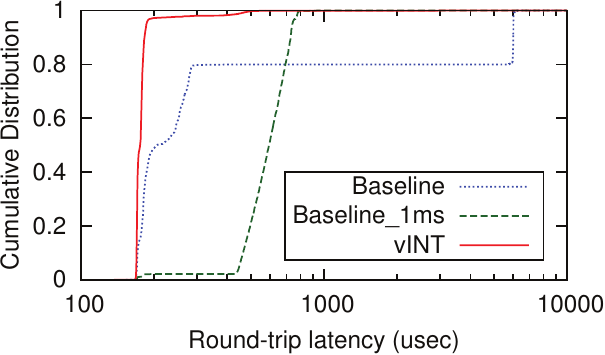}
}
\VS{-7pt}
\caption{Cumulative distribution of Netperf UDP round-trip latency}
\VS{-12pt}
\label{fig:INTR_netperf_latency}
\end{figure*}

\begin{figure}[t]
\centering
\VS{-6pt}
\subfloat{
\includegraphics[width=0.7\textwidth]{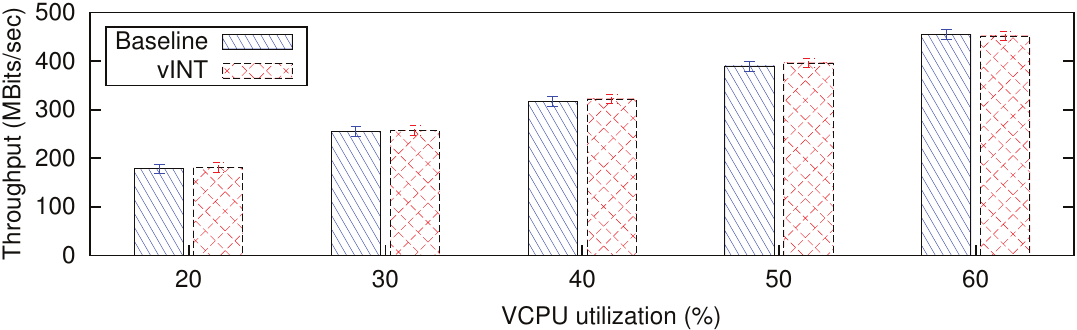}
}
\VS{-7pt}
\caption{Netperf TCP throughput}
\VS{-7pt}
\label{fig:INTR_netperf_throughput}
\end{figure}

\begin{figure}[t]
\centering
\VS{-9pt}
\subfloat{
\includegraphics[width=0.7\textwidth]{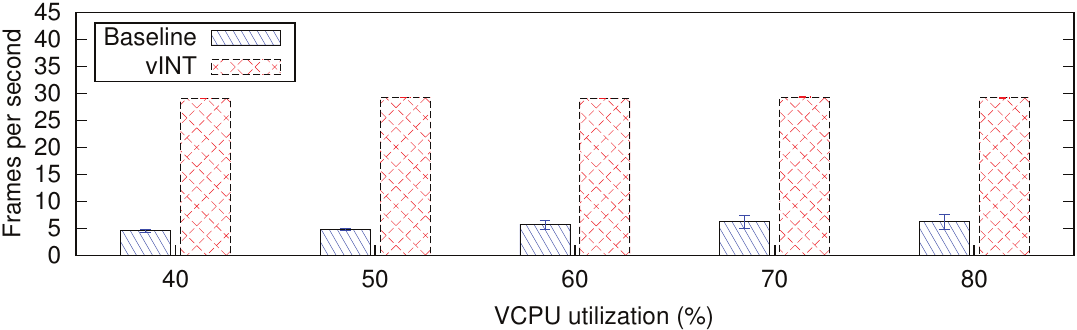}
}
\VS{-7pt}
\caption{MPlayer fps under virtual interrupt storms}
\VS{-13pt}
\label{fig:INTR_mplayer_fps}
\end{figure}

\smallskip
\noindent\textbf{Case Study:} 
The target system hosts one guest VM, which has four VCPUs: $\{v_1,v_2,v_3,v_4\}$. Each VCPU is statically assigned to a PCPU with the same index number, i.e. $v_i$ on Core $i$. The Gigabit NIC of the target system is assigned to the guest VM in pass-through mode. The physical interrupt of the NIC is statically pinned to Core 1 and the corresponding virtual interrupt is pinned to the VCPU $v_1$. The QEMU process is assigned the highest prioriry to prevent unexpected delays from QEMU device emulation, although it is not involved in the critical path of interrupt handling in pass-through mode. In our case study, we only focus on $v_1$ on Core 1 and other VCPUs on other cores are kept in idle. When vINT is not used, the VCPU $v_1$ is assigned 4~msec of budget and 10~msec of replenishment period (40\% VCPU utilization). When vINT is used, both $v_1$ and a pseudo-VCPU created for NIC interrupts are each assigned 2~msec of budget and 10~msec of replenishment period (total 40\% VCPU utilization). 

We use three applications in our case study: Netperf~\cite{netperf}, MPlayer~\cite{mplayer} and busyloop. {\em Netperf} is a network benchmark consisting of sender and receiver tasks. The Netperf sender task runs natively on a remote system, which has no other workload and is connected to the target system with a direct Ethernet connection. The Netperf receiver task runs in the VCPU $v_1$. When $v_1$ receives a virtual interrupt of the NIC, the ISR of the virtual interrupt activates the {\em softirq} task of the guest Linux kernel, which in turn activates the Netperf receiver task. Both the softirq and Netperf receiver tasks are assigned the highest real-time priority. {\em MPlayer} is an open-source movie player. MPlayer runs in $v_1$, with a real-time priority lower than the Netperf receiver task, and decodes a MPEG2 video stream with 1920x1080 (1080p) frame size and 29.97~fps. {\em Busyloop} is a background task that continuously consumes CPU time, and runs in $v_1$ with the lowest priority.

We first compare interrupt handling time with and without vINT. For this purpose, we use the UDP round-trip latency test of Netperf, which is highly affected by the system's interrupt handling time. \figref{INTR_netperf_latency} shows the cumulative distribution of the Netperf UDP round-trip latency. ``Baseline'' and ``Baseline\_1ms'' show the results without vINT, and ``vINT'' shows the results with vINT. Baseline and vINT use the aforementioned VCPU parameters for $v_1$. Baseline\_1ms uses 0.4~msec of budget and 1~msec of replenishment period for $v_1$, which results in the same VCPU utilization as Baseline. As shown in the figure, Baseline and Baseline\_1ms are significantly affected by the executions of lower-priority tasks within the same VCPU, but vINT is nearly unaffected. Especially, when both MPlayer and busyloop are running, vINT handles 95\% of round-trips in less than 200~$\mu$sec, while Baseline and Baseline\_1ms handle only 50\% and 2\% of round-trips in 200~$\mu$sec, respectively. Interestingly, Baseline\_1ms would be expected to outperform Baseline due to its shorter replenishment period, but the results are the opposite due to the higher overhead occurred.

Next, we identify the impact of vINT overhead on the throughput of NIC. \figref{INTR_netperf_throughput} shows the results of the TCP throughput test of Netperf with and without vINT as the VCPU utilization increases. The VCPU period is 10~msec in all cases. Only the budget varies from 2~msec to 6~msec. In case of vINT, each point on the x-axis represents the utilization of the pseudo-VCPU. As can be seen, there is no noticeable difference between Baseline and vINT in TCP throughput. This implies that the impact of the overhead induced by vINT is either negligible or acceptably small.

Lastly, we demonstrate the effect of vINT in protecting a real-time task against a virtual interrupt storm. \figref{INTR_mplayer_fps} compares the frame rate of MPlayer with and without vINT, in the presence of a virtual interrupt storm which is generated by the TCP throughput test of Netperf. In case of vINT, each point on the x-axis represents the total utilization of original and pseudo-VCPUs. Hence, the budget of the original VCPU varies from 4~msec to 8~msec for Baseline, and from 2~msec to 6~msec for vINT (the pseudo-VCPU budget is unchanged). When vINT is not used, the frame rate of Mplayer is severely degraded by a virtual interrupt storm, even when 80\% of VCPU utilization is assigned. In contrast, when vINT is used, the frame rate is very close to when there is no interrupt storm. This result shows that vINT can effectively protect the execution of a real-time task against the occurrence of a virtual interrupt storm in a virtualized environment.

\section{Summary}
\label{INTR_conclusions}

In this chapter, we presented vINT, an interrupt handling scheme to provide responsive and enforced interrupt handling in a virtualized environment. We introduced our analyses on interrupt handling time, and the schedulability of VCPUs and tasks with and without vINT. Experimental results show that vINT yields significant improvements in interrupt handling performance. For example, a system with vINT services 99\% of interrupt sets while a system without vINT cannot service any interrupt set. 
Our case study on the KVM hypervisor, chosen for convenience, also shows the effects of vINT in reducing interrupt handling time and protecting against interrupt storms. For example, a system with vINT handles 95\% of Ethernet round-trips in 200~$\mu$sec, and a system without vINT handles only 50\% of round-trips during that time. Under interrupt storms, the frame rate of MPlayer with vINT is nearly unaffected while the frame rate without vINT is dropped to one-fifth of the original one. 


\chapter{Predictable GPGPU Access Control}
\label{chapter_predictable_gpgpu_management}

In this chapter, we first review the use of a real-time synchronization protocol for tasks accessing a general-purpose GPU (graphics processing unit) on a multi-core platform, and characterize the limitations of this approach. Among a variety of real-time synchronization protocols, we focus on the multiprocessor priority ceiling protocol (MPCP)~\cite{MPCP,MPCP2} because it is designed for partitioned fixed-priority scheduling that we use in our work. 
Then, we present our new GPU access control technique, called a server-based approach. Our proposed server-based approach provides a dedicated {\em GPU server task} that receives GPU access requests from other tasks and handles the requests on behalf of them. Unlike the synchronization-based approach, the server-based approach allows tasks to suspend during their GPU executions, yielding significant CPU utilization benefits. The server-based approach can also reduce the response time of a task using a GPU, compared the synchronization-based approach. Although we have focused on a GPU in this work, our approach can be used for other types of computational accelerators, such as DSPs. 

We provide the schedulability analysis of tasks under our server-based approach, which accounts for the overhead of the use of the GPU server task. Experimental results indicate that, when the overhead is reasonable, the server-based approach significantly outperforms the synchronization-based approach, with as much as 66\% more tasksets being schedulable. 

The background and related earlier work on GPGPU management were presented in Section~\ref{backgroun_gpu_management}. The system model including assumptions and notation for tasks and GPU access segments can be found in Chapter~\ref{system_model}.

The rest of this chapter is organized as follows. Section~\ref{GPU_synchronization_approach} reviews the use of the synchronization-based approach for GPU access control. Section~\ref{GPU_server_approach} presents our proposed server-based approach. Section~\ref{GPU_evaluation} provides detailed evaluation. Section~\ref{GPU_conclusions} summarizes this chapter.

\section{Synchronization-based GPU Access Control}
\label{GPU_synchronization_approach}

The synchronization-based approach models the GPU as a global mutually-exclusive resource and the GPU access segments of tasks as critical sections. A single mutex is used for protecting such GPU critical sections. Hence, under the synchronization-based approach, a task can only enter its GPU access segment when the mutex for the GPU is not held by any other task. If the mutex is already held by another task, the task is inserted into the waiting list of the mutex and waits until the mutex can be held by that task.

Since our focus is on the multiprocessor priority ceiling protocol (MPCP), we shall briefly review the definition of MPCP below. More details on MPCP can be found in \cite{MPCP,MPCP2,Lakshmanan_ECRTS09}. 

\begin{enumerate}
	\item When a task $\tau_i$ requests an access to a global resource $R_k$, the resource $R_k$ can be granted to the task $\tau_i$, if it is not held by another task.
	\item While a task $\tau_i$ is holding a resource for its global critical section (gcs), the priority of $\tau_i$ is raised to $\pi_{B}+\pi_{i}$, where $\pi_{B}$ is a base task-priority level greater than that of any task in the system, and $\pi_{i}$ is the normal priority of $\tau_i$. This priority boosting is referred to as the {\em global priority ceiling} of the gcs of $\tau_i$.
	\item When a task $\tau_i$ requests access to a resource $R_k$, the resource $R_k$ cannot be granted to $\tau_i$, if it is already held by another task. In this case, the task $\tau_i$ is inserted to the waiting list of the mutex for $R_k$. 
	\item When a global resource $R_k$ is released and the waiting list of the mutex for $R_k$ is not empty, the highest-priority task in the waiting list is dequeued from the list and is granted the resource $R_k$. 
\end{enumerate}

\subsection{Limitations of Synchronization-based Approach}
\label{GPU_limitations_of_synchronization_approach}

\begin{figure}[t]
	\centering
	\subfloat{
		\includegraphics[width=0.7\textwidth]{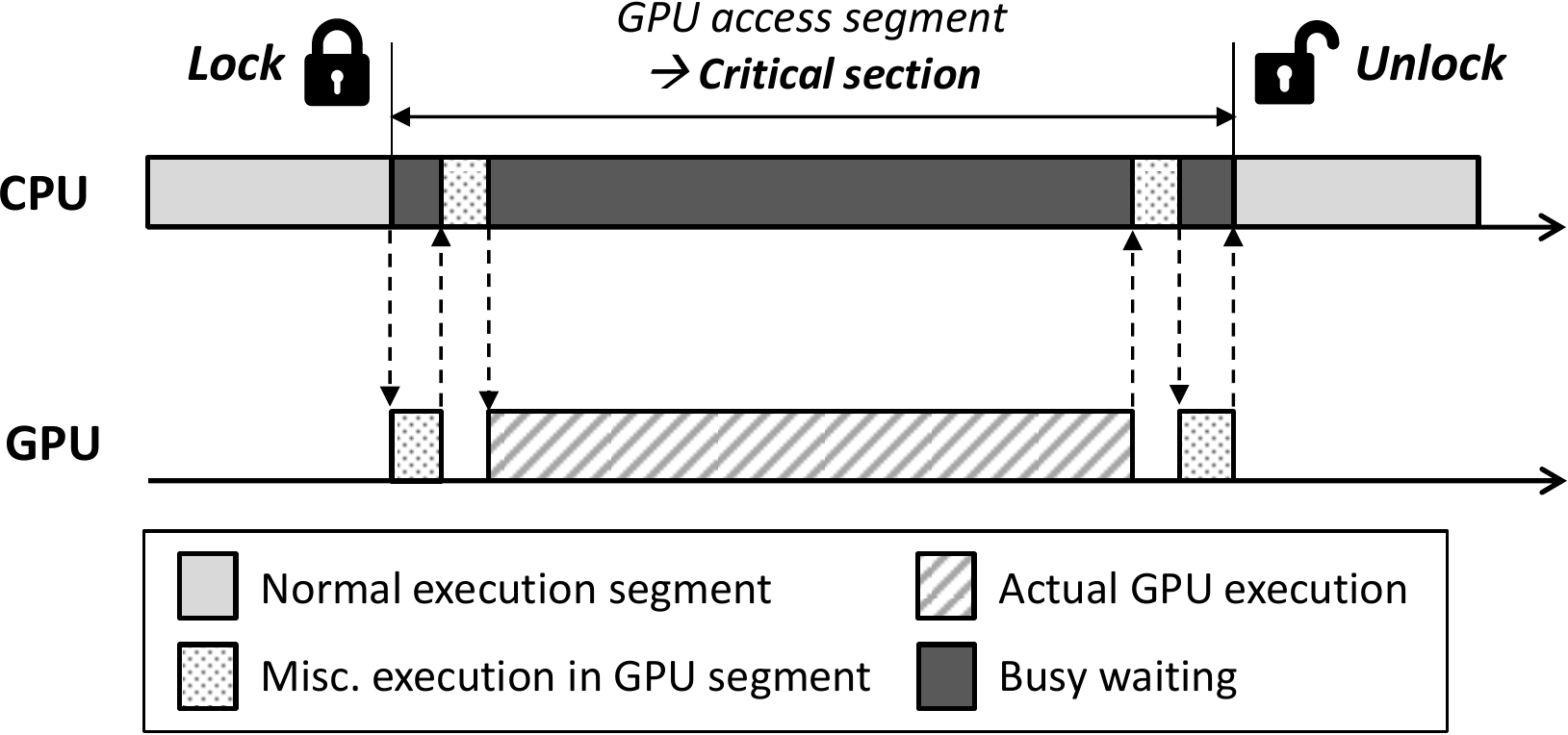}
	}
	\caption{Task execution pattern under the synchronization-based approach}
	\label{fig:GPU_synch_based_approach}
\end{figure}

As discussed in Section~\ref{backgroun_gpu_execution_pattern}, each GPU access segment contains various operations, including data copies, notifications, and the actual GPU code execution. Specifically, a task may suspend during the GPU code execution to save CPU utilization. However, under the synchronization-based approach, any task in its GPU access segment  should busy-wait for any operation conducted on the GPU in order to ensure predictability. This is because each GPU access segment is modeled as a critical section, and real-time synchronization protocols including MPCP commonly assume that (i) a critical section is executed entirely on the CPU, and (ii) there is no suspension during the execution of the critical section. \figref{GPU_synch_based_approach} shows the execution pattern of a GPU-using task under the synchronization-based approach. The entire GPU access segment is protected by a mutex. Hence, the task should hold the mutex to enter its GPU access segment. The task releases the mutex when it leaves the GPU access segment. While the GPU code execution happens on the GPU, the task consumes CPU time because of the busy-waiting requirement of the synchronization-based approach. As the GPU execution time increases, the CPU utilization loss under the synchronization-based approach is therefore expected to increase.

\begin{figure}[t]
	\centering
	\subfloat{
		\includegraphics[width=1\textwidth]{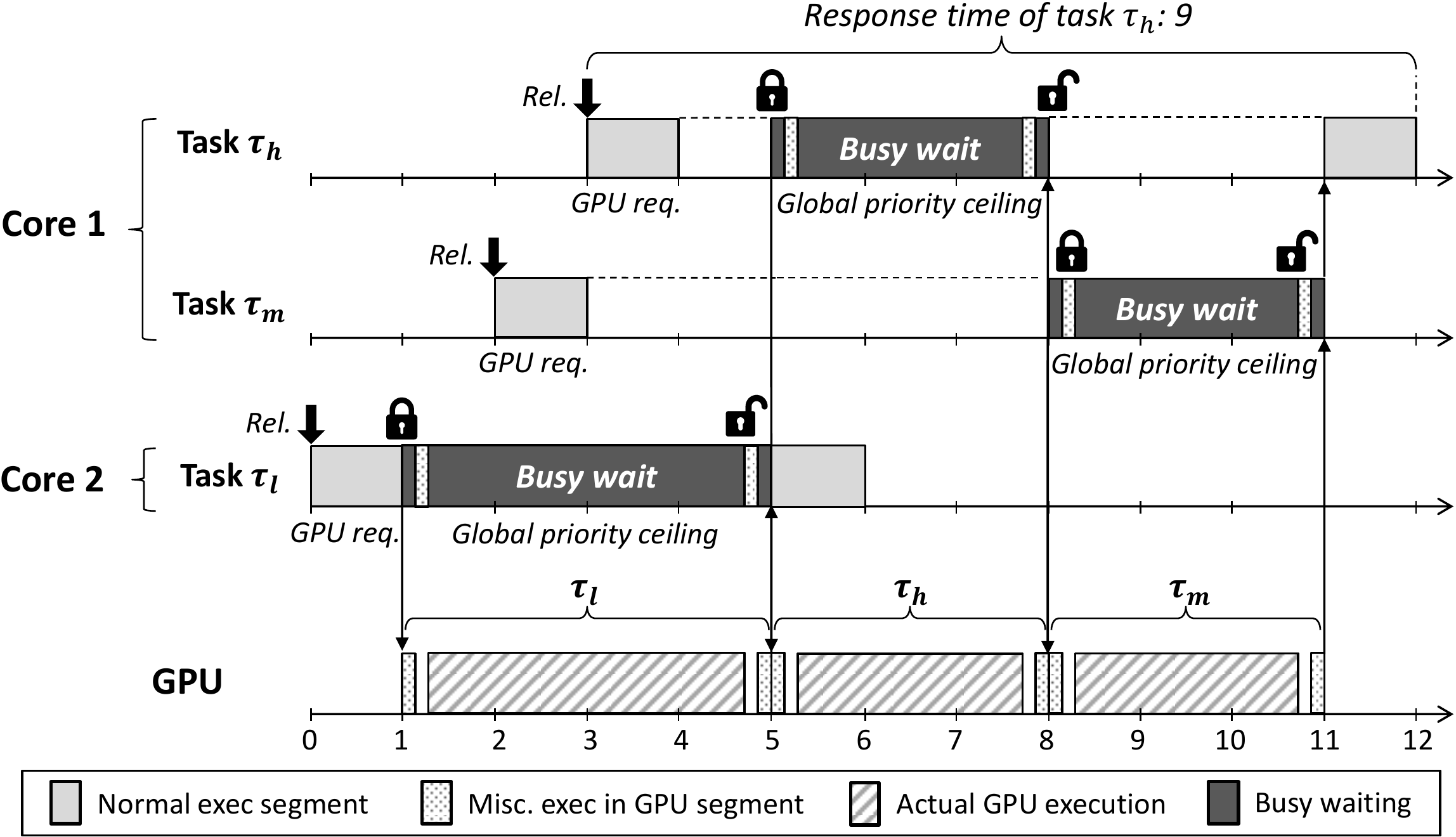}
	}
	\caption{Example of the synchronization-based approach}
	\label{fig:GPU_synch_based_approach_example}
\end{figure}

There is another issue with the synchronization-based approach. That is, the priority boosting mechanism of the synchronization-based approach introduces priority inversion that is unnecessarily long for tasks accessing a GPU. We describe this issue with the example illustrated in \figref{GPU_synch_based_approach_example}. In this figure, there are three tasks, $\tau_h$, $\tau_m$, and $\tau_l$, allocated to two CPU cores, Cores 1 and 2. Task $\tau_h$ is a high-priority task, $\tau_m$ is a medium-priority task, and $\tau_l$ is a low-priority task. Each task has one GPU access segment that is executed after a normal execution segment of one time unit. Each of $\tau_h$ and $\tau_m$ has a GPU access segment of three time units, $\tau_l$ has a GPU access segment of four time units. The GPU access segment of each task is followed by another normal execution segment of one time unit. In this figure, $\tau_l$ is released at time 0 and makes a GPU request at time 1. Since there is no other task using the GPU at that point, $\tau_l$ can hold the mutex for the GPU and enter its GPU access segment. Then, $\tau_l$ executes with the global priority ceiling associated with the mutex. Tasks $\tau_m$ and $\tau_h$ are released at time 2 and 3, respectively. They make GPU requests at time 3 and 4, but the GPU cannot be granted to any of them because it is already held by $\tau_l$. At time 5, $\tau_l$ releases the GPU and $\tau_h$ holds the GPU because it has higher priority than $\tau_m$. At time 8, $\tau_h$ finishes its GPU access segment and releases the GPU. Then, the task $\tau_m$ holds the GPU and enters its GPU access segment with the global priority ceiling. This makes $\tau_m$ to preempt the normal execution segment of $\tau_h$. Hence, although the majority of $\tau_m$'s GPU access segment merely performs busy-waiting, the execution of the normal segment of $\tau_h$ is delayed until the GPU access segment of $\tau_m$ finishes. Finally, $\tau_h$ completes its normal execution segment at time 12 and the response time of $\tau_h$ is 9 in this example. In the next section, we will present our new approach to address these issues.

\subsection{Schedulability Analysis}

We review the task schedulability analysis under the synchronization-based approach with MPCP. The analysis described here is originally developed by Lakshmanan et al.\cite{Lakshmanan_RTSS09}, and combined with a correction given by Bletsas et al.\cite{Bletsas_TR15} and Huang et al.~\cite{Huang_DAC15}. 
	
The worst-case response time of a task $\tau_i$ under the synchronization-based approach with MPCP is given by the following recurrence equation:
\begin{equation} \label{eq:GPU_synch_task_sched}
\begin{split}
W_i^{n+1}=&C_i+G_i+B_i^r+\sum_{\tau_h\in \mathbb{P}(\tau_i)\land \pi_h>\pi_i}\!\!\!\bigg\lceil{W_i^n\!+\!\{W_h\!-\!(C_h+G_h)\} \over T_h}\bigg\rceil (C_h+G_h)\\
&+(\eta_i + 1)\bigg(\sum_{\substack{\tau_l \in \mathbb{P}(\tau_i) \land \pi_l < \pi_i \land \eta_l > 0}} \max_{1\le u \le \eta_l} G_{l,u}\bigg)
\end{split}
\end{equation}
where $B_i^r$ is the remote blocking time for $\tau_i$, $\mathbb{P}(\tau_i)$ is the CPU core where $\tau_i$ is allocated, $\pi_i$ is the priority of $\tau_i$. It terminates when $W_i^{n+1} = W_i^n$, and the task $\tau_i$ is schedulable if its response time does not exceed its deadline:
$W_i^n \le D_i$. Since the task $\tau_i$ should busy-wait during its GPU access, the entire GPU access segment, $G_i$, is captured as the CPU usage of $\tau_i$, along with its WCET $C_i$.

The remote blocking time $B_i^r$ is given by $B_i^r=\sum_{1\le j \le \eta_i} B_{i,j}^r$, where $B_{i,j}^r$ is the remote blocking time for the $j$-th GPU access segment of $\tau_i$ to acquire the GPU. The term $B_{i,j}^r$ is bounded by the following recurrence:
\begin{equation} \label{eq:GPU_remote_blocking_time}
\begin{split}
B_{i,j}^{r,n+1}=\max_{\pi_l < \pi_i \land 1 \le u \le \eta_l} W^{gpu}_{l,u}+\sum_{\pi_h>\pi_i \land 1 \le u \le \eta_h} \bigg(\bigg\lceil{B_{i,j}^{r,n}\over T_h}\bigg\rceil+1\bigg)W_{h,u}^{gpu}
\end{split}
\end{equation}
where $B_{i,j}^{r,0}=\max_{\pi_l < \pi_i \land 1 \le u \le \eta_l} W^{gpu}_{l,u}$ (the first term of the equation), and $W_{l,u}^{gpu}$ represents the worst-case response time of a GPU access segment $G_{l,u}$. The first term of Eq.~\eqref{eq:GPU_remote_blocking_time}  captures the time for a lower-priority task to finish its GPU access segment. The second term represents the time for the GPU access segments of higher-priority tasks.

The worst-case response time of a GPU access segment $G_{l,u}$, namely $W_{l,u}^{gpu}$, is given by:
\begin{equation} \label{eq:GPU_resp_gpu_access_segment}
\begin{split}
W_{l,u}^{gpu}=G_{l,u}+\sum_{\tau_x\in \mathbb{P}(\tau_l)} \max_{1\le y \le \eta_x \land \pi_x > \pi_l }G_{x,y}
\end{split}
\end{equation}
This equation captures the length of $G_{l,u}$ and the lengths of GPU access segments of higher-priority tasks on the same core. It considers only one GPU access segment from each task, because every GPU access segment is associated with a global priority ceiling and $G_{l,u}$ will never be preempted by normal execution segments.

\section{Server-based GPU Access Control}
\label{GPU_server_approach}

\begin{figure}[t]
	\centering
	\subfloat{
		\includegraphics[width=0.75\textwidth]{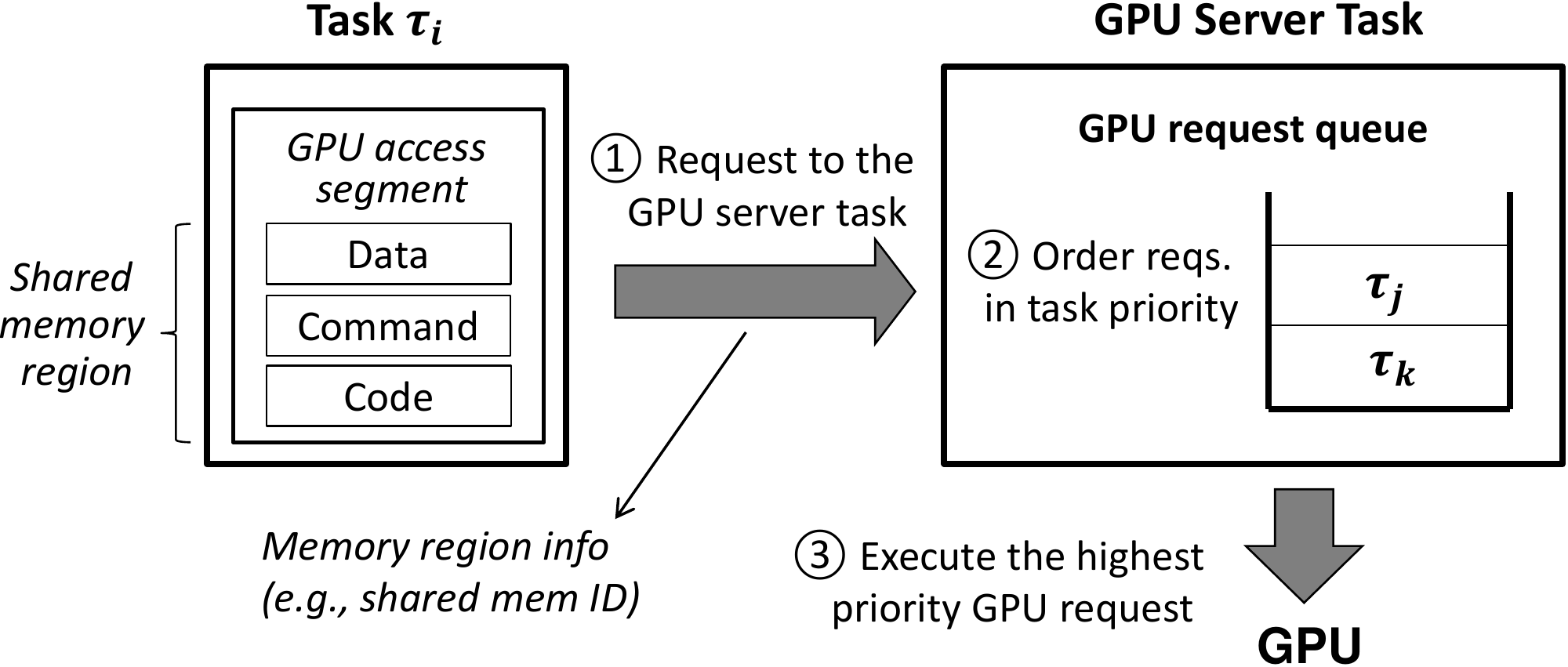}
	}
	\caption{Overview of the server-based approach}
	\label{fig:GPU_server_based_approach}
\end{figure}

In this section, we present our server-based approach for predictable GPU access control. This approach addresses the two main limitations of the synchronization-based approach: busy waiting and long priority inversion. To do so, our approach creates a {\em GPU server task} that handles GPU access requests from other tasks on behalf of them. The GPU server is assigned the highest priority in the system, which is to prevent preemptions by other tasks. \figref{GPU_server_based_approach} shows the sequence of GPU request handling under our approach. First, when a task $\tau_i$ enters its GPU access segment, it makes a GPU access request to the GPU server, not to the GPU device driver. The request is sent to the server by sending the memory region information for the GPU access segment, including in/output data, commands and code for GPU execution. This requires the memory regions to be configured as shared memory regions so that the GPU server can access them with their identifiers, e.g., \texttt{shmid}. After sending the request to the GPU server, the task $\tau_i$ can suspend, allowing other tasks to execute. Second, the GPU server enqueues the received request into the GPU request queue, if the GPU is being used by another request. The GPU request queue is a priority queue, where elements are ordered in their task priorities. Third, once the GPU becomes free, the GPU server dequeues a request from the head of the queue and executes the GPU access segment corresponding to that request. Finally, when the request finishes, the GPU server notifies the completion of the request to the task $\tau_i$. Then, $\tau_i$ resumes its execution.

\begin{figure}[t]
	\centering
	\subfloat{
		\includegraphics[width=1\textwidth]{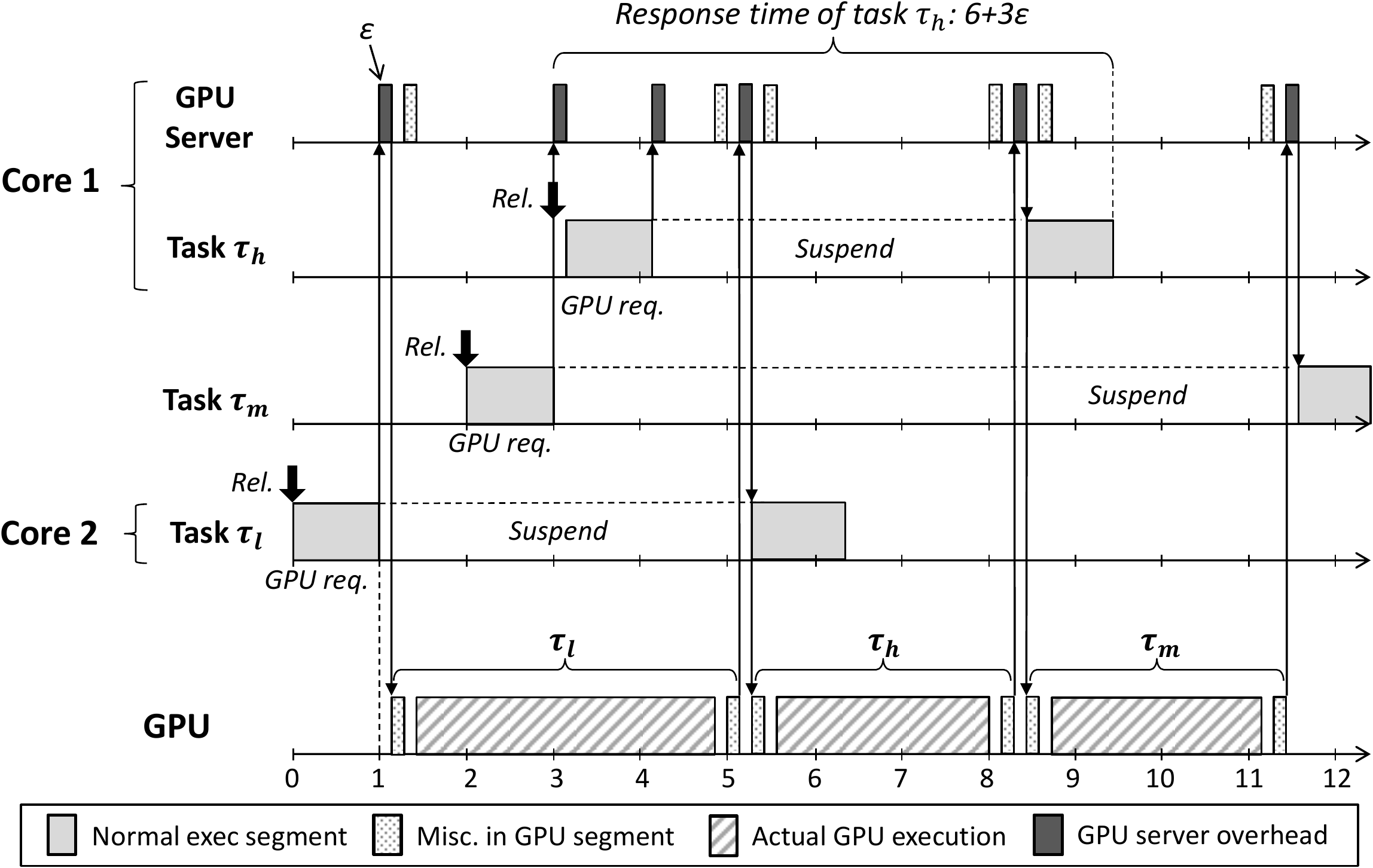}
	}
	\caption{Example of the server-based approach}
	\label{fig:GPU_server_based_approach_example}
\end{figure}

\figref{GPU_server_based_approach_example} shows an example of task scheduling under our server-based approach. This example has the same configuration as the one in \figref{GPU_synch_based_approach_example}. Hence, $\tau_h$ and $\tau_m$ are allocated to Core 1, and $\tau_l$ is allocated to Core 3. The GPU server, which the server-based approach creates, is allocated to Core 1. At time 1, the task $\tau_l$ makes a GPU access request to the GPU server. Then, the GPU server receives the request and executes the corresponding GPU access segment at time $1+\epsilon$, where the term $\epsilon$ is the amount of the overhead that the GPU server introduces. Since the server-based approach does not require tasks to busy-wait, $\tau_l$ can suspend until the completion of its GPU request. The GPU request of $\tau_m$ at time 3 is enqueued into the request queue of the GPU server. Since the GPU server executes with the highest priority in the system, it delays the execution of $\tau_h$ released at time 3 by $\epsilon$. Hence, $\tau_h$ starts execution at time $3+\epsilon$ and makes a GPU request at time $4+\epsilon$. When the GPU access segment of $\tau_l$ finishes, it is notified to the GPU server. Then, the GPU server notifies the completion of the GPU request to $\tau_l$, and executes the GPU access segment of $\tau_h$ at time $5+2\epsilon$. The task $\tau_h$ suspends until its GPU request finishes. The GPU access segment of $\tau_h$ finishes at time $8+2\epsilon$ and that of $\tau_m$ starts at time $8+3\epsilon$. Unlike the case under the synchronization-based approach, $\tau_h$ can continue to execute its normal execution segment from time $8+3\epsilon$, because $\tau_m$ suspends and the priority of $\tau_m$ is not boosted. The task $\tau_h$ finishes its normal execution segment at time $9+3\epsilon$ and the response time of $\tau_h$ is $6+3\epsilon$. Recall that the response time of $\tau_h$ is 9 under the synchronization-based approach, as shown in \figref{GPU_synch_based_approach_example}. Therefore, this example shows that, if $\epsilon<1$, the server-based approach can provide shorter response time than the synchronization-based approach.

\subsection{Schedulability Analysis}

We analyze task schedulability under our server-based approach. Since the GPU server handles the GPU requests of tasks on their behalf, we first identify the GPU request handling time of the GPU server. The maximum handling time of $\tau_i$'s GPU request is given by:
\begin{equation} \label{eq:GPU_server_handling_delay}
\begin{split}
B_i^{gpu}=\left\{
\begin{array}{lr}
B_i^{w}+(G_i+2\epsilon) &: \eta_i > 0\\
0 &: \eta_i = 0\\
\end{array}
\right.
\end{split}
\end{equation}
where $B_i^{w}$ is the maximum time the $\tau_i$'s GPU request has to wait, $G_i$ is the length of $\tau_i$'s GPU request, $\epsilon$ is the overhead of the GPU server, and $\eta_i$ is the number of GPU access segments of $\tau_i$. Obviously, in case of $\eta_i=0$, $B_i^{gpu}$ is zero. In case of $\eta_i> 0$, the reason for adding $2\epsilon$ to $G_i$ is that the GPU server intervenes before and after the execution of $\tau_i$'s GPU request. 

The maximum waiting time of $\tau_i$'s GPU request, $B_i^{w}$, is bounded by the following recurrence equation:
\begin{equation} \label{eq:GPU_server_waiting_delay}
\begin{split}
B_i^{w,n+1}=\max_{\tau_j \in \Gamma \land \tau_j \ne \tau_i}(G_l+\epsilon) + \sum_{\tau_h\in \Gamma \land \pi_h > \pi_i}\bigg\lceil{B_i^{w,n}\over T_h} \bigg\rceil (G_h+\epsilon)
\end{split}
\end{equation}
where $B_i^{w,0}=\max_{\tau_j \in \Gamma \land \tau_j \ne \tau_i}(G_l+\epsilon)$ (the first term of the equation), $\Gamma$ is the set of all tasks in the system, $\pi_h$ is the priority of a task $\tau_h$. The first term of this equation captures the longest GPU access segment among all other tasks, because GPU execution happens in a non-preemptive manner. Here, we add only one $\epsilon$ to $G_l$, because other GPU requests will be followed and the the GPU server needs to be invoked only once between two consecutive GPU requests, as depicted in \figref{GPU_server_based_approach_example}. The second term captures the fact that the GPU server prioritizes requests from higher-priority tasks. 

The response time of a task $\tau_i$ is affected by the presence of the GPU server on $\tau_i$'s core. If $\tau_i$ is allocated to a core different from the GPU server, the worst-case response time of $\tau_i$ under the server-based approach is given as follows:
\begin{equation} \label{eq:GPU_server_task_sched_no_server}
\begin{split}
W_i^{n+1}=&C_i+B^{gpu}_i+\sum_{\tau_h\in \mathbb{P}(\tau_i)\land \pi_h>\pi_i}\!\!\!\bigg\lceil{{W_i^n\!+\!(W_h\!-\!C_h)} \over {T_h}}\bigg\rceil C_h
\end{split}
\end{equation}
where $\mathbb{P}$ is the CPU core where $\tau_i$ is allocated. It terminates when $W_i^{n+1} = W_i^n$, and $\tau_i$ is schedulable if its
response time does not exceed its deadline: $W_i^n \le D_i$. Unlike the response-time analysis given in Eq.~\eqref{eq:GPU_synch_task_sched}, the GPU access segment of each task is not accounted for in this equation. This is because GPU access segments are executed by the GPU server  under the server-based approach. 

If $\tau_i$ is allocated to the same core as the GPU server, the worst-case response time of $\tau_i$ under the server-based approach is given as follows:
\begin{equation} \label{eq:GPU_server_task_sched}
\begin{split}
W_i^{n+1}=&C_i+B^{gpu}_i+\sum_{\tau_h\in \mathbb{P}(\tau_i)\land \pi_h>\pi_i}\!\!\!\bigg\lceil{{W_i^n\!+\!(W_h\!-\!C_h)} \over {T_h}}\bigg\rceil C_h\\
&+\sum_{\tau_j\in\Gamma \land \eta_j > 0} \left\lceil{{W_i^{n}+\{D_j\!-\!(X^m_j+2\epsilon)\}} \over{T_j}} \right\rceil(X^m_j+2\epsilon)
\end{split}
\end{equation}
where $X_j^m$ is the sum of the WCETs of miscellaneous operations in $\tau_i$'s GPU access segments, i.e., $X_j^m=\sum_{k=1}^{\eta_j}X_{j,k}^m$. The main difference of this equation from Eq.~\eqref{eq:GPU_server_task_sched_no_server} is the last term in this equation. The last term captures the execution time of the GPU server task. We capture this by summing up the miscellaneous operations and the server overhead ($X^m_j+2\epsilon$) caused by GPU requests from all other tasks. In this way, we can upper-bound task response time in the presence of the GPU server.

\section{Evaluation}
\label{GPU_evaluation}

\begin{table}[t]
	\centering
	{
		\footnotesize
		\caption[Base parameters for GPGPU access control experiments]{Base parameters for taskset generation}\label{tab:GPU_taskset_param}
		\begin{tabular}{l|c}
			\hline
			Parameters & Values\\\hline
			Number of CPU cores ($N_P$) & 4, 8\\
			Number of tasks per core & 4\\
			Ratio of GPU-using tasks & 20\%\\
			Task period ($T_i$) & [100, 500] msec\\
			Taskset utilization per core & 50\% \\
			Number of GPU access segment per task ($\eta_i$) & 1\\
			Length of GPU access segment ($G_i$) & [5, 10] msec\\
			WCET of misc. operations in GPU access segment ($X_i^m$)& [0.5, 2] msec\\
			GPU server overhead ($\epsilon$) & 100 $\mu$sec\\
			\hline
		\end{tabular}
	}
\end{table}

This section provides our experimental evaluation of two different approaches for GPU access control: the synchronization-based and server-based approaches. Our focus here is to explore the impact of those approaches on task schedulability. To do this, we use randomly-generated tasksets and capture the percentage of schedulable tasksets as the metric.

\begin{figure}[t]
	\centering
	\subfloat[$N_P=4$]{\label{fig:GPU_expr_gpu_segment_length_pcpu_4}
		\includegraphics[width=0.7\textwidth]{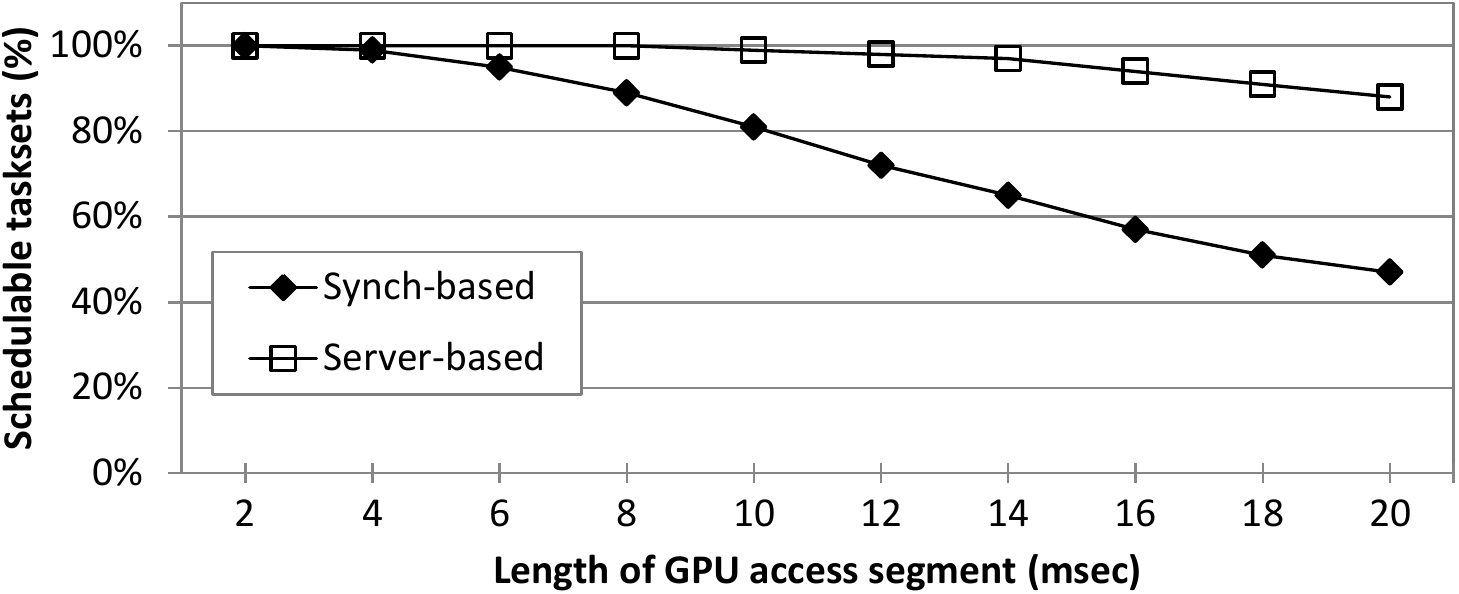}
	}\vspace{-5pt}\\
	\subfloat[$N_P=8$]{\label{fig:GPU_expr_gpu_segment_length_pcpu_8}
		\includegraphics[width=0.7\textwidth]{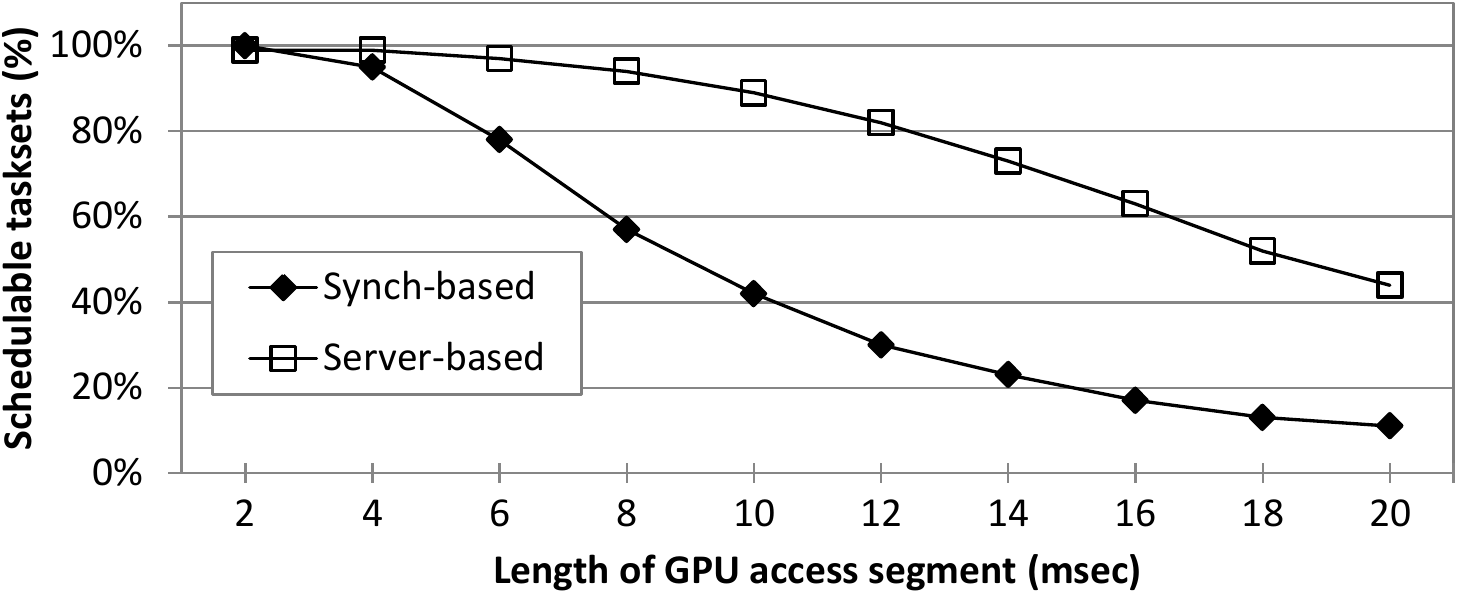}
	}\vspace{-5pt}
	\caption{Percentage of schedulable tasksets as the length of GPU access segment increases}
	\label{fig:GPU_expr_gpu_segment_length}
\end{figure}

\subsection{Experimental Setup}

We generate 10,000 tasksets with the parameters given in \tableref{GPU_taskset_param} for each experimental setting. We consider two different system configurations: systems with four cores ($N_P=4$) and eight cores ($N_P=8$). 
For each taskset, we first generate the defined number of CPU cores in the system and tasks for each core. 
Task periods are randomly selected within the defined minimum and maximum task period range. Task deadlines are set equal to their periods. On each core, the taskset utilization is split into $k$ random-sized pieces, where $k$ is the number of tasks per core. The size of each piece represents the utilization of the corresponding task. Then, the WCET of each task is calculated by dividing its utilization by its period. Task priorities are assigned by the Rate-Monotonic policy~\cite{Liu_Layland}, with arbitrary tie-breaking. A subset of the generated tasks is randomly chosen according to the defined ratio of GPU-using tasks, and each task in that subset is assigned a GPU access segment. Under the server-based approach, the GPU server is randomly allocated to one of the cores in the system.

\subsection{Results}
\figref{GPU_expr_gpu_segment_length} shows the percentage of schedulable tasksets as the length of the GPU access segment increases. In general, the percentage of schedulable tasksets is higher when $N_P=4$, compared to when $N_P=8$. This is because the GPU is contended for by more tasks as the number of cores increases.
In both $N_P=4$ and $N_P=8$, the server-based approach performs much better than the synchronization-based approach. Especially, when the GPU access segment is 12 msec and $N_P=8$, the difference in the percentage of schedulable tasksets between the two approaches is about 50\%. This big difference is mainly due to the fact that the server-based approach allows a task to suspend while its GPU segment is being executed on the GPU.

\begin{figure}[t]
	\centering
	\subfloat[$N_P=4$]{\label{fig:GPU_expr_gpu_task_ratio_pcpu_4}
		\includegraphics[width=0.7\textwidth]{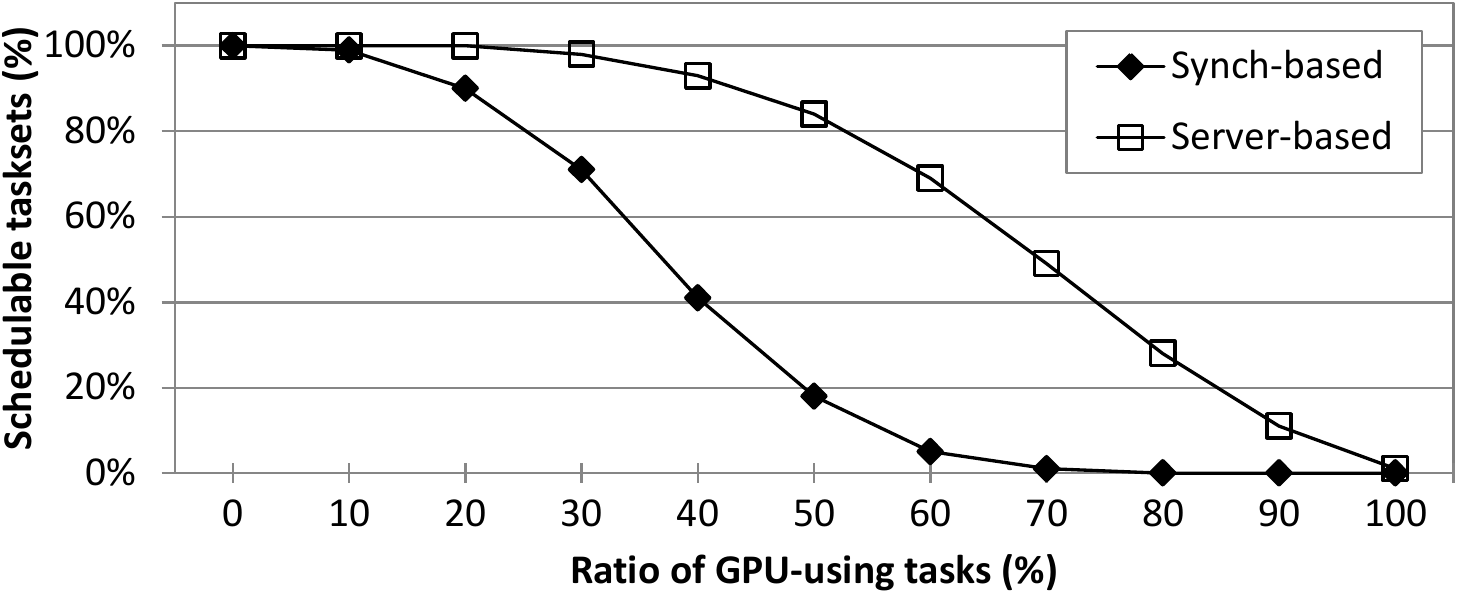}
	}\vspace{-5pt}\\
	\subfloat[$N_P=8$]{\label{fig:GPU_expr_gpu_task_ratio_pcpu_8}
		\includegraphics[width=0.7\textwidth]{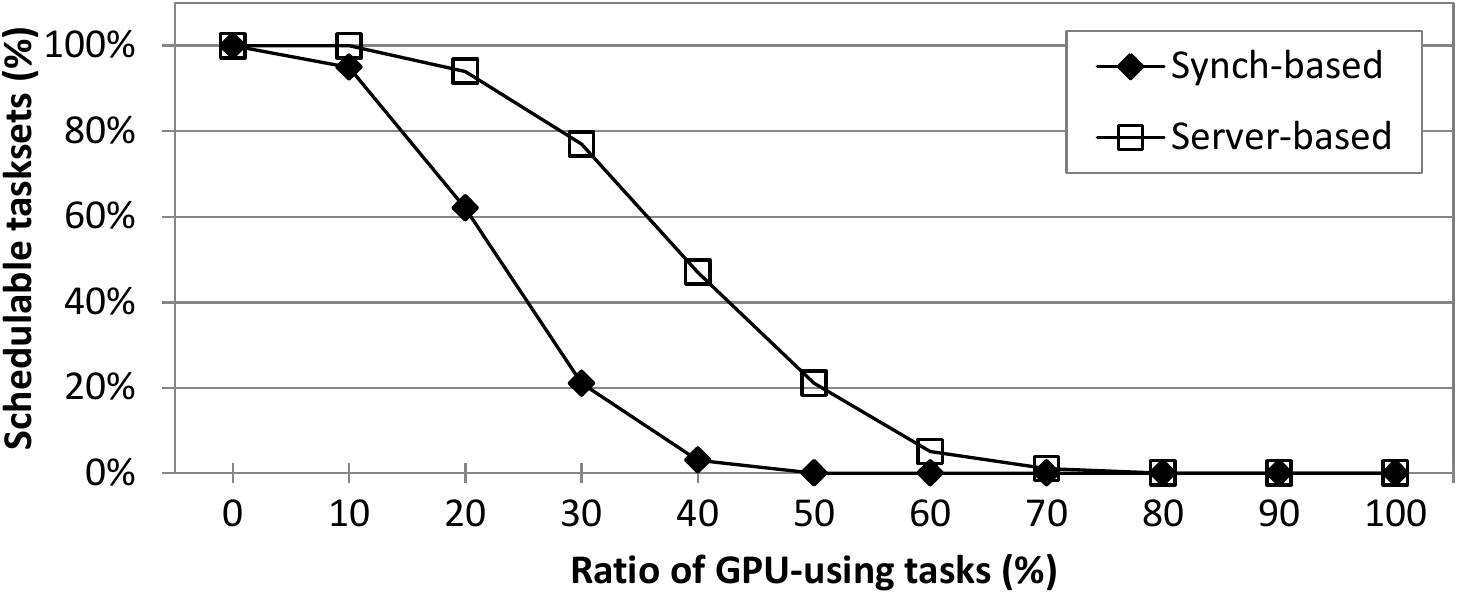}
	}\vspace{-5pt}
	\caption{Percentage of schedulable tasksets as the ratio of GPU-using tasks increases}
	\label{fig:GPU_expr_gpu_task_ratio}
\end{figure}

\begin{figure}[t]
	\centering
	\subfloat[$N_P=4$]{\label{fig:GPU_expr_misc_oper_time_pcpu_4}
		\includegraphics[width=0.7\textwidth]{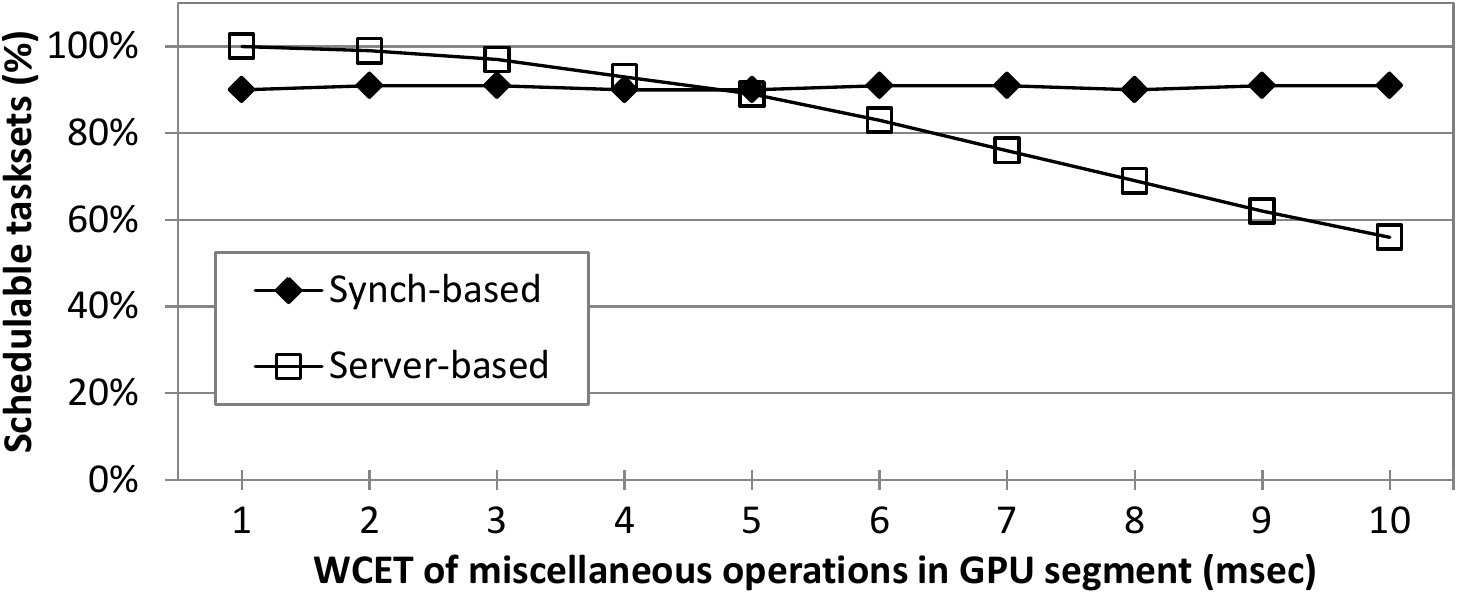}
	}\vspace{-5pt}\\
	\subfloat[$N_P=8$]{\label{fig:GPU_expr_misc_oper_time_pcpu_8}
		\includegraphics[width=0.7\textwidth]{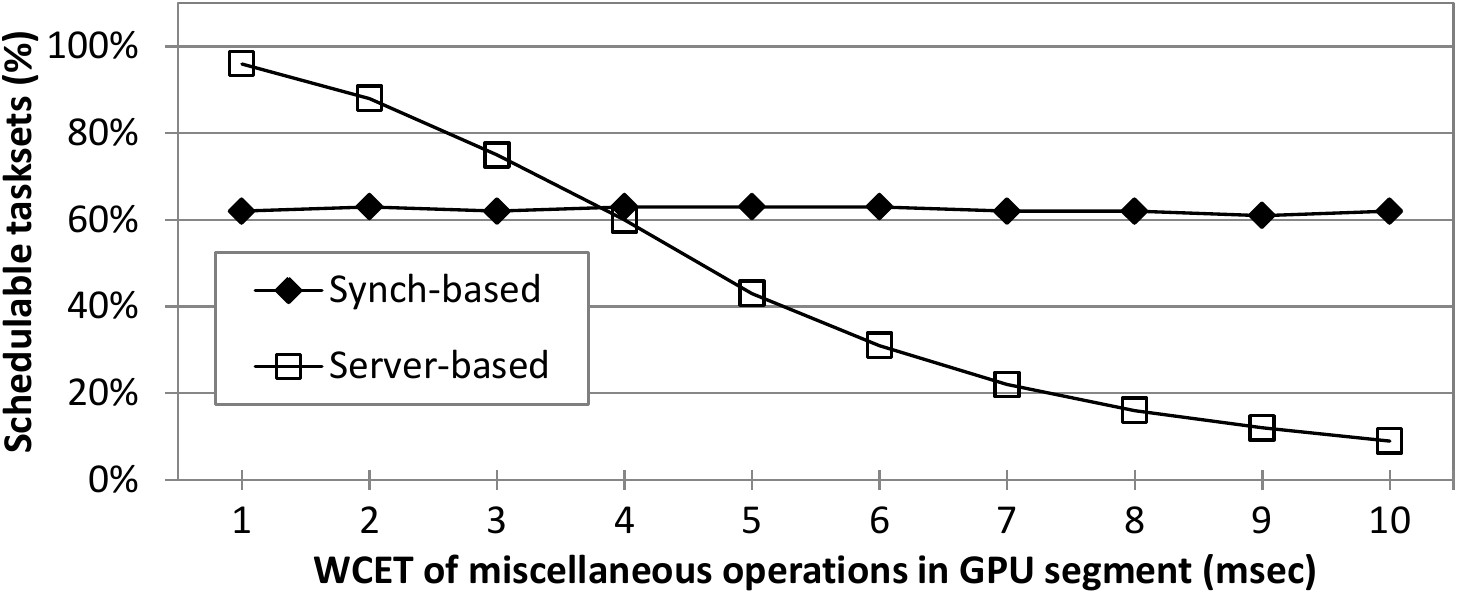}
	}\vspace{-5pt}
	\caption{Percentage of schedulable tasksets as the ratio of GPU-using tasks increases}
	\label{fig:GPU_expr_misc_oper_time}
\end{figure}

\figref{GPU_expr_gpu_task_ratio} shows the percentage of schedulable tasksets as the ratio of GPU-using tasks increases. The left-most point on the x-axis of each graph represents that all tasks are CPU-only tasks, and the right-most point represents that all tasks access the GPU. Under both approaches, the percentage of schedulable tasksets  reduces as the ratio of GPU-using tasks increases. However, there are many cases where the server-based approach significantly outperforms the synchronization-based approach, with as much as 66\% more tasksets being schedulable when the ratio is 50\% and $N_P=4$.

The benefit of the server-based approach is adversely affected by the amount of miscellaneous operations in GPU access segments, because such operations require the GPU server to consume CPU time. \figref{GPU_expr_misc_oper_time} shows the percentage of schedulable tasksets as the WCET of miscellaneous operations in GPU access segments increases. Since the synchronization-based approach makes tasks to busy-wait during their entire GPU access segments, its performance is not affected by the WCET of miscellaneous operations. However, as expected, the performance of the server-based approach reduces as the WCET of miscellaneous GPU operations increases.

\begin{figure}[t]
	\centering
	\subfloat[$N_P=4$]{\label{fig:GPU_expr_server_overhead_pcpu_4}
		\includegraphics[width=0.7\textwidth]{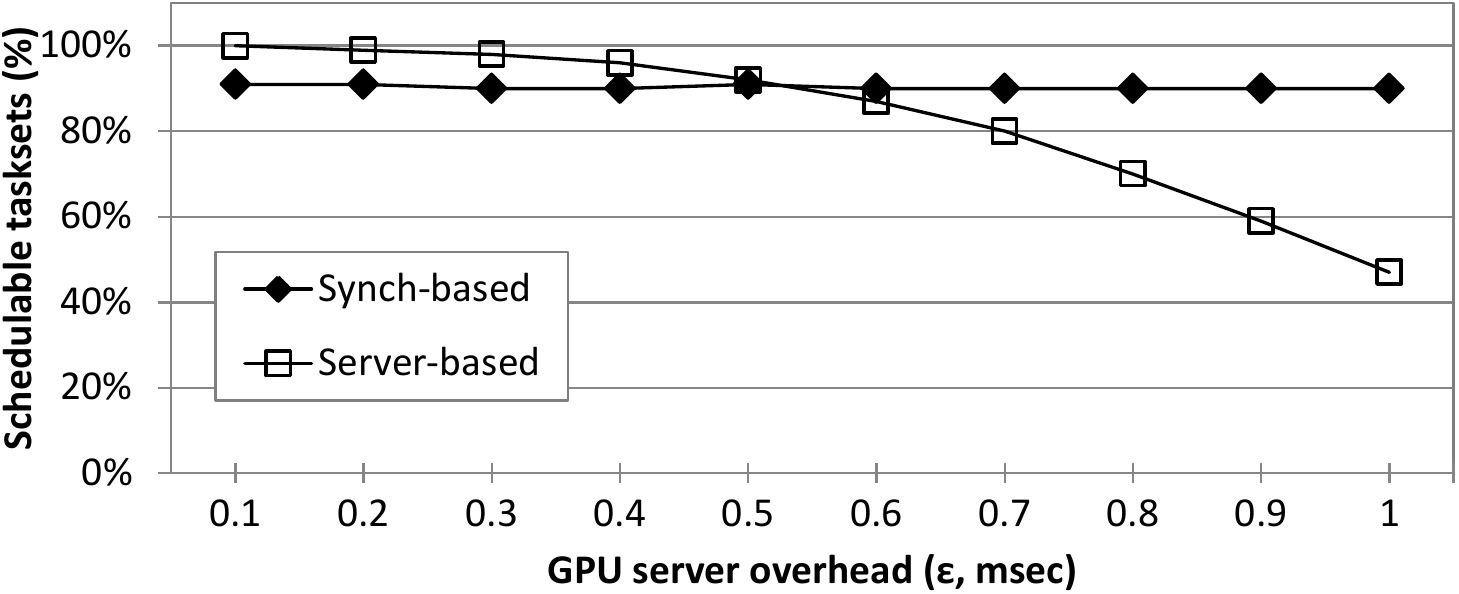}
	}\vspace{-5pt}\\
	\subfloat[$N_P=8$]{\label{fig:GPU_expr_server_overhead_pcpu_8}
		\includegraphics[width=0.7\textwidth]{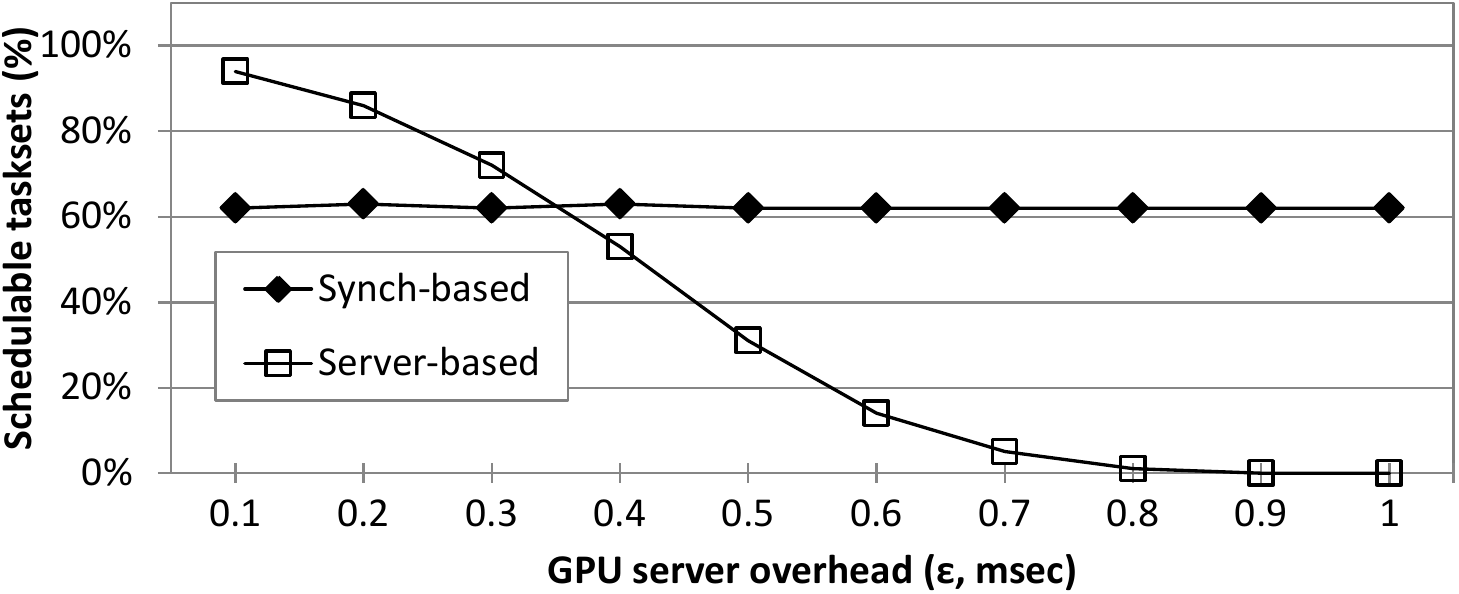}
	}\vspace{-5pt}
	\caption{Percentage of schedulable tasksets as the overhead of the GPU server ($\epsilon$) increases}
	\label{fig:GPU_expr_server_overhead}
\end{figure}

The performance of the server-based approach is also affected by the amount of the overhead $\epsilon$ that the GPU server introduces. Although $\epsilon$ of 100 $\mu$sec we have used in prior experiments is sufficient enough to upper-bound the GPU server overhead in practical systems, we further investigate with larger $\epsilon$ values.  \figref{GPU_expr_server_overhead} shows the percentage of schedulable tasksets as the GPU server overhead increases. Since such overhead only exists under the server-based approach, the performance of the synchronization-based approach is unaffected by this factor. On the other hand, the performance of the server-based approach reduces as the amount of the overhead increases. 

In summary, the server-based approach outperforms the synchronization-based approach in many cases. Especially, the benefit of the server-based approach can be significant when the length of GPU access segments is long or the ratio of GPU-using tasks is high. However, we find that the server-based approach does not dominate the synchronization-based approach. The use of the synchronization-based approach may be a better choice than that of the server-based approach, when miscellaneous operations, e.g., data copy between GPU and CPU, take the majority time of the GPU access segment.

\section{Summary}
\label{GPU_conclusions}

In this chapter, we presented our server-based approach to control GPU access requests from tasks in a predictable manner. Our approach is motivated by the limitations of the synchronization-based approach, namely busy-waiting and long priority inversion. By introducing a dedicated server task for GPU request handling, our approach addresses those limitations, while ensuring the analyzability and predictability of the system. We also described our analysis on task schedulability under the server-based approach. Experimental results show that the server-based approach yields significant improvements in task schedulability over the synchronization-based approach. For example, the system with the server-based approach schedules 66\% more randomly-generated tasksets than the one with the synchronization-based approach. Future work involves extending the GPU access control approaches to a virtualized environment.

\chapter{Guidelines for Future Computer Architecture Designs}
\label{chapter_guidelines_for_future_computer_architectures}

The analytical and systems techniques proposed in this dissertation provide predictable real-time performance on commodity multi-core platforms. 
We believe that, however, our techniques can benefit from new computer architecture support. 
In this chapter, we discuss hardware features that are as yet unavailable on most of today's platforms but can help the development of predictable and efficient systems. Our discussions may serve as guidelines to design future computer architectures for cyber-physical systems. In the following sections, we describe architecture support desired for each type of shared resource.

\section{Architecture Support for Concurrent Resources}

In Chapters~\ref{chapter_coordinated_cache_management}, \ref{chapter_bounding_and_reducing_memory_interference}, and \ref{chapter_cache_management_for_virtualization}, we proposed analytical and systems techniques to address cache and memory interference issues. In this section, we describe hardware features that can be used with our techniques to improve the degree of efficiency and predictability.

\begin{itemize}
\item \textbf{Fine-grained Hardware Cache Partitioning:} 
Our work uses a software-based cache partitioning technique to manage a last-level cache in software. As discussed in Section~\ref{problems_with_page_coloring}, software cache partitioning has two main problems: (i) the memory co-partitioning problem, which results in page swapping or waste of memory, and (ii) the availability of a limited number of cache partitions, which causes degraded performance. Our techniques proposed in Chapter~\ref{chapter_coordinated_cache_management} can significantly mitigate these problems, but cannot eliminate them. We believe that a fine-grained hardware cache partitioning feature, such as the one proposed in \cite{Kirk_RTSS89}, can further reduce the negative impact of those problems. If cache allocation to a task becomes independent of physical page allocation to the task, the memory co-partitioning problem will disappear, and the problem of finding a feasible cache allocation will be reduced to finding a cache allocation for guaranteeing timing constraints. Also, if more number of cache partitions are provided in the system, the problem of limited cache partitions will be easily resolved. Our analysis techniques and cache allocation algorithm proposed in Chapter~\ref{chapter_coordinated_cache_management} can be used together with future hardware cache partitioning features, since our techniques are independent of a specific cache partitioning technique used in the system. 

\item\textbf{Software-controllable Miss Status Holding Registers:}
Recent work in \cite{Valsan_RTAS16} reported that the contention on Miss Status Holding Registers (MSHR), which are employed in many of today's shared caches to support memory-level parallelism, may cause significant performance interference among tasks, even when tasks are assigned private cache partitions. The authors of that work also proposed an MSHR partitioning technique as a solution to this problem. We are a proponent of such a technique because it can be used along with cache partitioning techniques to improve the performance isolation capability of the system. 

\item\textbf{Task-aware Memory Scheduling:} Our work in Chapter~\ref{chapter_bounding_and_reducing_memory_interference} analyzes memory interference delay on a COTS DRAM system, where a memory controller handles memory requests without considering the priority of tasks that have generated those requests. In the computer architecture community, thread prioritization approaches~\cite{Kim_MICRO10, Subramanian_MICRO15, Kim_HPCA10, Mutlu_MICRO07, Mutlu_ISCA08, Subramanian_IEEE16, Subramanian_ICCD14} have been proposed to achieve high memory throughput and fairness. The key idea of those approaches is to make the memory controller be aware of threads so that the memory controller can prioritize memory requests based on the priorities of their origin threads. This idea can be used to improve the predictability and schedulability of the system. For instance, if the memory controller makes use of task priorities on memory request scheduling, memory requests from higher-priority tasks can be prioritized, thereby reducing their response times. Also, the idea of thread clustering in the memory controller~\cite{Kim_MICRO10} can be used to protect the performance of critical tasks in a safety-critical system.

\item\textbf{Slowdown Estimation Techniques:} Recent work in the computer architecture community on the design of memory controllers and memory systems has proposed techniques for dynamically estimating application slowdowns~\cite{Subramanian_HPCA13,Moscibroda_2007,Ebrahimi_ASPLOS10,Subramanian_MICRO15,Mutlu_MICRO07}. Although these techniques are not designed to find worst-case bounds, they can provide various metrics to understand the performance characteristics of application tasks running on the target hardware. Having such knowledge would help develop measurement-based WCET analysis tools for modern multi-core platforms. 

\end{itemize}

\section{Architecture Support for Mutually-Exclusive Resources}

In Chapters~\ref{chapter_synchronization} and \ref{chapter_interrupt_handling}, we proposed analytical and systems techniques to address challenges on accessing mutually-exclusive resources. In this section, we describe hardware features that can be utilized with our techniques to improve system performance while preserving predictability.

\begin{itemize}
\item \textbf{Critical Section Acceleration:} Our work in Chapter~\ref{chapter_synchronization} provides a synchronization mechanism to provide predictable access to mutually-exclusive resources. Although our proposed technique bounds and minimizes blocking time on accessing such resources, the blocking time can be further reduced with architecture support. In the computer architecture community, there has been recent work on accelerating the execution of critical sections with high-performance cores in an asymmetric multi-core processor~\cite{Suleman_ASPLOS09, Joao_ASPLOS12,Joao_ISCA13}. These approaches migrate a critical section to a dedicated high-performance core that executes the critical section faster than the other cores. Hence, they can reduce the blocking time imposed on other tasks waiting for the corresponding resource. There is also a memory scheduling technique that prioritizes the memory requests of tasks executing critical sections~\cite{Ebrahimi_MICRO11}. 
We believe that the effect of such techniques can be incorporated into our analysis for synchronization.

\item \textbf{Interrupt Throttling and Enforcement:} To protect the execution of tasks from interrupt storms, our work in Chapter~\ref{chapter_interrupt_handling} uses a software-based interrupt enforcement mechanism which incurs a small but measurable overhead. The interrupt throttling and enforcement mechanism can be implemented in hardware, and in fact, some of today's I/O devices employ these mechanisms, e.g., Intel Gigabit Ethernet controllers~\cite{IntelGigabitEthernet}. However, there exist many I/O devices that do not support such a mechanism. If the interrupt throttling and enforcement mechanism is implemented as part of the interrupt controller of the processor, instead of in individual I/O devices, all I/O devices equipped in the system can benefit from it. Also, the system can use a common approach to control the rates of interrupts coming from different types of I/O devices, simplifying the design of systems software.
\end{itemize}

\section{Architecture Support for Computational Accelerators}

In Chapter~\ref{chapter_predictable_gpgpu_management}, we proposed analytical and systems techniques to provide predictable access to a general-purpose GPU, which is a computational accelerator recently receiving much attention. In this section, we describe hardware features desirable for future GPU architectures with respect to predictability.

\begin{itemize}
\item \textbf{GPU Partitioning:} Although a GPU is viewed as a single accelerator device by application tasks, it consists of many GPU cores that execute a given parallel workload in an aggregate manner. Hence, depending on the characteristics of workloads, only some of the GPU cores may be utilized~\cite{Vijaykumar_ISCA15}. To address this GPU underutilization problem, some of recent GPU architectures, e.g., NVIDIA Kepler~\cite{NvidiaKepler}, introduce a feature to execute multiple GPU functions concurrently. However, this feature is limited only to GPU functions from threads sharing the same process context. More importantly, this feature may cause performance interference among concurrent GPU executions, which can hamper real-time predictability. Therefore, we believe that the partitioning of GPU cores is a desirable feature to improve GPU utilization while preserving predictability. For example, with GPU partitioning, each application task can specify its required number of GPU cores and the requested number of GPU cores can be granted to the task by the admission control mechanism of a GPU device driver. Then, multiple tasks can execute GPU functions concurrently by using their assigned GPU cores. GPU partitioning would open interesting research directions that could build upon our work.

\item \textbf{GPU Context Switching:} Today's GPU architectures do not support preemptive execution due to the high overhead expected on GPU context switching. However, recent work~\cite{Tanasic_ISCA14} reports that the overhead of GPU context switching is not as high as expected and the GPU preemption mechanism improves the throughput and fairness of the system, even in the presence of the GPU context switching overhead. The use of the GPU preemption mechanism can eliminate priority inversion issues caused by the non-preemptivity of today's GPUs. Therefore, we expect that it can significantly improve the schedulability of tasks using GPUs.

\end{itemize}

\section{Summary}
In this chapter, we discussed future architecture support that could be utilized with our proposed techniques in the design of cyber-physical systems. The addition of the discussed hardware features can improve the degree of the predictability and efficiency of cyber-physical systems significantly. These hardware features are also beneficial to a wide range of general-purpose systems, such as smartphones, video game consoles, web servers and cloud services, where fairness, responsiveness, and quality-of-service are the key performance requirements. Therefore, we strongly encourage hardware manufacturers to adopt these architecture techniques in their future products so that both cyber-physical and general-purpose systems can enjoy the benefit of them.

\chapter{Conclusions}
\label{conclusions}

In this dissertation, we have presented novel analytical and systems techniques to address the predictability issues associated with shared resources in multi-core platforms. 
Our work categorizes the shared resources into three types: (i) {\em concurrent resources} (Chapters~\ref{chapter_coordinated_cache_management}, \ref{chapter_bounding_and_reducing_memory_interference}, and \ref{chapter_cache_management_for_virtualization}), which allow concurrent access from multiple tasks executing on different cores, e.g., a last-level cache, a memory controller, and DRAM, (ii) {\em mutually-exclusive resources} (Chapters~\ref{chapter_synchronization} and \ref{chapter_interrupt_handling}), which require no more than one
task to access them at a time to prevent race conditions, e.g., sensors,
actuators, network interfaces, and shared data regions, and (iii) {\em computational accelerators} (Chapter~\ref{chapter_predictable_gpgpu_management}), which supplement the computational capacity, e.g., general-purpose graphics processing units (GPGPUs). Our proposed techniques in this work provide predictable real-time performance on accessing these three types of shared resources in modern multi-core platforms and guarantee the predictability of the entire system in an efficient way.


\section{Contributions of Our Work}

The work in this dissertation addresses the challenges faced by cache interference, memory interference, synchronization, interrupt handling, and GPGPU access control issues. While each chapter in this dissertation has dealt with a different issue, our analysis presented in each chapter can be easily combined with those in other chapters in an additive manner, e.g., combined cache and memory interference analysis given in Section~\ref{combining_with_cache_interference_analysis}. 
The following summarizes our contributions.

\begin{itemize} 
	
\item \textbf{Coordinated Approach for Predictable Cache Management:}
Chapter~\ref{chapter_coordinated_cache_management} presents our coordinated OS-level cache management scheme to address cache interference. Our scheme provides predictable cache performance through tight coordination of cache reservation, reserved cache sharing, and cache-aware task allocation. Our scheme mitigates the two major challenges of page coloring:
the memory co-partitioning problem and the availability of limited number of cache partitions. We provide a condition that checks the feasibility of cache sharing while guaranteeing the allocation of the required memory space to tasks. We also provide a response-time based schedulability analysis in the presence of cache interference. We have implemented and evaluated our scheme in Linux/RK running on an Intel Core i7 quad-core processor. Experimental results with our implementation indicate that, compared to the traditional approaches, our scheme yields a significant utilization benefit that increases with the number of tasks. 

\item \textbf{Bounding and Reducing Memory Interference:}
Chapter~\ref{chapter_bounding_and_reducing_memory_interference} describes our proposed techniques to bound and reduce memory interference on a multi-core platform with DRAM-based main memory. 
Our analysis is based on a realistic memory model, which considers the Joint Electron Device Engineering Council (JEDEC) DDR3 SDRAM standard, the FR-FCFS policy of the memory controller, and shared/private DRAM banks. To provide a tighter upper-bound on the memory interference delay, our analysis uses the combination of the request-driven and job-driven approaches. We find that memory interference can be significantly reduced by (i) partitioning DRAM banks, and (ii) co-locating memory-access-intensive tasks on the same processing core. Based on these observations, we develop a memory interference-aware task allocation algorithm. Experimental results from a real hardware platform show that our analysis can closely
estimate the memory interference delay under workloads with both high and low memory contention. Also, our memory interference-aware task allocation algorithm provides a significant improvement in task schedulability over previous work, with as much as 96\% more tasksets being schedulable.

\item \textbf{Predictable Cache Management for Virtualization:} Chapter~\ref{chapter_cache_management_for_virtualization} focuses on predictable cache management in a virtualized environment. We develop two hypervisor-level techniques, vLLC and vColoring, that enable the cache allocation of individual tasks running in a virtual machine (VM), which is not achievable by prior work. We have implemented vLLC
and vColoring on the KVM hypervisor running on x86 and ARM platforms. Experimental results with three different guest OSs show that both vLLC and vColoring can effectively control the cache allocation of tasks in a VM. vColoring can also be used for DRAM bank partitioning in a virtualized environment. In addition, we develop a cache management scheme that determines cache allocation to tasks, designs VMs in a cache-aware manner, and minimizes the aggregated utilization of VMs to be consolidated. Experimental results with randomly-generated tasksets show that our scheme yields a significant utilization benefit
compared to other approaches.

\item \textbf{Synchronization for Multi-Core Virtual Machines:} Chapter~\ref{chapter_synchronization} presents vMPCP, a synchronization framework to provide bounded blocking time on accessing mutually-exclusive resources in a virtualized environment. vMPCP extends the well-known multiprocessor priority ceiling protocol to the multi-core two-level hierarchical scheduling context. vMPCP reduces the major timing penalties caused by resources shared among tasks on virtual CPUs allocated to different physical cores, by exposing the executions of global critical sections to the hypervisor. 
We have presented the schedulability analysis under vMPCP, with the periodic and deferrable server policies with and without the budget overrun mechanism. 
Experimental results indicate that, under vMPCP, deferrable server outperforms periodic server when overrun is used, with as much as 80\% more tasksets being schedulable. 
We also have implemented vMPCP on the KVM hypervisor and demonstrated the effect of vMPCP in reducing task response times by an average of 29\% in our case study.

\item \textbf{Responsive and Enforced Interrupt Handling:}
Chapter~\ref{chapter_interrupt_handling} presents vINT, an interrupt handling scheme to provide responsive and enforced interrupt handling in a virtualized environment. vINT provides a pseudo-VCPU abstraction dedicated for interrupt handling, which overcomes the limits imposed by the timing parameters of virtual CPUs in an analyzable way. vINT also accounts for and enforces interrupt handling and resulting execution flows within a guest VM. We have presented our analyses on interrupt handling time, and the schedulability of VCPUs and tasks with and without vINT. Experimental results show that vINT yields significant improvements in interrupt handling performance while providing as good task schedulability as when it is not used. For example, a system with vINT services 99\% of interrupt sets while a system without vINT cannot service any interrupt set. 
Our case study based on a prototype implementation on the KVM hypervisor also shows that vINT yields significant benefits in reducing interrupt handling time and in protecting tasks against interrupt storms permeating into the VM.

\item \textbf{Predictable GPGPU Access Control:} In Chapter~\ref{chapter_predictable_gpgpu_management}, we first review a synchronization-based GPU access control approach that uses a real-time synchronization protocol for tasks accessing a GPU. We characterize the two major limitations of the synchronization-based approach: busy-waiting and long priority inversion. Then, we present our proposed server-based approach to control GPU access requests from tasks in a predictable manner. Our approach introduces a dedicated server task that handles GPU requests from other tasks with respect to their priority order. Our approach addresses the limitations of the synchronization-based approach. Although we focus on a GPU in this work, our approach can be used for other types of computational accelerators, such as DSPs. Experimental results show that our server-based approach yields significant improvements in task schedulability over the synchronization-based approach. For example, a quad-core system with the server-based approach schedules 66\% more randomly-generated tasksets than the one with the synchronization-based approach.

\end{itemize}

\section{Future Research Directions}

Cyber-physical systems (CPS) are expected to become more pervasive in various safety-critical application domains in near future. We believe that the work in this dissertation can be effectively used for designing predictable CPS. There still exists plenty of future work in this area. We describe some of future research topics in the following. 
\begin{itemize}

\item \textbf{Handling Variations in CPS Workloads:} To provide predictable real-time performance without sacrificing efficiency, program execution times should be bounded with acceptable margins. However, it is hard to find such bounds on programs that vary their execution times, e.g., the size and arrival rate of input determined by physical environmental factors. A potential direction to address this issue would be developing an analytical model of program execution times as a function of features. The features may be extracted from possible inputs and system conditions. Then, one can develop a systems framework that leverages the analytical model. For example, the system may control the CPU usage by limiting input arrival rate, while preserving predictability. The system may also offload some workloads to other computing resources before execution by checking their inputs. We believe that this idea of taking physical environmental factors into performance analysis and control is essential for the development of scalable and resilient CPS.

\item \textbf{Large-scale CPS with Distributed, Non-Uniform Multi-Core Platforms:}  Large-scale CPS will likely be implemented by a coordination of distributed, non-uniform multi-core platforms. Each platform may be equipped with a different set of I/O devices like sensors, which may need to be accessible by other platforms. Of course, there already exist some approaches for sharing I/O devices in a distributed environment, e.g., IP cameras and network speakers. However, they are specific to one I/O device type and require modifications to application software so that existing applications cannot use remote devices directly. For the ease development and integration of large-scale safety-critical CPS, we believe that OS support for seamless resource sharing is strongly required. Distributed real-time synchronization protocols should also be revisited and enhanced for the predictability of the entire system. In addition, each platform may be equipped with heterogeneous multi-core processors, such as various types of CPU cores and GPUs. Providing predictable performance on such processors is a challenging issue.

\item \textbf{Cloudlet Computing for CPS:} The use of cloud computing in CPS has recently gained considerable interest due to its many benefits, e.g., rapid provisioning and flexibility in software deployment. However, cloud computing may not be appropriate for mobile CPS like autonomous vehicles. Although some recent cars are equipped with cellular data communication capabilities, network connectivity, bandwidth and latency may not be sufficient for timing-sensitive applications. Hence, we believe that an in-vehicle cloudlet, which operates as a small-scale cloud system for CPS applications, will play an important role. Unlike regular cloudlets, the in-vehicle cloudlet should be able to provide quantifiable and predictable performance. A solution to this issue can be built upon our contributions for multi-core virtualization, described in Chapters~\ref{chapter_cache_management_for_virtualization}, \ref{chapter_synchronization}, and \ref{chapter_interrupt_handling}.

\end{itemize}

\backmatter

\renewcommand{\bibsection}{\chapter{\bibname}}
\bibliographystyle{unsrt}
\bibliography{paper} 
\end{document}